\documentclass[fleqn,usenatbib]{mnras}
 
\usepackage{newtxtext,newtxmath}
\usepackage[T1]{fontenc}
 
\DeclareRobustCommand{\VAN}[3]{#2}
\let\VANthebibliography\thebibliography
\def\thebibliography{\DeclareRobustCommand{\VAN}[3]{##3}\VANthebibliography}

\usepackage{graphicx}	% Including figure files
\usepackage{amsmath}	% Advanced maths commands
 
\usepackage{graphicx}
\usepackage{amsmath}
\usepackage{pifont}
\usepackage{makecell}
\usepackage{comment}
\usepackage{lscape}
\usepackage{placeins}
\usepackage{xcolor}
\usepackage{tabularx}
 
\hypersetup{colorlinks=true,linkcolor=blue,citecolor=blue,urlcolor=blue,filecolor=blue}
\usepackage{newunicodechar}
\newunicodechar{⁡}{}
\newcommand{\um}{$\mu$m}

\newcommand{\hst}{{\it HST}}
\newcommand{\jwst}{{\it JWST}}

\providecommand{\lesssim}{\mathrel{\raise.3ex\hbox{$<$}\mkern-14mu\lower0.6ex\hbox{$\sim$}}}
\providecommand{\gtrsim}{\mathrel{\raise.3ex\hbox{$>$}\mkern-14mu\lower0.6ex\hbox{$\sim$}}}
\title[Where the red rise of QSO1 stops]{Where the red rise stops: JWST/MIRI reveals a near-infrared turnover in the Little Red Dot Abell~2744-QSO1} 
 
\author[E. Iani et al.]{
E. Iani$^{1}$\thanks{E-mail: edoardo.iani@ist.ac.at},
P. Rinaldi$^{2, 3}$,
A. Torralba$^{1}$,
J. Matthee$^{1}$,
F. D'Eugenio$^{4,5}$,
A. Gupta$^{2,3}$,
S. Mascia$^{1}$,
\newauthor
C. Di Cesare$^{1}$,
G. \"Ostlin$^{6}$, 
G. Kotiwale$^{1}$,
M. Catone$^{7,8}$,
G. Rodighiero$^{7,8,9}$
\\
$^{1}$Institute of Science and Technology Austria (ISTA), Am Campus 1, 3400 Klosterneuburg, Austria\\
$^{2}$ Department of Astronomy, The University of Texas at Austin, Austin, TX 78712, USA \\
$^3$ Cosmic Frontier Center, The University of Texas at Austin, Austin, TX 78712, USA\\
$^4$ Kavli Institute for Cosmology, University of Cambridge, Madingley Road, Cambridge, CB3 0HA, United Kingdom\\
$^5$ Cavendish Laboratory - Astrophysics Group, University of Cambridge, 19 JJ Thomson Avenue, Cambridge, CB3 0HE, United
Kingdom\\
$^6$ Department of Astronomy, Oskar Klein Centre, Stockholm University, AlbaNova University Center, 10691 Stockholm, Sweden\\
$^7$ Dipartimento di Fisica e Astronomia “G. Galilei”, Universit\`a di Padova, Vicolo dell’Osservatorio 3, 35122 Padova, Italy\\
$^8$ INAF, Osservatorio Astronomico di Padova, Vicolo dell’Osservatorio 5, 35122 Padova, Italy\\
$^9$ Centro di Ateneo di Studi e Attivit\`a spaziali “Giuseppe Colombo” (CISAS), via Venezia 1, 35131 Padova, Italy\\
}
 
\date{Accepted XXX. Received YYY; in original form ZZZ}
 
\pubyear{\the\year{}}

\begin{document}
\label{firstpage}
\pagerange{\pageref{firstpage}--\pageref{lastpage}}
\maketitle
 
\begin{abstract}
Little Red Dots (LRDs) are compact sources with a blue rest-frame UV, a steep red optical continuum and, often, broad hydrogen lines. They have been interpreted as dense stellar systems, dust-reddened active galactic nuclei, or black holes embedded in dense, dust-poor gas, scenarios that are hard to tell apart in the rest-frame UV and optical. We present JWST/MIRI F770W and F1000W imaging of the triply lensed Abell~2744-QSO1 at $z=7.04$, probing rest-frame $0.96$ and $1.24$\um. QSO1 is unresolved, and its steep optical rise stops near rest-frame $1$\um. Up to $1.24$\um, a $T\simeq4450$~K blackbody, or a hotter one reddened by dust, describes the continuum, with no sign of the hot dust seen in most type-1 quasars. Our statistically preferred SED fits produce the red continuum with $(3$--$4)\times10^{9}~\rm M_\odot$ of stars, over 100 times the dynamical limit on the host mass, and are therefore disfavoured. An accretion disc in a dense gas cocoon respects this limit but over-predicts the $1.24$\um\ flux by $5\sigma$. Super-Eddington accretion models with a thermalising envelope reproduce the turnover, but only for a black hole $\approx2.5$~dex less massive than the dynamical estimate. For a dust-reddened accretion disc ($A_V\simeq2$~mag), dust at $\geq1200$~K can re-emit at most $\simeq2$ per cent of the absorbed luminosity, and most of it would have to emerge at $\simeq70$--600~K, unless the dust covers only a small fraction of the sky. The dynamics favour an accretion-powered source, but our data cannot distinguish dust reddening from processing by dense gas. MIRI imaging at $\gtrsim18$\um\ could separate these scenarios.
\end{abstract}

 
\section{Introduction}
\label{sec:intro}
Among the many discoveries enabled by JWST, Little Red Dots \citep[LRDs;][]{Matthee+24a,Kocevski+23,Labbe+23} have emerged as one of the most debated populations at intermediate and high redshift \citep{Harikane+23,Greene+24,Inayoshi+25a,Ubler+26}. They are surprisingly common at $z\sim5$--7 \citep{Kokorev+24b}, and they combine properties with no clear precedent among known galaxies or active galactic nuclei (AGN). LRDs are compact and often unresolved in the rest-frame optical \citep{Kocevski+25}. Their spectral energy distributions (SEDs) are ``V-shaped'', with a faint blue UV continuum and a steep red optical rise \citep{Furtak+24,Greene+24,Kocevski+25,Setton+25b}. Their broad hydrogen lines, with a full width at half maximum (FWHM) above $1000~\rm km~s^{-1}$, point to gas moving around a black hole \citep{Matthee+24a,Treiber+25,Hviding+25}. Other properties, however, are unusual for an AGN. The broad lines often have large Balmer decrements and non-Gaussian, exponential-like wings \citep{Nikopoulos+26,Rusakov+26,Torralba+26b,Chang+26,Brazzini+25,Scholtz+26,Sneppen+26c}. Many LRDs also show Balmer absorption on top of the broad emission \citep{Juodbalis+24a,Deugenio+26a,Matthee+26,Torralba+26c}, a feature seen in only a handful of quasars \citep[e.g.][]{Hall07,Zhang+15}. Most LRDs are X-ray undetected \citep{Ananna+24,Yue+24,KokuboHarikane25}, faint in the mid- and far-infrared \citep{Williams+24,Setton+25a,Akins+25b,Casey+25} and show little or no variability \citep{KokuboHarikane25,Tee+25,Secunda+26}. Type-1 AGN, instead, are typically bright in X-rays \citep[e.g.][]{Steffen+06,Just+07,Lusso+16}, show hot-dust emission in the near- and mid-infrared \citep[e.g.][]{Richards+06,Mor+12}, and vary on time-scales of months to years \citep[e.g.][]{VandenBerk+04,MacLeod+10}.

This combination of observed physical properties makes the nature of LRDs hard to establish. Proposed interpretations have ranged from massive compact galaxies \citep{Labbe+23,Baggen+24} and composite AGN--host systems \citep{Barro+24,Greene+24}, to dust-reddened AGN with high accretion rates or a large covering factor \citep{MadauMaiolino26}, to accreting black holes embedded in dense, hydrogen-rich gas \citep{Inayoshi+25b,Kido+25,Naidu+25,Yanagisawa+26,deGraaff+26}. More exotic options include supermassive stars, quasi-stars within proto-globular clusters \citep{Nandal+26,Begelman+26,Chisholm+26,Kokorev+26}.

Whichever of these scenarios is confirmed, it would change our understanding of early black-hole and galaxy growth, and each would change it in a different way. If the light of LRDs were dominated by stars, the stars would have to have formed rapidly in extraordinarily dense systems \citep{Labbe+23,Baggen+24}. Given the number densities of LRDs, they would also have converted a large fraction of the available baryons into stars by $z\gtrsim5$, possibly in tension with current cosmological models \citep[e.g.][]{BoylanKolchin23}. If instead LRDs were AGN and their broad lines traced virial motions, they would host black holes of $10^{6}$--$10^{9}~\rm M_\odot$, up to 1.5--2~dex above the local relation for their host stellar masses. This would challenge even the most extreme models of black-hole growth \citep{Maiolino+24,Juodzbalis+26b,Pacucci+23,Harikane+23}. Finally, if they were black holes embedded in dense, dust-poor gas, scattering could broaden the lines and modify their ratios, in which case virial masses would be overestimated thus easing the tension with black-hole growth models \citep{Rusakov+26,Chang+26,Ji+25,Sneppen+26c}. At the same time, the dense gas should also produce the observed red optical continuum. This would happen in two ways: it could imprint a Balmer break and absorption on an intrinsically blue AGN continuum \citep{Inayoshi+25b,Naidu+25,Torralba+26c,deGraaff+25B,Matthee+26}; or, if it is optically thick, it could thermalise the accretion luminosity and emit a cool, nearly blackbody spectrum, a ``pseudo-photosphere'' \citep{Kido+25,Liu+26,Begelman+26}. Both scenarios would reproduce the V-shaped SEDs and make LRDs an early, possibly short-lived phase of black-hole growth inside dense gas, with black holes less massive than their virial estimates and an emission very different from that of standard AGNs \citep{Inayoshi+25b,Kido+25,Naidu+25}. The origin, lifetime and geometry of this dense gas, however, remain unclear.

In the rest-frame UV and optical, studies have shown that these scenarios are hard to tell apart, since stars, a dust-reddened AGN and dense gas around the black hole can all reproduce the red optical continuum of LRDs \citep[e.g.][]{Greene+24,Barro+24,Inayoshi+25b}. This motivates observations in the rest-frame near-infrared. There, the light of a stellar population best traces its mass \cite[e.g.][]{BelldeJong01}. A dust-reddened AGN should show hot dust re-emitting the absorbed luminosity, which in type-1 quasars often produces a rise beyond rest-frame $\approx1$\um\ well described by a $T\approx1200$--1300~K blackbody \citep{Barvainis87,Richards+06,Glikman+06,Temple+21}. A pseudo-photosphere, instead, emits a nearly thermal spectrum with $T_{\rm eff}\approx4000$--6000~K, which turns over near rest-frame 1\um\ without hot dust \citep{Kido+25,Liu+25,Liu+26,Begelman+26,Sneppen+26c}.

At $z\gtrsim5$, the rest-frame near-infrared lies beyond observed 5\um\ and can only be reached with the Mid-Infrared Instrument (MIRI, \citealt{Rieke+15,Wright+23}) on board JWST. MIRI observations of LRDs in this regime so far give a mixed picture. Several bright LRDs show weak or absent hot-dust emission \citep{Williams+24,Setton+25a}, with mid-infrared colours bluer than those of AGN tori \citep{Perez-Gonzalez+24a}. In contrast, the \textit{Rosetta Stone} \citep{Juodbalis+24a} and stacks of fainter LRDs \citep{Delvecchio+25,Perez-Gonzalez+26,Barro+26b} show an excess beyond rest-frame $\approx2$\um, which could trace hot dust. At $z<1$, LRD analogues selected from the Dark Energy Spectroscopic Instrument (DESI, \citealt{DESI19}) survey show blackbody-like continua with $T\approx2000$--4700~K, peaking at rest-frame 0.6--1.5\um\ \citep[e.g.][]{Lin+26,Park+26}, as expected for a pseudo-photosphere. Some of them also show an excess at $\gtrsim3$\um, attributed to warm dust at $\approx500$--700~K \citep{Ashall+26}. Individual faint LRDs at the highest redshifts, where the population is most abundant, remain largely unexplored at these wavelengths.

In this work, we present deep JWST/MIRI F770W and F1000W imaging of Abell~2744-QSO1 (hereafter QSO1), a triply-imaged lensed source at $z=7.04$ (Section~\ref{sec:target}) and one of the best-studied LRDs to date \citep{Furtak+23,Furtak+24,Furtak+25,Ji+25,Ma+25,Maiolino+26b,Deugenio+26b,Ji+26c,Zhang+26,Tang+26a}.
At the redshift of QSO1, these filters probe rest-frame $\approx0.96$ and $1.24$\um, the immediate continuation of the observed optical rise.
We combine them with the existing UV-to-optical photometry and spectroscopy and with far-infrared and millimetre limits (Section~\ref{sec:data}) to address three open questions:
(i) how the optical rise of QSO1 continues into the near-infrared,
(ii) whether the red continuum comes from a massive stellar host or from dense gas around the black hole, and
(iii) whether QSO1 shows the hot-dust emission expected for a dust-reddened AGN, and where the luminosity absorbed by dust is re-emitted.

Throughout this paper, we assume a flat $\Lambda$-CDM cosmology with $\Omega_{\rm M} = 0.3$, $\Omega_\Lambda = 0.7$, and $H_0 = 70\rm~ km~ s^{-1}~ Mpc^{-1}$.
We adopt AB magnitudes \citep{Oke+83} and a \citet{Chabrier+03} initial mass function (IMF).

%%%%%%%%%%%%%%%%%%%%%%%%%%%%%%%%%%%%%%%%%%%%%%%%%%%%%%%%%%%%%%
\section{Target}
\label{sec:target}
QSO1 is an extremely red, compact source at $z_{\rm spec}=7.0364$ \citep{Tang+26a}, triply imaged by the galaxy cluster Abell~2744 (A2744, $z=0.308$). Its three images are Ima-A ($\rm RA=3.5798375$, $\rm Dec=-30.4015733$~deg), Ima-B ($3.5835458$, $-30.3966828$~deg) and Ima-C ($3.5972125$, $-30.3943442$~deg), with magnifications $\mu=5.4\pm1.1$, $7.5\pm1.5$ and $2.7\pm0.5$, respectively \citep{Furtak+25}. Thanks to the lensing magnification, we can study an intrinsically faint source ($M_{\rm UV}\simeq-16.3$) at different moments of its history. In fact, since the light of each image follows a different path through the cluster, it reaches us with a different delay. Image~B and image~C arrive $\approx2.3$ and $\approx19$~yr after image~A \citep{Furtak+25}, spanning up to $\approx2.4$~yr in the rest frame. Between these epochs, the equivalent widths of the broad H$\alpha$ and H$\beta$ lines have been found to change, revealing variability of the broad-line emission \citep{Furtak+25}.

Several observations point to an accreting black hole, with little room for a host galaxy. QSO1 is unresolved in JWST Near-Infrared Camera (NIRCam; \citealt{Rieke+23}) imaging, with $r_{\rm e}<30$~pc \citep{Furtak+24}, and even its rest-frame UV shows AGN features, namely broad Ly$\alpha$ and far-UV Fe~{\sc ii} emission \citep{Tang+26a}. Its narrow-H$\alpha$ velocity field indicates Keplerian rotation around a black hole of $\approx5\times10^{7}~\rm M_\odot$ \citep{Juodzbalis+26b}, consistent with the broad-line estimate \citep{Furtak+24}. This leaves at most $\approx2\times10^{7}~\rm M_\odot$ for the host stars, i.e. $M_{\rm BH}/M_\star>2$. The narrow-line widths also limit the total dynamical mass to $M_{\rm dyn}<4\times10^8~\rm M_\odot$ \citep{Ji+25}.

Despite these constraints, what produces the continuum of QSO1 is still debated. It shows one of the strongest and smoothest Balmer breaks among LRDs, comparable to that of \textit{the Cliff} \citep{deGraaff+25B}, followed by a steep optical rise. This shape has been reproduced by a massive, heavily attenuated stellar population \citep{Ma+25}, which however conflicts with the dynamical constraints above, by an accretion disc seen through nearly dust-free gas with $n_{\rm H}\sim10^{10}~{\rm cm^{-3}}$ and a large population of hydrogen in its $n=2$ level \citep{Ji+25,Deugenio+26b,Chang+26}, or by a pseudo-photosphere \cite[e.g.][]{Naidu+26}. If instead the broad Balmer decrement were due to dust, it would imply $A_V\approx2$--3~mag \citep{Ji+25,Deugenio+26b}. QSO1 is undetected by the Atacama Large Millimeter/submillimeter Array (ALMA; \citealt{Furtak+24,Labbe+25}), which constrains only cold dust. It is also undetected in deep \textit{Chandra} observations \citep{Furtak+24}, which at $z=7$ probe rest-frame energies of $\approx4$--56~keV. This X-ray faintness suggests either an intrinsically X-ray-weak source or a large gas column ($N_{\rm H}\gtrsim10^{24}~{\rm cm^{-2}}$; e.g. \citealt{Lambrides+26,Maiolino+25}). These constraints come from the rest-frame UV and optical and from far-infrared limits. The continuum beyond rest-frame $\approx0.6$\um, where the scenarios differ, had not been observed before this work.

%%%%%%%%%%%%%%%%%%%%%%%%%%%%%%%%%%%%%%%%%%%%%%%%%%%%%%%%%%%%%%
\section{Data}
\label{sec:data}
 
\subsection{Dataset}
\label{sec:dataset}
The core of our dataset is the deep MIRI imaging at 7.7 and 10~$\mu$m from the \jwst\ Cycle~3 program \textit{MIRI in Clusters} (MIRIC, GO-5578, PIs: Iani \& Rinaldi; Iani et al.\ in prep.), which covers $\sim$5.9~arcmin$^2$ of the central region of A2744, including all three images of QSO1.
The MIRI data were reduced with the same pipeline as in \citet{Rinaldi+25}, with further details to be presented in Iani et al. {\it (in prep.)}.
 
We complemented the MIRI data with all the available \hst\ imaging from the Advanced Camera for Surveys (ACS) and the Wide Field Camera~3 (WFC3), and with \jwst/NIRCam imaging.
We used the fully reduced, public mosaics of the UNCOVER DR3\footnote{\href{https://jwst-uncover.github.io/DR3.html\#Mosaics}{https://jwst-uncover.github.io/DR3.html}} \citep{Bezanson+24, Suess+24}.
These combine the NIRCam imaging from ERS-1324 (GLASS; PI: Treu), GO-2561 (UNCOVER; PIs: Labb\'e \& Bezanson), DD-2767 (PI: Kelly), GO-2883 (MAGNIF; PI: Sun), GO-3516 (ALT; PIs: Naidu \& Matthee), GO-3538 (PI: Iani) and GO-4111 (MegaScience; PI: Suess) with the \hst\ programs 11689 (PI: Dupke), 13386 (PI: Rodney), 13389 (PI: Siana), 13495 (HFF; PI: Lotz), 15117 (BUFFALO; PI: Steinhardt) and 17231 (PI: Treu).
 
We also used \jwst/NIRSpec Multi-Shutter Assembly (MSA) spectroscopy from SPURS (GO-9214, PIs: Mason \& Stark), obtained with the three medium-resolution ($R\simeq1000$) grating/filter combinations G140M/F100LP, G235M/F170LP and G395M/F290LP, which together cover $\sim$1--5.3~$\mu$m.
We retrieved the reduced spectra (v4) from the DAWN JWST Archive (DJA)\footnote{\url{https://dawn-cph.github.io/dja/}}.
The NIRSpec spectra were reduced via {\sc msaexp}\footnote{\url{https://github.com/gbrammer/msaexp}} \citep{Brammer+23} following the data reduction described in \cite{Heintz+24, deGraaff+25}.
All three gratings targeted Ima-B of QSO1.
 
To potentially constrain the dust emission, we also considered far-infrared and millimetre data.
The region covering the three images of QSO1 was observed by \textit{Spitzer}/MIPS at 24~$\mu$m and by \textit{Herschel}/PACS (100 and 160~$\mu$m) and SPIRE (250, 350 and 500~$\mu$m) \citep{Rawle+14, Rawle+16}.
These observations reach average $3\sigma$ depths of 53~$\mu$Jy, 2.6, 5.2, 16.8, 19.2 and 19.8~mJy, respectively, where for SPIRE we adopt the confusion-limited depths of \citet{Nguyen+10}.
QSO1 is not detected in any of these maps. We assumed uniform image-plane noise and adopt the same $3\sigma$ limit for each image.
 
QSO1 is also covered by ALMA Band~6 (1.2~mm) observations from DUALZ+ALCS+ALMA-HFF programs \citep{Fujimoto+24, Fujimoto+25, MunozArancibia+23}.
We downloaded the fully-reduced ALMA mosaic maps from the UNCOVER webpage\footnote{\href{https://jwst-uncover.github.io/DR2.html\#DUALZ}{https://jwst-uncover.github.io/DR2.html}}.
None of the three images is detected, with $3\sigma$ upper limits of 94.5, 90.9 and 84.0$\mu$Jy~beam$^{-1}$ for Ima-A, B and C, respectively.

\subsection{Photometry extraction}
\label{sec:photometry_extraction}
For each of the three multiple images of QSO1, we created $18''\times18''$ cutouts centred on the corresponding coordinates reported in Section~\ref{sec:target}.
Ima-B is significantly contaminated by a foreground object located at a small projected separation from the target ($\approx 1\arcsec$).
We modelled and removed this contaminating emission through an iterative deconvolution procedure based on the Richardson--Lucy algorithm \citep{Richardson+72,Lucy+74}, adopting filter-dependent model PSFs.
We present more details and examples of the contaminant modelling and subtraction in Appendix~\ref{app:gal-removal}.
 
We extracted the photometry using \textsc{sep} \citep{Barbary+16}, a \textsc{Python} implementation of \textsc{SourcExtractor} \citep{Bertin+96}.
For each filter and multiple image, we adopted the circular aperture that maximised the signal from the target, imposing a minimum radius corresponding to the radius enclosing $70\%$ of the PSF energy.
The measured fluxes and their uncertainties were subsequently corrected for the flux falling outside the adopted apertures using the corresponding PSF encircled-energy (EE) curves.
For MIRI, we adopted the in-flight EE curves of the JWST calibration reference files (CRDS \texttt{apcorr}, context \texttt{jwst\_1298.pmap}), which include the detector scattering and the extended wings of the PSF.
At our aperture radii ($0.40$--$0.56\arcsec$), they enclose $66$--$72\%$ of the flux, $\approx12\%$ (F770W) and $4$--$6\%$ (F1000W) less than an {\sc stpsf} model without detector effects, and agree within $1.5\%$ with the growth curve of a field star (Appendix~\ref{sec:psf}).

We estimated the photometric uncertainties empirically by placing 500 apertures, with the same area as that adopted for the target, at random positions within the surrounding sky region.
Neighbouring sources were masked before placing the apertures, and the standard deviation of the resulting sky-aperture fluxes was taken as the $1\sigma$ uncertainty on our target flux measurement.
To additionally account for systematic uncertainties in the absolute flux calibration of the \jwst\ data, we adopted a conservative minimum relative uncertainty of $5\%$ in all \jwst\ bands.
 
Finally, we corrected the photometry for Galactic foreground extinction.
We adopted $R_V=3.1$ and the colour excess reported by the {\sc irsa} dust maps\footnote{\url{https://irsa.ipac.caltech.edu/applications/DUST/}} at the position of QSO1, $E(B-V)=0.0112\pm0.0004$ \citep{Schlafly+11}.
The correction in each passband was calculated using the extinction curve of \citet{Gordon+23}, as implemented in the \texttt{dust\_extinction}\footnote{\url{https://dust-extinction.readthedocs.io/en/latest/}} \textsc{Python} package \citep{Gordon+24}.
The adopted multiplicative corrections, $f_{\rm ext}$, are reported in Table~\ref{tab:photometry}, together with the photometry of the three multiple images.
The individual-image measurements listed in the table are not corrected for gravitational-lensing magnification.
 
We then corrected the flux densities of the three multiple images for their magnification factors (Section~\ref{sec:target}; \citealt{Furtak+25}), propagating the magnification uncertainties, which include a 20\% floor for lensing systematics, into the uncertainty of each band.
Although lensing introduces time delays that allow us to observe our target at different intrinsic epochs, the resulting three SEDs show overall consistent colours in their broad-band photometry \citep{Furtak+25}, even in their new MIRI coverage, see Figure~\ref{fig:phot-rescaling-diagnostic}.
The bands that show the largest discrepancies are those affected by strong emission lines.
Despite their similar SEDs, Ima-A and Ima-B display approximately wavelength-independent offsets relative to Ima-C, with weighted mean flux-density ratios of
$f_{\rm Ima-A}/f_{\rm Ima-C}=1.87$ and
$f_{\rm Ima-B}/f_{\rm Ima-C}=1.22$.
These factors were derived from all detected bands (Appendix~\ref{app:phot-combination}). Excluding the two MIRI bands changes them by $\lesssim2\%$ ($1.86$ and $1.25$), so the MIRI data do not drive the normalisation.
In the medium bands closest to MIRI that are free of strong emission lines (F430M, F460M and F480M), the ratios are $1.92\pm0.08$ and $1.23\pm0.05$ (photometric uncertainties only). We use these values to test for chromatic variability in Section~\ref{sec:disc_variability}.
These offsets likely reflect residual uncertainties in the relative magnification factors, residual photometric systematics, or a combination of both.
We adopted Ima-C as the reference since it has the lowest magnification and is therefore expected to be less sensitive to uncertainties in the lens model.
We corrected the de-magnified photometry of Ima-A and Ima-B for the single wavelength-independent multiplicative factor found to match the normalisation of Ima-C.
We note that, because the combined SED is tied to the normalisation of Ima-C, the uncertainty on $\mu_{\rm Ima-C}$ acts as a common scaling of all bands, corresponding to a systematic uncertainty of $\pm0.08$~dex on all luminosities and masses derived in this work.
 
The final photometry of QSO1 was then obtained as the inverse-variance-weighted mean of the three rescaled measurements.
To account for any remaining differences between the multiple images, the uncertainty in each band was increased by adding in quadrature a systematic term derived from the inter-image scatter.
Further details of this procedure are provided in Appendix~\ref{app:phot-combination}.
The resulting combined SED is shown in Figure~\ref{fig:phot-rescaling-diagnostic}.
Following the relative rescaling, the measurements obtained from the three multiple images are mutually consistent within $3\sigma$ across the available passbands.
The magnification uncertainties included in these errors scale all bands of an image together and do not affect the shape of the SED. For the analysis of the continuum shape (Section~\ref{sec:nir_continuum}) we therefore combined the images with their photometric uncertainties only, keeping the inter-image systematic term and a $5\%$ calibration floor on the stack. The SED fits use the total uncertainties, which is conservative.

\subsection{NIRSpec spectrum}
\label{sec:nirspec_spectrum}
We constructed a continuous NIRSpec spectrum of QSO1 by combining the observations obtained with the G140M/F100LP, G235M/F170LP and G395M/F290LP grating--filter configurations (Section~\ref{sec:dataset}).
The individual spectra were resampled onto a common wavelength grid extending from $0.8$ to $5.6~\rm \mu m$, with a sampling of $\Delta\lambda=0.005~\rm \mu m$, using the flux-conserving \texttt{spectres} algorithm \citep{Carnall+17}.
In the wavelength regions covered by more than one configuration, the spectra were combined using inverse-variance weighting.
The corresponding wavelength-dependent spectral-resolution curves were derived following the parameterisation of the instrumental line-spread-function \citep{Shajib+25} and combined using the same weights.
We subsequently corrected for the overall normalisation and residual wavelength-dependent path losses by matching the spectrum to the combined imaging photometry derived in Section~\ref{sec:photometry_extraction}.
To this end, we calculated synthetic NIRCam photometry by convolving the spectrum with the corresponding filter-transmission curves retrieved from the Spanish Virtual Observatory Filter Profile Service \citep[SVO FPS;][]{Rodrigo+12,Rodrigo+20}.
The uncertainties in the synthetic photometry were propagated from the errors on the individual spectral elements.
For each filter covered by the spectrum, we measured the ratio between the observed and synthetic photometry.
Both the imaging and synthetic-photometry uncertainties were included in the corresponding error.
We modelled the wavelength dependence of these ratios using a second-order polynomial in logarithmic wavelength and flux-ratio space, and multiplied the spectrum by the resulting correction curve.
The uncertainty associated with the path-loss correction was estimated using 1000 Monte Carlo realisations of the photometric-to-synthetic flux-density ratios.
This uncertainty was propagated into the final spectral error together with the original uncertainty of the NIRSpec spectrum.
The comparison between the combined photometry and the spectrum before and after the path-loss correction is shown in Figure~\ref{fig:pathloss-diagnostic}, while further details of the procedure are presented in Appendix~\ref{app:pathloss}.

%%%%%%%%%%%%%%%%%%%%%%%%%%%%%%%%%%%%%%%%%%%%%%%%%%%%%%%%%%%%%%%
\section{Results}
\label{sec:results}
 
\subsection{MIRI morphology and variability}
\label{sec:miri_morph}
\label{sec:disc_variability}
We fitted each image in the contaminant-removed F770W and F1000W mosaics with a point source and with a circular Gaussian convolved with the PSF, accounting for the correlated noise of the drizzled mosaics, with {\sc socca} \citep{DiMascolo+26}.
To bracket the systematic uncertainty on the PSF, we repeated every fit with two effective PSFs: a model PSF built by resampling per-exposure {\sc stpsf} models in the same way as the mosaics (``ad-hoc'' PSF), and the same PSF broadened to match the core of a field star (``broadened'' PSF; Appendices~\ref{sec:psf} and \ref{app:morph}). The broadened PSF is calibrated empirically, but on a star much brighter than QSO1, whose core may be broadened by brightness-dependent detector effects, so the true PSF of QSO1 should lie between the two.
 
\begin{figure}
    \centering
    \includegraphics[width=\columnwidth, trim={0 0 9.5cm 0}, clip=True]{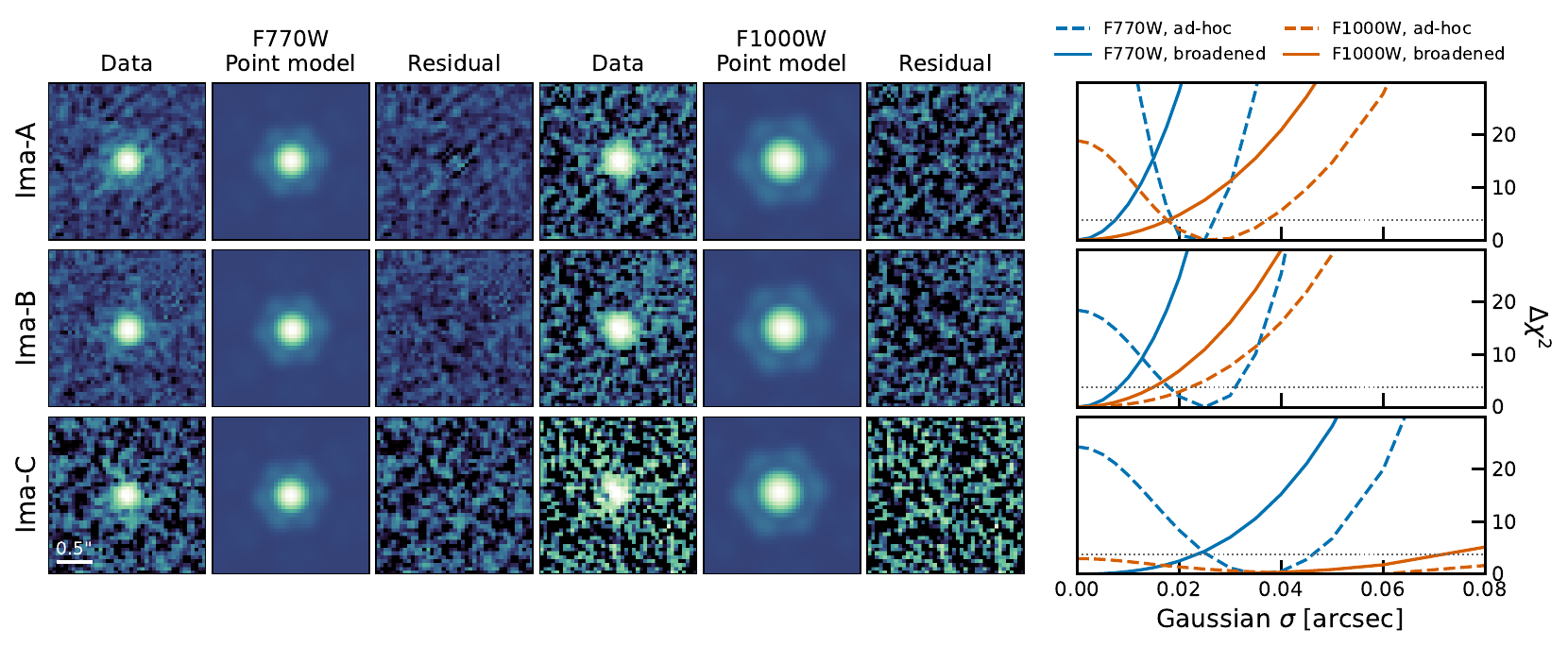}
    \caption{{\it QSO1 is unresolved in both MIRI bands in all three images.} MIRI morphology of the three images of QSO1 (rows: Ima-A, Ima-B, Ima-C). For F770W and F1000W, the panels show the data, the best-fit point-source model (broadened PSF) and the residuals, on $2.5\arcsec$ stamps. All images are consistent with a point source. The size limits quoted in the text are derived with the narrower ad-hoc PSF (Appendix~\ref{app:morph}).}
    \label{fig:miri_morph}
\end{figure}

With the broadened PSF, all six fits (three images, two filters) are best described by a point source (Figure~\ref{fig:miri_morph}).
With the ad-hoc PSF, some fits prefer a small extension ($\sigma\simeq0.025$--$0.035\arcsec$), but this has nearly the same angular size in all three images, despite their different magnifications, and is smaller than the core excess measured on the field star. It reflects a PSF systematic rather than intrinsic structure, and QSO1 is unresolved in MIRI in all three images.
We nevertheless derive the size limits with the ad-hoc PSF, since any mismatch then appears as extension and the limits are conservative.
The narrow H$\alpha$ and Ly$\alpha$ emission of QSO1 extends over a few hundred parsecs \citep{Maiolino+26b,Ji+26c}, but the emission lines contribute only $\approx3$--7\% of the MIRI flux densities, almost entirely from the broad component (Section~\ref{sec:nir_continuum}), so an extended narrow-line region is not expected to be detectable in MIRI.
Converted to the source plane (Appendix~\ref{app:morph}), the 95\% upper limits on the effective radius in F770W are $r_{\rm e}<64$, 70 and 173~pc for Ima-A, B and C. These limits are conservative. The tightest, $r_{\rm e}<64$~pc for Ima-A ($<51$~pc dividing by the tangential magnification alone, as in \citealt{Furtak+24}), is a factor of $\approx2$ weaker than the NIRCam limit of $r_{\rm e}<30$~pc, as expected from the broader MIRI PSF.
 
Ima-A and Ima-C sample epochs $\approx19$~yr apart in the observed frame ($\approx2.4$~yr in the rest frame; Section~\ref{sec:target}). Since the magnification uncertainties are common to all bands of an image, we compared the MIRI image ratios with those of the line-free NIRCam continuum (Section~\ref{sec:photometry_extraction}), using the photometric uncertainties only. For Ima-A, the cleanest image, the ratios relative to Ima-C ($1.89\pm0.13$ in F770W and $1.85\pm0.15$ in F1000W) are $0.98\pm0.08$ and $0.96\pm0.09$ times the NIRCam value. For Ima-B they are $12$--$16\%$ lower, within the scatter of its own NIRCam continuum ratios ($1.13$--$1.43$ in F430M--F480M), most likely because of residual contamination from the foreground object (Appendix~\ref{app:gal-removal}). Point-source photometry and different NIRCam reference bands change these ratios by at most $\approx15\%$, and no combination deviates from the NIRCam value by more than $3\sigma$. We find no evidence for wavelength-dependent variability above $\approx15\%$, as expected if the optical and near-infrared continua come from the same component. Smaller variations, such as those of a hot-dust component lagging the optical by $\sim0.7$~yr, the light-crossing time of the sublimation radius (Section~\ref{sec:disc_dust}), cannot be excluded.
The broad lines, instead, have been reported to vary. Their equivalent widths differ between the images by up to $\approx20\%$ \citep{Furtak+25}, and those of H$\beta$ and H$\alpha$ in Ima-C are about twice those of Ima-A and Ima-B \citep{Ji+25}, consistent with the $\approx10\%$ lower image ratios we measure in the line-dominated NIRCam bands. Paschen lines twice as strong in Ima-C would raise its MIRI flux densities by only $\approx3\%$ and $\approx7\%$ (Section~\ref{sec:nir_continuum}), within our uncertainties.

\subsection{The near-infrared turnover}
\label{sec:nir_continuum}
\label{sec:empirical_templates}
 
 \begin{figure}
     \centering
     \includegraphics[width=\linewidth]{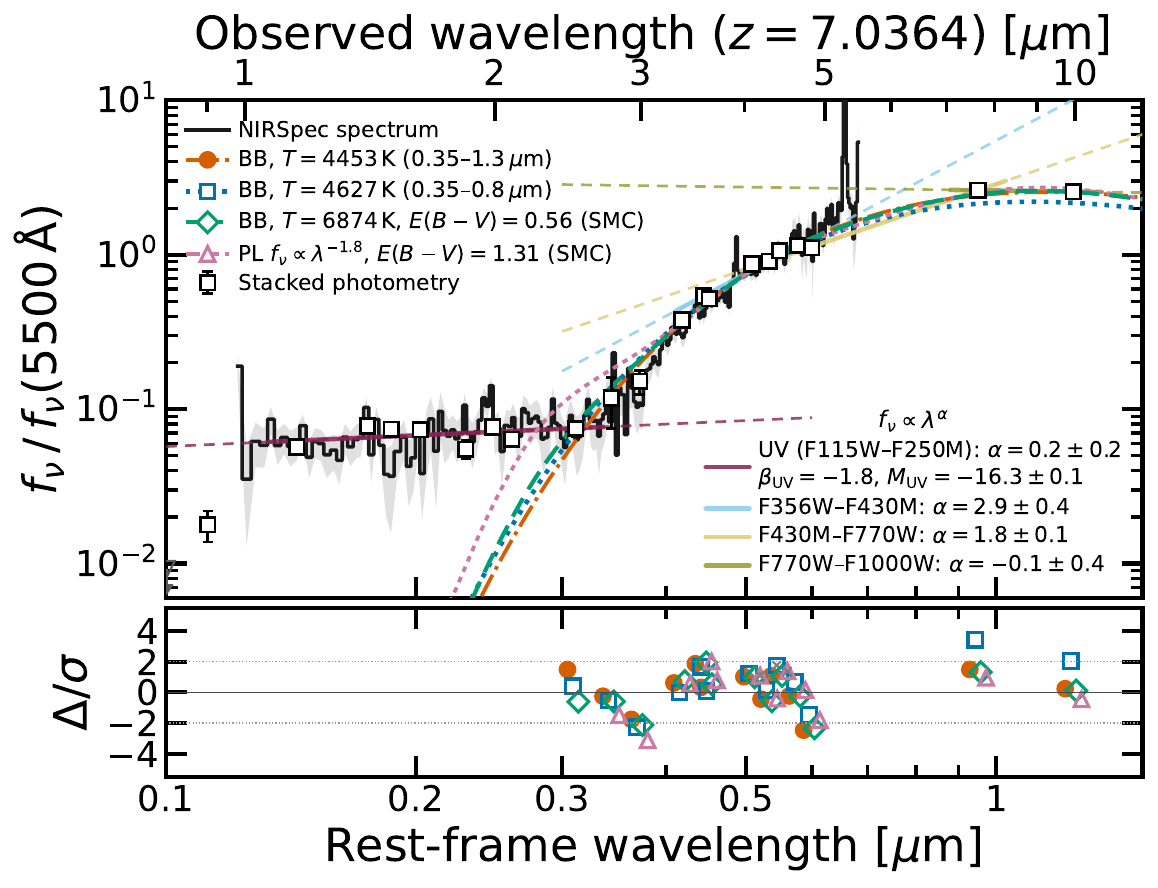}
     \caption{{\it The steep optical rise of QSO1 flattens between the two MIRI bands, and a single $T\approx4450$~K blackbody describes the continuum from rest-frame 0.35 to 1.24\um. A hotter blackbody reddened by dust fits equally well.} Rest-frame SED of QSO1 normalised at 5500\AA: path-loss-corrected NIRSpec combined spectrum (black) and stacked photometry (white squares). Lines show the blackbodies fitted to the photometry at rest-frame 0.35--1.3\um\ ($T_{\rm BB}=4453$~K, including MIRI; purple dash-dotted) and 0.35--0.8\um\ ($T_{\rm BB}=4627$~K, optical only; blue dotted), the green dashed line the $T_{\rm BB}=6874$~K blackbody attenuated by an SMC screen with $E(B-V)=0.56$ ($A_V\approx1.5$~mag), and the pink densely dotted line the reddened power law $f_\nu\propto\lambda^{-1.84}$ with $E(B-V)=1.31$ ($A_V\approx3.6$~mag; see Section~\ref{sec:nir_continuum}). Error bars and fits use the photometric uncertainties only (Section~\ref{sec:photometry_extraction}). The coloured segments show the local power-law slopes $f_\nu\propto\lambda^\alpha$ between F356W and F430M, F430M and F770W, and F770W and F1000W, and in the rest-frame UV (F115W--F250M), with their values in the legend. Bottom: residuals of the four continuum models, $(f_{\rm obs}-f_{\rm model})/\sigma$, at the bands between rest-frame 0.3 and 1.3\um, with the markers shown next to each model in the legend (crosses: F410M and F444W, excluded from the fits).}
     \label{fig:bb_slopes}
 \end{figure}
 
Between rest-frame $0.44$ and $0.53~$\um\ (F356W and F430M), the flux density rises as $f_\nu\propto\lambda^{\alpha}$ with $\alpha=2.9\pm0.4$, and the rise continues with $\alpha=1.83\pm0.13$ between $0.53$ and $0.95~$\um\ (F430M and F770W).
Between the two MIRI filters ($0.95$--$1.24~$\um), the continuum is instead flat in $f_\nu$, with $\alpha=-0.07\pm0.36$. Since $\nu f_\nu\propto\lambda^{\alpha-1}$, $\nu f_\nu$ peaks between rest-frame $0.95$ and $1.24$\um\ and declines at longer wavelengths.
Both MIRI filters contain hydrogen and helium recombination lines (the Paschen series and He{\sc i}$~\lambda10830$ in F770W, Pa$\beta$ in F1000W). Scaling the broad and narrow H$\alpha$ fluxes of \citet{Deugenio+26b} with Case-B line ratios and the broad-line attenuation (Appendix~\ref{app:nir}), lines contribute at most $\approx3\%$ and $\approx7\%$ of the F770W and F1000W flux densities, so the flattening is a property of the continuum.
 
A single blackbody fitted to the combined photometry at rest-frame wavelengths longer than $0.35~$\um\ (excluding F410M and F444W, which are contaminated by H$\beta$) has $T_{\rm BB}=4450\pm60$~K and $\chi^2_{\rm red}=\chi^2_{\rm min}/(N_{\rm bands}-N_{\rm par})=2.2$ (nine filters, two free parameters), and reproduces both MIRI points within $1.3\sigma$ (Figure~\ref{fig:bb_slopes}). The largest residuals are in two NIRCam bands (F356W and F480M, $\approx2$--$2.5\sigma$), not in MIRI.
Fitting only the rest-frame optical filters (F300M to F480M) gives $T_{\rm BB}=4630\pm110$~K, which under-predicts the F770W and F1000W flux densities by $18\%$ and $15\%$ ($3.3\sigma$ and $2.0\sigma$), whereas a power law fitted to the same filters ($\alpha=3.3$) over-predicts them by factors of $2.2$ and $5.3$.
The MIRI bands therefore trace the turnover of the same thermal component that produces the red optical continuum, and no additional component is needed up to $1.24$\um. The blackbody temperature is, however, degenerate with dust attenuation: a hotter, reddened blackbody ($T_{\rm BB}=6874$~K with $A_V \simeq 1.5$~mag, $\chi^2_{\rm red}=2.4$) describes the data equally well (Appendix~\ref{app:nir}).
 
\begin{figure*}
    \centering
    \includegraphics[width=\textwidth]{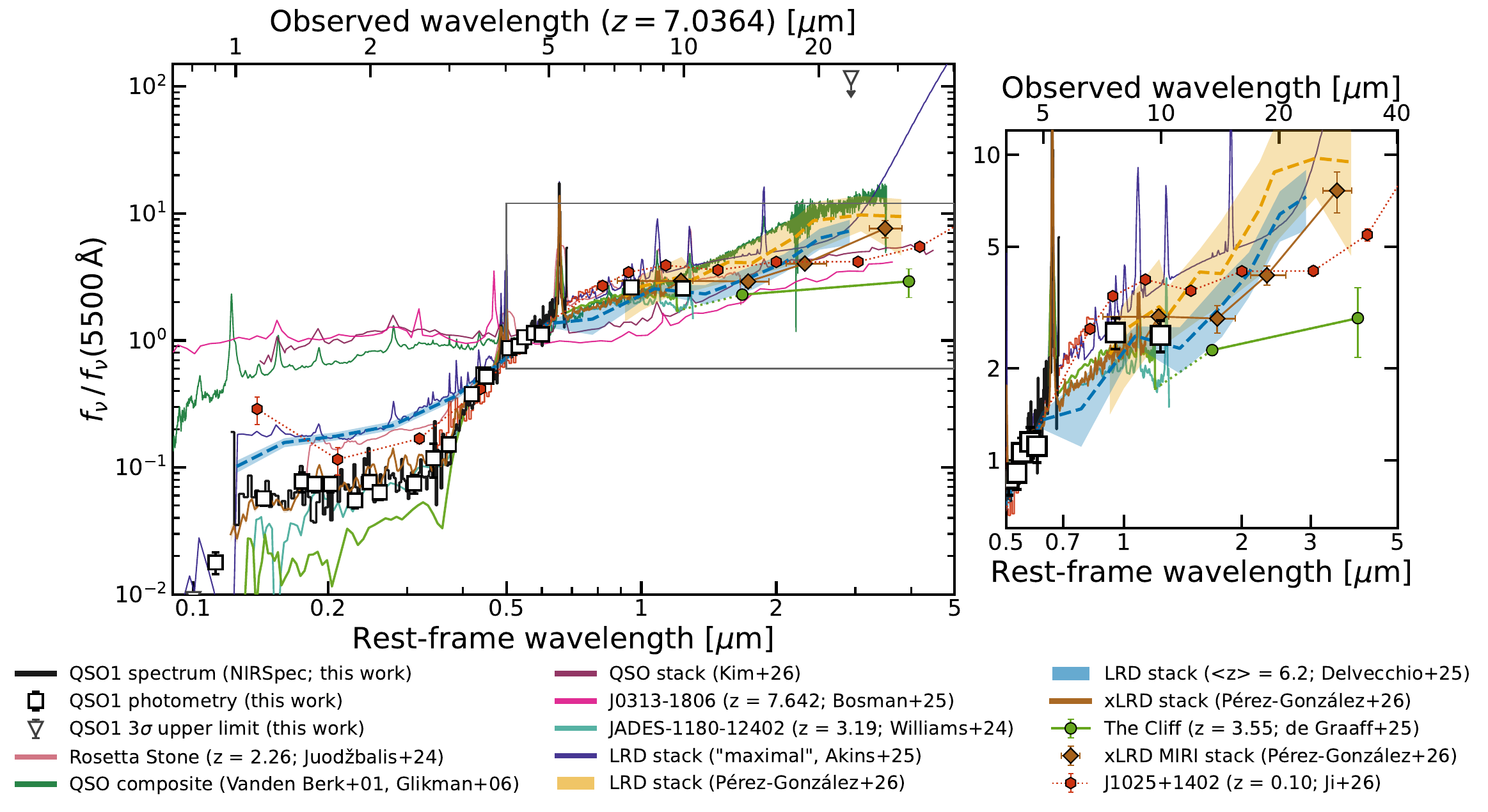}
    \caption{{\it Up to rest-frame 1.24\um\ the SED of QSO1 resembles those of other LRDs, and lacks the hot-dust rise of type-1 quasars. Beyond this wavelength, where our data stop, LRDs diverge.} Rest-frame SED of QSO1 (stacked photometry: white squares and $3\sigma$ upper limits; path-loss-corrected NIRSpec spectrum: black) compared with type-1 quasar templates \citep{VandenBerk+01,Glikman+06,Kim+26}, the $z\approx7.6$ quasar J0313$-$1806 \citep{Bosman+25}, the LRDs GN-28074 (the \textit{Rosetta Stone}; \citealt{Juodbalis+24a}) and JADES-1180-12402 \citep{Williams+24,Setton+25a}, \textit{the Cliff} (RUBIES-UDS-154183, $z=3.55$; \citealt{deGraaff+25B}; PRISM spectrum and NIRCam+MIRI photometry, circles), the local LRD J1025+1402 \citep[\textit{the Egg}, $z=0.10$;][]{Lin+25, Ji+26a}, the ``maximal'' LRD template of \citet{Akins+25b}, the LRD stacks of \citet{Delvecchio+25} and \citet{Perez-Gonzalez+26} (shaded: their uncertainties), and the stack of extreme LRDs (xLRDs) of \citet{Perez-Gonzalez+26} with its MIRI photometric stack (dark orange line and diamonds). All curves are normalised to their flux density at rest-frame 5500\AA. The right panel zooms on the rest-frame 0.5--5\um\ range (grey box in the left panel) and shows only QSO1, the LRD and xLRD stacks of \citet{Perez-Gonzalez+26}, {the Cliff}, the stacks of \citet{Delvecchio+25} and \citet{Akins+25b}, JADES-1180-12402 and {\it the Egg}. The QSO1 spectrum and photometry are from this work. The spectra of the \textit{Rosetta Stone} and JADES-1180-12402 are smoothed for readability.}
    \label{fig:sed_comparison}
\end{figure*}
 
Figure~\ref{fig:sed_comparison} compares QSO1 with empirical templates of type-1 quasars and LRDs, all normalised at rest-frame 5500\AA. Relative to the quasar templates \citep{VandenBerk+01,Glikman+06,Kim+26,Bosman+25}, QSO1 is 10--20 times fainter in the UV, but its steep optical rise brings it close to the quasar composite at $\approx1$\um. Only at $1.24$\um\ does the composite become brighter, as the hot-dust emission of type-1 quasars sets in.
Among LRDs, the stacks of \citet{Delvecchio+25} and \citet{Perez-Gonzalez+26} and the \textit{Rosetta Stone} \citep{Juodbalis+24a} closely reproduce the optical rise and the MIRI photometry of QSO1, although they are brighter in the UV. The closest match over the whole range is JADES-1180-12402 \citep[$z=3.19$;][]{Williams+24,Setton+25a}. With $L_{\nu}(5100~\textrm{\AA})/L_{\nu}(2500~\textrm{\AA})\approx9$--11, QSO1 falls among the extreme LRDs (xLRDs, with a ratio above 6.3) of \citet{Perez-Gonzalez+26}, whose stack agrees well with it. Up to $1.24$\um, QSO1 is therefore a typical LRD with the UV continuum of the most extreme ones.
The local LRD \textit{the Egg} \citep[$z=0.10$;][]{Lin+25,Ji+26a} and \textit{the Cliff} \citep[$z=3.55$;][]{deGraaff+25B}, whose extreme Balmer break is due to absorption in dense gas, have a similar optical rise and bracket QSO1 at 0.95--1.24\um, with \textit{the Egg} slightly above and \textit{the Cliff} slightly below.
Beyond $1.24$\um, where our data stop, LRDs diverge. The LRD stacks keep rising, whereas the \textit{the Egg} and \textit{the Cliff} stay nearly flat in $f_\nu$ up to rest-frame $\approx3$--4\um. \textit{The Egg} rises again only at longer wavelengths, where its WISE photometry traces warm dust, and \textit{the Cliff} shows no evidence of hot dust \citep{deGraaff+25B}.

\subsection{What produces the red continuum}
\label{sec:ed}
We compared the SED with three sets of models of increasing physical detail: the continuum models of \citet{Ma+25} and the accretion-flow envelopes of \citet{Liu+25} (Section~\ref{sec:ma25}), and our own models with the nebular emission of dense gas, fitted to the photometry and the spectrum (Section~\ref{sec:sedfit}).
 
\subsubsection{Continuum models}
\label{sec:ma25}
\begin{figure*}
    \centering
    \includegraphics[width=0.49\linewidth]{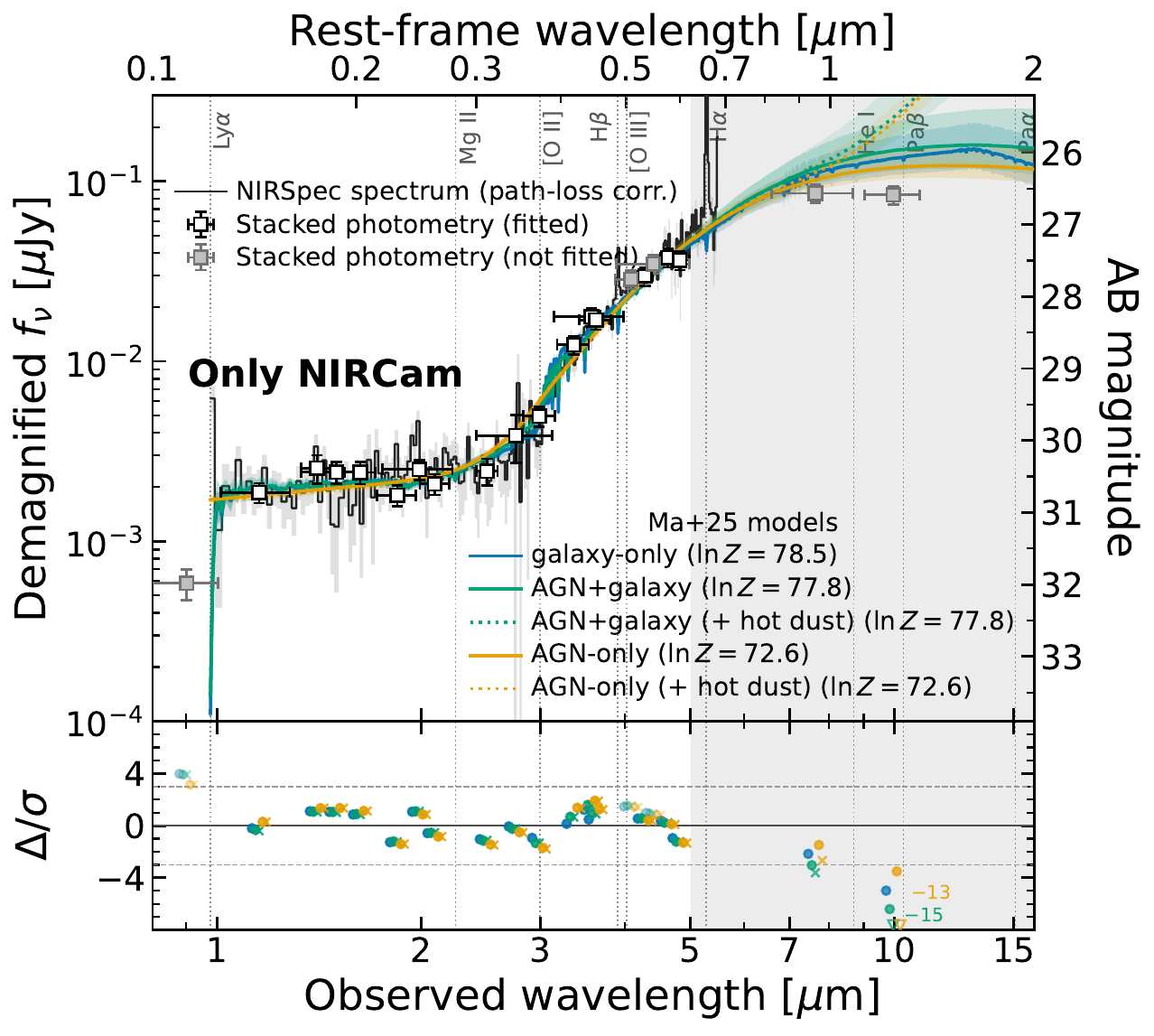}
    \includegraphics[width=0.49\linewidth]{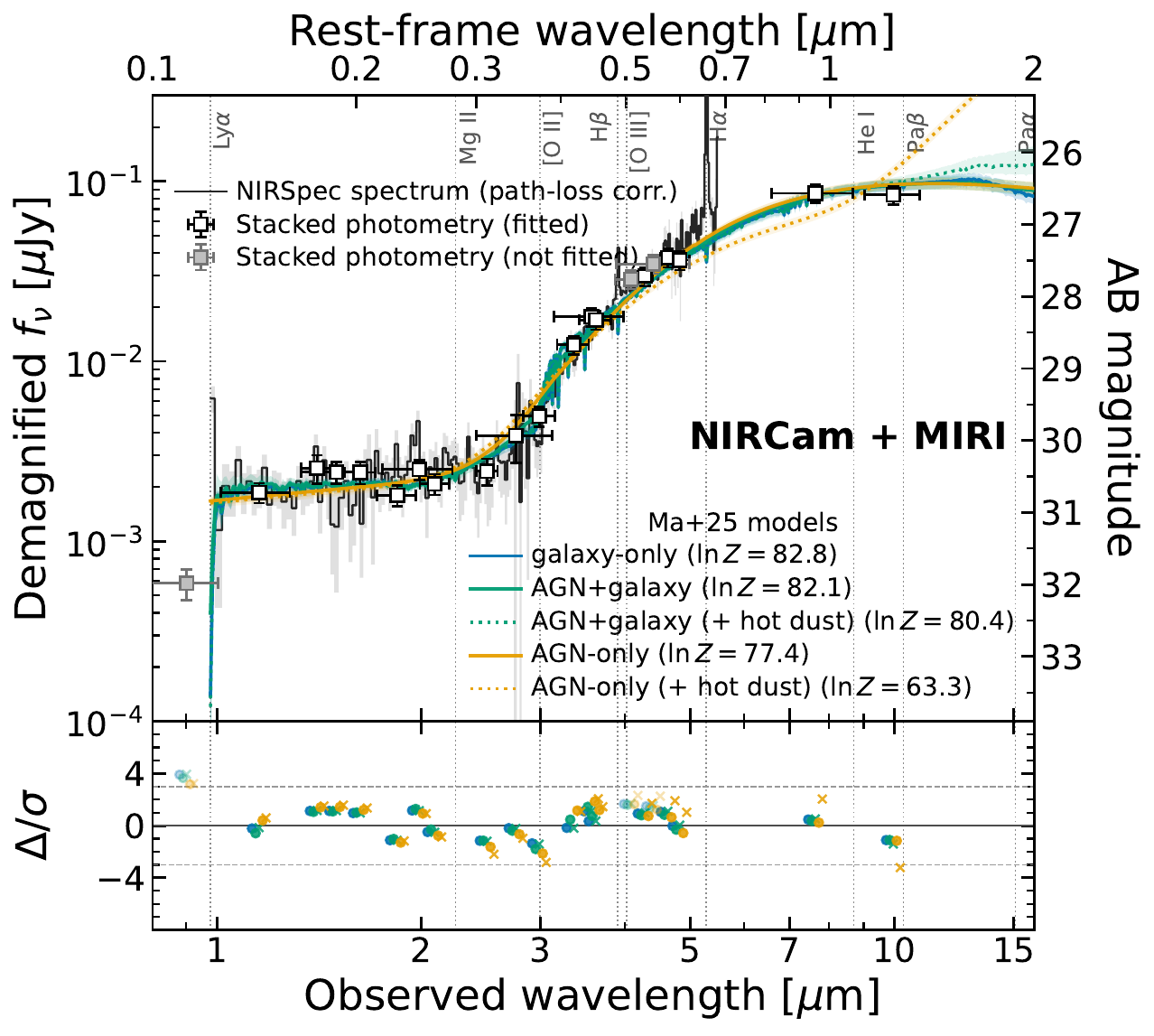}
    \caption{{\it MIRI reveals that the continuum models of \citet{Ma+25}, constrained by NIRCam alone, over-predict the near-infrared, and that quasar-like hot dust is disfavoured.}
    Posterior fits of the model families of \citet{Ma+25} to the stacked photometry of QSO1, fitted to the NIRCam bands only (left) and to NIRCam and MIRI (right). White squares are the fitted bands. Grey squares are bands not used in the fit (F090W, affected by the Ly$\alpha$ break, and F410M and F444W, dominated by H$\beta$). Horizontal bars show the filter coverage. In the left panel, the grey shaded region marks the MIRI range ($\lambda_{\rm obs}>5$\um), which is not fitted.
    The NIRSpec spectrum (path-loss corrected, black) is shown for reference only and is not fitted.
    Coloured lines and shaded regions show the posterior median and 16th--84th percentile range of each model family. Dotted lines include the hot-dust component of the \citet{Temple+21} quasar template.
    The log-evidence of each fit is given in the legend.
    Bottom panel: residuals, $(f_{\rm obs}-f_{\rm model})/\sigma$, of the synthetic photometry of each model.}
    \label{fig:ma25_posteriors}
\end{figure*}

\citet{Ma+25} modelled the rest-frame UV--optical continuum of QSO1 (up to $0.66$\um) with an attenuated, $\sim100$~Myr-old stellar population (galaxy-only), a reddened quasar continuum with a small scattered fraction (AGN-only) and a combination of the two (AGN+galaxy). All three described the data comparably well, but the solutions with a stellar component require a very steep attenuation curve \citep{Ji+25}. We first rebuilt their published best fits and extended them to the MIRI bands without refitting (Appendix~\ref{app:ma25}, Figure~\ref{fig:ma25_published}, Table~\ref{tab:ma25_miri}). We found that those with a stellar component over-predict F1000W by 48--60\% (79\% with the hot dust of the quasar template). We then refitted the same families to our photometry, with and without MIRI (Table~\ref{tab:ma25_posteriors}, Figure~\ref{fig:ma25_posteriors}).
 
Constrained by NIRCam alone, the models over-predict the MIRI flux densities by 27--76\% when a stellar component is present and by 18--43\% for the AGN-only model, over the whole posterior volume, because the attenuated continuum keeps rising beyond $1$\um. With MIRI, the galaxy-only model needs less attenuation than when fitted to NIRCam alone, and a surviving stellar mass of $\log_{10}(M_\star~[\rm M_\odot])=9.33^{+0.10}_{-0.09}$, still in tension with the upper limit on the host stellar mass (Section~\ref{sec:target}). The AGN+galaxy posterior splits into two modes, one close to this galaxy-only solution and one dominated by the reddened AGN with a minor stellar component ($\log_{10}(M_\star~[\rm M_\odot])\approx7.5$--8). The galaxy-only and AGN+galaxy models fit the data equally well ($\chi^2_{\rm red}=0.94$ and 1.20) and have indistinguishable Bayesian evidence\footnote{The Bayesian evidence, $\mathcal{Z}=\int\mathcal{L}(\theta)~\pi(\theta)~{\rm d}\theta$, is the likelihood $\mathcal{L}$ averaged over the prior $\pi$, and therefore penalises models with more parameters or wider priors. Differences $|\Delta\ln(\mathcal{Z})|>5$ are commonly regarded as strong evidence \citep{Trotta08}.} ($\Delta\ln(\mathcal{Z})=-0.7$ for AGN+galaxy relative to galaxy-only). The AGN-only model is disfavoured ($\Delta\ln(\mathcal{Z})=-5.4$, $\chi^2_{\rm red}=1.70$), and adding the hot-dust component of the \citet{Temple+21} quasar template disfavours it further ($\Delta\ln(\mathcal{Z})=-19.5$). In the AGN+galaxy model the hot dust instead suppresses the AGN, and the fit reverts to the galaxy-only solution (Appendix~\ref{app:ma25}).
 
\begin{figure}
    \centering
    \includegraphics[width=\columnwidth]{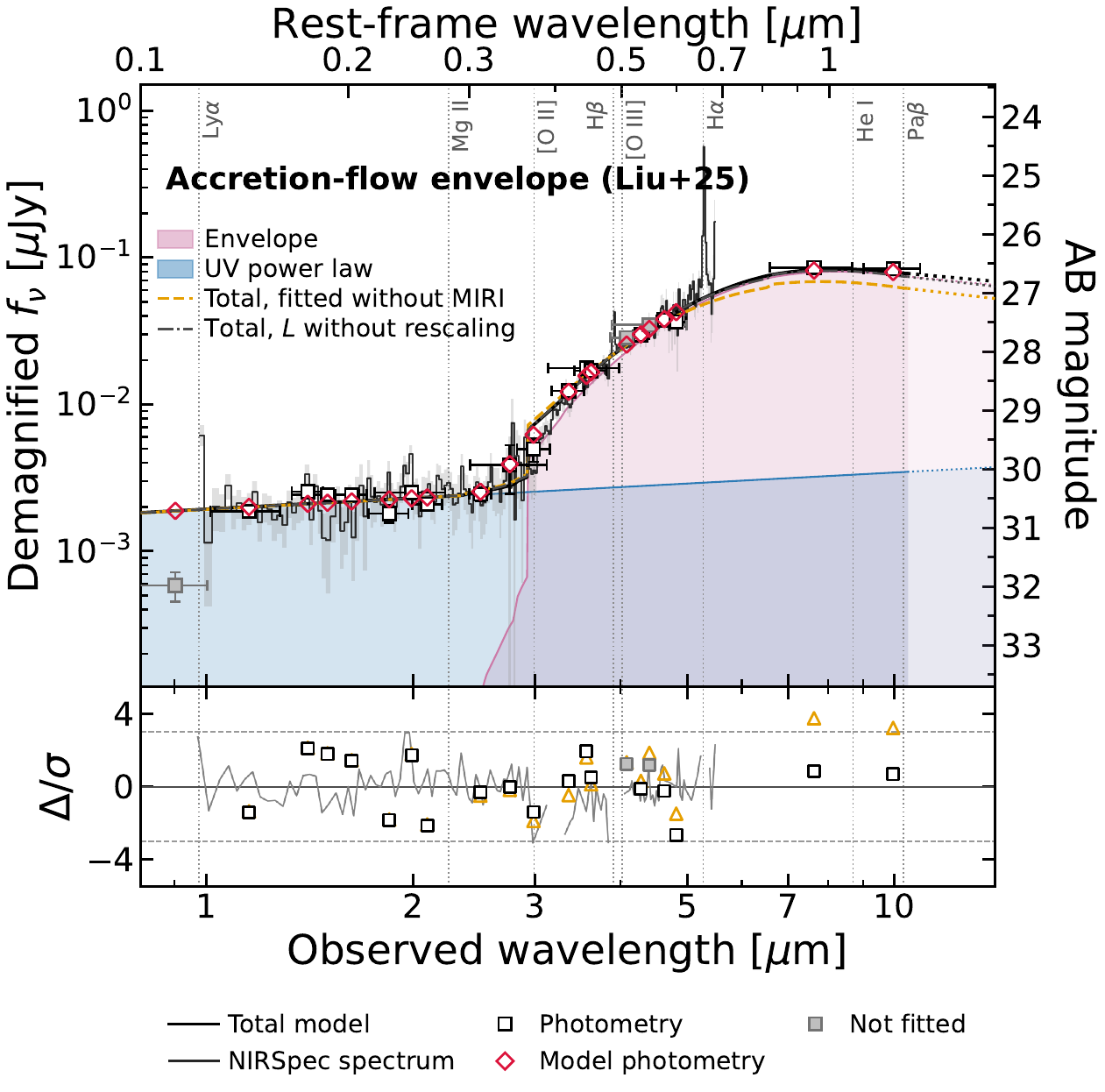}
    \caption{{\it A single accretion-flow envelope reproduces the continuum of QSO1 from the Balmer break to F1000W, but its luminosity matches the data only for a black-hole mass $\approx2.5$~dex below the dynamical estimate.} QSO1 compared with the accretion-flow envelope models of \citet{Liu+25} (Appendix~\ref{app:liu}). Top: photometry (white squares; grey: bands not fitted) and NIRSpec spectrum, with the fit to all bands (black), its envelope (pink) and UV power-law (blue) components and its synthetic photometry (diamonds), the fit without MIRI (dashed) and the solution whose luminosity matches the data without rescaling (dash-dotted). The model curves are dotted beyond rest-frame $1.3$\um, where the envelope spectra are extrapolated. Bottom: residuals, $(f_{\rm obs}-f_{\rm model})/\sigma$, of the photometry for the fits with (squares) and without MIRI (triangles), and of the spectrum for the fit to all bands (grey, averaged over six pixels, with emission lines masked since the models do not include them).}
    \label{fig:liu}
\end{figure}

The spherical accretion-flow models of \citet{Liu+25}, in which the emission of an accreting black hole thermalises in a dense envelope, behave differently (Appendix~\ref{app:liu}). With a power law added for the UV continuum, a single envelope spectrum reproduces the photometry from the Balmer break to F1000W ($\chi^2=14.3$ in the nine bands of Section~\ref{sec:nir_continuum}, against 15.4 for the 4450~K blackbody), with both MIRI bands within $1\sigma$ (Figure~\ref{fig:liu}), whereas the fit without MIRI under-predicts them by 3--4$\sigma$. In this picture the Balmer break, the turnover and the flat MIRI bands arise in the same thermalised gas, without hot dust or an evolved population. Two discrepancies remain. The model break is a sharp step at 3646~\AA, whereas the observed one is smoother, as expected if bulk or turbulent motions broaden the Balmer edge \citep{Ji+25,Naidu+25} or if the photosphere spans a range of densities and temperatures. The solution that matches the observed luminosity without rescaling has $L\approx2\times10^{43}$~erg~s$^{-1}$, i.e. $\log(M_{\rm BH}/{\rm M_\odot})\approx5.2$, and at the dynamical black-hole mass \citep{Juodzbalis+26b} these near-Eddington models would over-predict the observed luminosity by $\approx2.5$~dex (Section~\ref{sec:disc_photosphere}).
 
\subsubsection{SED modelling with dense gas}
\label{sec:sedfit}
Except for the envelopes, the models of Section~\ref{sec:ma25} include neither emission lines nor the dense gas that may shape the Balmer break and the broad lines of QSO1 \citep{Ji+25} and of LRDs in general \citep{Sun+26}. We therefore fitted the NIRSpec spectrum, emission lines included, and the photometry jointly with a Bayesian forward-modelling code based on the nested sampler \textsc{Nautilus} \citep{Lange+23}, using models with the nebular emission of stellar- and AGN-ionised gas over a wide range of densities (Appendix~\ref{app:sedfit}). The spectrum targets Ima-B and is flux-calibrated to the stacked photometry (Section~\ref{sec:nirspec_spectrum}), so the two data sets are not fully independent and its line equivalent widths may differ from those of the stack \citep{Furtak+25}.
 
All families contain {\sc bpass} binary stellar populations \citep{Eldridge+17,Stanway+18}, with a \citet{Chabrier+03}-like IMF extending to $300~\rm M_\odot$, metallicities of 0.05\% ({\tt zem5}), 0.5\% ({\tt zem4}), 5\% ({\tt z001}) and 10\% $Z_\odot$ ({\tt z002}) and nebular emission computed with \textsc{Cloudy} \citep[v25;][]{Gunasekera+25}. We considered three classes of models for the UV and red optical continuum:
(i) two stellar populations in gas of ordinary density ($n_{\rm H}<10^{6.0}~\rm cm^{-3}$);
(ii) an ordinary-density population plus a population coupled to dense photoionised gas ($n_{\rm H}=10^{6.5-12.0}~\rm cm^{-3}$), with a free covering fraction and lines with their own velocity dispersion;
(iii) an ordinary-density population plus an accretion disc embedded in a dense ($n_{\rm H}=10^{8.0-12.0}~\rm cm^{-3}$), dust-free cocoon\footnote{The \textsc{Cloudy} cocoon is an approximation in the regime favoured by the fits, because \textsc{Cloudy} solves the gas structure only along the radial direction of a spherically symmetric shell and does not follow scattering as a transfer process (Appendix~\ref{app:cocoon}).}, with a free covering fraction $C$ (Appendix~\ref{app:cocoon}).
The UV and optical components were attenuated either independently or by a single shared foreground screen with a \citet{Noll+09} law, and all families include dust emission tied to the absorbed luminosity (Section~\ref{sec:disc_dust}). Each family was fitted separately at each metallicity and compared through its Bayesian evidence (Tables~\ref{tab:sedfit_families} and \ref{tab:sedfit_evidence}). We also repeated class (iii) with the formed stellar mass limited to $\log_{10}(M_\star~[\mathrm{M}_\odot])\leq7.5$, i.e. a surviving mass of $\approx2\times10^7~\mathrm{M}_\odot$, the upper limit on the host stellar mass derived from the dynamics by \citet{Juodzbalis+26b} (``stellar-mass limit'' in the following).
 
\begin{figure*}
\centering
\includegraphics[width=\textwidth]{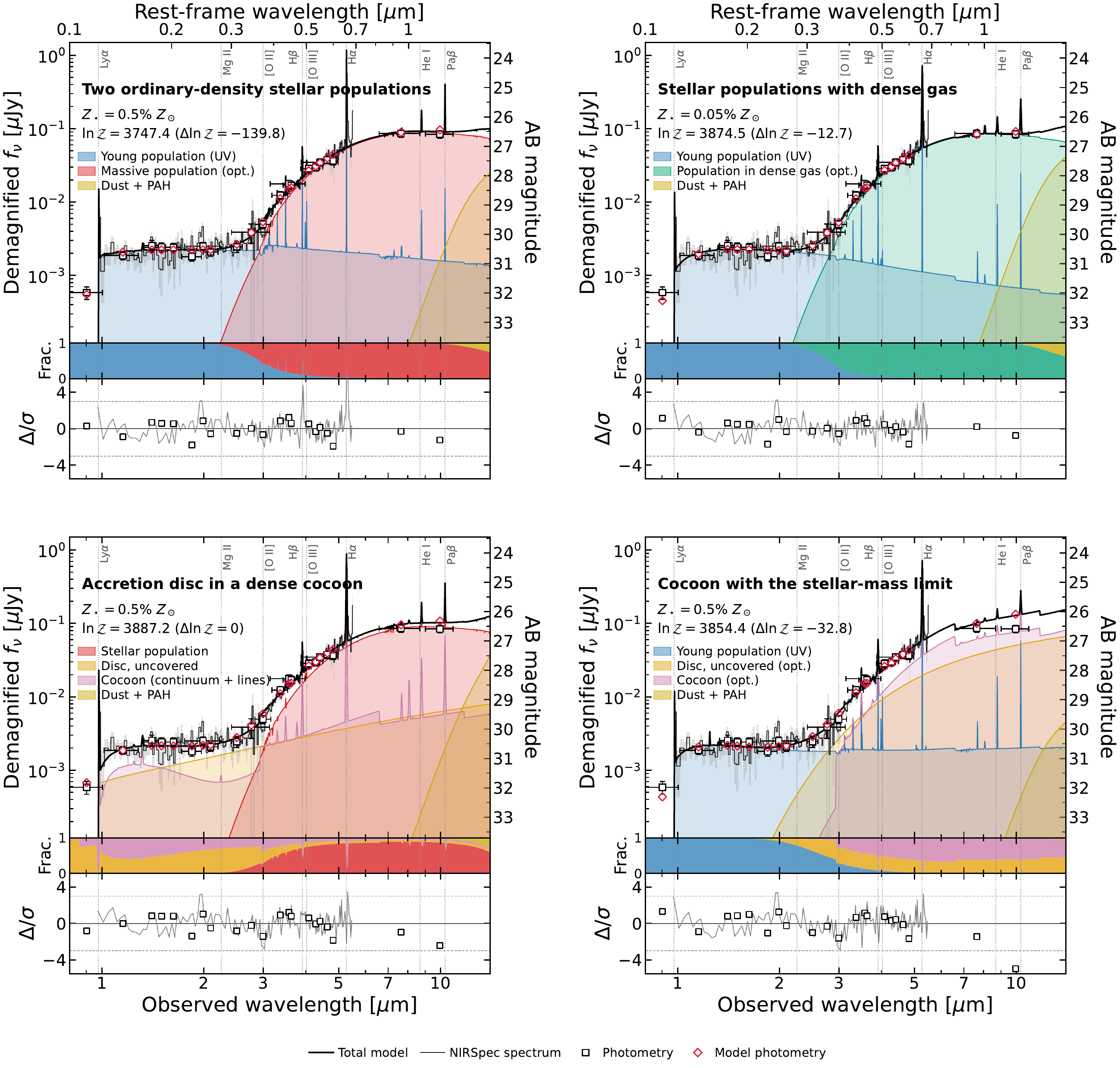}
\caption{{\it In our models, the highest-evidence solutions produce the red continuum with $\approx4\times10^9~\rm M_\odot$ of stars. With the stellar mass limited by the dynamics, the disc and dense cocoon produce it instead, but over-predict F1000W.} Maximum-likelihood solutions of the representative model families, each at the metallicity of its highest evidence, compared with the NIRSpec spectrum and the photometry: two ordinary-density stellar populations (\texttt{zem4}, top left), the highest-evidence stellar model including a dense photoionised component (\texttt{zem5}, top right), the highest-evidence accretion-disc cocoon model (\texttt{zem4}, bottom left) and the same model with the stellar mass limited to the upper limit on the host stellar mass, $M_\star\leq10^{7.5}~\mathrm{M}_\odot$ (\texttt{zem4}, bottom right). For each model, the upper panel shows the individual components (coloured), the total model (black), the photometry (squares, with horizontal bars showing the filter widths, and the model photometry as diamonds) and the NIRSpec spectrum (dark grey, with its uncertainty shaded). Dotted vertical lines mark the main emission lines. The middle panel gives the fractional contribution of each component, and the lower panel the residuals, $(f_{\rm obs}-f_{\rm model})/\sigma$, with the spectrum averaged over six pixels and its uncertainty propagated accordingly. Dashed lines mark $\pm3\sigma$.}
\label{fig:sed_models}
\end{figure*}
 
Figure~\ref{fig:sed_models} shows the highest-evidence maximum-likelihood solution of each class, Table~\ref{tab:sedfit_bestfit} their properties and Figure~\ref{fig:sedfit_corners} their posteriors. A single foreground screen cannot reproduce the V-shaped SED: with a shared screen, the dense-gas stellar solutions lose $\Delta\ln(\mathcal{Z})\simeq81$--274 and the cocoon solutions $\simeq94$--125, so the UV and the red continuum must come from regions with different effective attenuation.
 
\textit{Two ordinary-density stellar populations.} A young, almost unobscured population produces the UV and a massive, attenuated $\sim20$~Myr population the red continuum (formed $\log_{10}(M_\star~[\mathrm{M}_\odot])\simeq9.61$). This is the analogue of the galaxy-only model of \citet{Ma+25}, with a similar mass and the same steepest attenuation curve allowed ($\delta\simeq-0.95$), but a younger age ($\sim20$ against $\sim100$~Myr), favoured by the nebular emission and the Balmer region of the spectrum. It has by far the lowest evidence of the four.
 
\textit{Stellar populations with dense gas.} Coupling the red-continuum population to dense photoionised gas significantly improves the Bayesian evidence ($\Delta\ln(\mathcal{Z})\simeq127$). A $\sim16$~Myr population with $\log_{10}(M_\star~[\mathrm{M}_\odot])\simeq9.62$ produces the red continuum, $\approx140$--200 times the upper limit on the host stellar mass, and its lines need $\sigma_v\simeq800$~km~s$^{-1}$, difficult to reconcile with a stellar H~{\sc ii} region.
 
\textit{Accretion disc in a dense cocoon.} The highest-evidence run of all does not use the disc for the red continuum. The disc, seen directly and through a cocoon with $C\simeq0.26$, produces the UV, while a $\sim20$~Myr stellar population with $\log_{10}(M_\star~[\mathrm{M}_\odot])\simeq 9.57$ produces 75--85\% of the light at rest-frame 0.5--1.2\um, with the same tension with the dynamical mass.
 
\textit{Cocoon with the stellar-mass limit.} With the stellar-mass limit the roles are reversed. A young population ($\log_{10}(M_\star~[\mathrm{M}_\odot])\simeq6.96$) produces the UV, and the disc ($L_{\rm AGN}\approx2.4\times10^{44}$~erg~s$^{-1}$) and the cocoon ($C\simeq0.22$, $n_{\rm H}=10^{11}~\mathrm{cm^{-3}}$, $U=10^{-3}$ and $N_{\rm H}=10^{25.5}~\mathrm{cm^{-2}}$, the largest column of the grid) more than 90\% of the light beyond rest-frame 0.5\um, behind a screen with $A_V\simeq1.3$~mag and the steepest attenuation slope allowed. The same solution is found at all four metallicities. Its continuum keeps rising beyond rest-frame 1\um\ and over-predicts F1000W by $5\sigma$, whereas the solutions with a massive stellar population are within $1\sigma$ (stellar models) and $2.3\sigma$ (unconstrained cocoon).
 
Within the model families explored, the spectrum requires a dense, strongly reprocessing component, in line with the broad Balmer lines with a large decrement \citep{Deugenio+26b} and the broad Ly$\alpha$ and far-UV Fe~{\sc ii} emission of QSO1 \citep{Tang+26a}. The highest-evidence solutions produce the red continuum with $\approx4\times10^9~\rm M_\odot$ of stars, two orders of magnitude above the stellar-mass limit, and the solution within this limit has an evidence lower by $\Delta\ln(\mathcal{Z})\simeq20$--33. We regard this difference as indicative given the systematics of our modelling (Appendix~\ref{app:sedfit}), but it does not depend on how the images are combined (it holds for the photometry of each lensed image, with $\Delta\chi^2=31$--107; Appendix~\ref{app:images}) or on metallicity (Table~\ref{tab:sedfit_evidence}).

Two test models clarify the role of the stars (Appendix~\ref{app:extra_families}). In the first, there is no stellar population, and the disc and cocoon must also produce the UV. This model misses the Balmer break by $\approx3\sigma$ and over-predicts F1000W by $\approx5\sigma$ (Figure~\ref{fig:sedfit_nostars}). In the second, the stars are unattenuated and their mass is not limited. They then produce only the UV, and the fit converges to the solution with the stellar-mass limit. Within our grid, a separate UV component seems to be required, and the massive solutions rely on heavily attenuated stars. The UV may nevertheless be partly AGN light, as suggested by its far-UV Fe~{\sc ii} emission \citep{Tang+26a}, and radiative-transfer effects not included in our grid could produce the blue UV and the red continuum in the same gas \citep[e.g.][]{Mascia+26}.

The SED alone thus prefers a massive stellar population, whereas the dynamics, the broad lines and the UV features point to an accretion-powered source in dense gas.
 
\begin{table*}
    \centering
    \caption{Physical properties of the four representative solutions, each at the metallicity of its highest evidence (Figure~\ref{fig:sed_models}): posterior medians with 16th--84th percentile ranges. Ages are $\log_{10}(t~[\mathrm{yr}])$, stellar masses are formed masses $\log_{10}(M_\star~[\mathrm{M}_\odot])$, de-magnified. The surviving masses are at most $0.15$~dex lower at these ages. ``UV'' and ``opt.'' denote the stellar populations dominating the rest-frame UV and optical continua. $\delta$ is the slope offset of the attenuation curve (prior $[-1,0.4]$). $T_{\rm dust,0}$ is the intrinsic dust temperature, before CMB heating. $C$ is the covering fraction of the photoionised gas of each stellar population, or of the cocoon.}
    \label{tab:sedfit_bestfit}
    \small
    \begin{tabular}{lcccc}
    \hline\hline
     & Two ordinary-density & Stellar populations & Accretion disc in & Cocoon with the \\
 & stellar populations & with dense gas & a dense cocoon & stellar-mass limit \\
     & \texttt{zem4} & \texttt{zem5} & \texttt{zem4} & \texttt{zem4} \\
    \hline
    $\ln(\mathcal{Z})$ ($\Delta\ln(\mathcal{Z})$) & 3747.4 ($-139.8$) & 3874.5 ($-12.7$) & 3887.2 (0) & 3854.4 ($-32.8$) \\
    Optical continuum from & stars & stars (dense gas) & stars & disc + cocoon \\
    \hline
    \multicolumn{5}{l}{\textit{UV-emitting stellar population}}\\
    Age & $6.8^{+0.2}_{-0.1}$ & $6.8^{+0.1}_{-0.2}$ & -- & $6.9^{+0.1}_{-0.2}$ \\
    $\log_{10} M_\star$ & $6.69^{+0.19}_{-0.22}$ & $6.48^{+0.19}_{-0.18}$ & -- & $6.96^{+0.32}_{-0.35}$ \\
    $E(B-V)$ & $0.022^{+0.003}_{-0.004}$ & $0.019^{+0.004}_{-0.003}$ & -- & $0.25^{+0.12}_{-0.13}$ \\
    $C$ & $0.54^{+0.27}_{-0.16}$ & $0.44^{+0.23}_{-0.15}$ & -- & $0.63^{+0.26}_{-0.25}$ \\
    $\log_{10}(n_{\rm H}~[\mathrm{cm^{-3}}])$, $\log_{10}U$ & 5.5, $-2.5$ & 6.0, $-1.0$ & -- & 6.0, $-1.0$ \\
    \hline
    \multicolumn{5}{l}{\textit{Optical-continuum stellar population}}\\
    Age & $7.3^{+0.1}_{-0.1}$ & $7.2$ & $7.3^{+0.1}_{-0.1}$ & -- \\
    $\log_{10} M_\star$ & $9.61^{+0.08}_{-0.09}$ & $9.62^{+0.03}_{-0.03}$ & $9.57^{+0.10}_{-0.07}$ & -- \\
    $E(B-V)$ & $0.663^{+0.025}_{-0.019}$ & $0.650^{+0.018}_{-0.016}$ & $0.694^{+0.017}_{-0.016}$ & -- \\
    $C$ & $0.18^{+0.12}_{-0.06}$ & $0.38^{+0.03}_{-0.02}$ & $0.49^{+0.29}_{-0.23}$ & -- \\
    $\log_{10}(n_{\rm H}~[\mathrm{cm^{-3}}])$, $\log_{10}U$ & 6.0, $-1.5$ & 7.5, $-3.0$ & 3.0, $-1.5$ & -- \\
    $\sigma_{v}$ of dense-gas lines [km~s$^{-1}$] & -- & $798^{+53}_{-51}$ & -- & -- \\
    \hline
    \multicolumn{5}{l}{\textit{Accretion disc and cocoon}}\\
    $\log_{10}(L_{\rm AGN}~[L_\odot])$ & -- & -- & $9.97^{+0.21}_{-0.16}$ & $10.79^{+0.04}_{-0.04}$ \\
    $C$ & -- & -- & $0.26^{+0.04}_{-0.03}$ & $0.22^{+0.03}_{-0.02}$ \\
    $\log_{10}(n_{\rm H}~[\mathrm{cm^{-3}}])$, $\log_{10}U$ & -- & -- & 9.0, $-2.0$ & 11.0, $-3.0$ \\
    $\log_{10}(N_{\rm H}~[\mathrm{cm^{-2}}])$ & -- & -- & 25.0 & 25.5 \\
    $\sigma_v$ of cocoon lines [km~s$^{-1}$] & -- & -- & $855^{+74}_{-67}$ & $906^{+64}_{-64}$ \\
    $E(B-V)$ & -- & -- & $0.13^{+0.11}_{-0.07}$ & $0.315^{+0.012}_{-0.012}$ \\
    $\delta$ & -- & -- & $0.07^{+0.21}_{-0.38}$ & $-0.985^{+0.023}_{-0.011}$ \\
    \hline
    $\log_{10}(M_{\star,\rm tot}\rm ~[M_\odot])$ & 9.61 & 9.62 & 9.57 & 6.96 \\
    $\delta$ (stellar) & $-0.95^{+0.06}_{-0.04}$ & $-0.95^{+0.04}_{-0.03}$ & $-0.98^{+0.03}_{-0.02}$ & $0.30^{+0.07}_{-0.21}$ \\
    $T_{\rm dust,0}$ [K] & $228^{+168}_{-104}$ & $234^{+162}_{-95}$ & $226^{+175}_{-103}$ & $171^{+193}_{-101}$ \\
    $\log_{10}(L_{\rm dust}~[L_\odot])$ & $11.50^{+0.26}_{-0.29}$ & $11.62^{+0.23}_{-0.24}$ & $11.44^{+0.29}_{-0.28}$ & $10.29^{+0.29}_{-0.26}$ \\
    \hline\hline
    \end{tabular}
\end{table*}

%%%%%%%%%%%%%%%%%%%%%%%%%%%%%%%%%%%%%%%%%%%%%%%%%%%%%%%%%%%%%%
\section{Discussion}
\label{sec:discussion}
 
\subsection{A pseudo-photosphere?}
\label{sec:disc_photosphere}
The steep optical rise of QSO1 does not continue into the near-infrared, and its SED up to $1.24$\um\ is typical of LRDs (Section~\ref{sec:nir_continuum}). With $\nu L_\nu(1~\mu{\rm m})\approx1.8\times10^{43}$~erg~s$^{-1}$ and $M_{\rm UV}\simeq-16.3$, QSO1 extends this shape to one of the faintest and most distant LRDs observed at these wavelengths.
 
Two kinds of source can produce this shape. In the first, the engine (e.g. an accreting black hole) is enshrouded in dense gas that is optically thick enough to thermalise its emission, so that it radiates from a pseudo-photosphere with $T_{\rm eff}\approx4000$--6000~K, close to a blackbody \citep{Liu+26}. A single component then produces both the optical rise and the turnover, without the need of dust. Thermalisation is what makes this model work. In fact, gas that only absorbs the disc light can produce a Balmer break and can partially redden the optical continuum \citep{Inayoshi+25b}, but it cannot produce the steep optical rise and the turnover observed. In a pseudo-photosphere, these come from the thermal emission of the gas itself. Among our models, only the accretion-flow envelopes of \citet{Liu+25} are of this kind (Section~\ref{sec:ma25}).
However, while these envelope models reproduce the shape of the turnover, their luminosity is in tension with the dynamical black-hole mass (Section~\ref{sec:ma25}). These models, in fact, are powered by supercritical accretion, and photon trapping and outflows regulate their emergent luminosity to $\sim1$--10~$L_{\rm Edd}$ \citep{Kido+25,Liu+25,Jiang+14}. For the dynamical mass, $\log(M_{\rm BH}/{\rm M_\odot})=7.7\pm0.3$ \citep{Juodzbalis+26b}, $L_{\rm Edd}\approx6\times10^{45}$~erg~s$^{-1}$ (3--$12\times10^{45}$~erg~s$^{-1}$). QSO1 is far below this value, whichever bolometric luminosity is adopted. Integrating its observed SED gives $L_{\rm bol}\approx3\times10^{43}$~erg~s$^{-1}$ (Section~\ref{sec:disc_lbol}), i.e. $\lambda_{\rm Edd}\approx0.005$. The type-1 bolometric correction applied to H$\alpha$ gives $\approx2\times10^{44}$~erg~s$^{-1}$, i.e. $\lambda_{\rm Edd}\approx0.03$. Even the intrinsic luminosity of a reddened accretion disc, $(0.4$--$1.2)\times10^{45}$~erg~s$^{-1}$, would give only $\lambda_{\rm Edd}\approx0.06$--0.2. This tension would only ease if the black-hole mass were overestimated, the bolometric luminosity underestimated, or the emission strongly anisotropic, as expected for super-Eddington flows \citep{Jiang+14,Pacucci+24}.

In the second kind, a hotter source is reddened by dust, and the continuum turns over where the attenuation becomes small. The hotter the source, the more luminosity the dust must absorb. A 6900~K blackbody behind $A_V\approx1.5$~mag has $L_{\rm bol}\approx8\times10^{43}$~erg~s$^{-1}$, of which the dust absorbs $L_{\rm abs}\approx5\times10^{43}$~erg~s$^{-1}$. A $3\times10^4$~K blackbody behind $A_V\approx3.3$~mag would instead have $L_{\rm bol}\approx5\times10^{45}$~erg~s$^{-1}$, nearly all of it absorbed. This is close to the Eddington luminosity for the dynamical black-hole mass. All our other models are of this kind: the reddened accretion disc ($A_V\approx2.2$~mag), the reddened blackbody ($A_V\approx1.5$~mag), the massive stellar populations ($A_V\approx2.6$--2.8~mag) and, despite its dense gas envelope, the cocoon with the stellar-mass limit ($A_V\approx1.3$~mag, Section~\ref{sec:ed}).
In fact, while this cocoon reproduces the UV, the Balmer break and the broad lines with a black hole in dense gas, it does not reproduce the turnover, because its gas does not thermalise. The gas is optically thick in the Balmer continuum ($\tau\approx30$ below 3646~\AA), which produces the break, but only marginally in the Paschen continuum ($\tau\approx2$ below 8204~\AA) and nearly transparent beyond. Its near-infrared continuum is therefore the transmitted disc light and the nebular emission of the gas, roughly flat in $f_\nu$. The steep rise comes instead from its dust screen and the directly seen disc, which keep rising beyond 1\um\ and over-predict F1000W. The same mostly neutral gas also scatters little. In dense-gas models of LRDs, scattering by free electrons has been proposed to produce the broad wings of the Balmer lines \cite[e.g.][]{Chang+26}. In our cocoon, however, the fully ionised layer is thin and most free electrons come from the residual ionisation ($x_e\approx0.6$~\%) of the neutral gas. The electron-scattering optical depth is therefore only $\tau_{\rm es}\approx0.13$, $\approx4$--20 times below what is needed to produce the observed wings, and the line widths are set by the fitted velocity dispersion. So, for the cocoon model with stellar-mass limit, we attribute the MIRI mismatch mainly to the lack of thermalisation, although the single geometry, fixed illuminating spectrum and grid-edge column of our cocoon may also contribute (Appendix~\ref{app:cocoon}).

Although our cocoons do not reproduce the turnover, they still constrain how the dense gas is distributed around the black hole. Both cocoon solutions, with and without the stellar-mass limit, cover the disc only partially with dense gas ($C\simeq0.22$--0.26; Section~\ref{sec:sedfit}), which points to a clumpy or anisotropic medium. This is in line with the broad Ly$\alpha$ of QSO1, since high-column gas shapes the Balmer break while lower-opacity paths let Ly$\alpha$ and ionising photons escape \citep{Ji+26c,Tang+26a}. So does its UV Fe~{\sc ii} emission, which needs dense, partially ionised gas exposed to the central continuum \citep{Tang+26a}. The fitted covering fraction is therefore an effective quantity, which may trace openings in the cocoon, anisotropic escape or sightlines of different optical depth. We highlight that this covering fraction refers to the gas, and is distinct from the one of the dust discussed in Section~\ref{sec:disc_dust}.

Except for our cocoon, which misses the turnover, both kinds of source reproduce the observed SED shape, so the photometry alone cannot decide between them. They, however, rest on different physical requirements. The pseudo-photosphere models we tested need a luminosity close to Eddington, which is in tension with the dynamical black-hole mass. A reddened source instead requires dust that absorbs part of the intrinsic luminosity and re-emits it in the infrared. Among the reddened solutions, those dominated by stars must satisfy the dynamical limit on the host mass (Section~\ref{sec:disc_host}), and all must hide the re-emitted luminosity below our infrared limits (Section~\ref{sec:disc_dust}).

\subsection{A massive host?}
\label{sec:disc_host}
MIRI probes the rest-frame near-infrared, where the light of an evolved host best traces its mass \citep{BelldeJong01}, but in QSO1 this emission is unresolved ($r_{\rm e}<64$~pc) and follows the same thermal component as the optical. A surviving stellar mass of $\gtrsim3\times10^9~\rm M_\odot$ within $r_{\rm e}<30$--64~pc \citep{Furtak+24} would imply $\Sigma_\star=M_\star/(2\pi r_{\rm e}^2)\gtrsim1$--$5\times10^5~\rm M_\odot~pc^{-2}$, at or above the maximum observed in dense stellar systems \citep{Hopkins+10}. Such densities are extreme but not impossible, so they cannot exclude a stellar solution on their own. The dynamics can provide an additional test. Our inferred stellar mass is $\approx140$--200 times the dynamical upper limit on the host stellar mass \citep[see also \citealt{Ji+25}]{Juodzbalis+26b}, and even the galaxy-only model of \citet{Ma+25}, refitted with MIRI, still requires $\approx2\times10^9~\rm M_\odot$ (Section~\ref{sec:ma25}). One way to reconcile the inferred stellar mass and the dynamical upper limit would be if the dynamical mass were underestimated. In fact, since only the line-of-sight component of the rotation is measured, $v_{\rm obs}=v\sin i$, a nearly face-on disc would hide most of the true velocity thus biasing the estimated enclosed mass by a factor of $1/\sin^2 i$. However, \citet{Juodzbalis+26b} fit the inclination together with the mass and found $i=52^\circ\pm2^\circ$, and $\approx45^\circ\pm10^\circ$ with a non-parametric method. Hiding a factor of $\approx140$--200 would require the true inclination to be only $\approx3$--$4^\circ$, inconsistent with both estimates. Moreover, similar mass deficits have been found in many LRDs with measured line widths \citep{Ji+26a,Deugenio+26a,Wang+25a,Torralba+26b}, which are unlikely to all be seen face-on. 
If LRDs in general contained $\sim10^{9}$--$10^{10}~\rm M_\odot$ of stars within a few tens of parsecs, their high number densities \citep{Kokorev+24b} would reintroduce the baryon-budget tension raised by the first massive-galaxy interpretations of these sources \citep{Labbe+23,BoylanKolchin23}.
The stellar solution faces a further problem.
It needs dense gas moving at $\sim800$~km~s$^{-1}$, as expected near an accreting black hole rather than in H~{\sc ii} regions. 

A strong Balmer break does not necessarily require evolved stars but can arise from absorption in dense gas rather than from evolved stars \cite[e.g.][]{deGraaff+25B}.
In our modelling, however, even when a black hole in dense gas is available, the fit tends to prefer a solution with a strong contribution by stars. The highest-evidence cocoon produces the red continuum with $\approx4\times10^9~\rm M_\odot$ of stars and uses the disc only for the UV. Limiting the stellar mass to maximum values consistent with the dynamical mass disfavours the model ($\Delta\ln(\mathcal{Z})\simeq20$--33, see Section~\ref{sec:sedfit}). This preference does not come from the Balmer break or the optical continuum. It mainly comes from the MIRI bands, where the cocoon lacks a pseudo-photosphere (Section~\ref{sec:disc_photosphere}), and from the emission lines fit: the Balmer-line profiles and ratios, the high-order Balmer lines just redward of the break, and the UV lines. These are exactly the features our cocoon models describe least well, since they assume a single density, ionisation parameter and column, a single Gaussian line width, and escape-probability line transfer (Appendix~\ref{app:cocoon}). 
Models combining a range of densities and columns, or radiative-transfer models that follow line scattering and the thermalisation of the continuum \citep[e.g.][]{Matthews+25,Sneppen+26c,Liu+25}, may close the gap. 

We therefore regard the massive stellar solutions as unlikely, and the preference of the SED for them as a limitation of our dense-gas models rather than as evidence for a massive host.

\subsection{Where is the dust?}
\label{sec:disc_dust}
The dynamical mass mainly tests the models with stellar populations (Section~\ref{sec:disc_host}). The other reddened models, including the reddened accretion disc, can instead be tested through the amount of dust they require. In the dense-gas picture, the red optical continuum arises without any reddening (Section~\ref{sec:disc_photosphere}). In addition to the SED fits, two independent tracers can help establish whether QSO1 is reddened at all: the emission lines and the gas metallicity.
The Balmer decrement of the broad component is large and would imply $A_V\approx2$--3 mag if caused by dust (Section~\ref{sec:target}). However, dense gas near the black hole, inside the sublimation radius ($R_{\rm sub}\approx0.2$~pc for $L_{\rm bol}\approx2\times10^{44}$~erg~s$^{-1}$; \citealt{Barvainis87}) and therefore dust-free, can also produce large decrements without dust \citep{Ji+25,Chang+26,Torralba+26c,Matthee+26}. The Balmer decrement of the narrow component is instead consistent with Case~B \citep{Deugenio+26b}, suggesting little dust in the narrow-line gas. This gas, however, extends over a few hundred parsecs \citep{Maiolino+26b,Ji+26c} and does not necessarily probe dust in front of the broad-line region. The very low metallicity of QSO1 implies little dust per unit gas, but does not exclude $A_V\approx2$~mag. For the Galactic ratio, $N_{\rm H}/A_V=2.2\times10^{21}~{\rm cm^{-2}~mag^{-1}}$ \citep{GueverOezel09}, scaled linearly to the gas-phase metallicity of QSO1 ($\approx0.005~Z_\odot$; \citealt{Maiolino+26b}), $A_V\approx2$~mag requires $N_{\rm H}\approx10^{24}~{\rm cm^{-2}}$ outside the sublimation radius\footnote{We note that $N_{\rm H}\approx10^{24}~{\rm cm^{-2}}$ is a lower limit if the dust-to-gas ratio falls faster than linearly at low metallicity \citep{RemyRuyer+14}, and an upper limit if the nuclear gas is more enriched than the narrow-line gas.}. A column of this size is consistent with the X-ray faintness of QSO1 (Section~\ref{sec:target}). Neither tracer can therefore confirm or rule out dust reddening on its own. The energy budget offers a more direct test.

If QSO1 is reddened, the dust is substantial. Our best-fit reddened solutions have $A_V\approx2.6$--2.8~mag for the massive stellar populations, $\approx2.2$~mag for the reddened accretion disc and $\approx1.3$~mag for the screen of the cocoon with the stellar-mass limit (Section~\ref{sec:ed}). The energy absorbed by this dust must be re-emitted in the infrared, and our limits constrain how much of it can be hidden, and at which temperatures.
 
For the SED models of Section~\ref{sec:sedfit} the absorbed luminosity is a fitted quantity. Conversely, for the reddened accretion disc it has to be derived from the AGN-only fit to NIRCam and MIRI (Appendix~\ref{app:dust_budget}), whose intrinsic disc has $\log_{10}(\lambda L_{3000}~[{\rm erg~s^{-1}}])=44.22\pm0.04$. Between 912~\AA\ and $2.9$\um\ dust absorbs $L_{\rm abs}=(3.9\pm0.3)\times10^{44}$~erg~s$^{-1}$, 94\% of the intrinsic emission in this range. The template does not include the ionising continuum, and extending it below 912~\AA\ with the disc continuum of our cocoon models (Appendix~\ref{app:cocoon}) gives $L_{\rm AGN}\approx1.2\times10^{45}$~erg~s$^{-1}$, $\approx2$ times the value from the bolometric correction of \citet{Runnoe+12}. Below 912~\AA, the energy is absorbed first by the gas, but it is largely re-emitted as lines and recombination continua that the dust can then absorb. We therefore adopt $L_{\rm abs}\approx(0.4$--$1.2)\times10^{45}$~erg~s$^{-1}$ for the reddened accretion disc. The lower value counts only the non-ionising continuum, which the dust must absorb in any case, and the upper value assumes that all the ionising luminosity is eventually reprocessed by dust. The energy budget for the three model families spans a factor of $\approx30$: $(1.3$--$2.3)\times10^{45}$~erg~s$^{-1}$ for the solutions with a massive stellar population, $(0.4$--$1.2)\times10^{45}$~erg~s$^{-1}$ for the reddened accretion disc, and $(0.7$--$1.0)\times10^{44}$~erg~s$^{-1}$ for the cocoon with the stellar-mass limit.

\begin{figure}
    \centering
    \includegraphics[width=\columnwidth]{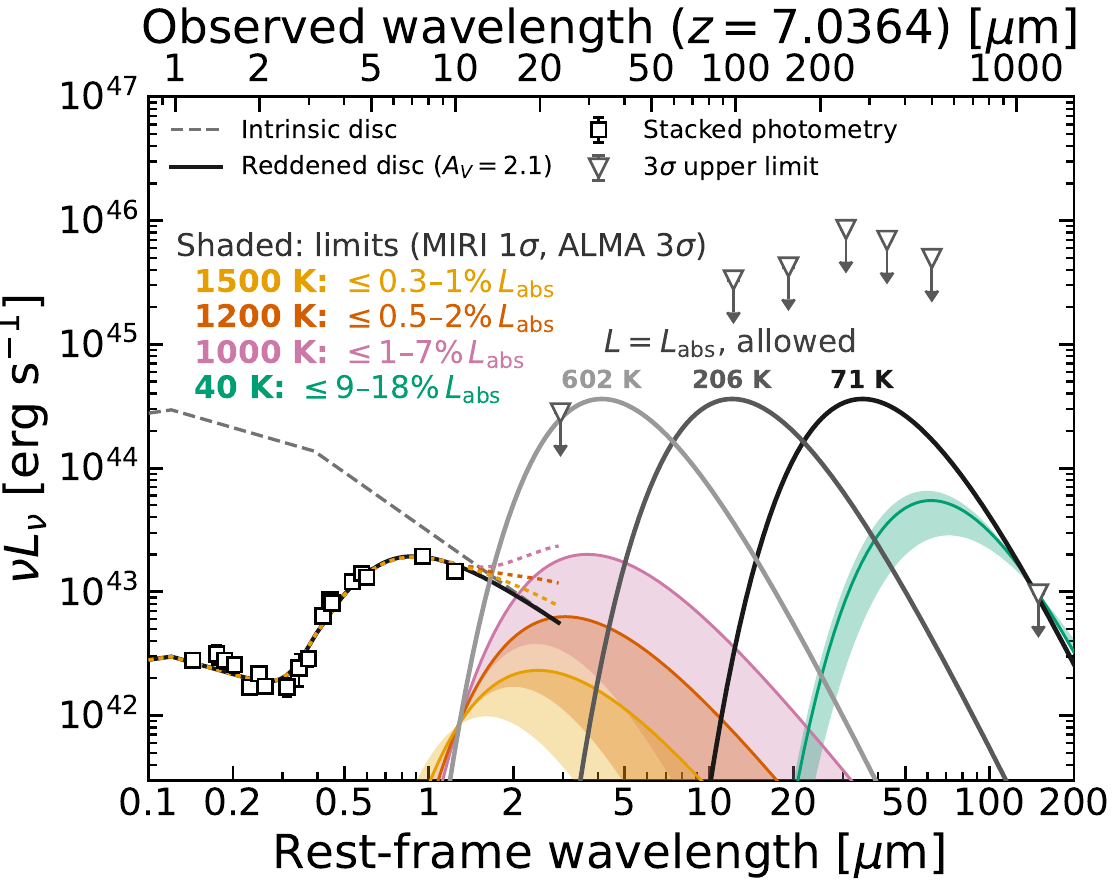}
    \caption{{\it For a reddened accretion disc, hot dust ($\gtrsim1000$~K) can re-emit at most a few per cent of the absorbed luminosity, and cold dust ($\approx40$~K) at most 15 per cent, whereas warm dust at $\approx70$--600~K could re-emit all of it without exceeding any limit.} Dust energy budget of QSO1 under the reddened-disc hypothesis (Appendix~\ref{app:dust_budget}): stacked NIRCam and MIRI photometry, $3\sigma$ upper limits from MIPS, \textit{Herschel} and ALMA, and the intrinsic and attenuated disc of the reddened-AGN model. Shaded bands show single-temperature dust components at their limits, labelled with the fraction of the absorbed luminosity $L_{\rm abs}$ they can re-emit. These limits are $1\sigma$ for the 1000, 1200 and 1500~K components, which are fitted jointly with the disc to the NIRCam and MIRI detections ($\beta=0$--2; line: blackbody), and $3\sigma$ for the 40~K component, which is set by ALMA ($\beta=1$--2; line: $\beta=1.8$). Dotted lines show the total of the re-fitted disc and each hot component. Solid grey curves show greybodies ($\beta=1.8$) that re-emit all of $L_{\rm abs}$ without violating any limit, at the lowest (71~K) and highest (602~K) such temperatures and at their logarithmic midpoint (206~K). All dust emission is shown as observed, including the CMB contrast.}
    \label{fig:dust_budget}
\end{figure}
 
Where this energy emerges depends on the dust temperature (Figure~\ref{fig:dust_budget}). Hot dust ($\gtrsim1000$~K) emits mostly near rest-frame 3\um, warm dust (100--600~K) at 6--40\um\ and cold dust ($\lesssim50$~K) near 100\um. Only hot dust is constrained directly by NIRCam and MIRI, through its Wien tail. Such temperatures are reached mainly when an AGN heats the dust, since dust heated by stars stays cooler. In the cocoon with the stellar-mass limit, which already over-predicts F1000W by $5\sigma$ (Section~\ref{sec:sedfit}), hot dust would only increase the tension. We therefore test hot dust for the reddened accretion disc only.
The fits of Section~\ref{sec:ma25} already disfavour quasar-like hot dust. To quantify how much is allowed, we fitted a single-temperature greybody ($L_\nu\propto\nu^\beta B_\nu(T_{\rm d})$, $\beta=0$--2) jointly with the reddened accretion disc to the NIRCam and MIRI photometry, and integrated it over all frequencies to obtain its luminosity (Appendix~\ref{app:dust_budget}). Based on this model, dust at 1200~K can re-emit at most 2\% of the lower value of $L_{\rm abs}$, and dust at 1000~K at most $\approx7\%$ (Table~\ref{tab:dust_limits})\footnote{Both values are for $\beta=0$, the least restrictive case. This case also approximates optically thick dust, whose surface emits close to a blackbody, as expected for the clouds of a clumpy AGN torus \citep{Nenkova+08b,Hoenig+10,Hoenig19}.}. The 1200~K limit corresponds to $L_{\rm hot}/L_{\rm AGN}\lesssim0.007$, a geometry-dependent proxy for the covering factor of the hot dust. This is more than ten times below the typical values of luminous quasars ($\approx0.1$) and Seyfert~1 galaxies ($\approx0.15$--0.35; \citealt{Mor+11,Mor+12}), although these values vary widely between objects, down to quasars with no detectable hot dust \citep{Hao+10,Jiang+10,Lyu+17}.
This comparison should be taken with care though. We convert the hot-dust flux into a luminosity assuming isotropic emission, whereas the near-infrared emission of a clumpy torus depends on the viewing angle by up to a factor of a few \citep{Nenkova+08b,Hoenig+10}. For a reddened accretion disc, our sightline crosses the dusty medium, so the hot inner face of the torus may be partly hidden and its intrinsic luminosity a few times higher than we infer. The limit also rests on a single band on the Wien tail (F1000W, rest-frame $1.24$\um), so it is sensitive to the calibration of this band, to its $\approx7\%$ line contribution (see Section~\ref{sec:nir_continuum}) and to the slope of the disc template beyond 1\um. With these caveats, QSO1 lies towards the hot-dust-poor end of type-1 AGN, but only up to rest-frame $1.24$\um. A torus whose emission rises only at longer wavelengths, as in the LRD stacks \citep{Perez-Gonzalez+26,Delvecchio+25}, is not constrained by our data (see Section~\ref{sec:disc_predictions}).

\begin{figure}
    \centering
    \includegraphics[width=\columnwidth]{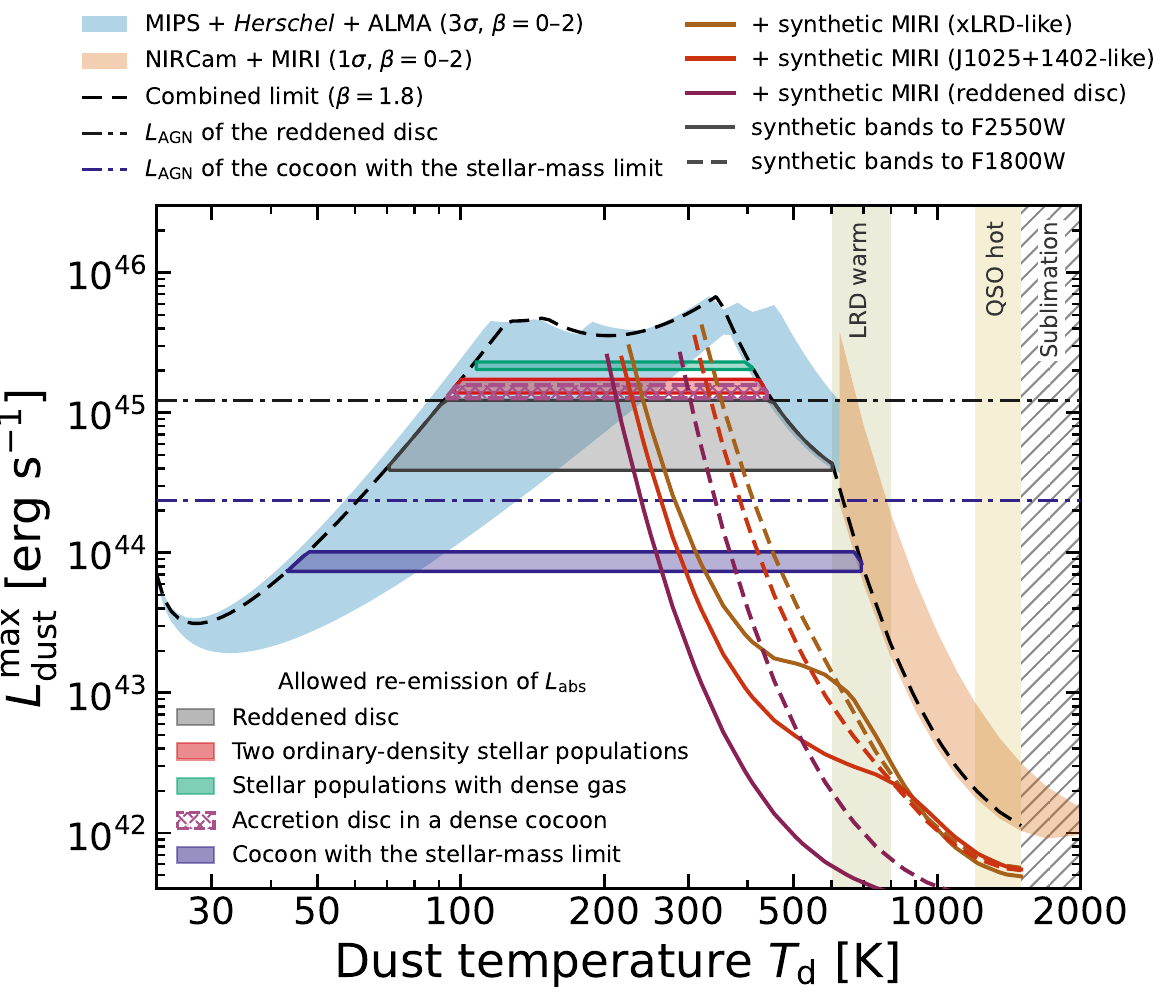}
    \caption{{\it No solution is excluded by energy balance, but the absorbed luminosity of the most attenuated solutions can be hidden only in a limited range of dust temperatures, which deeper MIRI imaging at longer wavelengths can test.} Maximum luminosity of dust emission at a single temperature allowed by the NIRCam and MIRI photometry (orange; fitted jointly with the reddened-disc model) and by the MIPS, \textit{Herschel} and ALMA limits (blue; including the CMB contrast), for $\beta=0$--2. Each limit is shown only where it is the tighter of the two. The dashed black curve is the combined limit for $\beta=1.8$. Shaded regions show the luminosity absorbed by dust in each model, plotted over the temperatures at which the combined limit allows it to be re-emitted in full as a single (modified) blackbody. For the reddened accretion disc (Section~\ref{sec:disc_dust}) the region extends from the absorption between 912~\AA\ and 2.9\um\ to the full intrinsic luminosity, and for the solutions of Section~\ref{sec:sedfit} it spans the 16th--84th percentiles of the absorbed luminosity. Dash-dotted lines mark the intrinsic disc luminosity of the reddened accretion disc and of the cocoon with the stellar-mass limit, both computed with the disc continuum of Appendix~\ref{app:cocoon}. The cocoon region lies below its $L_{\rm AGN}$ because most of the disc luminosity is emitted below 912~\AA\ and is absorbed by gas or escapes, not by dust. Vertical bands mark the temperatures of hot dust in type-1 quasars and of the warm-dust excess in low-redshift LRDs \citep{Ashall+26}, and the hatched band the sublimation range of silicate and graphite grains ($\approx1500$--2000~K; e.g. \citealt{Barvainis87}). Brown, red and purple curves show the combined limit on additional dust if QSO1 were also detected with S/N~$=10$ in F1280W, F1500W and F1800W (dashed) or in F1280W--F2550W (solid), with an SED that follows beyond F1000W the xLRD stack of \citet{Perez-Gonzalez+26}, \textit{the Egg} \citep{Lin+25,Ji+26a} or the reddened-disc model, normalised to QSO1 at rest-frame 1.24\um. Temperatures are those at $z=7.04$, including CMB heating, which is negligible at $T_{\rm d}\gtrsim40$~K.}
    \label{fig:dust_limits_T}
\end{figure}

The hot-dust limit weakens quickly towards cooler grains, from 2\% of $L_{\rm abs}$ at 1200~K to $\approx7\%$ at 1000~K and $\approx6$--50\% at 800~K (depending on $\beta$, see Table~\ref{tab:dust_limits}). Still, in the reddened accretion disc case, most of the absorbed energy must therefore emerge from cooler dust, which is loosely constrained only by the far-infrared and millimetre limits. Normalising a greybody with $\beta=1.8$ to these limits (Appendix~\ref{app:dust_budget}), the full absorbed luminosity of the reddened accretion disc can be hidden at $\approx70$--600~K, while ALMA limits dust at 40~K to at most 15\% of it (Figure~\ref{fig:dust_budget}). For the massive stellar solutions, the larger absorbed luminosity can be hidden only at $\approx100$--440~K. Dust can reach such temperatures in compact, low-metallicity starbursts, much warmer than in typical star-forming galaxies, and produce a steep mid-infrared rise without an AGN \citep[e.g.][]{Hainline+16,Sturm+25b,Rieke+25}. For the cocoon with the stellar-mass limit, the absorbed luminosity can be hidden at $\approx40$--700~K, almost any temperature allowed by our limits (Figure~\ref{fig:dust_limits_T}). No solution is therefore excluded on energy-balance grounds, but the dust of the reddened accretion disc and of the stellar solutions must lie, at least in part, at temperatures that longer-wavelength MIRI imaging can test (Section~\ref{sec:disc_predictions}).

Since grains in radiative equilibrium cool with distance from their heating source \citep{Barvainis87}, the estimated dust temperature constraints can also set where the dust can be. For the massive stellar solutions, dust at $\lesssim400$~K lies at $R\gtrsim25$~pc (applying the same scaling as an approximation), i.e. at or beyond the size limit of QSO1. For the cocoon with the stellar-mass limit, the small budget leaves the dust location unconstrained. For the reddened accretion disc, dust at $\lesssim600$~K lies at $R\gtrsim6$~pc for $L_{\rm AGN}\approx1.2\times10^{45}$~erg~s$^{-1}$, $\approx13$ times the sublimation radius at this luminosity. A shell at this distance, covering the whole sky with the column needed for $A_V\approx2$~mag at the metallicity of QSO1 ($N_{\rm H}\approx10^{24}~{\rm cm^{-2}}$), contains $M_{\rm gas}\approx3\times10^{7}~(R/14~{\rm pc})^2~\rm M_\odot$. At $R\approx50$~pc ($T_{\rm d}\approx280$~K), this approaches the dynamical mass \citep[$<4\times10^8~\rm M_\odot$;][]{Ji+25}. Both the large cavity and the large mass follow from assuming a single temperature. A torus with a range of temperatures can instead extend to the sublimation radius, provided its hottest ($\gtrsim1000$~K) dust carries no more than a few per cent of $L_{\rm abs}$, as in hot-dust-poor quasars \citep{Hao+10,Lyu+17}, and most of the luminosity is re-emitted by cooler dust further out.
We note that $L_{\rm abs}$ assumes that the dust covers all sightlines, like ours. If the dust of the reddened accretion disc covers only a fraction $C_f$ of the sky, it absorbs and re-emits only $C_f\,L_{\rm abs}$, and its gas mass is lower by the same factor. The rest of the disc light escapes unattenuated along the uncovered sightlines, towards other observers. For $C_f=0.3$ (0.1), the absorbed luminosity can then be hidden at $\approx50$--670~K ($\approx35$--750~K), with dust at 1200~K carrying at most $\approx7\%$ (22\%) of it. Such covering factors are at the low end of those of AGN. For the obscuring gas, typical values are $\approx0.85$ at $\lambda_{\rm Edd}\lesssim0.03$ and $\approx0.4$ above \citep{Ricci+17}, a range that includes QSO1 ($\lambda_{\rm Edd}\approx0.005$--0.2; Section~\ref{sec:disc_photosphere}). For the infrared-emitting dust, the covering factor decreases with luminosity, from about unity in low-luminosity AGN to $\approx0.1$ in luminous quasars \citep{Maiolino+07}, with a median of $\approx0.6$ in X-ray-selected type-1 AGN \citep{Lusso+13}. \citet{MadauMaiolino26} adopt $C_f\approx0.15$ in a model in which LRDs are blue broad-line AGN seen through dust at high inclination.

The covering factor also links QSO1 to the LRD population. In the reddened accretion disc interpretation, the fraction of systems seen through their dust, and hence selected as LRDs, equals the covering factor. For $C_f=0.3$ (0.1), each LRD would have $\approx2$ (9) counterparts seen without dust, as blue broad-line AGN of similar intrinsic luminosity. At $z\sim5$, LRDs are $\approx100$ times more abundant than UV-selected AGN of similar dust-corrected luminosity \citep{Kokorev+24b}, which would require dust covering most of the sky. Among the broad-line AGN identified spectroscopically with JWST, however, LRDs make up $\approx30$ per cent, a fraction that rises with luminosity and drops below 15 per cent under $L_{\rm bol}\approx1.5\times10^{44}$~erg~s$^{-1}$ \citep{Hainline+25}. At the intrinsic luminosity of QSO1, these small samples suggest $C_f\gtrsim0.3$, which disfavours the lowest covering factors but not the case $C_f\approx0.3$. The energy budget derived above for full covering is therefore the most demanding case.

The broad Ly$\alpha$ of QSO1 \citep{Tang+26a} constrains the geometry further. A uniform dust screen would suppress it by $\approx27$~mag for the reddened accretion disc ($A_V\approx2.2$~mag) and by $\approx17$~mag in the cocoon with the stellar-mass limit ($A_V\approx1.3$~mag). In such scenarios, the dust must therefore be clumpy, as typically found in AGN tori \citep[e.g.][]{Tristram+07,GRAVITY20b,Hoenig19}. It can then cover most sightlines, as required above, while a small fraction of the light leaks through gaps between the clumps. In the reddened-disc fit this unattenuated fraction is $\approx1\%$ (Appendix~\ref{app:dust_budget}), and it also produces the blue UV continuum. With Case-B or typical broad-line Ly$\alpha$/H$\alpha$ ratios and the broad-line attenuation, this fraction predicts a broad Ly$\alpha$ luminosity within a factor of a few of the observed one, before any absorption by the intergalactic medium. In the cocoon, the screen may instead partly stand in for reddening by the dense gas itself, in which case its dust emission is an upper limit.

\begin{figure}
    \centering
    \includegraphics[width=\linewidth]{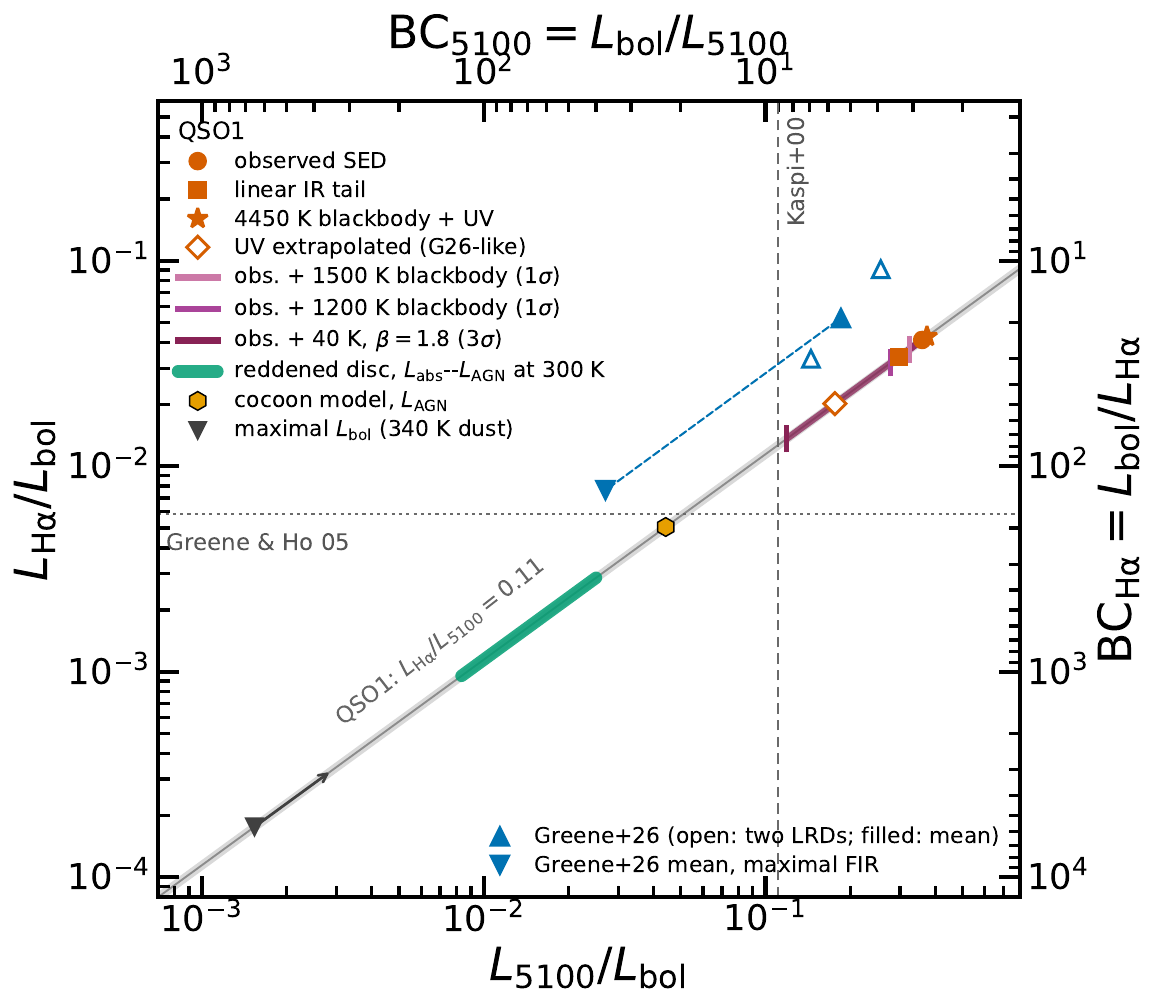}
    \caption{{\it Without warm dust, the bolometric corrections of QSO1 are comparable to those of other LRDs and well below those of type-1 AGN. They would exceed the type-1 values only if a reddened accretion disc re-emitted its absorbed luminosity as warm dust.} H$\alpha$ and 5100\AA\ luminosities of QSO1 relative to its bolometric luminosity, for the luminosity integrated over the observed SED (orange symbols, a lower limit on $L_{\rm bol}$), for the observed SED plus the dust components of Figure~\ref{fig:dust_budget} at their limits (segments: allowed range), for a reddened accretion disc whose absorbed luminosity, from $L_{\rm abs}$ to $L_{\rm AGN}$, is re-emitted at 300~K (green), for the intrinsic disc luminosity of the cocoon model with the stellar-mass limit (hexagon), and for the maximal luminosity allowed by the MIRI and far-infrared limits (triangle, upper limit on $L_{\rm bol}$). All QSO1 points lie on the line set by its $L_{\rm H\alpha}/L_{5100}$. Blue triangles: the two LRDs of \citet{Greene+25b}, their mean and the mean with their maximal far-infrared emission. Dashed and dotted lines: bolometric corrections of type-1 AGN \citep{Kaspi+00,GreeneHo05}. The top and right axes give the bolometric corrections.}
    \label{fig:lbol}
\end{figure}

\subsection{Bolometric luminosity}
\label{sec:disc_lbol}
The dust budget also sets the bolometric luminosity of QSO1. Integrating the de-magnified SED from Ly$\alpha$ to 1~mm, with a Rayleigh--Jeans tail beyond F1000W, gives $L_{\rm obs}=2.9\times10^{43}$~erg~s$^{-1}$ ($3.5\times10^{43}$~erg~s$^{-1}$ with a linear decline to the ALMA band as in \citealt{Greene+25b}), a lower limit on $L_{\rm bol}$ since it omits the ionising and X-ray emission. With $L_{5100}=1.05\times10^{43}$~erg~s$^{-1}$ and $L_{\rm H\alpha}=(1.20\pm0.07)\times10^{42}$~erg~s$^{-1}$ (measured on the Ima-B spectrum, whose broad-line equivalent widths are about half those of Ima-C; \citealt{Ji+25}), the bolometric corrections are ${\rm BC}_{5100}\approx2.8$--3.4 and ${\rm BC}_{\rm H\alpha}\approx24$--29 (Figure~\ref{fig:lbol}). They are comparable to those of the two LRDs of \citet{Greene+25b} (5.4 and 19) and well below those of type-1 AGN \citep[9 and 170;][]{Kaspi+00,GreeneHo05}, whose H$\alpha$ correction would give $L_{\rm bol}\approx2\times10^{44}$~erg~s$^{-1}$ \citep{Juodzbalis+26b}, $\approx6$ times the observed luminosity. Hot dust within our limits raises these corrections only slightly, to at most ${\rm BC}_{5100}\approx3.6$ and ${\rm BC}_{\rm H\alpha}\approx31$ ($\geq1200$~K), and cold dust by at most a factor of $\approx2.5$, to $\approx8$ and $\approx75$ ($\approx40$~K). Warm dust can raise them much further. A reddened accretion disc re-emitting its absorbed luminosity at $\approx300$~K would have $L_{\rm bol}\approx(0.4$--$1.3)\times10^{45}$~erg~s$^{-1}$, ${\rm BC}_{5100}\approx40$--120 and ${\rm BC}_{\rm H\alpha}\approx350$--1050, above even the maximal values of \citet{Greene+25b}, and the disc of the cocoon with the stellar-mass limit falls in between ($2.4\times10^{44}$~erg~s$^{-1}$, ${\rm BC}_{5100}\approx23$, ${\rm BC}_{\rm H\alpha}\approx200$). The largest luminosity compatible with all our limits, for dust at $\approx340$~K, is $L_{\rm bol}<6.8\times10^{45}$~erg~s$^{-1}$.

\subsection{Predictions for longer-wavelength MIRI imaging}
\label{sec:disc_predictions}
Our data stop at rest-frame $1.24$\um, while the LRD stacks and local LRDs depart most from a single thermal component at 2--3\um\ (Section~\ref{sec:empirical_templates}). We therefore predict what MIRI imaging beyond F1000W would see in QSO1 for three continuations of its SED: the xLRD stack of \citet{Perez-Gonzalez+26}, \textit{the Egg} \citep{Lin+25,Ji+26a} and the reddened-disc model of Section~\ref{sec:ma25}, each normalised to QSO1 at rest-frame $1.24$\um\ (Table~\ref{tab:miri_predictions}). The xLRD and {\it the Egg} continuations would exceed the $4450$~K blackbody by factors of $\approx2$--5 at rest-frame 2--3\um. The excess would carry $L_{\rm d}\approx(1$--$2)\times10^{43}$~erg~s$^{-1}$ up to rest-frame $\approx20$\um, consistent with the warm dust inferred for low-redshift LRDs \citep[$\approx500$--700~K;][]{Ashall+26}. This emission is allowed by our limits, as it lies $\approx20$--40 times below the MIPS limit. It would raise $L_{\rm obs}$ by $\approx40$--80\%, i.e. to ${\rm BC}_{5100}\approx4$--5, comparable to the values of \citet{Greene+25b}. However, it would amount to only a few per cent of the absorbed luminosity of the reddened accretion disc or of the massive stellar solutions, so the bulk of that luminosity would still have to be re-emitted by cooler dust.

In the image plane of Ima-A, the three continuations give the same F1280W flux density, $\approx0.9$~$\mu$Jy ($\approx24.0$~AB). They separate at longer wavelengths. F1800W ranges from $0.81$~$\mu$Jy ($24.1$~AB) for the reddened accretion disc to $1.18$~$\mu$Jy ($23.7$~AB) for the xLRD stack, and F2550W from $0.69$~$\mu$Jy ($24.3$~AB) to $1.99$~$\mu$Jy ($23.2$~AB), with \textit{the Egg} in between ($0.98$ and $1.02$~$\mu$Jy, i.e. $23.9$~AB, in both bands). \textit{The Cliff} \citep{deGraaff+25B} predicts flux densities within 5\% of those of \textit{the Egg} in all five bands. If QSO1 followed the xLRD or {\it the Egg} SED and were detected at S/N $=$ 10 in F1280W--F2550W, the excess up to rest-frame $3.2$\um\ would be measured as dust at $\approx600$--650~K with $L_{\rm d}\approx(0.2$--$1)\times10^{43}$~erg~s$^{-1}$. The temperature at which the near- and far-infrared limits cross would drop from $\approx620$ to $\approx220$~K, and the absorbed luminosity of a reddened accretion disc could be hidden only at $\approx70$--250~K, compared with 70--600~K today (Figure~\ref{fig:dust_limits_T}). Imaging up to F1800W alone would already restrict it to $\lesssim300$--380~K.

\begin{table}
\centering
\caption{Expected flux densities of Ima-A (image plane, i.e. not corrected for magnification) in the MIRI bands beyond F1000W, if the SED of QSO1 continued like the xLRD stack of \citet{Perez-Gonzalez+26}, like \textit{the Egg} \citep{Lin+25,Ji+26a} or like the reddened accretion disc model of Section~\ref{sec:ma25}, each normalised to QSO1 at rest-frame 1.24\um. For reference, Ima-A has $0.91$ and $0.90~\mu$Jy (24.00 and 24.02~AB) in F770W and F1000W.}
\label{tab:miri_predictions}
\begin{tabular}{lccc}
\hline
\hline
Filter (rest-frame \um) & xLRD stack & \textit{the Egg} & reddened disc \\
 & [$\mu$Jy] (AB) & [$\mu$Jy] (AB) & [$\mu$Jy] (AB) \\
\hline
F1280W (1.59) & 0.89 (24.03) & 0.88 (24.04) & 0.90 (24.02) \\
F1500W (1.86) & 0.96 (23.94) & 0.95 (23.96) & 0.86 (24.06) \\
F1800W (2.24) & 1.18 (23.72) & 0.98 (23.92) & 0.81 (24.13) \\
F2100W (2.61) & 1.47 (23.48) & 0.99 (23.92) & 0.76 (24.20) \\
F2550W (3.17) & 1.99 (23.15) & 1.02 (23.87) & 0.69 (24.31) \\
\hline
\hline
\end{tabular}
\end{table}

%%%%%%%%%%%%%%%%%%%%%%%%%%%%%%%%%%%%%%%%%%%%%%%%%%%%%%%%%%%%%%
\section{Conclusions}
\label{sec:conclusions}
We have presented the first rest-frame near-infrared observations of the lensed LRD QSO1 at $z=7.04$, obtained with MIRI in F770W and F1000W. Our main results are:
\begin{enumerate}
    \item All three images are unresolved, with a conservative source-plane 95\% upper limit of $r_{\rm e}<64$~pc for Ima-A. The MIRI-to-NIRCam continuum ratios of Ima-A and Ima-C agree within $\approx5\%$, with no evidence for wavelength-dependent variability above $\approx15\%$ over the observer-frame $\approx19$~yr ($\approx2.4$~yr rest-frame) delay between Ima-A and Ima-C, whereas the broad lines have been reported to vary \citep{Furtak+25,Ji+25}.
    \item The steep optical rise turns over between $0.96$ and $1.24$\um. A single $T\simeq4450$~K blackbody describes the optical continuum and both MIRI bands, and a hotter, reddened thermal continuum fits equally well. Up to $1.24$\um, the SED of QSO1 is typical of LRDs, and its UV-to-optical ratio places it among the extreme LRDs of \citet{Perez-Gonzalez+26}. Without MIRI, models constrained by the UV-to-optical data mispredict the $1$\um\ continuum by 15--76\%.
    \item Within the model families explored, the spectrum requires a dense, strongly reprocessing component and differential attenuation ($\Delta\ln(\mathcal{Z})\gtrsim80$ against a shared screen). The highest-evidence solutions produce the red continuum with $\gtrsim3\times10^9~\rm M_\odot$ of stars, $\approx140$--200 times the upper limit on the host stellar mass. Within this limit, the disc and cocoon produce the red continuum but over-predict F1000W, with an evidence lower by $\Delta\ln(\mathcal{Z})\simeq20$--33 that comes from the MIRI bands and the emission lines, the features our cocoon models describe least well. Accretion-flow envelope models reproduce the shape of the turnover, but their luminosity matches QSO1 only for a black-hole mass $\approx2.5$~dex below the dynamical one, which implies $\lambda_{\rm Edd}\approx0.005$--0.03 rather than the near-Eddington accretion these models assume.
    \item There is no evidence for the standard quasar hot-dust upturn up to $1.24$\um. For a disc reddened by $A_V\approx2$~mag, which would absorb $L_{\rm abs}\approx(0.4$--$1.2)\times10^{45}$~erg~s$^{-1}$, at most $\approx2\%$ can be re-emitted by hot dust ($T_{\rm d}\geq1200$~K) and at most $\approx15\%$ by cold dust ($T_{\rm d}\approx40$~K). The remainder could be emitted undetected by warm dust at $\approx70$--600~K, and the larger budget of the massive stellar solutions only at $\approx100$--400~K. The cocoon with the stellar-mass limit absorbs $\approx20$ times less and is unconstrained. A reddened accretion disc therefore remains allowed, provided that most of its absorbed luminosity emerges at $\approx70$--600~K and dust hotter than $\approx1000$~K carries at most a few per cent of it, as in hot-dust-poor quasars. Without dust, the observed luminosity gives ${\rm BC}_{5100}\approx3$ and ${\rm BC}_{\rm H\alpha}\approx25$, comparable to other LRDs and well below type-1 AGN. If a reddened accretion disc re-emitted its absorbed luminosity as warm dust, these would rise to $\approx40$--120 and $\approx350$--1050.
\end{enumerate}
If the upper limit on the host stellar mass holds, the stellar solutions explored here are excluded, and our results favour an accretion-powered origin of the red continuum, either reddened by dust or processed by dense gas, which our data cannot yet distinguish. An accretion-powered source is also suggested by the broad Ly$\alpha$ and far-UV Fe~{\sc ii} emission of QSO1. Our \textsc{Cloudy} cocoons do not form a pseudo-photosphere and miss the turnover, whereas envelope models that do form one miss the luminosity expected for the dynamical black-hole mass. Reconciling the two, and exploring non-isotropic dust geometries that would relax the energy-balance constraints derived here, are needed before QSO1 can be modelled self-consistently.
 
Even with MIRI, our description of the continuum of QSO1 stops at rest-frame $1.24$\um, short of the range where LRDs depart most from a single thermal component. QSO1 belongs to the extreme LRDs of \citet{Perez-Gonzalez+26}, the subtype whose stack rises again beyond rest-frame $\approx2$\um, by factors of $\approx2$--5 over the $4450$~K blackbody at 2--3\um. For the most extreme LRDs, near-infrared data up to $\approx1.2$\um\ are therefore not sufficient to constrain the dust budget or the nature of the red continuum. At $z\gtrsim7$ this range lies beyond observed 16\um\ and can be reached only with deep MIRI imaging at $\gtrsim18$\um\ (F1800W, F2100W and F2550W). For Ima-A, the xLRD stack, \textit{the Egg} and a reddened accretion disc predict $0.8$--$1.2$~$\mu$Jy (23.7--24.1~AB) in F1800W and $0.7$--$2.0$~$\mu$Jy (23.2--24.3~AB) in F2550W (Section~\ref{sec:disc_predictions}). Detections at S/N~$=10$ up to F2550W would measure any warm-dust excess and confine the dust of a reddened accretion disc to $\approx70$--250~K, and F1800W alone would restrict it to $\lesssim300$--380~K. Such data would tighten the energy budget of every solution considered here and distinguish a true pseudo-photosphere from a reddened, hotter continuum. 
 
%%%%%%%%%%%%%%%%%%%%%%%%%%%%%%%%%%%%%%%%%%%%%%%%%%%%%%%%%%%%%%
\section*{Acknowledgements}
The authors thank R. Maiolino, K. Caputi, S. Finkelstein and A. Sneppen for the useful discussion. 
The authors thank X. Ji for sharing the data of J1025+1402 (the Egg).
This research is based in part on observations made with the NASA/ESA Hubble Space Telescope obtained from the Space Telescope Science Institute, which is operated by the Association of Universities for Research in Astronomy, Inc., under NASA contract NAS 5–26555.
This work is based in part on observations made with the NASA/ESA/CSA James Webb Space Telescope. The data were obtained from the Mikulski Archive for Space Telescopes at the Space Telescope Science Institute, which is operated by the Association of Universities for Research in Astronomy, Inc., under NASA contract NAS 5-03127 for JWST. These observations are associated with programs 1324, 2561, 2767, 2883, 3516, 3538, 4111, 5578 and 9214.
The authors acknowledge the team led by PIs F. Sun (\#2883) and E. Iani (\#3538) for developing their observing programs with a zero-exclusive-access period.
Some of the data products presented herein were retrieved from the Dawn JWST Archive (DJA). DJA is an initiative of the Cosmic Dawn Center (DAWN), which is funded by the Danish National Research Foundation under grant DNRF140.
PR acknowledges support from the University of Texas at Austin Cosmic Frontier Center. 
FDE acknowledges support by the Science and Technology Facilities Council (STFC), by the ERC through Advanced Grant 695671 ``QUENCH'', and by the UKRI Frontier Research grant RISEandFALL. Funded by the European Union (ERC, AGENTS,  101076224). Views and opinions expressed are however those of the author(s) only and do not necessarily reflect those of the European Union or the European Research Council. Neither the European Union nor the granting authority can be held responsible for them. 
Generative artificial-intelligence tools (Anthropic - Claude Opus 5.5) were used to assist with language polishing, improve the readability and presentation of the figures, and support the software implementation of the models described in Section~4.3.2 and Appendix I. All scientific choices, model validation, analyses, interpretations, and final presentation were reviewed and verified by the authors.

\section*{Data Availability}
The JWST and HST data used in this work are available from the Mikulski Archive for Space Telescopes (MAST). The NIRSpec spectra are available from the DAWN JWST Archive. The derived data underlying this article will be shared on reasonable request to the corresponding author.

\bibliographystyle{mnras}
\bibliography{paper}

@ARTICLE{Trotta08,
       author = {{Trotta}, Roberto},
        title = "{Bayes in the sky: Bayesian inference and model selection in cosmology}",
      journal = {Contemporary Physics},
         year = 2008,
        month = mar,
       volume = {49},
       number = {2},
        pages = {71-104},
          doi = {10.1080/00107510802066753},
archivePrefix = {arXiv},
       eprint = {0803.4089},
 primaryClass = {astro-ph},
       adsurl = {https://ui.adsabs.harvard.edu/abs/2008ConPh..49...71T}
}

@ARTICLE{Matthews+25,
       author = {{Matthews}, James H. and {Long}, Knox S. and {Knigge}, Christian and {Sim}, Stuart A. and {Parkinson}, Edward J. and {Higginbottom}, Nick and {Mangham}, Samuel W. and {Scepi}, Nicolas and {Wallis}, Austen and {Hewitt}, Henrietta A. and {Mosallanezhad}, Amin},
        title = "{SIROCCO: a publicly available Monte Carlo ionization and radiative transfer code for astrophysical outflows}",
      journal = {\mnras},
         year = 2025,
        month = jan,
       volume = {536},
       number = {1},
        pages = {879-904},
          doi = {10.1093/mnras/stae2677},
archivePrefix = {arXiv},
       eprint = {2410.19908},
 primaryClass = {astro-ph.HE},
       adsurl = {https://ui.adsabs.harvard.edu/abs/2025MNRAS.536..879M}
}

@ARTICLE{Kaspi+00,
       author = {{Kaspi}, Shai and {Smith}, Paul S. and {Netzer}, Hagai and {Maoz}, Dan and {Jannuzi}, Buell T. and {Giveon}, Uriel},
        title = "{Reverberation Measurements for 17 Quasars and the Size-Mass-Luminosity Relations in Active Galactic Nuclei}",
      journal = {\apj},
         year = 2000,
        month = apr,
       volume = {533},
       number = {2},
        pages = {631-649},
          doi = {10.1086/308704},
archivePrefix = {arXiv},
       eprint = {astro-ph/9911476},
 primaryClass = {astro-ph},
       adsurl = {https://ui.adsabs.harvard.edu/abs/2000ApJ...533..631K}
}

@ARTICLE{Greene+25b,
       author = {{Greene}, Jenny E. and {Setton}, David J. and {Furtak}, Lukas J. and {Naidu}, Rohan P. and {Volonteri}, Marta and {Dayal}, Pratika and {Labbe}, Ivo and {van Dokkum}, Pieter and {Bezanson}, Rachel and {Brammer}, Gabriel and {Cutler}, Sam E. and {Glazebrook}, Karl and {de Graaff}, Anna and {Hirschmann}, Michaela and {Hviding}, Raphael E. and {Kokorev}, Vasily and {Leja}, Joel and {Liu}, Hanpu and {Ma}, Yilun and {Matthee}, Jorryt and {Nanayakkara}, Themiya and {Oesch}, Pascal A. and {Pan}, Richard and {Price}, Sedona H. and {Spilker}, Justin S. and {Wang}, Bingjie and {Weaver}, John R. and {Whitaker}, Katherine E. and {Williams}, Christina C. and {Zitrin}, Adi},
        title = "{What You See Is What You Get: Empirically Measured Bolometric Luminosities of Little Red Dots}",
      journal = {\apj},
         year = 2026,
        month = jan,
       volume = {996},
       number = {2},
          eid = {129},
        pages = {129},
          doi = {10.3847/1538-4357/ae1836},
archivePrefix = {arXiv},
       eprint = {2509.05434},
 primaryClass = {astro-ph.GA},
       adsurl = {https://ui.adsabs.harvard.edu/abs/2026ApJ...996..129G}
}

@ARTICLE{GreeneHo05,
       author = {{Greene}, Jenny E. and {Ho}, Luis C.},
        title = "{Estimating Black Hole Masses in Active Galaxies Using the H{\ensuremath{\alpha}} Emission Line}",
      journal = {\apj},
         year = 2005,
        month = sep,
       volume = {630},
       number = {1},
        pages = {122-129},
          doi = {10.1086/431897},
archivePrefix = {arXiv},
       eprint = {astro-ph/0508335},
 primaryClass = {astro-ph},
       adsurl = {https://ui.adsabs.harvard.edu/abs/2005ApJ...630..122G}
}

@ARTICLE{Mascia+26,
       author = {{Mascia}, Sara and {Matthee}, Jorryt and {Torralba}, Alberto and {Greene}, Jenny E. and {Eilers}, Anna-Christina and {Iani}, Edoardo and {Naidu}, Rohan P.},
        title = "{The ionizing properties of JWST's compact broad-line emitters}",
      journal = {arXiv e-prints},
         year = 2026,
        month = aug,
          eid = {arXiv:2608.25021},
        pages = {arXiv:2608.25021},
          doi = {10.48550/arXiv.2608.25021},
archivePrefix = {arXiv},
       eprint = {2608.25021},
 primaryClass = {astro-ph.GA},
       adsurl = {https://ui.adsabs.harvard.edu/abs/2026arXiv260825021M}
}

@ARTICLE{Sneppen+26c,
       author = {{Sneppen}, A. and {Watson}, D. and {Matthews}, J.~H. and {Nikopoulos}, G. and {Allen}, N. and {Brammer}, G. and {Damgaard}, R. and {Heintz}, K.~E. and {Knigge}, C. and {Long}, K.~S. and {Rusakov}, V. and {Sim}, S.~A. and {Witstok}, J.},
        title = "{Inside the cocoon: a comprehensive explanation of the spectra of Little Red Dots}",
      journal = {arXiv e-prints},
         year = 2026,
        month = jan,
          eid = {arXiv:2601.18864},
        pages = {arXiv:2601.18864},
          doi = {10.48550/arXiv.2601.18864},
archivePrefix = {arXiv},
       eprint = {2601.18864},
 primaryClass = {astro-ph.GA},
       adsurl = {https://ui.adsabs.harvard.edu/abs/2026arXiv260118864S}
}

@ARTICLE{Tristram+07,
       author = {{Tristram}, K.~R.~W. and {Meisenheimer}, K. and {Jaffe}, W. and {Schartmann}, M. and {Rix}, H.-W. and {Leinert}, Ch. and {Morel}, S. and {Wittkowski}, M. and {R{\"o}ttgering}, H. and {Perrin}, G. and {Lopez}, B. and {Raban}, D. and {Cotton}, W.~D. and {Graser}, U. and {Paresce}, F. and {Henning}, Th.},
        title = "{Resolving the complex structure of the dust torus in the active nucleus of the Circinus galaxy}",
      journal = {\aap},
         year = 2007,
        month = nov,
       volume = {474},
       number = {3},
        pages = {837-850},
          doi = {10.1051/0004-6361:20078369},
archivePrefix = {arXiv},
       eprint = {0709.0209},
 primaryClass = {astro-ph},
       adsurl = {https://ui.adsabs.harvard.edu/abs/2007A&A...474..837T}
}

@ARTICLE{Pacucci+24,
       author = {{Pacucci}, Fabio and {Narayan}, Ramesh},
        title = "{Mildly Super-Eddington Accretion onto Slowly Spinning Black Holes Explains the X-Ray Weakness of the Little Red Dots}",
      journal = {\apj},
         year = 2024,
        month = nov,
       volume = {976},
       number = {1},
          eid = {96},
        pages = {96},
          doi = {10.3847/1538-4357/ad84f7},
archivePrefix = {arXiv},
       eprint = {2407.15915},
 primaryClass = {astro-ph.HE},
       adsurl = {https://ui.adsabs.harvard.edu/abs/2024ApJ...976...96P}
}

@ARTICLE{Jiang+14,
       author = {{Jiang}, Yan-Fei and {Stone}, James M. and {Davis}, Shane W.},
        title = "{A Global Three-dimensional Radiation Magneto-hydrodynamic Simulation of Super-Eddington Accretion Disks}",
      journal = {\apj},
         year = 2014,
        month = dec,
       volume = {796},
       number = {2},
          eid = {106},
        pages = {106},
          doi = {10.1088/0004-637X/796/2/106},
archivePrefix = {arXiv},
       eprint = {1410.0678},
 primaryClass = {astro-ph.HE},
       adsurl = {https://ui.adsabs.harvard.edu/abs/2014ApJ...796..106J}
}

@ARTICLE{Runnoe+12,
       author = {{Runnoe}, Jessie C. and {Brotherton}, Michael S. and {Shang}, Zhaohui},
        title = "{Updating quasar bolometric luminosity corrections - II. Infrared bolometric corrections}",
      journal = {\mnras},
         year = 2012,
        month = nov,
       volume = {426},
       number = {4},
        pages = {2677-2688},
          doi = {10.1111/j.1365-2966.2012.21644.x},
archivePrefix = {arXiv},
       eprint = {1207.2124},
 primaryClass = {astro-ph.CO},
       adsurl = {https://ui.adsabs.harvard.edu/abs/2012MNRAS.426.2677R}
}

@ARTICLE{ForemanMackey+13,
       author = {{Foreman-Mackey}, Daniel and {Hogg}, David W. and {Lang}, Dustin and {Goodman}, Jonathan},
        title = "{emcee: The MCMC Hammer}",
      journal = {\pasp},
         year = 2013,
        month = mar,
       volume = {125},
       number = {925},
        pages = {306},
          doi = {10.1086/670067},
archivePrefix = {arXiv},
       eprint = {1202.3665},
 primaryClass = {astro-ph.IM},
       adsurl = {https://ui.adsabs.harvard.edu/abs/2013PASP..125..306F}
}

@ARTICLE{Mor+11,
       author = {{Mor}, Rivay and {Trakhtenbrot}, Benny},
        title = "{Hot-dust Clouds with Pure-graphite Composition around type-I Active Galactic Nuclei}",
      journal = {\apjl},
         year = 2011,
        month = aug,
       volume = {737},
       number = {2},
        pages = {L36},
          doi = {10.1088/2041-8205/737/2/L36},
archivePrefix = {arXiv},
       eprint = {1105.3198},
 primaryClass = {astro-ph.CO},
       adsurl = {https://ui.adsabs.harvard.edu/abs/2011ApJ...737L..36M}
}

@ARTICLE{Mor+12,
       author = {{Mor}, Rivay and {Netzer}, Hagai},
        title = "{Hot graphite dust and the infrared spectral energy distribution of active galactic nuclei}",
      journal = {\mnras},
         year = 2012,
        month = feb,
       volume = {420},
       number = {1},
        pages = {526-541},
          doi = {10.1111/j.1365-2966.2011.20060.x},
archivePrefix = {arXiv},
       eprint = {1110.5326},
 primaryClass = {astro-ph.CO},
       adsurl = {https://ui.adsabs.harvard.edu/abs/2012MNRAS.420..526M}
}

@ARTICLE{Liu+25,
       author = {{Liu}, Hanpu and {Jiang}, Yan-Fei and {Quataert}, Eliot and {Greene}, Jenny E. and {Ma}, Yilun},
        title = "{The Balmer Break and Optical Continuum of Little Red Dots from Super-Eddington Accretion}",
      journal = {\apj},
         year = 2025,
        month = nov,
       volume = {994},
       number = {1},
          eid = {113},
        pages = {113},
          doi = {10.3847/1538-4357/ae0c19},
archivePrefix = {arXiv},
       eprint = {2507.07190},
 primaryClass = {astro-ph.GA},
       adsurl = {https://ui.adsabs.harvard.edu/abs/2025ApJ...994..113L}
}

@ARTICLE{Ji+26a,
       author = {{Ji}, Xihan and {D'Eugenio}, Francesco and {Juod{\v{z}}balis}, Ignas and {Walton}, Dominic J. and {Fabian}, Andrew C. and {Maiolino}, Roberto and {Ramos Almeida}, Cristina and {Acosta Pulido}, Jose A. and {Belokurov}, Vasily A. and {Isobe}, Yuki and {Jones}, Gareth and {Maraston}, Claudia and {Scholtz}, Jan and {Simmonds}, Charlotte and {Tacchella}, Sandro and {Terlevich}, Elena and {Terlevich}, Roberto},
        title = "{Lord of LRDs: insights into a 'Little Red Dot' with a low-ionization spectrum at z = 0.1}",
      journal = {\mnras},
         year = 2026,
        month = jan,
       volume = {545},
       number = {3},
          eid = {staf2235},
        pages = {staf2235},
          doi = {10.1093/mnras/staf2235},
archivePrefix = {arXiv},
       eprint = {2507.23774},
 primaryClass = {astro-ph.GA},
       adsurl = {https://ui.adsabs.harvard.edu/abs/2026MNRAS.545f2235J}
}

@ARTICLE{Choi+16,
       author = {{Choi}, Jieun and {Dotter}, Aaron and {Conroy}, Charlie and {Cantiello}, Matteo and {Paxton}, Bill and {Johnson}, Benjamin D.},
        title = "{Mesa Isochrones and Stellar Tracks (MIST). I. Solar-scaled Models}",
      journal = {\apj},
         year = 2016,
        month = jun,
       volume = {823},
       number = {2},
          eid = {102},
        pages = {102},
          doi = {10.3847/0004-637X/823/2/102},
archivePrefix = {arXiv},
       eprint = {1604.08592},
 primaryClass = {astro-ph.SR},
       adsurl = {https://ui.adsabs.harvard.edu/abs/2016ApJ...823..102C}
}

@ARTICLE{Dotter+16,
       author = {{Dotter}, Aaron},
        title = "{MESA Isochrones and Stellar Tracks (MIST) 0: Methods for the Construction of Stellar Isochrones}",
      journal = {\apjs},
         year = 2016,
        month = jan,
       volume = {222},
       number = {1},
          eid = {8},
        pages = {8},
          doi = {10.3847/0067-0049/222/1/8},
archivePrefix = {arXiv},
       eprint = {1601.05144},
 primaryClass = {astro-ph.SR},
       adsurl = {https://ui.adsabs.harvard.edu/abs/2016ApJS..222....8D}
}

@ARTICLE{Sanchez-Blazquez+06,
       author = {{S{\'a}nchez-Bl{\'a}zquez}, P. and {Peletier}, R.~F. and {Jim{\'e}nez-Vicente}, J. and {Cardiel}, N. and {Cenarro}, A.~J. and {Falc{\'o}n-Barroso}, J. and {Gorgas}, J. and {Selam}, S. and {Vazdekis}, A.},
        title = "{Medium-resolution Isaac Newton Telescope library of empirical spectra}",
      journal = {\mnras},
         year = 2006,
        month = sep,
       volume = {371},
       number = {2},
        pages = {703-718},
          doi = {10.1111/j.1365-2966.2006.10699.x},
archivePrefix = {arXiv},
       eprint = {astro-ph/0607009},
 primaryClass = {astro-ph},
       adsurl = {https://ui.adsabs.harvard.edu/abs/2006MNRAS.371..703S}
}

@ARTICLE{VandenBerk+01,
       author = {{Vanden Berk}, Daniel E. and {Richards}, Gordon T. and {Bauer}, Amanda and {Strauss}, Michael A. and {Schneider}, Donald P. and {Heckman}, Timothy M. and {York}, Donald G. and {Hall}, Patrick B. and {Fan}, Xiaohui and {Knapp}, G.~R. and {Anderson}, Scott F. and {Annis}, James and {Bahcall}, Neta A. and {Bernardi}, Mariangela and {Briggs}, John W. and {Brinkmann}, J. and {Brunner}, Robert and {Burles}, Scott and {Carey}, Larry and {Castander}, Francisco J. and {Connolly}, A.~J. and {Crocker}, J.~H. and {Csabai}, Istv{\'a}n and {Doi}, Mamoru and {Finkbeiner}, Douglas and {Friedman}, Scott and {Frieman}, Joshua A. and {Fukugita}, Masataka and {Gunn}, James E. and {Hennessy}, G.~S. and {Ivezi{\'c}}, {\v{Z}}eljko and {Kent}, Stephen and {Kunszt}, Peter Z. and {Lamb}, D.~Q. and {Leger}, R. French and {Long}, Daniel C. and {Loveday}, Jon and {Lupton}, Robert H. and {Meiksin}, Avery and {Merelli}, Aronne and {Munn}, Jeffrey A. and {Newberg}, Heidi Jo and {Newcomb}, Matt and {Nichol}, R.~C. and {Owen}, Russell and {Pier}, Jeffrey R. and {Pope}, Adrian and {Rockosi}, Constance M. and {Schlegel}, David J. and {Siegmund}, Walter A. and {Smee}, Stephen and {Snir}, Yehuda and {Stoughton}, Chris and {Stubbs}, Christopher and {SubbaRao}, Mark and {Szalay}, Alexander S. and {Szokoly}, Gyula P. and {Tremonti}, Christy and {Uomoto}, Alan and {Waddell}, Patrick and {Yanny}, Brian and {Zheng}, Wei},
        title = "{Composite Quasar Spectra from the Sloan Digital Sky Survey}",
      journal = {\aj},
         year = 2001,
        month = aug,
       volume = {122},
       number = {2},
        pages = {549-564},
          doi = {10.1086/321167},
archivePrefix = {arXiv},
       eprint = {astro-ph/0105231},
 primaryClass = {astro-ph},
       adsurl = {https://ui.adsabs.harvard.edu/abs/2001AJ....122..549V}
}

@ARTICLE{ShakuraSunyaev73,
       author = {{Shakura}, N.~I. and {Sunyaev}, R.~A.},
        title = "{Black holes in binary systems. Observational appearance.}",
      journal = {\aap},
         year = 1973,
        month = jan,
       volume = {24},
        pages = {337-355},
       adsurl = {https://ui.adsabs.harvard.edu/abs/1973A&A....24..337S}
}

@ARTICLE{GueverOezel09,
       author = {{G{\"u}ver}, Tolga and {{\"O}zel}, Feryal},
        title = "{The relation between optical extinction and hydrogen column density in the Galaxy}",
      journal = {\mnras},
         year = 2009,
        month = dec,
       volume = {400},
       number = {4},
        pages = {2050-2053},
          doi = {10.1111/j.1365-2966.2009.15598.x},
archivePrefix = {arXiv},
       eprint = {0903.2057},
 primaryClass = {astro-ph.GA},
       adsurl = {https://ui.adsabs.harvard.edu/abs/2009MNRAS.400.2050G}
}

@ARTICLE{Lusso+13,
       author = {{Lusso}, E. and {Hennawi}, J.~F. and {Comastri}, A. and {Zamorani}, G. and {Richards}, G.~T. and {Vignali}, C. and {Treister}, E. and {Schawinski}, K. and {Salvato}, M. and {Gilli}, R.},
        title = "{The Obscured Fraction of Active Galactic Nuclei in the XMM-COSMOS Survey: A Spectral Energy Distribution Perspective}",
      journal = {\apj},
         year = 2013,
        month = nov,
       volume = {777},
       number = {2},
          eid = {86},
        pages = {86},
          doi = {10.1088/0004-637X/777/2/86},
archivePrefix = {arXiv},
       eprint = {1309.0814},
 primaryClass = {astro-ph.CO},
       adsurl = {https://ui.adsabs.harvard.edu/abs/2013ApJ...777...86L}
}

@ARTICLE{Maiolino+07,
       author = {{Maiolino}, R. and {Shemmer}, O. and {Imanishi}, M. and {Netzer}, H. and {Oliva}, E. and {Lutz}, D. and {Sturm}, E.},
        title = "{Dust covering factor, silicate emission, and star formation in luminous QSOs}",
      journal = {\aap},
         year = 2007,
        month = jun,
       volume = {468},
       number = {3},
        pages = {979-992},
          doi = {10.1051/0004-6361:20077252},
archivePrefix = {arXiv},
       eprint = {0704.1559},
 primaryClass = {astro-ph},
       adsurl = {https://ui.adsabs.harvard.edu/abs/2007A&A...468..979M}
}

@ARTICLE{Park+26,
       author = {{Park}, Kevin and {Torralba}, Alberto and {Matthee}, Jorryt and {Mascia}, Sara and {Haiman}, Zolt{\'a}n and {Naidu}, Rohan P. and {de Graaff}, Anna},
        title = "{A new sample of Little Red Dots at $z<0.45$ in DESI DR1: Broad Balmer lines, low ionization spectrum and no variability}",
      journal = {arXiv e-prints},
         year = 2026,
        month = may,
          eid = {arXiv:2605.14233},
        pages = {arXiv:2605.14233},
          doi = {10.48550/arXiv.2605.14233},
archivePrefix = {arXiv},
       eprint = {2605.14233},
 primaryClass = {astro-ph.GA},
       adsurl = {https://ui.adsabs.harvard.edu/abs/2026arXiv260514233P}
}

@ARTICLE{Brazzini+25,
       author = {{Brazzini}, Matilde and {D'Eugenio}, Francesco and {Maiolino}, Roberto and {Juod{\v{z}}balis}, Ignas and {Ji}, Xihan and {Scholtz}, Jan and {Chang}, Seok-Jun},
        title = "{Ruling out dominant electron scattering in Little Red Dots' Rosetta Stone using multiple hydrogen lines}",
      journal = {\mnras},
         year = 2025,
        month = nov,
       volume = {544},
       number = {1},
        pages = {L167-L173},
          doi = {10.1093/mnrasl/slaf116},
archivePrefix = {arXiv},
       eprint = {2507.08929},
 primaryClass = {astro-ph.GA},
       adsurl = {https://ui.adsabs.harvard.edu/abs/2025MNRAS.544L.167B}
}

@ARTICLE{Jiang+10,
       author = {{Jiang}, Linhua and {Fan}, Xiaohui and {Brandt}, W.~N. and {Carilli}, Chris L. and {Egami}, Eiichi and {Hines}, Dean C. and {Kurk}, Jaron D. and {Richards}, Gordon T. and {Shen}, Yue and {Strauss}, Michael A. and {Vestergaard}, Marianne and {Walter}, Fabian},
        title = "{Dust-free quasars in the early Universe}",
      journal = {\nat},
         year = 2010,
        month = mar,
       volume = {464},
       number = {7287},
        pages = {380-383},
          doi = {10.1038/nature08877},
archivePrefix = {arXiv},
       eprint = {1003.3432},
 primaryClass = {astro-ph.CO},
       adsurl = {https://ui.adsabs.harvard.edu/abs/2010Natur.464..380J}
}

@ARTICLE{Yanagisawa+26,
       author = {{Yanagisawa}, Hiroto and {Ouchi}, Masami and {Golubchik}, Miriam and {Oguri}, Masamune and {Fujimoto}, Seiji and {Kokorev}, Vasily and {Brammer}, Gabriel and {Sun}, Fengwu and {Nakane}, Minami and {Harikane}, Yuichi and {Umeda}, Hiroya and {Akins}, Hollis B. and {Atek}, Hakim and {Bauer}, Franz E. and {Brada{\v{c}}}, Maru{\v{s}}a and {Chisholm}, John and {Coe}, Dan and {Diego}, Jose M. and {Ferguson}, Henry C. and {Finkelstein}, Steven L. and {Furtak}, Lukas J. and {Inayoshi}, Kohei and {Koekemoer}, Anton M. and {Matthee}, Jorryt and {Naidu}, Rohan P. and {Ono}, Yoshiaki and {Pan}, Richard and {Richard}, Johan and {Robbins}, Luke and {Willott}, Chris and {Zitrin}, Adi and {Amor{\'\i}n}, Ricardo O. and {Bradley}, Larry D. and {Bromm}, Volker and {Conselice}, Christopher J. and {Dayal}, Pratika and {Kartaltepe}, Jeyhan S. and {Lopes}, Paulo A.~A. and {Lucas}, Ray A. and {Magdis}, Georgios E. and {Martis}, Nicholas S. and {Papovich}, Casey and {Schaerer}, Daniel and {Valentino}, Francesco and {Vanzella}, Eros and {Allingham}, Joseph F.~V. and {Grogin}, Norman A. and {Gonz{\'a}lez-Otero}, Mauro and {Ricotti}, Massimo and {Windhorst}, Rogier A.},
        title = "{VENUS: Two Faint Little Red Dots Separated by $\sim70\,\mathrm{pc}$ Hidden in a Single Lensed Galaxy at $z\sim7$}",
      journal = {arXiv e-prints},
         year = 2026,
        month = jan,
          eid = {arXiv:2601.06015},
        pages = {arXiv:2601.06015},
          doi = {10.48550/arXiv.2601.06015},
archivePrefix = {arXiv},
       eprint = {2601.06015},
 primaryClass = {astro-ph.GA},
       adsurl = {https://ui.adsabs.harvard.edu/abs/2026arXiv260106015Y}
}

@ARTICLE{Nenkova+08b,
       author = {{Nenkova}, Maia and {Sirocky}, Matthew M. and {Nikutta}, Robert and {Ivezi{\'c}}, {\v{Z}}eljko and {Elitzur}, Moshe},
        title = "{AGN Dusty Tori. II. Observational Implications of Clumpiness}",
      journal = {\apj},
         year = 2008,
        month = sep,
       volume = {685},
       number = {1},
        pages = {160-180},
          doi = {10.1086/590483},
archivePrefix = {arXiv},
       eprint = {0806.0512},
 primaryClass = {astro-ph},
       adsurl = {https://ui.adsabs.harvard.edu/abs/2008ApJ...685..160N}
}

@ARTICLE{Hoenig+10,
       author = {{H{\"o}nig}, S.~F. and {Kishimoto}, M.},
        title = "{The dusty heart of nearby active galaxies. II. From clumpy torus models to physical properties of dust around AGN}",
      journal = {\aap},
         year = 2010,
        month = nov,
       volume = {523},
          eid = {A27},
        pages = {A27},
          doi = {10.1051/0004-6361/200912676},
archivePrefix = {arXiv},
       eprint = {0909.4539},
 primaryClass = {astro-ph.CO},
       adsurl = {https://ui.adsabs.harvard.edu/abs/2010A&A...523A..27H}
}

@ARTICLE{Kokorev+26,
       author = {{Kokorev}, V. and {Chisholm}, J. and {Naidu}, R.~P. and {Gieles}, M. and {Finkelstein}, S. and {Berg}, D. and {Akins}, H. and {Taylor}, A. and {Fujimoto}, S. and {Furtak}, L.~J. and et al.},
        title = "{The Ashes of Supermassive Stars: Globular Cluster-like Aluminum Enhancement in Little Red Dots}",
      journal = {arXiv e-prints},
         year = 2026,
        month = sep,
          eid = {arXiv:2609.09271},
        pages = {arXiv:2609.09271},
          doi = {10.48550/arXiv.2609.09271},
archivePrefix = {arXiv},
       eprint = {2609.09271},
 primaryClass = {astro-ph.GA},
       adsurl = {https://ui.adsabs.harvard.edu/abs/2026arXiv260909271K}
}

@ARTICLE{Naidu+26,
       author = {{Naidu}, Rohan P. and {Matthee}, Jorryt and {de Graaff}, Anna and {Torralba}, Alberto and {Ashall}, Chris and {Katz}, Harley and {Chisholm}, John and {Brammer}, Gabriel and {Dessart}, Luc and {Eilers}, Anna-Christina and {Hviding}, Raphael E. and {Jones}, David O. and {Kokorev}, Vasily and {Leja}, Joel and {Liu}, Hanpu and {Liu}, Zhaoran and {Nandal}, Devesh and {Oesch}, Pascal A. and {Ransome}, Conor L. and {Simcoe}, Robert A. and {Sun}, Wendy Q. and {Weibel}, Andrea and {Xiao}, Mengyuan},
        title = "{Little Red Dots as Intermediate Mass, Super-Eddington Engines: Insights from Type IIn Supernovae and The 1837-1856 Great Eruption of $η$ Carinae}",
      journal = {arXiv e-prints},
         year = 2026,
        month = jun,
          eid = {arXiv:2606.30711},
        pages = {arXiv:2606.30711},
          doi = {10.48550/arXiv.2606.30711},
archivePrefix = {arXiv},
       eprint = {2606.30711},
 primaryClass = {astro-ph.GA},
       adsurl = {https://ui.adsabs.harvard.edu/abs/2026arXiv260630711N}
}

@ARTICLE{Kocevski+23,
       author = {{Kocevski}, Dale D. and {Onoue}, Masafusa and {Inayoshi}, Kohei and {Trump}, Jonathan R. and {Arrabal Haro}, Pablo and {Grazian}, Andrea and {Dickinson}, Mark and {Finkelstein}, Steven L. and {Kartaltepe}, Jeyhan S. and {Hirschmann}, Michaela and et al.},
        title = "{Hidden Little Monsters: Spectroscopic Identification of Low-mass, Broad-line AGNs at z > 5 with CEERS}",
      journal = {\apjl},
         year = 2023,
        month = sep,
       volume = {954},
       number = {1},
          eid = {L4},
        pages = {L4},
          doi = {10.3847/2041-8213/ace5a0},
archivePrefix = {arXiv},
       eprint = {2302.00012},
 primaryClass = {astro-ph.GA},
       adsurl = {https://ui.adsabs.harvard.edu/abs/2023ApJ...954L...4K}
}

@ARTICLE{Hoenig19,
       author = {{H{\"o}nig}, Sebastian F.},
        title = "{Redefining the Torus: A Unifying View of AGNs in the Infrared and Submillimeter}",
      journal = {\apj},
         year = 2019,
        month = oct,
       volume = {884},
       number = {2},
          eid = {171},
        pages = {171},
          doi = {10.3847/1538-4357/ab4591},
archivePrefix = {arXiv},
       eprint = {1909.08639},
 primaryClass = {astro-ph.GA},
       adsurl = {https://ui.adsabs.harvard.edu/abs/2019ApJ...884..171H}
}

@ARTICLE{Hao+10,
       author = {{Hao}, Heng and {Elvis}, Martin and {Civano}, Francesca and {Lanzuisi}, Giorgio and {Brusa}, Marcella and {Lusso}, Elisabeta and {Zamorani}, Gianni and {Comastri}, Andrea and {Bongiorno}, Angela and {Impey}, Chris D. and {Koekemoer}, Anton M. and {Le Floc'h}, Emeric and {Salvato}, Mara and {Sanders}, David and {Trump}, Jonathan R. and {Vignali}, Cristian},
        title = "{Hot-dust-poor Type 1 Active Galactic Nuclei in the COSMOS Survey}",
      journal = {\apjl},
         year = 2010,
        month = nov,
       volume = {724},
       number = {1},
        pages = {L59-L63},
          doi = {10.1088/2041-8205/724/1/L59},
archivePrefix = {arXiv},
       eprint = {1009.3276},
 primaryClass = {astro-ph.CO},
       adsurl = {https://ui.adsabs.harvard.edu/abs/2010ApJ...724L..59H}
}

@ARTICLE{Lyu+17,
       author = {{Lyu}, Jianwei and {Rieke}, George H.},
        title = "{The Intrinsic Far-infrared Continua of Type-1 Quasars}",
      journal = {\apj},
         year = 2017,
        month = jun,
       volume = {841},
       number = {2},
          eid = {76},
        pages = {76},
          doi = {10.3847/1538-4357/aa7051},
archivePrefix = {arXiv},
       eprint = {1704.06987},
 primaryClass = {astro-ph.GA},
       adsurl = {https://ui.adsabs.harvard.edu/abs/2017ApJ...841...76L}
}

@ARTICLE{BoylanKolchin23,
       author = {{Boylan-Kolchin}, Michael},
        title = "{Stress testing {\ensuremath{\Lambda}}CDM with high-redshift galaxy candidates}",
      journal = {Nature Astronomy},
         year = 2023,
        month = jun,
       volume = {7},
        pages = {731-735},
          doi = {10.1038/s41550-023-01937-7},
archivePrefix = {arXiv},
       eprint = {2208.01611},
 primaryClass = {astro-ph.CO},
       adsurl = {https://ui.adsabs.harvard.edu/abs/2023NatAs...7..731B}
}

@ARTICLE{Scholtz+26,
       author = {{Scholtz}, J. and {D'Eugenio}, F. and {Maiolino}, R. and {Brazzini}, M. and {{\"U}bler}, H. and {Ji}, X. and {Perna}, M. and {Sun}, F. and {Brocchi}, G. and {Carniani}, S. and {Cresci}, G. and {Ivey}, L.~R. and {Juod{\v{z}}balis}, I. and {Marconi}, A. and {Mazzolari}, G. and {Risaliti}, G. and {Trefoloni}, B.},
        title = "{Little Red and Blue Dots: simply stratified Broad Line Regions}",
      journal = {arXiv e-prints},
         year = 2026,
        month = mar,
          eid = {arXiv:2603.22277},
        pages = {arXiv:2603.22277},
          doi = {10.48550/arXiv.2603.22277},
archivePrefix = {arXiv},
       eprint = {2603.22277},
 primaryClass = {astro-ph.GA},
       adsurl = {https://ui.adsabs.harvard.edu/abs/2026arXiv260322277S}
}

@ARTICLE{Sun+26,
       author = {{Sun}, Wendy Q. and {Naidu}, Rohan P. and {Matthee}, Jorryt and {de Graaff}, Anna and {Chisholm}, John and {Greene}, Jenny E. and {Oesch}, Pascal A. and {Torralba}, Alberto and {Hviding}, Raphael E. and {Brammer}, Gabriel and {Simcoe}, Robert A. and {Bose}, Sownak and {Bouwens}, Rychard and {Dayal}, Pratika and {Eilers}, Anna-Christina and {Fei}, Qinyue and {Furtak}, Lukas J. and {Gottumukkala}, Rashmi and {Goulding}, Andy and {Heintz}, Kasper E. and {Hirschmann}, Michaela and {Kokorev}, Vasily and {Leja}, Joel and {Liu}, Zhaoran and {Natarajan}, Priyamvada and {Santarelli}, Andrew D. and {Setton}, David J. and {Smith}, Aaron and {Tacchella}, Sandro and {Volonteri}, Marta and {Walter}, Fabian and {Weibel}, Andrea and {Williams}, Christina C.},
        title = "{Little Red Dot {\ensuremath{-}} Host Galaxy = Black Hole Star: A Gas-Enshrouded Heart at the Center of Every Little Red Dot}",
      journal = {The Open Journal of Astrophysics},
         year = 2026,
        month = may,
       volume = {9},
        pages = {62505},
          doi = {10.33232/001c.162505},
archivePrefix = {arXiv},
       eprint = {2601.20929},
 primaryClass = {astro-ph.GA},
       adsurl = {https://ui.adsabs.harvard.edu/abs/2026OJAp....962505S}
}

@ARTICLE{MadauMaiolino26,
       author = {{Madau}, Piero and {Maiolino}, Roberto},
        title = "{Little red dots as obscured little blue dots: A super-Eddington unification model}",
      journal = {\aap},
         year = 2026,
        month = oct,
       volume = {714},
          eid = {A276},
        pages = {A276},
          doi = {10.1051/0004-6361/202659606},
       adsurl = {https://ui.adsabs.harvard.edu/abs/2026A&A...714A.276M}
}

@ARTICLE{Ubler+26,
       author = {{{\"U}bler}, Hannah},
        title = "{Massive black holes in the first billion years with JWST}",
      journal = {Nature Astronomy},
         year = 2026,
        month = sep,
       volume = {10},
       number = {9},
        pages = {1273-1282},
          doi = {10.1038/s41550-026-02966-8},
       adsurl = {https://ui.adsabs.harvard.edu/abs/2026NatAs..10.1273U}
}

@ARTICLE{Wang+25a,
       author = {{Wang}, Bingjie and {de Graaff}, Anna and {Davies}, Rebecca L. and {Greene}, Jenny E. and {Leja}, Joel and {Brammer}, Gabriel B. and {Goulding}, Andy D. and {Miller}, Tim B. and {Suess}, Katherine A. and {Weibel}, Andrea and {Williams}, Christina C. and {Bezanson}, Rachel and {Boogaard}, Leindert A. and {Cleri}, Nikko J. and {Hirschmann}, Michaela and {Katz}, Harley and {Labb{\'e}}, Ivo and {Maseda}, Michael V. and {Matthee}, Jorryt and {McConachie}, Ian and {Naidu}, Rohan P. and {Oesch}, Pascal A. and {Rix}, Hans-Walter and {Setton}, David J. and {Whitaker}, Katherine E.},
        title = "{RUBIES: JWST/NIRSpec Confirmation of an Infrared-luminous, Broad-line Little Red Dot with an Ionized Outflow}",
      journal = {\apj},
         year = 2025,
        month = may,
       volume = {984},
       number = {2},
          eid = {121},
        pages = {121},
          doi = {10.3847/1538-4357/adc1ca},
archivePrefix = {arXiv},
       eprint = {2403.02304},
 primaryClass = {astro-ph.GA},
       adsurl = {https://ui.adsabs.harvard.edu/abs/2025ApJ...984..121W}
}

@ARTICLE{Lin+26,
       author = {{Lin}, Xiaojing and {Fan}, Xiaohui and {Cai}, Zheng and {Liu}, Yichen and {Sun}, Fengwu and {Bian}, Fuyan and {Li}, Mingyu and {Mao}, Junjie and {Greene}, Jenny E. and {Liu}, Hanpu and {Li}, Jiaxuan and {Liu}, Weizhe and {Ma}, Yilun and {Sun}, Zechang and {Zhang}, Zijian},
        title = "{(LRDs)$^2$: The Low-ReDshift Little Red Dots Survey. II. DESI DR1 Sample}",
      journal = {arXiv e-prints},
         year = 2026,
        month = may,
          eid = {arXiv:2605.21574},
        pages = {arXiv:2605.21574},
          doi = {10.48550/arXiv.2605.21574},
archivePrefix = {arXiv},
       eprint = {2605.21574},
 primaryClass = {astro-ph.GA},
       adsurl = {https://ui.adsabs.harvard.edu/abs/2026arXiv260521574L}
}

@ARTICLE{Torralba+26b,
       author = {{Torralba}, Alberto and {Matthee}, Jorryt and {Pezzulli}, Gabriele and {Naidu}, Rohan P. and {Ishikawa}, Yuzo and {Brammer}, Gabriel B. and {Chang}, Seok-Jun and {Chisholm}, John and {de Graaff}, Anna and {D'Eugenio}, Francesco and {Di Cesare}, Claudia and {Eilers}, Anna-Christina and {Greene}, Jenny E. and {Gronke}, Max and {Iani}, Edoardo and {Kokorev}, Vasily and {Kotiwale}, Gauri and {Kramarenko}, Ivan and {Ma}, Yilun and {Mascia}, Sara and {Navarrete}, Benjam{\'\i}n and {Nelson}, Erica and {Oesch}, Pascal and {Simcoe}, Robert A. and {Wuyts}, Stijn},
        title = "{The warm outer layer of a little red dot as the source of [Fe II] and collisional Balmer lines with scattering wings}",
      journal = {\aap},
         year = 2026,
        month = feb,
       volume = {707},
          eid = {A75},
        pages = {A75},
          doi = {10.1051/0004-6361/202557537},
archivePrefix = {arXiv},
       eprint = {2510.00103},
 primaryClass = {astro-ph.GA},
       adsurl = {https://ui.adsabs.harvard.edu/abs/2026A&A...707A..75T}
}

@ARTICLE{Deugenio+26a,
       author = {{D'Eugenio}, Francesco and {Juod{\v{z}}balis}, Ignas and {Ji}, Xihan and {Scholtz}, Jan and {Maiolino}, Roberto and {Carniani}, Stefano and {Perna}, Michele and {Mazzolari}, Giovanni and {{\"U}bler}, Hannah and {Arribas}, Santiago and {Bhatawdekar}, Rachana and {Bunker}, Andrew J. and {Cresci}, Giovanni and {Curtis-Lake}, Emma and {Hainline}, Kevin and {Inayoshi}, Kohei and {Isobe}, Yuki and {Ji}, Zhiyuan and {Johnson}, Benjamin D. and {Jones}, Gareth C. and {Looser}, Tobias J. and {Nelson}, Erica J. and {Parlanti}, Eleonora and {Pusk{\'a}s}, D{\'a}vid and {Rinaldi}, Pierluigi and {Robertson}, Brant and {Rodr{\'\i}guez Del Pino}, Bruno and {Shivaei}, Irene and {Sun}, Fengwu and {Tacchella}, Sandro and {Venturi}, Giacomo and {Volonteri}, Marta and {Williams}, Christina C. and {Willmer}, Christopher N.~A. and {Willott}, Chris and {Witstok}, Joris},
        title = "{JADES and BlackTHUNDER: rest-frame Balmer-line absorption and the local environment in a Little Red Dot at z = 5}",
      journal = {\mnras},
         year = 2026,
        month = jan,
       volume = {545},
       number = {3},
          eid = {staf2117},
        pages = {staf2117},
          doi = {10.1093/mnras/staf2117},
archivePrefix = {arXiv},
       eprint = {2506.14870},
 primaryClass = {astro-ph.GA},
       adsurl = {https://ui.adsabs.harvard.edu/abs/2026MNRAS.545f2117D}
}

@ARTICLE{deGraaff+26,
       author = {{de Graaff}, Anna and {Hviding}, Raphael E. and {Naidu}, Rohan P. and {Greene}, Jenny E. and {Miller}, Tim B. and {Leja}, Joel and {Matthee}, Jorryt and {Brammer}, Gabriel and {Katz}, Harley and {Bezanson}, Rachel and {Boogaard}, Leindert A. and {Bose}, Sownak and {Chisholm}, John and {Cleri}, Nikko J. and {Dayal}, Pratika and {Feldmann}, Robert and {Fudamoto}, Yoshinobu and {Fujimoto}, Seiji and {Furtak}, Lukas J. and {Glazebrook}, Karl and {Gottumukkala}, Rashmi and {Heintz}, Kasper E. and {Kokorev}, Vasily and {Labbe}, Ivo and {Maseda}, Michael V. and {McConachie}, Ian and {Nanayakkara}, Themiya and {Nelson}, Erica and {Nowaczyk}, Przemys{\l}aw and {Oesch}, Pascal A. and {Rix}, Hans-Walter and {Setton}, David J. and {Torralba}, Alberto and {Walter}, Fabian and {Wang}, Bingjie and {Weibel}, Andrea and {van der Wel}, Arjen},
        title = "{Little red dots host black hole stars: a unified family of gas-reddened AGN revealed by JWST/NIRSpec spectroscopy}",
      journal = {\mnras},
         year = 2026,
        month = sep,
       volume = {551},
       number = {3},
          eid = {stag1567},
        pages = {stag1567},
          doi = {10.1093/mnras/stag1567},
archivePrefix = {arXiv},
       eprint = {2511.21820},
 primaryClass = {astro-ph.GA},
       adsurl = {https://ui.adsabs.harvard.edu/abs/2026MNRAS.551g1567D}
}

@article{Calzetti+00,
  author = {{Calzetti}, Daniela and {Armus}, Lee and {Bohlin}, Ralph C. and {Kinney}, Anne L. and {Koornneef}, Jan and {Storchi-Bergmann}, Thaisa},
  title = {{The Dust Content and Opacity of Actively Star-forming Galaxies}},
  journal = {\apj},
  year = {2000},
  month = apr,
  volume = {533},
  number = {2},
  pages = {682-695},
  doi = {10.1086/308692},
  archiveprefix = {arXiv},
  eprint = {astro-ph/9911459},
  primaryclass = {astro-ph},
  adsurl = {https://ui.adsabs.harvard.edu/abs/2000ApJ...533..682C},
}

@ARTICLE{MunozArancibia+23,
       author = {{Mu{\~n}oz Arancibia}, A.~M. and {Gonz{\'a}lez-L{\'o}pez}, J. and {Ibar}, E. and {Bauer}, F.~E. and {Anguita}, T. and {Aravena}, M. and {Demarco}, R. and {Kneissl}, R. and {Koekemoer}, A.~M. and {Troncoso-Iribarren}, P. and {Zitrin}, A.},
        title = "{The ALMA Frontier Fields Survey. VI. Lensing-corrected 1.1 mm number counts in Abell 2744, MACSJ0416.1-2403, MACSJ1149.5+2223, Abell 370, and Abell S1063}",
      journal = {\aap},
         year = 2023,
        month = jul,
       volume = {675},
          eid = {A85},
        pages = {A85},
          doi = {10.1051/0004-6361/202243528},
archivePrefix = {arXiv},
       eprint = {2203.06195},
 primaryClass = {astro-ph.GA},
       adsurl = {https://ui.adsabs.harvard.edu/abs/2023A&A...675A..85M}
}

@ARTICLE{Fujimoto+24,
       author = {{Fujimoto}, Seiji and {Kohno}, Kotaro and {Ouchi}, Masami and {Oguri}, Masamune and {Kokorev}, Vasily and {Brammer}, Gabriel and {Sun}, Fengwu and {Gonz{\'a}lez-L{\'o}pez}, Jorge and {Bauer}, Franz E. and {Caminha}, Gabriel B. and {Hatsukade}, Bunyo and {Richard}, Johan and {Smail}, Ian and {Tsujita}, Akiyoshi and {Ueda}, Yoshihiro and {Uematsu}, Ryosuke and {Zitrin}, Adi and {Coe}, Dan and {Kneib}, Jean-Paul and {Postman}, Marc and {Umetsu}, Keiichi and {Lagos}, Claudia del P. and {Popping}, Gerg{\"o} and {Ao}, Yiping and {Bradley}, Larry and {Caputi}, Karina and {Dessauges-Zavadsky}, Miroslava and {Egami}, Eiichi and {Espada}, Daniel and {Ivison}, R.~J. and {Jauzac}, Mathilde and {Knudsen}, Kirsten K. and {Koekemoer}, Anton M. and {Magdis}, Georgios E. and {Mahler}, Guillaume and {Mu{\~n}oz Arancibia}, A.~M. and {Rawle}, Timothy and {Shimasaku}, Kazuhiro and {Toft}, Sune and {Umehata}, Hideki and {Valentino}, Francesco and {Wang}, Tao and {Wang}, Wei-Hao},
        title = "{ALMA Lensing Cluster Survey: Deep 1.2 mm Number Counts and Infrared Luminosity Functions at z ≃ 1─8}",
      journal = {\apjs},
         year = 2024,
        month = dec,
       volume = {275},
       number = {2},
          eid = {36},
        pages = {36},
          doi = {10.3847/1538-4365/ad5ae2},
archivePrefix = {arXiv},
       eprint = {2303.01658},
 primaryClass = {astro-ph.GA},
       adsurl = {https://ui.adsabs.harvard.edu/abs/2024ApJS..275...36F}
}

@ARTICLE{Fujimoto+25,
       author = {{Fujimoto}, Seiji and {Bezanson}, Rachel and {Labbe}, Ivo and {Brammer}, Gabriel and {Price}, Sedona H. and {Wang}, Bingjie and {Weaver}, John R. and {Fudamoto}, Yoshinobu and {Oesch}, Pascal A. and {Williams}, Christina C. and {Dayal}, Pratika and {Feldmann}, Robert and {Greene}, Jenny E. and {Leja}, Joel and {Whitaker}, Katherine E. and {Zitrin}, Adi and {Cutler}, Sam E. and {Furtak}, Lukas J. and {Pan}, Richard and {Chemerynska}, Iryna and {Kokorev}, Vasily and {Miller}, Tim B. and {Atek}, Hakim and {van Dokkum}, Pieter and {Juneau}, St{\'e}phanie and {Kassin}, Susan and {Khullar}, Gourav and {Marchesini}, Danilo and {Maseda}, Michael and {Nelson}, Erica J. and {Setton}, David J. and {Smit}, Renske},
        title = "{DUALZ{\textemdash}Deep UNCOVER-ALMA Legacy High-Z Survey}",
      journal = {\apjs},
         year = 2025,
        month = jun,
       volume = {278},
       number = {2},
          eid = {45},
        pages = {45},
          doi = {10.3847/1538-4365/adc677},
archivePrefix = {arXiv},
       eprint = {2309.07834},
 primaryClass = {astro-ph.GA},
       adsurl = {https://ui.adsabs.harvard.edu/abs/2025ApJS..278...45F}
}

@ARTICLE{Zhang+26,
       author = {{Zhang}, Saiyang and {Liu}, Boyuan and {Bromm}, Volker and {K{\"u}hnel}, Florian},
        title = "{Primordial Black Holes as Seeds for Extremely Overmassive Active Galactic Nuclei Observed by JWST}",
      journal = {\apjl},
         year = 2026,
        month = mar,
       volume = {1000},
       number = {1},
          eid = {L19},
        pages = {L19},
          doi = {10.3847/2041-8213/ae4bd0},
archivePrefix = {arXiv},
       eprint = {2512.14066},
 primaryClass = {astro-ph.GA},
       adsurl = {https://ui.adsabs.harvard.edu/abs/2026ApJ..1000L..19Z}
}

@ARTICLE{Yue+24,
       author = {{Yue}, Minghao and {Eilers}, Anna-Christina and {Ananna}, Tonima Tasnim and {Panagiotou}, Christos and {Kara}, Erin and {Miyaji}, Takamitsu},
        title = "{Stacking X-Ray Observations of ``Little Red Dots'': Implications for Their Active Galactic Nucleus Properties}",
      journal = {\apjl},
         year = 2024,
        month = oct,
       volume = {974},
       number = {2},
          eid = {L26},
        pages = {L26},
          doi = {10.3847/2041-8213/ad7eba},
archivePrefix = {arXiv},
       eprint = {2404.13290},
 primaryClass = {astro-ph.GA},
       adsurl = {https://ui.adsabs.harvard.edu/abs/2024ApJ...974L..26Y}
}

@ARTICLE{Ananna+24,
       author = {{Ananna}, Tonima Tasnim and {Bogd{\'a}n}, {\'A}kos and {Kov{\'a}cs}, Orsolya E. and {Natarajan}, Priyamvada and {Hickox}, Ryan C.},
        title = "{X-Ray View of Little Red Dots: Do They Host Supermassive Black Holes?}",
      journal = {\apjl},
         year = 2024,
        month = jul,
       volume = {969},
       number = {1},
          eid = {L18},
        pages = {L18},
          doi = {10.3847/2041-8213/ad5669},
archivePrefix = {arXiv},
       eprint = {2404.19010},
 primaryClass = {astro-ph.GA},
       adsurl = {https://ui.adsabs.harvard.edu/abs/2024ApJ...969L..18A}
}

@ARTICLE{Hviding+25,
       author = {{Hviding}, Raphael E. and {de Graaff}, Anna and {Miller}, Tim B. and {Setton}, David J. and {Greene}, Jenny E. and {Labb{\'e}}, Ivo and {Brammer}, Gabriel and {Bezanson}, Rachel and {Boogaard}, Leindert A. and {Cleri}, Nikko J. and {Leja}, Joel and {Maseda}, Michael V. and {McConachie}, Ian and {Matthee}, Jorryt and {Naidu}, Rohan P. and {Oesch}, Pascal A. and {Wang}, Bingjie and {Whitaker}, Katherine E. and {Williams}, Christina C.},
        title = "{RUBIES: A spectroscopic census of little red dots: All point sources with v-shaped continua have broad lines}",
      journal = {\aap},
         year = 2025,
        month = oct,
       volume = {702},
          eid = {A57},
        pages = {A57},
          doi = {10.1051/0004-6361/202555816},
archivePrefix = {arXiv},
       eprint = {2506.05459},
 primaryClass = {astro-ph.GA},
       adsurl = {https://ui.adsabs.harvard.edu/abs/2025A&A...702A..57H}
}

@ARTICLE{Matthee+24a,
       author = {{Matthee}, Jorryt and {Naidu}, Rohan P. and {Brammer}, Gabriel and {Chisholm}, John and {Eilers}, Anna-Christina and {Goulding}, Andy and {Greene}, Jenny and {Kashino}, Daichi and {Labbe}, Ivo and {Lilly}, Simon J. and {Mackenzie}, Ruari and {Oesch}, Pascal A. and {Weibel}, Andrea and {Wuyts}, Stijn and {Xiao}, Mengyuan and {Bordoloi}, Rongmon and {Bouwens}, Rychard and {van Dokkum}, Pieter and {Illingworth}, Garth and {Kramarenko}, Ivan and {Maseda}, Michael V. and {Mason}, Charlotte and {Meyer}, Romain A. and {Nelson}, Erica J. and {Reddy}, Naveen A. and {Shivaei}, Irene and {Simcoe}, Robert A. and {Yue}, Minghao},
        title = "{Little Red Dots: An Abundant Population of Faint Active Galactic Nuclei at z {\ensuremath{\sim}} 5 Revealed by the EIGER and FRESCO JWST Surveys}",
      journal = {\apj},
         year = 2024,
        month = mar,
       volume = {963},
       number = {2},
          eid = {129},
        pages = {129},
          doi = {10.3847/1538-4357/ad2345},
archivePrefix = {arXiv},
       eprint = {2306.05448},
 primaryClass = {astro-ph.GA},
       adsurl = {https://ui.adsabs.harvard.edu/abs/2024ApJ...963..129M}
}

@ARTICLE{Kokorev+24b,
       author = {{Kokorev}, Vasily and {Caputi}, Karina I. and {Greene}, Jenny E. and {Dayal}, Pratika and {Trebitsch}, Maxime and {Cutler}, Sam E. and {Fujimoto}, Seiji and {Labb{\'e}}, Ivo and {Miller}, Tim B. and {Iani}, Edoardo and {Navarro-Carrera}, Rafael and {Rinaldi}, Pierluigi},
        title = "{A Census of Photometrically Selected Little Red Dots at 4 < z < 9 in JWST Blank Fields}",
      journal = {\apj},
         year = 2024,
        month = jun,
       volume = {968},
       number = {1},
          eid = {38},
        pages = {38},
          doi = {10.3847/1538-4357/ad4265},
archivePrefix = {arXiv},
       eprint = {2401.09981},
 primaryClass = {astro-ph.GA},
       adsurl = {https://ui.adsabs.harvard.edu/abs/2024ApJ...968...38K}
}

@ARTICLE{Oke+83,
       author = {{Oke}, J.~B. and {Gunn}, J.~E.},
        title = "{Secondary standard stars for absolute spectrophotometry.}",
      journal = {\apj},
         year = 1983,
        month = mar,
       volume = {266},
        pages = {713-717},
          doi = {10.1086/160817},
       adsurl = {https://ui.adsabs.harvard.edu/abs/1983ApJ...266..713O}
}

@ARTICLE{Chabrier+03,
       author = {{Chabrier}, Gilles},
        title = "{Galactic Stellar and Substellar Initial Mass Function}",
      journal = {\pasp},
         year = 2003,
        month = jul,
       volume = {115},
       number = {809},
        pages = {763-795},
          doi = {10.1086/376392},
archivePrefix = {arXiv},
       eprint = {astro-ph/0304382},
 primaryClass = {astro-ph},
       adsurl = {https://ui.adsabs.harvard.edu/abs/2003PASP..115..763C}
}

@ARTICLE{Matthee+26,
       author = {{Matthee}, Jorryt and {Torralba}, Alberto and {Pezzulli}, Gabriele and {Naidu}, Rohan P. and {Chisholm}, John and {Mascia}, Sara and {Greene}, Jenny E. and {Ishikawa}, Yuzo and {Gronke}, Max and {Wuyts}, Stijn and {Bordoloi}, Rongmon and {Brammer}, Gabriel and {Chang}, Seok-Jun and {Eilers}, Anna-Christina and {de Graaff}, Anna and {Hviding}, Raphael E. and {Iani}, Edoardo and {Illingworth}, Garth and {Kashino}, Daichi and {Labbe}, Ivo and {Ma}, Yilun and {Maseda}, Michael V. and {Meyer}, Romain and {Nelson}, Erica and {Oesch}, Pascal and {Xiao}, Mengyuan},
        title = "{The Engine and its Flows: Little Red Dot spectra are shaped by the column densities of their gas envelopes}",
      journal = {arXiv e-prints},
         year = 2026,
        month = mar,
          eid = {arXiv:2603.17667},
        pages = {arXiv:2603.17667},
          doi = {10.48550/arXiv.2603.17667},
archivePrefix = {arXiv},
       eprint = {2603.17667},
 primaryClass = {astro-ph.GA},
       adsurl = {https://ui.adsabs.harvard.edu/abs/2026arXiv260317667M}
}

@ARTICLE{Ji+26c,
       author = {{Ji}, Xihan and {Pezzulli}, Gabriele and {D'Eugenio}, Francesco and {Maiolino}, Roberto and {Carniani}, Stefano and {Tacchella}, Sandro and {Jones}, Gareth and {Smith}, Aaron and {Witstok}, Joris and {Fabian}, Andrew C. and {Geris}, Sophia and {Harshan}, Anishya and {Isobe}, Yuki and {Ivey}, Lucy R. and {Juod{\v{z}}balis}, Ignas and {Pascalau}, Robert and {Scholtz}, Jan and {Witten}, Callum},
        title = "{Holes in the BH$^\star$? AGN signatures in the FUV spectrum of a black-hole dominated Little Red Dot at $z=7.04$}",
      journal = {arXiv e-prints},
         year = 2026,
        month = apr,
          eid = {arXiv:2604.03370},
        pages = {arXiv:2604.03370},
          doi = {10.48550/arXiv.2604.03370},
archivePrefix = {arXiv},
       eprint = {2604.03370},
 primaryClass = {astro-ph.GA},
       adsurl = {https://ui.adsabs.harvard.edu/abs/2026arXiv260403370J}
}

@ARTICLE{Ji+25,
       author = {{Ji}, Xihan and {Maiolino}, Roberto and {{\"U}bler}, Hannah and {Scholtz}, Jan and {D'Eugenio}, Francesco and {Sun}, Fengwu and {Perna}, Michele and {Turner}, Hannah and {Carniani}, Stefano and {Arribas}, Santiago and {Bennett}, Jake S. and {Bunker}, Andrew and {Charlot}, St{\'e}phane and {Cresci}, Giovanni and {Curti}, Mirko and {Egami}, Eiichi and {Fabian}, Andy and {Inayoshi}, Kohei and {Isobe}, Yuki and {Jones}, Gareth and {Juod{\v{z}}balis}, Ignas and {Kumari}, Nimisha and {Lyu}, Jianwei and {Mazzolari}, Giovanni and {Parlanti}, Eleonora and {Robertson}, Brant and {Rodr{\'\i}guez Del Pino}, Bruno and {Schneider}, Raffaella and {Sijacki}, Debora and {Tacchella}, Sandro and {Trinca}, Alessandro and {Valiante}, Rosa and {Venturi}, Giacomo and {Volonteri}, Marta and {Willott}, Chris and {Witten}, Callum and {Witstok}, Joris},
        title = "{BlackTHUNDER ─ A non-stellar Balmer break in a black hole-dominated little red dot at z = 7.04}",
      journal = {\mnras},
         year = 2025,
        month = dec,
       volume = {544},
       number = {4},
        pages = {3900-3935},
          doi = {10.1093/mnras/staf1867},
archivePrefix = {arXiv},
       eprint = {2501.13082},
 primaryClass = {astro-ph.GA},
       adsurl = {https://ui.adsabs.harvard.edu/abs/2025MNRAS.544.3900J}
}

@ARTICLE{Deugenio+26b,
       author = {{D'Eugenio}, Francesco and {Maiolino}, Roberto and {Perna}, Michele and {{\"U}bler}, Hannah and {Ji}, Xihan and {McClymont}, William and {Koudmani}, Sophie and {Sijacki}, Debora and {Juod{\v{z}}balis}, Ignas and {Scholtz}, Jan and {Bennett}, Jake S. and {Bunker}, Andrew J. and {Carniani}, Stefano and {Charlot}, St{\'e}phane and {Cresci}, Giovanni and {Curtis-Lake}, Emma and {Dalla Bont{\`a}}, Elena and {Inayoshi}, Kohei and {Jones}, Gareth C. and {Lyu}, Jianwei and {Marconi}, Alessandro and {Mazzolari}, Giovanni and {Nelson}, Erica J. and {Parlanti}, Eleonora and {Robertson}, Brant E. and {Schneider}, Raffaella and {Simmonds}, Charlotte and {Tacchella}, Sandro and {Venturi}, Giacomo and {Willott}, Chris and {Witstok}, Joris and {Witten}, Callum},
        title = "{BlackTHUNDER strikes twice: Balmer-line absorption in an overmassive Little Red Dot at z = 7.04}",
      journal = {\mnras},
         year = 2026,
        month = apr,
       volume = {547},
       number = {4},
          eid = {stag401},
        pages = {stag401},
          doi = {10.1093/mnras/stag401},
archivePrefix = {arXiv},
       eprint = {2503.11752},
 primaryClass = {astro-ph.GA},
       adsurl = {https://ui.adsabs.harvard.edu/abs/2026MNRAS.547ag401D}
}

@ARTICLE{Maiolino+26b,
       author = {{Maiolino}, Roberto and {{\"U}bler}, Hannah and {D'Eugenio}, Francesco and {Scholtz}, Jan and {Juod{\v{z}}balis}, Ignas and {Ji}, Xihan and {Perna}, Michele and {Bromm}, Volker and {Dayal}, Pratika and {Koudmani}, Sophie and {Liu}, Boyuan and {Schneider}, Raffaella and {Sijacki}, Debora and {Valiante}, Rosa and {Trinca}, Alessandro and {Zhang}, Saiyang and {Volonteri}, Marta and {Inayoshi}, Kohei and {Carniani}, Stefano and {Nakajima}, Kimihiko and {Isobe}, Yuki and {Witstok}, Joris and {Jones}, Gareth C. and {Tacchella}, Sandro and {Arribas}, Santiago and {Bunker}, Andrew and {Cataldi}, Elisa and {Charlot}, Stephane and {Curti}, Giovanni Cresci Mirko and {Fabian}, Andrew C. and {Katz}, Harley and {Kumari}, Nimisha and {Laporte}, Nicolas and {Mazzolari}, Giovanni and {Robertson}, Brant and {Sun}, Fengwu and {Rodriguez Del Pino}, Bruno and {Venturi}, Giacomo},
        title = "{A black hole in a near pristine galaxy 700 Myr after the big bang}",
      journal = {\mnras},
         year = 2026,
        month = may,
       volume = {548},
       number = {1},
          eid = {staf2109},
        pages = {staf2109},
          doi = {10.1093/mnras/staf2109},
archivePrefix = {arXiv},
       eprint = {2505.22567},
 primaryClass = {astro-ph.GA},
       adsurl = {https://ui.adsabs.harvard.edu/abs/2026MNRAS.548f2109M}
}

@ARTICLE{Tang+26a,
       author = {{Tang}, Mengtao and {Stark}, Daniel P. and {Mason}, Charlotte A. and {Chen}, Zuyi and {Katz}, Harley and {Gronke}, Max and {Furtak}, Lukas J. and {Chang}, Seok-Jun and {Matthee}, Jorryt and {Whitler}, Lily and {Zitrin}, Adi and {Endsley}, Ryan and {Gelli}, Viola and {Roychowdhury}, Tamojeet and {Senchyna}, Peter and {Topping}, Michael W. and {Zhang}, Meng},
        title = "{SPURS: Evidence for Clumpy Neutral Envelopes and Ionized IGM Surrounding Little Red Dots in Abell 2744 from Ultra-Deep Rest-UV Spectroscopy}",
      journal = {arXiv e-prints},
         year = 2026,
        month = apr,
          eid = {arXiv:2604.03563},
        pages = {arXiv:2604.03563},
          doi = {10.48550/arXiv.2604.03563},
archivePrefix = {arXiv},
       eprint = {2604.03563},
 primaryClass = {astro-ph.GA},
       adsurl = {https://ui.adsabs.harvard.edu/abs/2026arXiv260403563T}
}

@ARTICLE{Ma+25,
       author = {{Ma}, Yilun and {Greene}, Jenny E. and {Setton}, David J. and {Volonteri}, Marta and {Leja}, Joel and {Wang}, Bingjie and {Bezanson}, Rachel and {Brammer}, Gabriel and {Cutler}, Sam E. and {Dayal}, Pratika and {van Dokkum}, Pieter and {Furtak}, Lukas J. and {Glazebrook}, Karl and {Goulding}, Andy D. and {de Graaff}, Anna and {Kokorev}, Vasily and {Labbe}, Ivo and {Pan}, Richard and {Price}, Sedona H. and {Weaver}, John R. and {Williams}, Christina C. and {Whitaker}, Katherine E. and {Zitrin}, Adi},
        title = "{UNCOVER: 404 Error{\textemdash}Models Not Found for the Triply Imaged Little Red Dot A2744-QSO1}",
      journal = {\apj},
         year = 2025,
        month = mar,
       volume = {981},
       number = {2},
          eid = {191},
        pages = {191},
          doi = {10.3847/1538-4357/ada613},
archivePrefix = {arXiv},
       eprint = {2410.06257},
 primaryClass = {astro-ph.GA},
       adsurl = {https://ui.adsabs.harvard.edu/abs/2025ApJ...981..191M}
}

@ARTICLE{Stalevski+16,
       author = {{Stalevski}, Marko and {Ricci}, Claudio and {Ueda}, Yoshihiro and {Lira}, Paulina and {Fritz}, Jacopo and {Baes}, Maarten},
        title = "{The dust covering factor in active galactic nuclei}",
      journal = {\mnras},
         year = 2016,
        month = may,
       volume = {458},
       number = {3},
        pages = {2288-2302},
          doi = {10.1093/mnras/stw444},
archivePrefix = {arXiv},
       eprint = {1602.06954},
 primaryClass = {astro-ph.GA},
       adsurl = {https://ui.adsabs.harvard.edu/abs/2016MNRAS.458.2288S}
}

@ARTICLE{Stalevski+12a,
       author = {{Stalevski}, Marko and {Fritz}, Jacopo and {Baes}, Maarten and {Nakos}, Theodoros and {Popovi{\'c}}, Luka {\v{C}}.},
        title = "{3D radiative transfer modelling of the dusty tori around active galactic nuclei as a clumpy two-phase medium}",
      journal = {\mnras},
         year = 2012,
        month = mar,
       volume = {420},
       number = {4},
        pages = {2756-2772},
          doi = {10.1111/j.1365-2966.2011.19775.x},
archivePrefix = {arXiv},
       eprint = {1109.1286},
 primaryClass = {astro-ph.CO},
       adsurl = {https://ui.adsabs.harvard.edu/abs/2012MNRAS.420.2756S}
}

@ARTICLE{daCunha+13b,
       author = {{da Cunha}, Elisabete and {Groves}, Brent and {Walter}, Fabian and {Decarli}, Roberto and {Weiss}, Axel and {Bertoldi}, Frank and {Carilli}, Chris and {Daddi}, Emanuele and {Elbaz}, David and {Ivison}, Rob and {Maiolino}, Roberto and {Riechers}, Dominik and {Rix}, Hans-Walter and {Sargent}, Mark and {Smail}, Ian},
        title = "{On the Effect of the Cosmic Microwave Background in High-redshift (Sub-)millimeter Observations}",
      journal = {\apj},
         year = 2013,
        month = mar,
       volume = {766},
       number = {1},
          eid = {13},
        pages = {13},
          doi = {10.1088/0004-637X/766/1/13},
archivePrefix = {arXiv},
       eprint = {1302.0844},
 primaryClass = {astro-ph.CO},
       adsurl = {https://ui.adsabs.harvard.edu/abs/2013ApJ...766...13D}
}

@ARTICLE{Leitherer+02a,
       author = {{Leitherer}, Claus and {Li}, I.-Hui and {Calzetti}, Daniela and {Heckman}, Timothy M.},
        title = "{Global Far-Ultraviolet (912-1800 {\r{A}}) Properties of Star-forming Galaxies}",
      journal = {\apjs},
         year = 2002,
        month = jun,
       volume = {140},
       number = {2},
        pages = {303-329},
          doi = {10.1086/342486},
       adsurl = {https://ui.adsabs.harvard.edu/abs/2002ApJS..140..303L}
}

@ARTICLE{Perez-Gonzalez+26,
       author = {{P{\'e}rez-Gonz{\'a}lez}, Pablo G. and {Barro}, Guillermo and {Carniani}, Stefano and {D'Eugenio}, Francesco and {Rieke}, George H. and {Tripodi}, Roberta and {Bunker}, Andrew J. and {Ji}, Xihan and {Marques-Chaves}, Rui and {Schaerer}, Daniel and {Venturi}, Giacomo and {Ar{\'e}valo-Gonz{\'a}lez}, Flor and {Arribas}, Santiago and {Rinaldi}, Pierluigi and {Rodr{\'\i}guez Del Pino}, Bruno and {Witstok}, Joris and {Bhatawdekar}, Rachana and {Boogaard}, Leindert A. and {Charlot}, Stephane and {Chevallard}, Jacopo and {Costantin}, Luca and {Curti}, Mirko and {Curtis-Lake}, Emma and {Daddi}, Emanuele and {Davis}, Kelcey and {Dickinson}, Mark and {Donnan}, Callum T. and {Donnan}, Fergus R. and {Dunlop}, James S. and {Eisenstein}, Daniel J. and {Ferguson}, Henry C. and {Fern{\'a}ndez Aranda}, Rom{\'a}n and {Finkelstein}, Steven L. and {Fujimoto}, Seiji and {Gandolfi}, Giovanni and {Giavalisco}, Mauro and {Grogin}, Norman A. and {Hamed}, Mahmoud and {Hirschmann}, Michaela and {Kartaltepe}, Jeyhan S. and {Kocevski}, Dale D. and {Leung}, Gene C.~K. and {Lofaro}, Cristina M. and {Lucas}, Ray A. and {McLeod}, Derek J. and {Melinder}, Jens and {{\"O}stlin}, Goran and {Papovich}, Casey and {Pentericci}, Laura and {P{\'e}rez-D{\'\i}az}, Borja and {Rieke}, Marcia and {Scholtz}, Jan and {Somerville}, Rachel S. and {Stanton}, Thomas M. and {Stevenson}, Struan D. and {Shivaei}, Irene and {Tacchella}, Sandro and {Trump}, Jonathan R. and {{\"U}bler}, Hannah and {Wang}, Xin and {Williams}, Christina C. and {Willmer}, Christopher N.~A. and {Yung}, L.~Y. Aaron and {Zhu}, Yongda},
        title = "{Little Red Dots: One Photometric Tag Concealing Diverse Spectroscopic Flavors of Massive Star Formation and Black Hole Activity}",
      journal = {arXiv e-prints},
         year = 2026,
        month = feb,
          eid = {arXiv:2602.20247},
        pages = {arXiv:2602.20247},
          doi = {10.48550/arXiv.2602.20247},
archivePrefix = {arXiv},
       eprint = {2602.20247},
 primaryClass = {astro-ph.GA},
       adsurl = {https://ui.adsabs.harvard.edu/abs/2026arXiv260220247P}
}

@ARTICLE{Delvecchio+25,
       author = {{Delvecchio}, I. and {Daddi}, E. and {Magnelli}, B. and {Elbaz}, D. and {Giavalisco}, M. and {Traina}, A. and {Lanzuisi}, G. and {Akins}, H.~B. and {Belli}, S. and {Casey}, C.~M. and {Gentile}, F. and {Gruppioni}, C. and {Pozzi}, F. and {Zamorani}, G.},
        title = "{Active galactic nuclei-heated dust revealed in ``little red dots''}",
      journal = {\aap},
         year = 2025,
        month = dec,
       volume = {704},
          eid = {A313},
        pages = {A313},
          doi = {10.1051/0004-6361/202557164},
archivePrefix = {arXiv},
       eprint = {2509.07100},
 primaryClass = {astro-ph.GA},
       adsurl = {https://ui.adsabs.harvard.edu/abs/2025A&A...704A.313D}
}

@ARTICLE{Akins+25b,
       author = {{Akins}, Hollis B. and {Casey}, Caitlin M. and {Lambrides}, Erini and {Allen}, Natalie and {Andika}, Irham T. and {Brinch}, Malte and {Champagne}, Jaclyn B. and {Cooper}, Olivia and {Ding}, Xuheng and {Drakos}, Nicole E. and {Faisst}, Andreas and {Finkelstein}, Steven L. and {Franco}, Maximilien and {Fujimoto}, Seiji and {Gentile}, Fabrizio and {Gillman}, Steven and {Gozaliasl}, Ghassem and {Harish}, Santosh and {Hayward}, Christopher C. and {Hirschmann}, Michaela and {Ilbert}, Olivier and {Kartaltepe}, Jeyhan S. and {Kocevski}, Dale D. and {Koekemoer}, Anton M. and {Kokorev}, Vasily and {Liu}, Daizhong and {Long}, Arianna S. and {McCracken}, Henry Joy and {McKinney}, Jed and {Onoue}, Masafusa and {Paquereau}, Louise and {Renzini}, Alvio and {Rhodes}, Jason and {Robertson}, Brant E. and {Shuntov}, Marko and {Silverman}, John D. and {Tanaka}, Takumi S. and {Toft}, Sune and {Trakhtenbrot}, Benny and {Valentino}, Francesco and {Zavala}, Jorge},
        title = "{COSMOS-Web: The Overabundance and Physical Nature of ``Little Red Dots''{\textemdash}Implications for Early Galaxy and SMBH Assembly}",
      journal = {\apj},
         year = 2025,
        month = sep,
       volume = {991},
       number = {1},
          eid = {37},
        pages = {37},
          doi = {10.3847/1538-4357/ade984},
archivePrefix = {arXiv},
       eprint = {2406.10341},
 primaryClass = {astro-ph.GA},
       adsurl = {https://ui.adsabs.harvard.edu/abs/2025ApJ...991...37A}
}

@ARTICLE{Barro+26b,
       author = {{Barro}, Guillermo and {P{\'e}rez-Gonz{\'a}lez}, Pablo G. and {Kocevski}, Dale and {Trump}, Jonathan R. and {Dickinson}, Mark and {Arrabal Haro}, Pablo and {Brooks}, Madisyn and {Donnan}, Callum T. and {Dunlop}, James S. and {Finkelstein}, Steven L. and {Franco}, Maximilien and {Gandolfi}, Giovanni and {Giavalisco}, Mauro and {Grogin}, Norman A. and {Hirschmann}, Michaela and {Kartaltepe}, Jeyhan S. and {Koekemoer}, Anton M. and {Larson}, Rebecca L. and {Leung}, Gene C.~K. and {Lucas}, Ray A. and {McGrath}, Elizabeth J. and {Papovich}, Casey and {P{\'e}rez-D{\'\i}az}, Borja and {Somerville}, Rachel S. and {Taylor}, Elizabeth and {Taylor}, Anthony J. and {Tripodi}, Roberta and {Yung}, L.~Y. Aaron and {Wang}, Xin},
        title = "{From ``The Cliff'' to ``Virgil'': Mapping the Spectral Diversity of Little Red Dots with JWST/NIRSpec}",
      journal = {\apj},
         year = 2026,
        month = may,
       volume = {1003},
       number = {1},
          eid = {96},
        pages = {96},
          doi = {10.3847/1538-4357/ae5d2c},
archivePrefix = {arXiv},
       eprint = {2512.15853},
 primaryClass = {astro-ph.GA},
       adsurl = {https://ui.adsabs.harvard.edu/abs/2026ApJ..1003...96B}
}

@ARTICLE{Juodbalis+24a,
       author = {{Juod{\v{z}}balis}, Ignas and {Ji}, Xihan and {Maiolino}, Roberto and {D'Eugenio}, Francesco and {Scholtz}, Jan and {Risaliti}, Guido and {Fabian}, Andrew C. and {Mazzolari}, Giovanni and {Gilli}, Roberto and {Prandoni}, Isabella and {Arribas}, Santiago and {Bunker}, Andrew J. and {Carniani}, Stefano and {Charlot}, St{\'e}phane and {Curtis-Lake}, Emma and {de Graaff}, Anna and {Hainline}, Kevin and {Parlanti}, Eleonora and {Perna}, Michele and {P{\'e}rez-Gonz{\'a}lez}, Pablo G. and {Robertson}, Brant and {Tacchella}, Sandro and {{\"U}bler}, Hannah and {Williams}, Christina C. and {Willott}, Chris and {Witstok}, Joris},
        title = "{JADES - the Rosetta stone of JWST-discovered AGN: deciphering the intriguing nature of early AGN}",
      journal = {\mnras},
         year = 2024,
        month = nov,
       volume = {535},
       number = {1},
        pages = {853-873},
          doi = {10.1093/mnras/stae2367},
archivePrefix = {arXiv},
       eprint = {2407.08643},
 primaryClass = {astro-ph.GA},
       adsurl = {https://ui.adsabs.harvard.edu/abs/2024MNRAS.535..853J}
}

@ARTICLE{Williams+24,
       author = {{Williams}, Christina C. and {Alberts}, Stacey and {Ji}, Zhiyuan and {Hainline}, Kevin N. and {Lyu}, Jianwei and {Rieke}, George and {Endsley}, Ryan and {Suess}, Katherine A. and {Sun}, Fengwu and {Johnson}, Benjamin D. and {Florian}, Michael and {Shivaei}, Irene and {Rujopakarn}, Wiphu and {Baker}, William M. and {Bhatawdekar}, Rachana and {Boyett}, Kristan and {Bunker}, Andrew J. and {Cameron}, Alex J. and {Carniani}, Stefano and {Charlot}, Stephane and {Curtis-Lake}, Emma and {DeCoursey}, Christa and {de Graaff}, Anna and {Egami}, Eiichi and {Eisenstein}, Daniel J. and {Gibson}, Justus L. and {Hausen}, Ryan and {Helton}, Jakob M. and {Maiolino}, Roberto and {Maseda}, Michael V. and {Nelson}, Erica J. and {P{\'e}rez-Gonz{\'a}lez}, Pablo G. and {Rieke}, Marcia J. and {Robertson}, Brant E. and {Saxena}, Aayush and {Tacchella}, Sandro and {Willmer}, Christopher N.~A. and {Willott}, Chris J.},
        title = "{The Galaxies Missed by Hubble and ALMA: The Contribution of Extremely Red Galaxies to the Cosmic Census at 3 < z < 8}",
      journal = {\apj},
         year = 2024,
        month = jun,
       volume = {968},
       number = {1},
          eid = {34},
        pages = {34},
          doi = {10.3847/1538-4357/ad3f17},
archivePrefix = {arXiv},
       eprint = {2311.07483},
 primaryClass = {astro-ph.GA},
       adsurl = {https://ui.adsabs.harvard.edu/abs/2024ApJ...968...34W}
}

@ARTICLE{Rawle+14,
       author = {{Rawle}, T.~D. and {Altieri}, B. and {Egami}, E. and {P{\'e}rez-Gonz{\'a}lez}, P.~G. and {Richard}, J. and {Santos}, J.~S. and {Valtchanov}, I. and {Walth}, G. and {Bouy}, H. and {Haines}, C.~P. and {Okabe}, N.},
        title = "{Star formation in the massive cluster merger Abell 2744}",
      journal = {\mnras},
         year = 2014,
        month = jul,
       volume = {442},
       number = {1},
        pages = {196-206},
          doi = {10.1093/mnras/stu868},
archivePrefix = {arXiv},
       eprint = {1405.1046},
 primaryClass = {astro-ph.GA},
       adsurl = {https://ui.adsabs.harvard.edu/abs/2014MNRAS.442..196R}
}

@ARTICLE{Nguyen+10,
       author = {{Nguyen}, H.~T. and {Schulz}, B. and {Levenson}, L. and {Amblard}, A. and {Arumugam}, V. and {Aussel}, H. and {Babbedge}, T. and {Blain}, A. and {Bock}, J. and {Boselli}, A. and {Buat}, V. and {Castro-Rodriguez}, N. and {Cava}, A. and {Chanial}, P. and {Chapin}, E. and {Clements}, D.~L. and {Conley}, A. and {Conversi}, L. and {Cooray}, A. and {Dowell}, C.~D. and {Dwek}, E. and {Eales}, S. and {Elbaz}, D. and {Fox}, M. and {Franceschini}, A. and {Gear}, W. and {Glenn}, J. and {Griffin}, M. and {Halpern}, M. and {Hatziminaoglou}, E. and {Ibar}, E. and {Isaak}, K. and {Ivison}, R.~J. and {Lagache}, G. and {Lu}, N. and {Madden}, S. and {Maffei}, B. and {Mainetti}, G. and {Marchetti}, L. and {Marsden}, G. and {Marshall}, J. and {O'Halloran}, B. and {Oliver}, S.~J. and {Omont}, A. and {Page}, M.~J. and {Panuzzo}, P. and {Papageorgiou}, A. and {Pearson}, C.~P. and {Perez Fournon}, I. and {Pohlen}, M. and {Rangwala}, N. and {Rigopoulou}, D. and {Rizzo}, D. and {Roseboom}, I.~G. and {Rowan-Robinson}, M. and {Scott}, D. and {Seymour}, N. and {Shupe}, D.~L. and {Smith}, A.~J. and {Stevens}, J.~A. and {Symeonidis}, M. and {Trichas}, M. and {Tugwell}, K.~E. and {Vaccari}, M. and {Valtchanov}, I. and {Vigroux}, L. and {Wang}, L. and {Ward}, R. and {Wiebe}, D. and {Wright}, G. and {Xu}, C.~K. and {Zemcov}, M.},
        title = "{HerMES: The SPIRE confusion limit}",
      journal = {\aap},
         year = 2010,
        month = jul,
       volume = {518},
          eid = {L5},
        pages = {L5},
          doi = {10.1051/0004-6361/201014680},
archivePrefix = {arXiv},
       eprint = {1005.2207},
 primaryClass = {astro-ph.CO},
       adsurl = {https://ui.adsabs.harvard.edu/abs/2010A&A...518L...5N}
}

@ARTICLE{Rawle+16,
       author = {{Rawle}, T.~D. and {Altieri}, B. and {Egami}, E. and {P{\'e}rez-Gonz{\'a}lez}, P.~G. and {Boone}, F. and {Clement}, B. and {Ivison}, R.~J. and {Richard}, J. and {Rujopakarn}, W. and {Valtchanov}, I. and {Walth}, G. and {Weiner}, B.~J. and {Blain}, A.~W. and {Dessauges-Zavadsky}, M. and {Kneib}, J.-P. and {Lutz}, D. and {Rodighiero}, G. and {Schaerer}, D. and {Smail}, I.},
        title = "{A complete census of Herschel-detected infrared sources within the HST Frontier Fields}",
      journal = {\mnras},
         year = 2016,
        month = jun,
       volume = {459},
       number = {2},
        pages = {1626-1645},
          doi = {10.1093/mnras/stw712},
archivePrefix = {arXiv},
       eprint = {1508.00586},
 primaryClass = {astro-ph.GA},
       adsurl = {https://ui.adsabs.harvard.edu/abs/2016MNRAS.459.1626R}
}

@ARTICLE{Rinaldi+25,
       author = {{Rinaldi}, Pierluigi and {P{\'e}rez-Gonz{\'a}lez}, Pablo G. and {Rieke}, George H. and {Lyu}, Jianwei and {D'Eugenio}, Francesco and {Wu}, Zihao and {Carniani}, Stefano and {Looser}, Tobias J. and {Shivaei}, Irene and {Boogaard}, Leindert A. and {Diaz-Santos}, Tanio and {Colina}, Luis and {{\"O}stlin}, G{\"o}ran and {Alberts}, Stacey and {{\'A}lvarez-M{\'a}rquez}, Javier and {Annuziatella}, Marianna and {Aravena}, Manuel and {Bhatawdekar}, Rachana and {Bunker}, Andrew J. and {Caputi}, Karina I. and {Charlot}, St{\'e}phane and {Crespo G{\'o}mez}, Alejandro and {Curti}, Mirko and {Eckart}, Andreas and {Gillman}, Steven and {Hainline}, Kevin and {Kumari}, Nimisha and {Hjorth}, Jens and {Iani}, Edoardo and {Inami}, Hanae and {Ji}, Zhiyuan and {Johnson}, Benjamin D. and {Jones}, Gareth C. and {Labiano}, {\'A}lvaro and {Maiolino}, Roberto and {Melinder}, Jens and {Moutard}, Thibaud and {Peissker}, Florian and {Rieke}, Marcia and {Robertson}, Brant and {Scholtz}, Jan and {Tacchella}, Sandro and {van der Werf}, Paul P. and {Walter}, Fabian and {Williams}, Christina C. and {Willott}, Chris and {Witstok}, Joris and {{\"U}bler}, Hannah and {Zhu}, Yongda},
        title = "{Deciphering the Nature of Virgil: An Obscured Active Galactic Nucleus Lurking within an Apparently Normal Ly{\ensuremath{\alpha}} Emitter during Cosmic Reionization}",
      journal = {\apj},
         year = 2025,
        month = nov,
       volume = {994},
       number = {1},
          eid = {86},
        pages = {86},
          doi = {10.3847/1538-4357/ae089c},
archivePrefix = {arXiv},
       eprint = {2504.01852},
 primaryClass = {astro-ph.GA},
       adsurl = {https://ui.adsabs.harvard.edu/abs/2025ApJ...994...86R}
}

@ARTICLE{Suess+24,
       author = {{Suess}, Katherine A. and {Weaver}, John R. and {Price}, Sedona H. and {Pan}, Richard and {Wang}, Bingjie and {Bezanson}, Rachel and {Brammer}, Gabriel and {Cutler}, Sam E. and {Labb{\'e}}, Ivo and {Leja}, Joel and {Williams}, Christina C. and {Whitaker}, Katherine E. and {Atek}, Hakim and {Dayal}, Pratika and {de Graaff}, Anna and {Feldmann}, Robert and {Franx}, Marijn and {Fudamoto}, Yoshinobu and {Fujimoto}, Seiji and {Furtak}, Lukas J. and {Goulding}, Andy D. and {Greene}, Jenny E. and {Khullar}, Gourav and {Kokorev}, Vasily and {Kriek}, Mariska and {Lorenz}, Brian and {Marchesini}, Danilo and {Maseda}, Michael V. and {Matthee}, Jorryt and {Miller}, Tim B. and {Mitsuhashi}, Ikki and {Mowla}, Lamiya A. and {Muzzin}, Adam and {Naidu}, Rohan P. and {Nanayakkara}, Themiya and {Nelson}, Erica J. and {Oesch}, Pascal A. and {Setton}, David J. and {Shipley}, Heath and {Smit}, Renske and {Spilker}, Justin S. and {van Dokkum}, Pieter and {Zitrin}, Adi},
        title = "{Medium Bands, Mega Science: A JWST/NIRCam Medium-band Imaging Survey of A2744}",
      journal = {\apj},
         year = 2024,
        month = nov,
       volume = {976},
       number = {1},
          eid = {101},
        pages = {101},
          doi = {10.3847/1538-4357/ad75fe},
archivePrefix = {arXiv},
       eprint = {2404.13132},
 primaryClass = {astro-ph.GA},
       adsurl = {https://ui.adsabs.harvard.edu/abs/2024ApJ...976..101S}
}

@ARTICLE{Bezanson+24,
       author = {{Bezanson}, Rachel and {Labbe}, Ivo and {Whitaker}, Katherine E. and {Leja}, Joel and {Price}, Sedona H. and {Franx}, Marijn and {Brammer}, Gabriel and {Marchesini}, Danilo and {Zitrin}, Adi and {Wang}, Bingjie and {Weaver}, John R. and {Furtak}, Lukas J. and {Atek}, Hakim and {Coe}, Dan and {Cutler}, Sam E. and {Dayal}, Pratika and {van Dokkum}, Pieter and {Feldmann}, Robert and {F{\"o}rster Schreiber}, Natascha M. and {Fujimoto}, Seiji and {Geha}, Marla and {Glazebrook}, Karl and {de Graaff}, Anna and {Greene}, Jenny E. and {Juneau}, St{\'e}phanie and {Kassin}, Susan and {Kriek}, Mariska and {Khullar}, Gourav and {Maseda}, Michael and {Mowla}, Lamiya A. and {Muzzin}, Adam and {Nanayakkara}, Themiya and {Nelson}, Erica J. and {Oesch}, Pascal A. and {Pacifici}, Camilla and {Pan}, Richard and {Papovich}, Casey and {Setton}, David J. and {Shapley}, Alice E. and {Smit}, Renske and {Stefanon}, Mauro and {Taylor}, Edward N. and {Williams}, Christina C.},
        title = "{The JWST UNCOVER Treasury Survey: Ultradeep NIRSpec and NIRCam Observations before the Epoch of Reionization}",
      journal = {\apj},
         year = 2024,
        month = oct,
       volume = {974},
       number = {1},
          eid = {92},
        pages = {92},
          doi = {10.3847/1538-4357/ad66cf},
archivePrefix = {arXiv},
       eprint = {2212.04026},
 primaryClass = {astro-ph.GA},
       adsurl = {https://ui.adsabs.harvard.edu/abs/2024ApJ...974...92B}
}

@ARTICLE{Juodzbalis+26b,
       author = {{Juod{\v{z}}balis}, Ignas and {Marconcini}, Cosimo and {D'Eugenio}, Francesco and {Maiolino}, Roberto and {Marconi}, Alessandro and {{\"U}bler}, Hannah and {Scholtz}, Jan and {Ji}, Xihan and {Jones}, Gareth C. and {Perna}, Michele and {Arribas}, Santiago and {Bennett}, Jake S. and {Bromm}, Volker and {Bunker}, Andrew J. and {Carniani}, Stefano and {Charlot}, St{\'e}phane and {Cresci}, Giovanni and {Dayal}, Pratika and {Egami}, Eiichi and {Fabian}, Andrew and {Inayoshi}, Kohei and {Isobe}, Yuki and {Ivey}, Lucy R. and {Koudmani}, Sophie and {Laporte}, Nicolas and {Liu}, Boyuan and {Lyu}, Jianwei and {Mazzolari}, Giovanni and {Monty}, Stephanie and {Parlanti}, Eleonora and {P{\'e}rez-Gonz{\'a}lez}, Pablo G. and {Robertson}, Brant and {Schneider}, Raffaella and {Sijacki}, Debora and {Tacchella}, Sandro and {Trinca}, Alessandro and {Valiante}, Rosa and {Volonteri}, Marta and {Witstok}, Joris and {Zhang}, Saiyang},
        title = "{A direct black-hole mass measurement in a little red dot at high redshift}",
      journal = {\nat},
         year = 2026,
        month = may,
       volume = {653},
       number = {8116},
        pages = {1017-1021},
          doi = {10.1038/s41586-026-10579-4},
archivePrefix = {arXiv},
       eprint = {2508.21748},
 primaryClass = {astro-ph.GA},
       adsurl = {https://ui.adsabs.harvard.edu/abs/2026Natur.653.1017J}
}

@ARTICLE{Torralba+26c,
       author = {{Torralba}, Alberto and {Matthee}, Jorryt and {Weibel}, Andrea and {Naidu}, Rohan P. and {Ma}, Yilun and {Cloonan}, Aidan P. and {Desai}, Aayush and {de Graaff}, Anna and {Greene}, Jenny E. and {Jespersen}, Christian Kragh and {Kramarenko}, Ivan G. and {Mascia}, Sara and {Oesch}, Pascal A. and {Sun}, Wendy Q. and {Williams}, Christina C.},
        title = "{A Black Hole Star at Cosmic Noon: Extreme Balmer Break, Photospheric Continuum, and Broad Absorption by Thick Winds in a Little Red Dot at z = 1.7}",
      journal = {\apjl},
         year = 2026,
        month = jul,
       volume = {1005},
       number = {2},
          eid = {L37},
        pages = {L37},
          doi = {10.3847/2041-8213/ae7bfd},
archivePrefix = {arXiv},
       eprint = {2603.28335},
 primaryClass = {astro-ph.GA},
       adsurl = {https://ui.adsabs.harvard.edu/abs/2026ApJ..1005L..37T}
}

@ARTICLE{Inoue+14,
       author = {{Inoue}, Akio K. and {Shimizu}, Ikkoh and {Iwata}, Ikuru and {Tanaka}, Masayuki},
        title = "{An updated analytic model for attenuation by the intergalactic medium}",
      journal = {\mnras},
         year = 2014,
        month = aug,
       volume = {442},
       number = {2},
        pages = {1805-1820},
          doi = {10.1093/mnras/stu936},
archivePrefix = {arXiv},
       eprint = {1402.0677},
 primaryClass = {astro-ph.CO},
       adsurl = {https://ui.adsabs.harvard.edu/abs/2014MNRAS.442.1805I}
}

@ARTICLE{Perez-Gonzalez+24a,
       author = {{P{\'e}rez-Gonz{\'a}lez}, Pablo G. and {Barro}, Guillermo and {Rieke}, George H. and {Lyu}, Jianwei and {Rieke}, Marcia and {Alberts}, Stacey and {Williams}, Christina C. and {Hainline}, Kevin and {Sun}, Fengwu and {Pusk{\'a}s}, D{\'a}vid and {Annunziatella}, Marianna and {Baker}, William M. and {Bunker}, Andrew J. and {Egami}, Eiichi and {Ji}, Zhiyuan and {Johnson}, Benjamin D. and {Robertson}, Brant and {Rodr{\'\i}guez Del Pino}, Bruno and {Rujopakarn}, Wiphu and {Shivaei}, Irene and {Tacchella}, Sandro and {Willmer}, Christopher N.~A. and {Willott}, Chris},
        title = "{What Is the Nature of Little Red Dots and what Is Not, MIRI SMILES Edition}",
      journal = {\apj},
         year = 2024,
        month = jun,
       volume = {968},
       number = {1},
          eid = {4},
        pages = {4},
          doi = {10.3847/1538-4357/ad38bb},
archivePrefix = {arXiv},
       eprint = {2401.08782},
 primaryClass = {astro-ph.GA},
       adsurl = {https://ui.adsabs.harvard.edu/abs/2024ApJ...968....4P}
}

@ARTICLE{Storey+95,
       author = {{Storey}, P.~J. and {Hummer}, D.~G.},
        title = "{Recombination line intensities for hydrogenic ions-IV. Total recombination coefficients and machine-readable tables for Z=1 to 8}",
      journal = {\mnras},
         year = 1995,
        month = jan,
       volume = {272},
       number = {1},
        pages = {41-48},
          doi = {10.1093/mnras/272.1.41},
       adsurl = {https://ui.adsabs.harvard.edu/abs/1995MNRAS.272...41S}
}

@ARTICLE{Shajib+25,
       author = {{Shajib}, Anowar J. and {Treu}, Tommaso and {Melo}, Alejandra and {Roberts-Borsani}, Guido and {Knabel}, Shawn and {Cappellari}, Michele and {Frieman}, Joshua A.},
        title = "{An accurate measurement of the spectral resolution of the JWST Near Infrared Spectrograph}",
      journal = {\aap},
         year = 2025,
        month = oct,
       volume = {702},
          eid = {L12},
        pages = {L12},
          doi = {10.1051/0004-6361/202556281},
archivePrefix = {arXiv},
       eprint = {2507.03746},
 primaryClass = {astro-ph.IM},
       adsurl = {https://ui.adsabs.harvard.edu/abs/2025A&A...702L..12S}
}

@ARTICLE{Carnall+17,
       author = {{Carnall}, A.~C.},
        title = "{SpectRes: A Fast Spectral Resampling Tool in Python}",
      journal = {arXiv e-prints},
         year = 2017,
        month = may,
          eid = {arXiv:1705.05165},
        pages = {arXiv:1705.05165},
          doi = {10.48550/arXiv.1705.05165},
archivePrefix = {arXiv},
       eprint = {1705.05165},
 primaryClass = {astro-ph.IM},
       adsurl = {https://ui.adsabs.harvard.edu/abs/2017arXiv170505165C}
}

@ARTICLE{Gordon+24,
       author = {{Gordon}, Karl},
        title = "{dust\_extinction: Interstellar Dust Extinction Models}",
      journal = {The Journal of Open Source Software},
         year = 2024,
        month = aug,
       volume = {9},
       number = {100},
          eid = {7023},
        pages = {7023},
          doi = {10.21105/joss.07023},
       adsurl = {https://ui.adsabs.harvard.edu/abs/2024JOSS....9.7023G}
}

@ARTICLE{Gordon+23,
       author = {{Gordon}, Karl D. and {Clayton}, Geoffrey C. and {Decleir}, Marjorie and {Fitzpatrick}, E.~L. and {Massa}, Derck and {Misselt}, Karl A. and {Tollerud}, Erik J.},
        title = "{One Relation for All Wavelengths: The Far-ultraviolet to Mid-infrared Milky Way Spectroscopic R(V)-dependent Dust Extinction Relationship}",
      journal = {\apj},
         year = 2023,
        month = jun,
       volume = {950},
       number = {2},
          eid = {86},
        pages = {86},
          doi = {10.3847/1538-4357/accb59},
archivePrefix = {arXiv},
       eprint = {2304.01991},
 primaryClass = {astro-ph.GA},
       adsurl = {https://ui.adsabs.harvard.edu/abs/2023ApJ...950...86G}
}

@ARTICLE{Schlafly+11,
       author = {{Schlafly}, Edward F. and {Finkbeiner}, Douglas P.},
        title = "{Measuring Reddening with Sloan Digital Sky Survey Stellar Spectra and Recalibrating SFD}",
      journal = {\apj},
         year = 2011,
        month = aug,
       volume = {737},
       number = {2},
          eid = {103},
        pages = {103},
          doi = {10.1088/0004-637X/737/2/103},
archivePrefix = {arXiv},
       eprint = {1012.4804},
 primaryClass = {astro-ph.GA},
       adsurl = {https://ui.adsabs.harvard.edu/abs/2011ApJ...737..103S}
}

@ARTICLE{Bertin+96,
       author = {{Bertin}, E. and {Arnouts}, S.},
        title = "{SExtractor: Software for source extraction.}",
      journal = {\aaps},
         year = 1996,
        month = jun,
       volume = {117},
        pages = {393-404},
          doi = {10.1051/aas:1996164},
       adsurl = {https://ui.adsabs.harvard.edu/abs/1996A&AS..117..393B}
}

@ARTICLE{Barbary+16,
       author = {{Barbary}, Kyle},
        title = "{SEP: Source Extractor as a library}",
      journal = {The Journal of Open Source Software},
         year = 2016,
        month = oct,
       volume = {1},
       number = {6},
          eid = {58},
        pages = {58},
          doi = {10.21105/joss.00058},
       adsurl = {https://ui.adsabs.harvard.edu/abs/2016JOSS....1...58B}
}

@ARTICLE{Lucy+74,
       author = {{Lucy}, L.~B.},
        title = "{An iterative technique for the rectification of observed distributions}",
      journal = {\aj},
         year = 1974,
        month = jun,
       volume = {79},
        pages = {745},
          doi = {10.1086/111605},
       adsurl = {https://ui.adsabs.harvard.edu/abs/1974AJ.....79..745L}
}

@ARTICLE{Richardson+72,
       author = {{Richardson}, William Hadley},
        title = "{Bayesian-Based Iterative Method of Image Restoration}",
      journal = {Journal of the Optical Society of America (1917-1983)},
         year = 1972,
        month = jan,
       volume = {62},
       number = {1},
        pages = {55},
          doi = {10.1364/JOSA.62.000055},
       adsurl = {https://ui.adsabs.harvard.edu/abs/1972JOSA...62...55R}
}

@INPROCEEDINGS{Rodrigo+20,
       author = {{Rodrigo}, C. and {Solano}, E.},
        title = "{The SVO Filter Profile Service}",
    booktitle = {XIV.0 Scientific Meeting (virtual) of the Spanish Astronomical Society},
         year = 2020,
        month = jul,
          eid = {182},
        pages = {182},
       adsurl = {https://ui.adsabs.harvard.edu/abs/2020sea..confE.182R}
}

@MISC{Rodrigo+12,
       author = {{Rodrigo}, Carlos and {Solano}, Enrique and {Bayo}, Amelia},
        title = "{SVO Filter Profile Service Version 1.0}",
 howpublished = {IVOA Working Draft 15 October 2012},
         year = 2012,
        month = oct,
        pages = {1015},
          doi = {10.5479/ADS/bib/2012ivoa.rept.1015R},
       adsurl = {https://ui.adsabs.harvard.edu/abs/2012ivoa.rept.1015R}
}

@ARTICLE{Kapoor+26,
       author = {{Kapoor}, Shrriya and {Matthee}, Jorryt and {Torralba}, Alberto and {Kramarenko}, Ivan G. and {Bordoloi}, Rongmon and {Greene}, Jenny E. and {Iani}, Edoardo and {Kashino}, Daichi and {Liu}, Zhaoran and {Mackenzie}, Ruari and {Mascia}, Sara and {Naidu}, Rohan P. and {Simcoe}, Rob},
        title = "{Little Red Dots at z\raisebox{-0.5ex}\textasciitilde2 in EIGER reveal a gentle decline with respect to their peak number density at z\raisebox{-0.5ex}\textasciitilde5}",
      journal = {arXiv e-prints},
         year = 2026,
        month = jun,
          eid = {arXiv:2607.00084},
        pages = {arXiv:2607.00084},
          doi = {10.48550/arXiv.2607.00084},
archivePrefix = {arXiv},
       eprint = {2607.00084},
 primaryClass = {astro-ph.GA},
       adsurl = {https://ui.adsabs.harvard.edu/abs/2026arXiv260700084K}
}

@ARTICLE{Chang+26,
       author = {{Chang}, Seok-Jun and {Gronke}, Max and {Matthee}, Jorryt and {Mason}, Charlotte},
        title = "{Impact of resonance, Raman, and Thomson scattering on hydrogen line formation in Little Red Dots}",
      journal = {\mnras},
         year = 2026,
        month = feb,
       volume = {545},
       number = {4},
          eid = {staf2131},
        pages = {staf2131},
          doi = {10.1093/mnras/staf2131},
archivePrefix = {arXiv},
       eprint = {2508.08768},
 primaryClass = {astro-ph.GA},
       adsurl = {https://ui.adsabs.harvard.edu/abs/2026MNRAS.545f2131C}
}

@ARTICLE{Casey+25,
       author = {{Casey}, Caitlin M. and {Akins}, Hollis B. and {Finkelstein}, Steven L. and {Franco}, Maximilien and {Fujimoto}, Seiji and {Liu}, Daizhong and {Long}, Arianna S. and {Magdis}, Georgios and {Manning}, Sinclaire M. and {McKinney}, Jed and {Shuntov}, Marko and {Tanaka}, Takumi S.},
        title = "{An Upper Limit of {}10$^{6}$ M$_{{\ensuremath{\odot}}}$ in Dust from ALMA Observations in 60 Little Red Dots}",
      journal = {\apjl},
         year = 2025,
        month = sep,
       volume = {990},
       number = {2},
          eid = {L61},
        pages = {L61},
          doi = {10.3847/2041-8213/adfa91},
archivePrefix = {arXiv},
       eprint = {2505.18873},
 primaryClass = {astro-ph.GA},
       adsurl = {https://ui.adsabs.harvard.edu/abs/2025ApJ...990L..61C}
}

@ARTICLE{Setton+25a,
       author = {{Setton}, David J. and {Greene}, Jenny E. and {Spilker}, Justin S. and {Williams}, Christina C. and {Labb{\'e}}, Ivo and {Ma}, Yilun and {Wang}, Bingjie and {Whitaker}, Katherine E. and {Leja}, Joel and {de Graaff}, Anna and {Alberts}, Stacey and {Bezanson}, Rachel and {Boogaard}, Leindert A. and {Brammer}, Gabriel and {Cutler}, Sam E. and {Cleri}, Nikko J. and {Cooper}, Olivia R. and {Dayal}, Pratika and {Fujimoto}, Seiji and {Furtak}, Lukas J. and {Goulding}, Andy D. and {Hirschmann}, Michaela and {Kokorev}, Vasily and {Maseda}, Michael V. and {McConachie}, Ian and {Matthee}, Jorryt and {Miller}, Tim B. and {Naidu}, Rohan P. and {Oesch}, Pascal A. and {Pan}, Richard and {Price}, Sedona H. and {Suess}, Katherine A. and {Weaver}, John R. and {Xiao}, Mengyuan and {Zhang}, Yunchong and {Zitrin}, Adi},
        title = "{A Confirmed Deficit of Hot and Cold Dust Emission in the Most Luminous Little Red Dots}",
      journal = {\apjl},
         year = 2025,
        month = sep,
       volume = {991},
       number = {1},
          eid = {L10},
        pages = {L10},
          doi = {10.3847/2041-8213/ade78b},
archivePrefix = {arXiv},
       eprint = {2503.02059},
 primaryClass = {astro-ph.GA},
       adsurl = {https://ui.adsabs.harvard.edu/abs/2025ApJ...991L..10S}
}

@ARTICLE{Setton+25b,
       author = {{Setton}, David J. and {Greene}, Jenny E. and {de Graaff}, Anna and {Ma}, Yilun and {Leja}, Joel and {Matthee}, Jorryt and {Bezanson}, Rachel and {Boogaard}, Leindert A. and {Cleri}, Nikko J. and {Katz}, Harley and {Labbe}, Ivo and {Maseda}, Michael V. and {McConachie}, Ian and {Miller}, Tim B. and {Price}, Sedona H. and {Suess}, Katherine A. and {van Dokkum}, Pieter and {Wang}, Bingjie and {Weibel}, Andrea and {Whitaker}, Katherine E. and {Williams}, Christina C.},
        title = "{Little Red Dots at an Inflection Point: Ubiquitous V-shaped Turnover Consistently Occurs at the Balmer Limit}",
      journal = {\apj},
         year = 2025,
        month = dec,
       volume = {995},
       number = {1},
          eid = {118},
        pages = {118},
          doi = {10.3847/1538-4357/ae1500},
archivePrefix = {arXiv},
       eprint = {2411.03424},
 primaryClass = {astro-ph.GA},
       adsurl = {https://ui.adsabs.harvard.edu/abs/2025ApJ...995..118S}
}

@ARTICLE{Glikman+06,
       author = {{Glikman}, Eilat and {Helfand}, David J. and {White}, Richard L.},
        title = "{A Near-Infrared Spectral Template for Quasars}",
      journal = {\apj},
         year = 2006,
        month = apr,
       volume = {640},
       number = {2},
        pages = {579-591},
          doi = {10.1086/500098},
archivePrefix = {arXiv},
       eprint = {astro-ph/0511640},
 primaryClass = {astro-ph},
       adsurl = {https://ui.adsabs.harvard.edu/abs/2006ApJ...640..579G}
}

@ARTICLE{Rieke+23,
       author = {{Rieke}, Marcia J. and {Kelly}, Douglas M. and {Misselt}, Karl and {Stansberry}, John and {Boyer}, Martha and {Beatty}, Thomas and {Egami}, Eiichi and {Florian}, Michael and {Greene}, Thomas P. and {Hainline}, Kevin and {Leisenring}, Jarron and {Roellig}, Thomas and {Schlawin}, Everett and {Sun}, Fengwu and {Tinnin}, Lee and {Williams}, Christina C. and {Willmer}, Christopher N.~A. and {Wilson}, Debra and {Clark}, Charles R. and {Rohrbach}, Scott and {Brooks}, Brian and {Canipe}, Alicia and {Correnti}, Matteo and {DiFelice}, Audrey and {Gennaro}, Mario and {Girard}, Julien H. and {Hartig}, George and {Hilbert}, Bryan and {Koekemoer}, Anton M. and {Nikolov}, Nikolay K. and {Pirzkal}, Norbert and {Rest}, Armin and {Robberto}, Massimo and {Sunnquist}, Ben and {Telfer}, Randal and {Wu}, Chi Rai and {Ferry}, Malcolm and {Lewis}, Dan and {Baum}, Stefi and {Beichman}, Charles and {Doyon}, Ren{\'e} and {Dressler}, Alan and {Eisenstein}, Daniel J. and {Ferrarese}, Laura and {Hodapp}, Klaus and {Horner}, Scott and {Jaffe}, Daniel T. and {Johnstone}, Doug and {Krist}, John and {Martin}, Peter and {McCarthy}, Donald W. and {Meyer}, Michael and {Rieke}, George H. and {Trauger}, John and {Young}, Erick T.},
        title = "{Performance of NIRCam on JWST in Flight}",
      journal = {\pasp},
         year = 2023,
        month = feb,
       volume = {135},
       number = {1044},
          eid = {028001},
        pages = {028001},
          doi = {10.1088/1538-3873/acac53},
archivePrefix = {arXiv},
       eprint = {2212.12069},
 primaryClass = {astro-ph.IM},
       adsurl = {https://ui.adsabs.harvard.edu/abs/2023PASP..135b8001R}
}

@INPROCEEDINGS{DESI19,
       author = {{Levi}, Michael and {Allen}, Lori E. and {Raichoor}, Anand and {Baltay}, Charles and {BenZvi}, Segev and {Beutler}, Florian and {Bolton}, Adam and {Castander}, Francisco J. and {Chuang}, Chia-Hsun and {Cooper}, Andrew and et al.},
        title = "{The Dark Energy Spectroscopic Instrument (DESI)}",
    booktitle = {Bulletin of the American Astronomical Society},
         year = 2019,
       volume = {51},
        month = sep,
    publisher = {AIP},
          eid = {57},
        pages = {57},
          doi = {10.48550/arXiv.1907.10688},
archivePrefix = {arXiv},
       eprint = {1907.10688},
 primaryClass = {astro-ph.IM},
       adsurl = {https://ui.adsabs.harvard.edu/abs/2019BAAS...51g..57L}
}

@ARTICLE{GRAVITY20b,
       author = {{GRAVITY Collaboration} and {Pfuhl}, O. and {Davies}, R. and {Dexter}, J. and {Netzer}, H. and {H{\"o}nig}, S. and {Lutz}, D. and {Schartmann}, M. and {Sturm}, E. and {Amorim}, A. and {Brandner}, W. and {Cl{\'e}net}, Y. and {de Zeeuw}, P.~T. and {Eckart}, A. and {Eisenhauer}, F. and {F{\"o}rster Schreiber}, N.~M. and {Gao}, F. and {Garcia}, P.~J.~V. and {Genzel}, R. and {Gillessen}, S. and {Gratadour}, D. and {Kishimoto}, M. and {Lacour}, S. and {Millour}, F. and {Ott}, T. and {Paumard}, T. and {Perraut}, K. and {Perrin}, G. and {Peterson}, B.~M. and {Petrucci}, P.~O. and {Prieto}, M.~A. and {Rouan}, D. and {Shangguan}, J. and {Shimizu}, T. and {Sternberg}, A. and {Straub}, O. and {Straubmeier}, C. and {Tacconi}, L.~J. and {Tristram}, K.~R.~W. and {Vermot}, P. and {Waisberg}, I. and {Widmann}, F. and {Woillez}, J.},
        title = "{An image of the dust sublimation region in the nucleus of NGC 1068}",
      journal = {\aap},
         year = 2020,
        month = feb,
       volume = {634},
          eid = {A1},
        pages = {A1},
          doi = {10.1051/0004-6361/201936255},
archivePrefix = {arXiv},
       eprint = {1912.01361},
 primaryClass = {astro-ph.GA},
       adsurl = {https://ui.adsabs.harvard.edu/abs/2020A&A...634A...1G}
}

@ARTICLE{Kim+26,
       author = {{Kim}, Minjin and {Kim}, Yongjung and {Jeong}, Woong-Seob and {Yang}, Yujin and {Pyo}, Jeonghyun and {Feder}, Richard M. and {Lee}, Bomee and {Kim}, Dohyeong and {Oh}, Kyuseok and {Woo}, Jong-Hak and {Lee}, Jeong Hwan and {Cheng}, Yun-Ting and {Chiang}, Yi-Kuan and {Cooray}, Asantha and {Crill}, Brendan P. and {Dore}, Olivier and {Faisst}, Andreas L. and {Huai}, Zhaoyu and {Hui}, Howard and {Masters}, Daniel C. and {Nguyen}, Chi H. and {Zemcov}, Michael},
        title = "{A UV-to-Near-infrared QSO Composite Spectrum from the SPHEREx All-Sky Survey}",
      journal = {arXiv e-prints},
         year = 2026,
        month = jul,
          eid = {arXiv:2607.10282},
        pages = {arXiv:2607.10282},
          doi = {10.48550/arXiv.2607.10282},
archivePrefix = {arXiv},
       eprint = {2607.10282},
 primaryClass = {astro-ph.GA},
       adsurl = {https://ui.adsabs.harvard.edu/abs/2026arXiv260710282K}
}

@ARTICLE{Bosman+25,
       author = {{Bosman}, Sarah E.~I. and {{\'A}lvarez-M{\'a}rquez}, Javier and {Davies}, Frederick B. and {Protu{\v{s}}ov{\'a}}, Klaudia and {Hennawi}, Joseph F. and {Yang}, Jinyi and {Spina}, Benedetta and {Colina}, Luis and {Fan}, Xiaohui and {{\"O}stlin}, G{\"o}ran and {Walter}, Fabian and {Wang}, Feige and {Ward}, Martin and {Alonso Herrero}, Almudena and {Barth}, Aaron J. and {Belladitta}, Silvia and {Boogaard}, Leindert and {Caputi}, Karina I. and {Connor}, Thomas and {{\v{D}}urov{\v{c}}{\'\i}kov{\'a}}, Dominika and {Eilers}, Anna-Christina and {Crespo G{\'o}mez}, Alejandro and {Hjorth}, Jens and {Jun}, Hyunsung D. and {Langeroodi}, Danial and {Liu}, Weizhe and {Lupi}, Alessandro and {Mazzucchelli}, Chiara and {Pye}, John P. and {Rinaldi}, Pierluigi and {van der Werf}, Paul and {Volonteri}, Marta},
        title = "{A close look at the black hole masses and hot dusty toruses of the first quasars with MIRI-MRS}",
      journal = {arXiv e-prints},
         year = 2025,
        month = nov,
          eid = {arXiv:2511.02902},
        pages = {arXiv:2511.02902},
          doi = {10.48550/arXiv.2511.02902},
archivePrefix = {arXiv},
       eprint = {2511.02902},
 primaryClass = {astro-ph.GA},
       adsurl = {https://ui.adsabs.harvard.edu/abs/2025arXiv251102902B}
}

@ARTICLE{Stanway+18,
       author = {{Stanway}, E.~R. and {Eldridge}, J.~J.},
        title = "{Re-evaluating old stellar populations}",
      journal = {\mnras},
         year = 2018,
        month = sep,
       volume = {479},
       number = {1},
        pages = {75-93},
          doi = {10.1093/mnras/sty1353},
archivePrefix = {arXiv},
       eprint = {1805.08784},
 primaryClass = {astro-ph.GA},
       adsurl = {https://ui.adsabs.harvard.edu/abs/2018MNRAS.479...75S}
}

@ARTICLE{Eldridge+17,
       author = {{Eldridge}, J.~J. and {Stanway}, E.~R. and {Xiao}, L. and {McClelland}, L.~A.~S. and {Taylor}, G. and {Ng}, M. and {Greis}, S.~M.~L. and {Bray}, J.~C.},
        title = "{Binary Population and Spectral Synthesis Version 2.1: Construction, Observational Verification, and New Results}",
      journal = {\pasa},
         year = 2017,
        month = nov,
       volume = {34},
          eid = {e058},
        pages = {e058},
          doi = {10.1017/pasa.2017.51},
archivePrefix = {arXiv},
       eprint = {1710.02154},
 primaryClass = {astro-ph.SR},
       adsurl = {https://ui.adsabs.harvard.edu/abs/2017PASA...34...58E}
}

@ARTICLE{Furtak+24,
       author = {{Furtak}, Lukas J. and {Labb{\'e}}, Ivo and {Zitrin}, Adi and {Greene}, Jenny E. and {Dayal}, Pratika and {Chemerynska}, Iryna and {Kokorev}, Vasily and {Miller}, Tim B. and {Goulding}, Andy D. and {de Graaff}, Anna and {Bezanson}, Rachel and {Brammer}, Gabriel B. and {Cutler}, Sam E. and {Leja}, Joel and {Pan}, Richard and {Price}, Sedona H. and {Wang}, Bingjie and {Weaver}, John R. and {Whitaker}, Katherine E. and {Atek}, Hakim and {Bogd{\'a}n}, {\'A}kos and {Charlot}, St{\'e}phane and {Curtis-Lake}, Emma and {van Dokkum}, Pieter and {Endsley}, Ryan and {Feldmann}, Robert and {Fudamoto}, Yoshinobu and {Fujimoto}, Seiji and {Glazebrook}, Karl and {Juneau}, St{\'e}phanie and {Marchesini}, Danilo and {Maseda}, Micheal V. and {Nelson}, Erica and {Oesch}, Pascal A. and {Plat}, Ad{\`e}le and {Setton}, David J. and {Stark}, Daniel P. and {Williams}, Christina C.},
        title = "{A high black-hole-to-host mass ratio in a lensed AGN in the early Universe}",
      journal = {\nat},
         year = 2024,
        month = apr,
       volume = {628},
       number = {8006},
        pages = {57-61},
          doi = {10.1038/s41586-024-07184-8},
archivePrefix = {arXiv},
       eprint = {2308.05735},
 primaryClass = {astro-ph.GA},
       adsurl = {https://ui.adsabs.harvard.edu/abs/2024Natur.628...57F}
}

@ARTICLE{Furtak+23,
       author = {{Furtak}, Lukas J. and {Zitrin}, Adi and {Plat}, Ad{\`e}le and {Fujimoto}, Seiji and {Wang}, Bingjie and {Nelson}, Erica J. and {Labb{\'e}}, Ivo and {Bezanson}, Rachel and {Brammer}, Gabriel B. and {van Dokkum}, Pieter and {Endsley}, Ryan and {Glazebrook}, Karl and {Greene}, Jenny E. and {Leja}, Joel and {Price}, Sedona H. and {Smit}, Renske and {Stark}, Daniel P. and {Weaver}, John R. and {Whitaker}, Katherine E. and {Atek}, Hakim and {Chevallard}, Jacopo and {Curtis-Lake}, Emma and {Dayal}, Pratika and {Feltre}, Anna and {Franx}, Marijn and {Fudamoto}, Yoshinobu and {Marchesini}, Danilo and {Mowla}, Lamiya A. and {Pan}, Richard and {Suess}, Katherine A. and {Vidal-Garc{\'\i}a}, Alba and {Williams}, Christina C.},
        title = "{JWST UNCOVER: Extremely Red and Compact Object at z $_{phot}$ ≃ 7.6 Triply Imaged by A2744}",
      journal = {\apj},
         year = 2023,
        month = aug,
       volume = {952},
       number = {2},
          eid = {142},
        pages = {142},
          doi = {10.3847/1538-4357/acdc9d},
archivePrefix = {arXiv},
       eprint = {2212.10531},
 primaryClass = {astro-ph.GA},
       adsurl = {https://ui.adsabs.harvard.edu/abs/2023ApJ...952..142F}
}

@ARTICLE{Furtak+25,
       author = {{Furtak}, Lukas J. and {Secunda}, Amy R. and {Greene}, Jenny E. and {Zitrin}, Adi and {Labb{\'e}}, Ivo and {Golubchik}, Miriam and {Bezanson}, Rachel and {Kokorev}, Vasily and {Atek}, Hakim and {Brammer}, Gabriel B. and {Chemerynska}, Iryna and {Cutler}, Sam E. and {Dayal}, Pratika and {Feldmann}, Robert and {Fujimoto}, Seiji and {Glazebrook}, Karl and {Leja}, Joel and {Ma}, Yilun and {Matthee}, Jorryt and {Naidu}, Rohan P. and {Nelson}, Erica J. and {Oesch}, Pascal A. and {Pan}, Richard and {Price}, Sedona H. and {Suess}, Katherine A. and {Wang}, Bingjie and {Weaver}, John R. and {Whitaker}, Katherine E.},
        title = "{Investigating photometric and spectroscopic variability in the multiply imaged little red dot A2744-QSO1}",
      journal = {\aap},
         year = 2025,
        month = jun,
       volume = {698},
          eid = {A227},
        pages = {A227},
          doi = {10.1051/0004-6361/202554110},
archivePrefix = {arXiv},
       eprint = {2502.07875},
 primaryClass = {astro-ph.GA},
       adsurl = {https://ui.adsabs.harvard.edu/abs/2025A&A...698A.227F}
}

@ARTICLE{Lange+23,
       author = {{Lange}, Johannes U.},
        title = "{NAUTILUS: boosting Bayesian importance nested sampling with deep learning}",
      journal = {\mnras},
         year = 2023,
        month = oct,
       volume = {525},
       number = {2},
        pages = {3181-3194},
          doi = {10.1093/mnras/stad2441},
archivePrefix = {arXiv},
       eprint = {2306.16923},
 primaryClass = {astro-ph.IM},
       adsurl = {https://ui.adsabs.harvard.edu/abs/2023MNRAS.525.3181L}
}

@software{DiMascolo+26,
       author = {{Di Mascolo}, Luca},
        title = "{socca: Source Characterization using a Composable Analysis}",
 howpublished = {Astrophysics Source Code Library, record ascl:2602.005},
         year = 2026,
        month = feb,
          eid = {ascl:2602.005},
archivePrefix = {ascl},
       eprint = {2602.005},
       adsurl = {https://ui.adsabs.harvard.edu/abs/2026ascl.soft02005D}
}

@article{Gordon+03,
  author = {{Gordon}, Karl D. and {Clayton}, Geoffrey C. and {Misselt}, K.~A. and {Landolt}, Arlo U. and {Wolff}, Michael J.},
  title = {{A Quantitative Comparison of the Small Magellanic Cloud, Large Magellanic Cloud, and Milky Way Ultraviolet to Near-Infrared Extinction Curves}},
  journal = {\apj},
  year = {2003},
  month = sep,
  volume = {594},
  number = {1},
  pages = {279-293},
  doi = {10.1086/376774},
  archiveprefix = {arXiv},
  eprint = {astro-ph/0305257},
  primaryclass = {astro-ph},
  adsurl = {https://ui.adsabs.harvard.edu/abs/2003ApJ...594..279G},
}

@article{Naidu+25,
  author = {{Naidu}, Rohan P. and {Matthee}, Jorryt and {Katz}, Harley and {de Graaff}, Anna and {Oesch}, Pascal and {Smith}, Aaron and {Greene}, Jenny E. and {Brammer}, Gabriel and {Weibel}, Andrea and {Hviding}, Raphael and {Chisholm}, John and {Labb{\'e}}, Ivo and {Simcoe}, Robert A. and {Witten}, Callum and {Atek}, Hakim and {Baggen}, Josephine F.~W. and {Belli}, Sirio and {Bezanson}, Rachel and {Boogaard}, Leindert A. and {Bose}, Sownak and {Covelo-Paz}, Alba and {Dayal}, Pratika and {Fudamoto}, Yoshinobu and {Furtak}, Lukas J. and {Giovinazzo}, Emma and {Goulding}, Andy and {Gronke}, Max and {Heintz}, Kasper E. and {Hirschmann}, Michaela and {Illingworth}, Garth and {Inoue}, Akio K. and {Johnson}, Benjamin D. and {Leja}, Joel and {Leonova}, Ecaterina and {McConachie}, Ian and {Maseda}, Michael V. and {Natarajan}, Priyamvada and {Nelson}, Erica and {Setton}, David J. and {Shivaei}, Irene and {Sobral}, David and {Stefanon}, Mauro and {Tacchella}, Sandro and {Toft}, Sune and {Torralba}, Alberto and {van Dokkum}, Pieter and {van der Wel}, Arjen and {Volonteri}, Marta and {Walter}, Fabian and {Wang}, Bingjie and {Watson}, Darach},
  title = {{A ``Black Hole Star'' Reveals the Remarkable Gas-Enshrouded Hearts of the Little Red Dots}},
  journal = {arXiv e-prints},
  year = {2025},
  month = mar,
  eid = {arXiv:2503.16596},
  pages = {arXiv:2503.16596},
  doi = {10.48550/arXiv.2503.16596},
  archiveprefix = {arXiv},
  eprint = {2503.16596},
  primaryclass = {astro-ph.GA},
  adsurl = {https://ui.adsabs.harvard.edu/abs/2025arXiv250316596N},
}

@article{Noll+09,
  author = {{Noll}, S. and {Burgarella}, D. and {Giovannoli}, E. and {Buat}, V. and {Marcillac}, D. and {Mu{\~n}oz-Mateos}, J.~C.},
  title = {{Analysis of galaxy spectral energy distributions from far-UV to far-IR with CIGALE: studying a SINGS test sample}},
  journal = {\aap},
  year = {2009},
  month = dec,
  volume = {507},
  number = {3},
  pages = {1793-1813},
  doi = {10.1051/0004-6361/200912497},
  archiveprefix = {arXiv},
  eprint = {0909.5439},
  primaryclass = {astro-ph.CO},
  adsurl = {https://ui.adsabs.harvard.edu/abs/2009A&A...507.1793N},
}

@INPROCEEDINGS{Perrin+14,
       author = {{Perrin}, Marshall D. and {Sivaramakrishnan}, Anand and {Lajoie}, Charles-Philippe and {Elliott}, Erin and {Pueyo}, Laurent and {Ravindranath}, Swara and {Albert}, Lo{\"\i}c.},
        title = "{Updated point spread function simulations for JWST with WebbPSF}",
    booktitle = {Space Telescopes and Instrumentation 2014: Optical, Infrared, and Millimeter Wave},
         year = 2014,
       editor = {{Oschmann}, Jr., Jacobus M. and {Clampin}, Mark and {Fazio}, Giovanni G. and {MacEwen}, Howard A.},
       series = {Society of Photo-Optical Instrumentation Engineers (SPIE) Conference Series},
       volume = {9143},
        month = aug,
          eid = {91433X},
        pages = {91433X},
          doi = {10.1117/12.2056689},
       adsurl = {https://ui.adsabs.harvard.edu/abs/2014SPIE.9143E..3XP}
}

@ARTICLE{Ricci+17,
       author = {{Ricci}, Claudio and {Trakhtenbrot}, Benny and {Koss}, Michael J. and {Ueda}, Yoshihiro and {Schawinski}, Kevin and {Oh}, Kyuseok and {Lamperti}, Isabella and {Mushotzky}, Richard and {Treister}, Ezequiel and {Ho}, Luis C. and {Weigel}, Anna and {Bauer}, Franz E. and {Paltani}, Stephane and {Fabian}, Andrew C. and {Xie}, Yanxia and {Gehrels}, Neil},
        title = "{The close environments of accreting massive black holes are shaped by radiative feedback}",
      journal = {\nat},
         year = 2017,
        month = sep,
       volume = {549},
       number = {7673},
        pages = {488-491},
          doi = {10.1038/nature23906},
archivePrefix = {arXiv},
       eprint = {1709.09651},
 primaryClass = {astro-ph.HE},
       adsurl = {https://ui.adsabs.harvard.edu/abs/2017Natur.549..488R}
}

@ARTICLE{Just+07,
       author = {{Just}, D.~W. and {Brandt}, W.~N. and {Shemmer}, O. and {Steffen}, A.~T. and {Schneider}, D.~P. and {Chartas}, G. and {Garmire}, G.~P.},
        title = "{The X-Ray Properties of the Most Luminous Quasars from the Sloan Digital Sky Survey}",
      journal = {\apj},
         year = 2007,
        month = aug,
       volume = {665},
       number = {2},
        pages = {1004-1022},
          doi = {10.1086/519990},
archivePrefix = {arXiv},
       eprint = {0705.3059},
 primaryClass = {astro-ph},
       adsurl = {https://ui.adsabs.harvard.edu/abs/2007ApJ...665.1004J}
}

@ARTICLE{Steffen+06,
       author = {{Steffen}, A.~T. and {Strateva}, I. and {Brandt}, W.~N. and {Alexander}, D.~M. and {Koekemoer}, A.~M. and {Lehmer}, B.~D. and {Schneider}, D.~P. and {Vignali}, C.},
        title = "{The X-Ray-to-Optical Properties of Optically Selected Active Galaxies over Wide Luminosity and Redshift Ranges}",
      journal = {\aj},
         year = 2006,
        month = jun,
       volume = {131},
       number = {6},
        pages = {2826-2842},
          doi = {10.1086/503627},
archivePrefix = {arXiv},
       eprint = {astro-ph/0602407},
 primaryClass = {astro-ph},
       adsurl = {https://ui.adsabs.harvard.edu/abs/2006AJ....131.2826S}
}

@ARTICLE{Lusso+16,
       author = {{Lusso}, E. and {Risaliti}, G.},
        title = "{The Tight Relation between X-Ray and Ultraviolet Luminosity of Quasars}",
      journal = {\apj},
         year = 2016,
        month = mar,
       volume = {819},
       number = {2},
          eid = {154},
        pages = {154},
          doi = {10.3847/0004-637X/819/2/154},
archivePrefix = {arXiv},
       eprint = {1602.01090},
 primaryClass = {astro-ph.GA},
       adsurl = {https://ui.adsabs.harvard.edu/abs/2016ApJ...819..154L}
}

@ARTICLE{Tananbaum+79,
       author = {{Tananbaum}, H. and {Avni}, Y. and {Branduardi}, G. and {Elvis}, M. and {Fabbiano}, G. and {Feigelson}, E. and {Giacconi}, R. and {Henry}, J.~P. and {Pye}, J.~P. and {Soltan}, A. and {Zamorani}, G.},
        title = "{X-ray studies of quasars with the Einstein Observatory.}",
      journal = {\apjl},
         year = 1979,
        month = nov,
       volume = {234},
        pages = {L9-L13},
          doi = {10.1086/183100},
       adsurl = {https://ui.adsabs.harvard.edu/abs/1979ApJ...234L...9T}
}

@ARTICLE{Gunasekera+25,
       author = {{Gunasekera}, C.~M. and {van Hoof}, P.~A.~M. and {Dehghanian}, M. and {Chakraborty}, P. and {Shaw}, G. and {Bianchi}, S. and {Chatzikos}, M. and {Tsujimoto}, M. and {Ferland}, G.~J.},
        title = "{The 2025 release of Cloudy}",
      journal = {\rmxaa},
         year = 2025,
        month = nov,
       volume = {61},
        pages = {120-133},
          doi = {10.22201/ia.01851101p.2025.61.03.01},
archivePrefix = {arXiv},
       eprint = {2508.01102},
 primaryClass = {astro-ph.GA},
       adsurl = {https://ui.adsabs.harvard.edu/abs/2025RMxAA..61c.120G}
}

@ARTICLE{Yang+20,
       author = {{Yang}, G. and {Boquien}, M. and {Buat}, V. and {Burgarella}, D. and {Ciesla}, L. and {Duras}, F. and {Stalevski}, M. and {Brandt}, W.~N. and {Papovich}, C.},
        title = "{X-CIGALE: Fitting AGN/galaxy SEDs from X-ray to infrared}",
      journal = {\mnras},
         year = 2020,
        month = jan,
       volume = {491},
       number = {1},
        pages = {740-757},
          doi = {10.1093/mnras/stz3001},
archivePrefix = {arXiv},
       eprint = {2001.08263},
 primaryClass = {astro-ph.GA},
       adsurl = {https://ui.adsabs.harvard.edu/abs/2020MNRAS.491..740Y}
}

@ARTICLE{Zhang+26b,
       author = {{Zhang}, Hao and {Theule}, Patrice and {Buat}, Veronique and {Burgarella}, Denis and {Pons}, Estelle and {Boquien}, Mederic},
        title = "{Nebular\_AGN: A CIGALE module for fitting active galactic nucleus emission lines}",
      journal = {arXiv e-prints},
         year = 2026,
        month = jun,
          eid = {arXiv:2606.14643},
        pages = {arXiv:2606.14643},
          doi = {10.48550/arXiv.2606.14643},
archivePrefix = {arXiv},
       eprint = {2606.14643},
 primaryClass = {astro-ph.GA},
       adsurl = {https://ui.adsabs.harvard.edu/abs/2026arXiv260614643Z}
}

@ARTICLE{Conroy+09,
       author = {{Conroy}, Charlie and {Gunn}, James E. and {White}, Martin},
        title = "{The Propagation of Uncertainties in Stellar Population Synthesis Modeling. I. The Relevance of Uncertain Aspects of Stellar Evolution and the Initial Mass Function to the Derived Physical Properties of Galaxies}",
      journal = {\apj},
         year = 2009,
        month = jul,
       volume = {699},
       number = {1},
        pages = {486-506},
          doi = {10.1088/0004-637X/699/1/486},
archivePrefix = {arXiv},
       eprint = {0809.4261},
 primaryClass = {astro-ph},
       adsurl = {https://ui.adsabs.harvard.edu/abs/2009ApJ...699..486C}
}

@ARTICLE{Nicholls+17,
       author = {{Nicholls}, David C. and {Sutherland}, Ralph S. and {Dopita}, Michael A. and {Kewley}, Lisa J. and {Groves}, Brent A.},
        title = "{Abundance scaling in stars, nebulae and galaxies}",
      journal = {\mnras},
         year = 2017,
        month = apr,
       volume = {466},
       number = {4},
        pages = {4403-4422},
          doi = {10.1093/mnras/stw3235},
archivePrefix = {arXiv},
       eprint = {1612.03546},
 primaryClass = {astro-ph.GA},
       adsurl = {https://ui.adsabs.harvard.edu/abs/2017MNRAS.466.4403N}
}

@ARTICLE{Temple+21,
       author = {{Temple}, Matthew J. and {Hewett}, Paul C. and {Banerji}, Manda},
        title = "{Modelling type 1 quasar colours in the era of Rubin and Euclid}",
      journal = {\mnras},
         year = 2021,
        month = nov,
       volume = {508},
       number = {1},
        pages = {737-754},
          doi = {10.1093/mnras/stab2586},
archivePrefix = {arXiv},
       eprint = {2109.04472},
 primaryClass = {astro-ph.GA},
       adsurl = {https://ui.adsabs.harvard.edu/abs/2021MNRAS.508..737T}
}

@ARTICLE{Nikopoulos+26,
       author = {{Nikopoulos}, G.~P. and {Watson}, D. and {Sneppen}, A. and {Rusakov}, V. and {Heintz}, K.~E. and {Witstok}, J. and {Brammer}, G.},
        title = "{Evidence of violation of case B recombination in little red dots}",
      journal = {\aap},
         year = 2026,
        month = jun,
       volume = {710},
          eid = {A136},
        pages = {A136},
          doi = {10.1051/0004-6361/202557604},
archivePrefix = {arXiv},
       eprint = {2510.06362},
 primaryClass = {astro-ph.GA},
       adsurl = {https://ui.adsabs.harvard.edu/abs/2026A&A...710A.136N}
}

@ARTICLE{Inayoshi+25a,
       author = {{Inayoshi}, Kohei},
        title = "{Little Red Dots as the Very First Activity of Black Hole Growth}",
      journal = {\apjl},
         year = 2025,
        month = jul,
       volume = {988},
       number = {1},
          eid = {L22},
        pages = {L22},
          doi = {10.3847/2041-8213/adea66},
archivePrefix = {arXiv},
       eprint = {2503.05537},
 primaryClass = {astro-ph.GA},
       adsurl = {https://ui.adsabs.harvard.edu/abs/2025ApJ...988L..22I}
}

@ARTICLE{Heintz+24,
       author = {{Heintz}, Kasper E. and {Watson}, Darach and {Brammer}, Gabriel and {Vejlgaard}, Simone and {Hutter}, Anne and {Strait}, Victoria B. and {Matthee}, Jorryt and {Oesch}, Pascal A. and {Jakobsson}, P{\'a}ll and {Tanvir}, Nial R. and {Laursen}, Peter and {Naidu}, Rohan P. and {Mason}, Charlotte A. and {Killi}, Meghana and {Jung}, Intae and {Hsiao}, Tiger Yu-Yang and {Abdurro'uf} and {Coe}, Dan and {Arrabal Haro}, Pablo and {Finkelstein}, Steven L. and {Toft}, Sune},
        title = "{Strong damped Lyman-{\ensuremath{\alpha}} absorption in young star-forming galaxies at redshifts 9 to 11}",
      journal = {Science},
         year = 2024,
        month = may,
       volume = {384},
       number = {6698},
        pages = {890-894},
          doi = {10.1126/science.adj0343},
archivePrefix = {arXiv},
       eprint = {2306.00647},
 primaryClass = {astro-ph.GA},
       adsurl = {https://ui.adsabs.harvard.edu/abs/2024Sci...384..890H}
}

@ARTICLE{Hopkins+10,
       author = {{Hopkins}, Philip F. and {Murray}, Norman and {Quataert}, Eliot and {Thompson}, Todd A.},
        title = "{A maximum stellar surface density in dense stellar systems}",
      journal = {\mnras},
         year = 2010,
        month = jan,
       volume = {401},
       number = {1},
        pages = {L19-L23},
          doi = {10.1111/j.1745-3933.2009.00777.x},
archivePrefix = {arXiv},
       eprint = {0908.4088},
 primaryClass = {astro-ph.CO},
       adsurl = {https://ui.adsabs.harvard.edu/abs/2010MNRAS.401L..19H}
}

@ARTICLE{Ashall+26,
       author = {{Ashall}, Chris and {Naidu}, Rohan P. and {Torralba}, Alberto and {Shivaei}, Irene and {Chisholm}, John and {Liu}, Hanpu and {Liu}, Zhaoran and {Matthee}, Jorryt and {Medler}, Kyle and {Mera}, Tyco and {Moriya}, Takashi J. and {Nandal}, Devesh and {Simcoe}, Robert A. and {Sun}, Wendy Q.},
        title = "{Can Little Red Dots Contribute To The Early Universe Cosmic Dust Budget?}",
      journal = {arXiv e-prints},
         year = 2026,
        month = aug,
          eid = {arXiv:2608.29011},
        pages = {arXiv:2608.29011},
          doi = {10.48550/arXiv.2608.29011},
archivePrefix = {arXiv},
       eprint = {2608.29011},
 primaryClass = {astro-ph.GA},
       adsurl = {https://ui.adsabs.harvard.edu/abs/2026arXiv260829011A}
}

@ARTICLE{Barvainis87,
       author = {{Barvainis}, Richard},
        title = "{Hot Dust and the Near-Infrared Bump in the Continuum Spectra of Quasars and Active Galactic Nuclei}",
      journal = {\apj},
         year = 1987,
        month = sep,
       volume = {320},
        pages = {537},
          doi = {10.1086/165571},
       adsurl = {https://ui.adsabs.harvard.edu/abs/1987ApJ...320..537B}
}

@ARTICLE{Richards+06,
       author = {{Richards}, Gordon T. and {Lacy}, Mark and {Storrie-Lombardi}, Lisa J. and {Hall}, Patrick B. and {Gallagher}, S.~C. and {Hines}, Dean C. and {Fan}, Xiaohui and {Papovich}, Casey and {Vanden Berk}, Daniel E. and {Trammell}, George B. and {Schneider}, Donald P. and {Vestergaard}, Marianne and {York}, Donald G. and {Jester}, Sebastian and {Anderson}, Scott F. and {Budav{\'a}ri}, Tam{\'a}s and {Szalay}, Alexander S.},
        title = "{Spectral Energy Distributions and Multiwavelength Selection of Type 1 Quasars}",
      journal = {\apjs},
         year = 2006,
        month = oct,
       volume = {166},
       number = {2},
        pages = {470-497},
          doi = {10.1086/506525},
archivePrefix = {arXiv},
       eprint = {astro-ph/0601558},
 primaryClass = {astro-ph},
       adsurl = {https://ui.adsabs.harvard.edu/abs/2006ApJS..166..470R}
}

@ARTICLE{deGraaff+25B,
       author = {{de Graaff}, Anna and {Rix}, Hans-Walter and {Naidu}, Rohan P. and {Labb{\'e}}, Ivo and {Wang}, Bingjie and {Leja}, Joel and {Matthee}, Jorryt and {Katz}, Harley and {Greene}, Jenny E. and {Hviding}, Raphael E. and {Baggen}, Josephine and {Bezanson}, Rachel and {Boogaard}, Leindert A. and {Brammer}, Gabriel and {Dayal}, Pratika and {van Dokkum}, Pieter and {Goulding}, Andy D. and {Hirschmann}, Michaela and {Maseda}, Michael V. and {McConachie}, Ian and {Miller}, Tim B. and {Nelson}, Erica and {Oesch}, Pascal A. and {Setton}, David J. and {Shivaei}, Irene and {Weibel}, Andrea and {Whitaker}, Katherine E. and {Williams}, Christina C.},
        title = "{A remarkable ruby: Absorption in dense gas, rather than evolved stars, drives the extreme Balmer break of a little red dot at z = 3.5}",
      journal = {\aap},
         year = 2025,
        month = sep,
       volume = {701},
          eid = {A168},
        pages = {A168},
          doi = {10.1051/0004-6361/202554681},
archivePrefix = {arXiv},
       eprint = {2503.16600},
 primaryClass = {astro-ph.GA},
       adsurl = {https://ui.adsabs.harvard.edu/abs/2025A&A...701A.168D}
}

@software{Brammer+23,
       author = {{Brammer}, Gabriel},
        title = "{msaexp: NIRSpec analysis tools}",
         year = 2023,
        month = aug,
          eid = {10.5281/zenodo.8233475},
          doi = {10.5281/zenodo.8233475},
      version = {0.6.13},
    publisher = {Zenodo},
       adsurl = {https://ui.adsabs.harvard.edu/abs/2023zndo...8233475B}
}

@ARTICLE{Secunda+26,
       author = {{Secunda}, Amy and {Somerville}, Rachel S. and {Jiang}, Yan-Fei and {Greene}, Jenny E. and {Furtak}, Lukas J. and {Zitrin}, Adi},
        title = "{Do Little Red Dots Vary?}",
      journal = {\apj},
         year = 2026,
        month = jan,
       volume = {996},
       number = {1},
          eid = {6},
        pages = {6},
          doi = {10.3847/1538-4357/ae1f08},
archivePrefix = {arXiv},
       eprint = {2509.03571},
 primaryClass = {astro-ph.GA},
       adsurl = {https://ui.adsabs.harvard.edu/abs/2026ApJ...996....6S}
}

@ARTICLE{Maiolino+25,
       author = {{Maiolino}, Roberto and {Risaliti}, Guido and {Signorini}, Matilde and {Trefoloni}, Bartolomeo and {Juod{\v{z}}balis}, Ignas and {Scholtz}, Jan and {{\"U}bler}, Hannah and {D'Eugenio}, Francesco and {Carniani}, Stefano and {Fabian}, Andy and {Ji}, Xihan and {Mazzolari}, Giovanni and {Bertola}, Elena and {Brusa}, Marcella and {Bunker}, Andrew J. and {Charlot}, Stephane and {Comastri}, Andrea and {Cresci}, Giovanni and {DeCoursey}, Christa Noel and {Egami}, Eiichi and {Fiore}, Fabrizio and {Gilli}, Roberto and {Perna}, Michele and {Tacchella}, Sandro and {Venturi}, Giacomo},
        title = "{JWST meets Chandra: a large population of Compton thick, feedback-free, and intrinsically X-ray weak AGN, with a sprinkle of SNe}",
      journal = {\mnras},
         year = 2025,
        month = apr,
       volume = {538},
       number = {3},
        pages = {1921-1943},
          doi = {10.1093/mnras/staf359},
archivePrefix = {arXiv},
       eprint = {2405.00504},
 primaryClass = {astro-ph.GA},
       adsurl = {https://ui.adsabs.harvard.edu/abs/2025MNRAS.538.1921M}
}

@ARTICLE{Rieke+25,
       author = {{Rieke}, George H. and {Sun}, Yang and {Lyu}, Jianwei and {Willmer}, Christopher N.~A. and {Zhu}, Yongda and {Rinaldi}, Pierluigi and {Stone}, Meredith A. and {Hainline}, Kevin N. and {P{\'e}rez-Gonz{\'a}lez}, Pablo G.},
        title = "{Confirming Near- to Mid-infrared Photometrically Identified Obscured AGNs in the JWST Era}",
      journal = {\apj},
         year = 2025,
        month = nov,
       volume = {994},
       number = {1},
          eid = {35},
        pages = {35},
          doi = {10.3847/1538-4357/adff79},
archivePrefix = {arXiv},
       eprint = {2510.07303},
 primaryClass = {astro-ph.GA},
       adsurl = {https://ui.adsabs.harvard.edu/abs/2025ApJ...994...35R}
}

@ARTICLE{Hainline+16,
       author = {{Hainline}, Kevin N. and {Reines}, Amy E. and {Greene}, Jenny E. and {Stern}, Daniel},
        title = "{Mid-infrared Colors of Dwarf Galaxies: Young Starbursts Mimicking Active Galactic Nuclei}",
      journal = {\apj},
         year = 2016,
        month = dec,
       volume = {832},
       number = {2},
          eid = {119},
        pages = {119},
          doi = {10.3847/0004-637X/832/2/119},
archivePrefix = {arXiv},
       eprint = {1609.06721},
 primaryClass = {astro-ph.GA},
       adsurl = {https://ui.adsabs.harvard.edu/abs/2016ApJ...832..119H}
}

@ARTICLE{Sturm+25b,
       author = {{Sturm}, Megan R. and {Hayes}, Bayli and {Reines}, Amy E.},
        title = "{Star-Forming Nuclear Clusters in Dwarf Galaxies Mimicking AGN Signatures in the Mid-Infrared}",
      journal = {arXiv e-prints},
         year = 2025,
        month = jan,
          eid = {arXiv:2501.09791},
        pages = {arXiv:2501.09791},
          doi = {10.48550/arXiv.2501.09791},
archivePrefix = {arXiv},
       eprint = {2501.09791},
 primaryClass = {astro-ph.GA},
       adsurl = {https://ui.adsabs.harvard.edu/abs/2025arXiv250109791S}
}

@ARTICLE{RemyRuyer+14,
       author = {{R{\'e}my-Ruyer}, A. and {Madden}, S.~C. and {Galliano}, F. and {Galametz}, M. and {Takeuchi}, T.~T. and {Asano}, R.~S. and {Zhukovska}, S. and {Lebouteiller}, V. and {Cormier}, D. and {Jones}, A. and {Bocchio}, M. and {Baes}, M. and {Bendo}, G.~J. and {Boquien}, M. and {Boselli}, A. and {DeLooze}, I. and {Doublier-Pritchard}, V. and {Hughes}, T. and {Karczewski}, O. {\L}. and {Spinoglio}, L.},
        title = "{Gas-to-dust mass ratios in local galaxies over a 2 dex metallicity range}",
      journal = {\aap},
         year = 2014,
        month = mar,
       volume = {563},
          eid = {A31},
        pages = {A31},
          doi = {10.1051/0004-6361/201322803},
archivePrefix = {arXiv},
       eprint = {1312.3442},
 primaryClass = {astro-ph.GA},
       adsurl = {https://ui.adsabs.harvard.edu/abs/2014A&A...563A..31R}
}

@ARTICLE{Lambrides+26,
       author = {{Lambrides}, Erini and {Larson}, Rebecca L. and {Garofali}, Kristen and {Ptak}, Andrew and {Chiaberge}, Marco and {Long}, Arianna S. and {Hutchison}, Taylor A. and {Norman}, Colin and {McKinney}, Jed and {Akins}, Hollis B. and {Berg}, Danielle A. and {Chisholm}, John and {Civano}, Francesca and {Cloonan}, Aidan P. and {Endsley}, Ryan and {Faisst}, Andreas L. and {Gilli}, Roberto and {Gillman}, Steven and {Hirschmann}, Michaela and {Kartaltepe}, Jeyhan S. and {Kocevski}, Dale D. and {Kokorev}, Vasily and {Pacucci}, Fabio and {Richardson}, Chris T. and {Stiavelli}, Massimo and {Whalen}, Kelly E.},
        title = "{The case for super-Eddington accretion in JWST broad-line active galactic nuclei during the first billion years}",
      journal = {Nature Astronomy},
         year = 2026,
        month = jun,
       volume = {10},
        pages = {868-879},
          doi = {10.1038/s41550-026-02813-w},
archivePrefix = {arXiv},
       eprint = {2409.13047},
 primaryClass = {astro-ph.HE},
       adsurl = {https://ui.adsabs.harvard.edu/abs/2026NatAs..10..868L}
}

@ARTICLE{Labbe+25,
       author = {{Labbe}, Ivo and {Greene}, Jenny E. and {Bezanson}, Rachel and {Fujimoto}, Seiji and {Furtak}, Lukas J. and {Goulding}, Andy D. and {Matthee}, Jorryt and {Naidu}, Rohan P. and {Oesch}, Pascal A. and {Atek}, Hakim and {Brammer}, Gabriel and {Chemerynska}, Iryna and {Coe}, Dan and {Cutler}, Sam E. and {Dayal}, Pratika and {Feldmann}, Robert and {Franx}, Marijn and {Glazebrook}, Karl and {Leja}, Joel and {Maseda}, Michael and {Marchesini}, Danilo and {Nanayakkara}, Themiya and {Nelson}, Erica J. and {Pan}, Richard and {Papovich}, Casey and {Price}, Sedona H. and {Suess}, Katherine A. and {Wang}, Bingjie and {Weaver}, John R. and {Whitaker}, Katherine E. and {Williams}, Christina C. and {Zitrin}, Adi},
        title = "{UNCOVER: Candidate Red Active Galactic Nuclei at 3 < z < 7 with JWST and ALMA}",
      journal = {\apj},
         year = 2025,
        month = jan,
       volume = {978},
       number = {1},
          eid = {92},
        pages = {92},
          doi = {10.3847/1538-4357/ad3551},
archivePrefix = {arXiv},
       eprint = {2306.07320},
 primaryClass = {astro-ph.GA},
       adsurl = {https://ui.adsabs.harvard.edu/abs/2025ApJ...978...92L}
}

@ARTICLE{KokuboHarikane25,
       author = {{Kokubo}, Mitsuru and {Harikane}, Yuichi},
        title = "{Challenging the Active Galactic Nucleus Scenario for JWST/NIRSpec Little Red Dot and Non─Little Red Dot Broad H{\ensuremath{\alpha}} Emitters in Light of Nondetection of NIRCam Photometric Variability and X-Ray}",
      journal = {\apj},
         year = 2025,
        month = dec,
       volume = {995},
       number = {1},
          eid = {24},
        pages = {24},
          doi = {10.3847/1538-4357/ae119e},
archivePrefix = {arXiv},
       eprint = {2407.04777},
 primaryClass = {astro-ph.GA},
       adsurl = {https://ui.adsabs.harvard.edu/abs/2025ApJ...995...24K}
}

@ARTICLE{Tee+25,
       author = {{Tee}, Wei Leong and {Fan}, Xiaohui and {Wang}, Feige and {Yang}, Jinyi},
        title = "{Lack of Rest-frame Ultraviolet Variability in Little Red Dots Based on HST and JWST Observations}",
      journal = {\apjl},
         year = 2025,
        month = apr,
       volume = {983},
       number = {1},
          eid = {L26},
        pages = {L26},
          doi = {10.3847/2041-8213/adc5e3},
archivePrefix = {arXiv},
       eprint = {2412.05242},
 primaryClass = {astro-ph.GA},
       adsurl = {https://ui.adsabs.harvard.edu/abs/2025ApJ...983L..26T}
}

@ARTICLE{deGraaff+25,
       author = {{de Graaff}, Anna and {Brammer}, Gabriel and {Weibel}, Andrea and {Lewis}, Zach and {Maseda}, Michael V. and {Oesch}, Pascal A. and {Bezanson}, Rachel and {Boogaard}, Leindert A. and {Cleri}, Nikko J. and {Cooper}, Olivia R. and {Gottumukkala}, Rashmi and {Greene}, Jenny E. and {Hirschmann}, Michaela and {Hviding}, Raphael E. and {Katz}, Harley and {Labb{\'e}}, Ivo and {Leja}, Joel and {Matthee}, Jorryt and {McConachie}, Ian and {Miller}, Tim B. and {Naidu}, Rohan P. and {Price}, Sedona H. and {Rix}, Hans-Walter and {Setton}, David J. and {Suess}, Katherine A. and {Wang}, Bingjie and {Whitaker}, Katherine E. and {Williams}, Christina C.},
        title = "{RUBIES: A complete census of the bright and red distant Universe with JWST/NIRSpec}",
      journal = {\aap},
         year = 2025,
        month = may,
       volume = {697},
          eid = {A189},
        pages = {A189},
          doi = {10.1051/0004-6361/202452186},
archivePrefix = {arXiv},
       eprint = {2409.05948},
 primaryClass = {astro-ph.GA},
       adsurl = {https://ui.adsabs.harvard.edu/abs/2025A&A...697A.189D}
}

@ARTICLE{Chisholm+26,
       author = {{Chisholm}, John and {Berg}, Danielle A. and {Boylan-Kolchin}, Michael and {de Graaff}, Anna and {Furtak}, Lukas J. and {Kokorev}, Vasily and {Matthee}, Jorryt and {Mu{\~n}oz}, Julian B. and {Naidu}, Rohan P. and {Sander}, Andreas A.~C.},
        title = "{Little Red Dots as Globular Clusters in Formation}",
      journal = {\apjl},
         year = 2026,
        month = jun,
       volume = {1004},
       number = {1},
          eid = {L4},
        pages = {L4},
          doi = {10.3847/2041-8213/ae6dae},
archivePrefix = {arXiv},
       eprint = {2602.15935},
 primaryClass = {astro-ph.GA},
       adsurl = {https://ui.adsabs.harvard.edu/abs/2026ApJ..1004L...4C}
}

@ARTICLE{Rusakov+26,
       author = {{Rusakov}, V. and {Watson}, D. and {Nikopoulos}, G.~P. and {Brammer}, G. and {Gottumukkala}, R. and {Harvey}, T. and {Heintz}, K.~E. and {Damgaard}, R. and {Sim}, S.~A. and {Sneppen}, A. and {Vijayan}, A.~P. and {Adams}, N. and {Austin}, D. and {Conselice}, C.~J. and {Goolsby}, C.~M. and {Toft}, S. and {Witstok}, J.},
        title = "{Little red dots as young supermassive black holes in dense ionized cocoons}",
      journal = {\nat},
         year = 2026,
        month = jan,
       volume = {649},
       number = {8097},
        pages = {574-579},
          doi = {10.1038/s41586-025-09900-4},
archivePrefix = {arXiv},
       eprint = {2503.16595},
 primaryClass = {astro-ph.GA},
       adsurl = {https://ui.adsabs.harvard.edu/abs/2026Natur.649..574R}
}

@ARTICLE{Harikane+23,
       author = {{Harikane}, Yuichi and {Zhang}, Yechi and {Nakajima}, Kimihiko and {Ouchi}, Masami and {Isobe}, Yuki and {Ono}, Yoshiaki and {Hatano}, Shun and {Xu}, Yi and {Umeda}, Hiroya},
        title = "{A JWST/NIRSpec First Census of Broad-line AGNs at z = 4-7: Detection of 10 Faint AGNs with M $_{BH}$ {}10$^{6}$-{}10$^{8}$ M $_{☉}$ and Their Host Galaxy Properties}",
      journal = {\apj},
         year = 2023,
        month = dec,
       volume = {959},
       number = {1},
          eid = {39},
        pages = {39},
          doi = {10.3847/1538-4357/ad029e},
archivePrefix = {arXiv},
       eprint = {2303.11946},
 primaryClass = {astro-ph.GA},
       adsurl = {https://ui.adsabs.harvard.edu/abs/2023ApJ...959...39H}
}

@ARTICLE{Treiber+25,
       author = {{Treiber}, Helena and {Greene}, Jenny E. and {Weaver}, John R. and {Miller}, Tim B. and {Furtak}, Lukas J. and {Setton}, David J. and {Wang}, Bingjie and {de Graaff}, Anna and {Bezanson}, Rachel and {Brammer}, Gabriel and {Cutler}, Sam E. and {Dayal}, Pratika and {Feldmann}, Robert and {Fujimoto}, Seiji and {Goulding}, Andy D. and {Kokorev}, Vasily and {Labbe}, Ivo and {Leja}, Joel and {Marchesini}, Danilo and {Nanayakkara}, Themiya and {Nelson}, Erica and {Pan}, Richard and {Price}, Sedona H. and {Siegel}, Jared and {Suess}, Katherine A. and {Whitaker}, Katherine E.},
        title = "{UNCOVERing the High-redshift AGN Population among Extreme UV Line Emitters}",
      journal = {\apj},
         year = 2025,
        month = may,
       volume = {984},
       number = {1},
          eid = {93},
        pages = {93},
          doi = {10.3847/1538-4357/adc38f},
archivePrefix = {arXiv},
       eprint = {2409.12232},
 primaryClass = {astro-ph.GA},
       adsurl = {https://ui.adsabs.harvard.edu/abs/2025ApJ...984...93T}
}

@ARTICLE{Kocevski+25,
       author = {{Kocevski}, Dale D. and {Finkelstein}, Steven L. and {Barro}, Guillermo and {Taylor}, Anthony J. and {Calabr{\`o}}, Antonello and {Laloux}, Brivael and {Buchner}, Johannes and {Trump}, Jonathan R. and {Leung}, Gene C.~K. and {Yang}, Guang and {Dickinson}, Mark and {P{\'e}rez-Gonz{\'a}lez}, Pablo G. and {Pacucci}, Fabio and {Inayoshi}, Kohei and {Somerville}, Rachel S. and {McGrath}, Elizabeth J. and {Akins}, Hollis B. and {Bagley}, Micaela B. and {Bowler}, Rebecca A.~A. and {Bisigello}, Laura and {Carnall}, Adam and {Casey}, Caitlin M. and {Cheng}, Yingjie and {Cleri}, Nikko J. and {Costantin}, Luca and {Cullen}, Fergus and {Davis}, Kelcey and {Donnan}, Callum T. and {Dunlop}, James S. and {Ellis}, Richard S. and {Ferguson}, Henry C. and {Fujimoto}, Seiji and {Fontana}, Adriano and {Giavalisco}, Mauro and {Grazian}, Andrea and {Grogin}, Norman A. and {Hathi}, Nimish P. and {Hirschmann}, Michaela and {Huertas-Company}, Marc and {Holwerda}, Benne W. and {Illingworth}, Garth and {Juneau}, St{\'e}phanie and {Kartaltepe}, Jeyhan S. and {Koekemoer}, Anton M. and {Li}, Wenxiu and {Lucas}, Ray A. and {Magee}, Dan and {Mason}, Charlotte and {McLeod}, Derek J. and {McLure}, Ross J. and {Napolitano}, Lorenzo and {Papovich}, Casey and {Pirzkal}, Nor and {Rodighiero}, Giulia and {Santini}, Paola and {Wilkins}, Stephen M. and {Yung}, L.~Y. Aaron},
        title = "{The Rise of Faint, Red Active Galactic Nuclei at z > 4: A Sample of Little Red Dots in the JWST Extragalactic Legacy Fields}",
      journal = {\apj},
         year = 2025,
        month = jun,
       volume = {986},
       number = {2},
          eid = {126},
        pages = {126},
          doi = {10.3847/1538-4357/adbc7d},
archivePrefix = {arXiv},
       eprint = {2404.03576},
 primaryClass = {astro-ph.GA},
       adsurl = {https://ui.adsabs.harvard.edu/abs/2025ApJ...986..126K}
}

@ARTICLE{Greene+24,
       author = {{Greene}, Jenny E. and {Labbe}, Ivo and {Goulding}, Andy D. and {Furtak}, Lukas J. and {Chemerynska}, Iryna and {Kokorev}, Vasily and {Dayal}, Pratika and {Volonteri}, Marta and {Williams}, Christina C. and {Wang}, Bingjie and {Setton}, David J. and {Burgasser}, Adam J. and {Bezanson}, Rachel and {Atek}, Hakim and {Brammer}, Gabriel and {Cutler}, Sam E. and {Feldmann}, Robert and {Fujimoto}, Seiji and {Glazebrook}, Karl and {de Graaff}, Anna and {Khullar}, Gourav and {Leja}, Joel and {Marchesini}, Danilo and {Maseda}, Michael V. and {Matthee}, Jorryt and {Miller}, Tim B. and {Naidu}, Rohan P. and {Nanayakkara}, Themiya and {Oesch}, Pascal A. and {Pan}, Richard and {Papovich}, Casey and {Price}, Sedona H. and {van Dokkum}, Pieter and {Weaver}, John R. and {Whitaker}, Katherine E. and {Zitrin}, Adi},
        title = "{UNCOVER Spectroscopy Confirms the Surprising Ubiquity of Active Galactic Nuclei in Red Sources at z > 5}",
      journal = {\apj},
         year = 2024,
        month = mar,
       volume = {964},
       number = {1},
          eid = {39},
        pages = {39},
          doi = {10.3847/1538-4357/ad1e5f},
archivePrefix = {arXiv},
       eprint = {2309.05714},
 primaryClass = {astro-ph.GA},
       adsurl = {https://ui.adsabs.harvard.edu/abs/2024ApJ...964...39G}
}

@ARTICLE{Begelman+26,
       author = {{Begelman}, Mitchell C. and {Dexter}, Jason},
        title = "{Little Red Dots as Late-stage Quasi-stars}",
      journal = {\apj},
         year = 2026,
        month = jan,
       volume = {996},
       number = {1},
          eid = {48},
        pages = {48},
          doi = {10.3847/1538-4357/ae274a},
archivePrefix = {arXiv},
       eprint = {2507.09085},
 primaryClass = {astro-ph.GA},
       adsurl = {https://ui.adsabs.harvard.edu/abs/2026ApJ...996...48B}
}

@ARTICLE{Barro+24,
       author = {{Barro}, Guillermo and {P{\'e}rez-Gonz{\'a}lez}, Pablo G. and {Kocevski}, Dale D. and {McGrath}, Elizabeth J. and {Trump}, Jonathan R. and {Simons}, Raymond C. and {Somerville}, Rachel S. and {Yung}, L.~Y. Aaron and {Arrabal Haro}, Pablo and {Akins}, Hollis B. and {Bagley}, Michaela B. and {Cleri}, Nikko J. and {Costantin}, Luca and {Davis}, Kelcey and {Dickinson}, Mark and {Finkelstein}, Steve L. and {Giavalisco}, Mauro and {G{\'o}mez-Guijarro}, Carlos and {Hathi}, Nimish P. and {Hirschmann}, Michaela and {Holwerda}, Benne W. and {Huertas-Company}, Marc and {Kartaltepe}, Jeyhan S. and {Koekemoer}, Anton M. and {Lucas}, Ray A. and {Papovich}, Casey and {Pirzkal}, Nor and {Seill{\'e}}, Lise-Marie and {Tacchella}, Sandro and {Wuyts}, Stijn and {Wilkins}, Stephen M. and {de la Vega}, Alexander and {Yang}, Guang and {Zavala}, Jorge A.},
        title = "{Extremely Red Galaxies at z = 5─9 with MIRI and NIRSpec: Dusty Galaxies or Obscured Active Galactic Nuclei?}",
      journal = {\apj},
         year = 2024,
        month = mar,
       volume = {963},
       number = {2},
          eid = {128},
        pages = {128},
          doi = {10.3847/1538-4357/ad167e},
archivePrefix = {arXiv},
       eprint = {2305.14418},
 primaryClass = {astro-ph.GA},
       adsurl = {https://ui.adsabs.harvard.edu/abs/2024ApJ...963..128B}
}

@ARTICLE{Inayoshi+25b,
       author = {{Inayoshi}, Kohei and {Maiolino}, Roberto},
        title = "{Extremely Dense Gas around Little Red Dots and High-redshift Active Galactic Nuclei: A Nonstellar Origin of the Balmer Break and Absorption Features}",
      journal = {\apjl},
         year = 2025,
        month = feb,
       volume = {980},
       number = {2},
          eid = {L27},
        pages = {L27},
          doi = {10.3847/2041-8213/adaebd},
archivePrefix = {arXiv},
       eprint = {2409.07805},
 primaryClass = {astro-ph.GA},
       adsurl = {https://ui.adsabs.harvard.edu/abs/2025ApJ...980L..27I}
}

@ARTICLE{Baggen+24,
       author = {{Baggen}, Josephine F.~W. and {van Dokkum}, Pieter and {Brammer}, Gabriel and {de Graaff}, Anna and {Franx}, Marijn and {Greene}, Jenny and {Labb{\'e}}, Ivo and {Leja}, Joel and {Maseda}, Michael V. and {Nelson}, Erica J. and {Rix}, Hans-Walter and {Wang}, Bingjie and {Weibel}, Andrea},
        title = "{The Small Sizes and High Implied Densities of ``Little Red Dots'' with Balmer Breaks Could Explain Their Broad Emission Lines without an Active Galactic Nucleus}",
      journal = {\apjl},
         year = 2024,
        month = dec,
       volume = {977},
       number = {1},
          eid = {L13},
        pages = {L13},
          doi = {10.3847/2041-8213/ad90b8},
archivePrefix = {arXiv},
       eprint = {2408.07745},
 primaryClass = {astro-ph.GA},
       adsurl = {https://ui.adsabs.harvard.edu/abs/2024ApJ...977L..13B}
}

@ARTICLE{Liu+26,
       author = {{Liu}, Hanpu and {Jiang}, Yan-Fei and {Quataert}, Eliot and {Greene}, Jenny E. and {Ma}, Yilun and {Lin}, Xiaojing},
        title = "{Synthetic Spectral Library of Optically Thick Atmospheres for Little Red Dots}",
      journal = {arXiv e-prints},
         year = 2026,
        month = mar,
          eid = {arXiv:2603.02317},
        pages = {arXiv:2603.02317},
          doi = {10.48550/arXiv.2603.02317},
archivePrefix = {arXiv},
       eprint = {2603.02317},
 primaryClass = {astro-ph.GA},
       adsurl = {https://ui.adsabs.harvard.edu/abs/2026arXiv260302317L}
}

@ARTICLE{Labbe+23,
       author = {{Labb{\'e}}, Ivo and {van Dokkum}, Pieter and {Nelson}, Erica and {Bezanson}, Rachel and {Suess}, Katherine A. and {Leja}, Joel and {Brammer}, Gabriel and {Whitaker}, Katherine and {Mathews}, Elijah and {Stefanon}, Mauro and {Wang}, Bingjie},
        title = "{A population of red candidate massive galaxies  600 Myr after the Big Bang}",
      journal = {\nat},
         year = 2023,
        month = apr,
       volume = {616},
       number = {7956},
        pages = {266-269},
          doi = {10.1038/s41586-023-05786-2},
archivePrefix = {arXiv},
       eprint = {2207.12446},
 primaryClass = {astro-ph.GA},
       adsurl = {https://ui.adsabs.harvard.edu/abs/2023Natur.616..266L}
}

@ARTICLE{Kido+25,
       author = {{Kido}, Daisaburo and {Ioka}, Kunihito and {Hotokezaka}, Kenta and {Inayoshi}, Kohei and {Irwin}, Christopher M.},
        title = "{Black hole envelopes in Little Red Dots}",
      journal = {\mnras},
         year = 2025,
        month = dec,
       volume = {544},
       number = {4},
        pages = {3407-3416},
          doi = {10.1093/mnras/staf1898},
archivePrefix = {arXiv},
       eprint = {2505.06965},
 primaryClass = {astro-ph.HE},
       adsurl = {https://ui.adsabs.harvard.edu/abs/2025MNRAS.544.3407K}
}

@ARTICLE{Nandal+26,
       author = {{Nandal}, Devesh and {Loeb}, Abraham},
        title = "{Supermassive Stars Match the Spectral Signatures of JWST's Little Red Dots}",
      journal = {\apj},
         year = 2026,
        month = feb,
       volume = {998},
       number = {1},
          eid = {124},
        pages = {124},
          doi = {10.3847/1538-4357/ae32f3},
archivePrefix = {arXiv},
       eprint = {2507.12618},
 primaryClass = {astro-ph.GA},
       adsurl = {https://ui.adsabs.harvard.edu/abs/2026ApJ...998..124N}
}

@ARTICLE{Dong+08,
       author = {{Dong}, Xiaobo and {Wang}, Tinggui and {Wang}, Jianguo and {Yuan}, Weimin and {Zhou}, Hongyan and {Dai}, Haifeng and {Zhang}, Kai},
        title = "{Broad-line Balmer decrements in blue active galactic nuclei}",
      journal = {\mnras},
         year = 2008,
        month = jan,
       volume = {383},
       number = {2},
        pages = {581-592},
          doi = {10.1111/j.1365-2966.2007.12560.x},
archivePrefix = {arXiv},
       eprint = {0710.1458},
 primaryClass = {astro-ph},
       adsurl = {https://ui.adsabs.harvard.edu/abs/2008MNRAS.383..581D}
}

@article{Maiolino+24,
  author = {{Maiolino}, Roberto and {Scholtz}, Jan and {Witstok}, Joris and {Carniani}, Stefano and {D'Eugenio}, Francesco and {de Graaff}, Anna and {{\"U}bler}, Hannah and {Tacchella}, Sandro and {Curtis-Lake}, Emma and {Arribas}, Santiago and {Bunker}, Andrew and {Charlot}, St{\'e}phane and {Chevallard}, Jacopo and {Curti}, Mirko and {Looser}, Tobias J. and {Maseda}, Michael V. and {Rawle}, Timothy D. and {Rodr{\'\i}guez del Pino}, Bruno and {Willott}, Chris J. and {Egami}, Eiichi and {Eisenstein}, Daniel J. and {Hainline}, Kevin N. and {Robertson}, Brant and {Williams}, Christina C. and {Willmer}, Christopher N.~A. and {Baker}, William M. and {Boyett}, Kristan and {DeCoursey}, Christa and {Fabian}, Andrew C. and {Helton}, Jakob M. and {Ji}, Zhiyuan and {Jones}, Gareth C. and {Kumari}, Nimisha and {Laporte}, Nicolas and {Nelson}, Erica J. and {Perna}, Michele and {Sandles}, Lester and {Shivaei}, Irene and {Sun}, Fengwu},
  title = {{A small and vigorous black hole in the early Universe}},
  journal = {\nat},
  year = {2024},
  month = mar,
  volume = {627},
  number = {8002},
  pages = {59-63},
  doi = {10.1038/s41586-024-07052-5},
  archiveprefix = {arXiv},
  eprint = {2305.12492},
  primaryclass = {astro-ph.GA},
  adsurl = {https://ui.adsabs.harvard.edu/abs/2024Natur.627...59M},
}

@article{Pacucci+23,
  author = {{Pacucci}, Fabio and {Nguyen}, Bao and {Carniani}, Stefano and {Maiolino}, Roberto and {Fan}, Xiaohui},
  title = {{JWST CEERS and JADES Active Galaxies at z = 4-7 Violate the Local M $_{{\textbullet}}$-M $_{{\ensuremath{\star}}}$ Relation at >3{\ensuremath{\sigma}}: Implications for Low-mass Black Holes and Seeding Models}},
  journal = {\apjl},
  year = {2023},
  month = nov,
  volume = {957},
  number = {1},
  eid = {L3},
  pages = {L3},
  doi = {10.3847/2041-8213/ad0158},
  archiveprefix = {arXiv},
  eprint = {2308.12331},
  primaryclass = {astro-ph.GA},
  adsurl = {https://ui.adsabs.harvard.edu/abs/2023ApJ...957L...3P},
}

@article{Lin+25,
  author = {{Lin}, Ruqiu and {Zheng}, Zhen-Ya and {Jiang}, Chunyan and {Yuan}, Fang-Ting and {Ho}, Luis C. and {Wang}, Junxian and {Jiang}, Linhua and {Rhoads}, James E. and {Malhotra}, Sangeeta and {Barrientos}, L. Felipe and {Wold}, Isak and {Infante}, Leopoldo and {Zhu}, Shuairu and {Ji}, Xiang and {Fu}, Xiaodan},
  title = {{Discovery of Local Analogs to JWST's Little Red Dots}},
  journal = {\apjl},
  year = {2025},
  month = feb,
  volume = {980},
  number = {2},
  eid = {L34},
  pages = {L34},
  doi = {10.3847/2041-8213/adaaf1},
  archiveprefix = {arXiv},
  eprint = {2412.08396},
  primaryclass = {astro-ph.GA},
  adsurl = {https://ui.adsabs.harvard.edu/abs/2025ApJ...980L..34L},
}

@article{Rieke+15,
  author = {{Rieke}, G.~H. and {Wright}, G.~S. and {B{\"o}ker}, T. and {Bouwman}, J. and {Colina}, L. and {Glasse}, Alistair and {Gordon}, K.~D. and {Greene}, T.~P. and {G{\"u}del}, Manuel and {Henning}, Th. and {Justtanont}, K. and {Lagage}, P.-O. and {Meixner}, M.~E. and {N{\o}rgaard-Nielsen}, H.-U. and {Ray}, T.~P. and {Ressler}, M.~E. and {van Dishoeck}, E.~F. and {Waelkens}, C.},
  title = {{The Mid-Infrared Instrument for the James Webb Space Telescope, I: Introduction}},
  journal = {\pasp},
  year = {2015},
  month = jul,
  volume = {127},
  number = {953},
  pages = {584},
  doi = {10.1086/682252},
  archiveprefix = {arXiv},
  eprint = {1508.02294},
  primaryclass = {astro-ph.IM},
  adsurl = {https://ui.adsabs.harvard.edu/abs/2015PASP..127..584R},
}

@article{Wright+23,
  author = {{Wright}, Gillian S. and {Rieke}, George H. and {Glasse}, Alistair and {Ressler}, Michael and {Garc{\'\i}a Mar{\'\i}n}, Macarena and {Aguilar}, Jonathan and {Alberts}, Stacey and {{\'A}lvarez-M{\'a}rquez}, Javier and {Argyriou}, Ioannis and {Banks}, Kimberly and {Baudoz}, Pierre and {Boccaletti}, Anthony and {Bouchet}, Patrice and {Bouwman}, Jeroen and {Brandl}, Bernard R. and {Breda}, David and {Bright}, Stacey and {Cale}, Steven and {Colina}, Luis and {Cossou}, Christophe and {Coulais}, Alain and {Cracraft}, Misty and {De Meester}, Wim and {Dicken}, Daniel and {Engesser}, Michael and {Etxaluze}, Mireya and {Fox}, Ori D. and {Friedman}, Scott and {Fu}, Henry and {Gasman}, Danny and {G{\'a}sp{\'a}r}, Andr{\'a}s and {Gastaud}, Ren{\'e} and {Geers}, Vincent and {Glauser}, Adrian Michael and {Gordon}, Karl D. and {Greene}, Thomas and {Greve}, Thomas R. and {Grundy}, Timothy and {G{\"u}del}, Manuel and {Guillard}, Pierre and {Haderlein}, Peter and {Hashimoto}, Ryan and {Henning}, Thomas and {Hines}, Dean and {Holler}, Bryan and {Detre}, {\"O}rs Hunor and {Jahromi}, Amir and {James}, Bryan and {Jones}, Olivia C. and {Justtanont}, Kay and {Kavanagh}, Patrick and {Kendrew}, Sarah and {Klaassen}, Pamela and {Krause}, Oliver and {Labiano}, Alvaro and {Lagage}, Pierre-Olivier and {Lambros}, Scott and {Larson}, Kirsten and {Law}, David and {Lee}, David and {Libralato}, Mattia and {Lorenzo Alverez}, Jose and {Meixner}, Margaret and {Morrison}, Jane and {Mueller}, Migo and {Murray}, Katherine and {Mycroft}, Matthew and {Myers}, Richard and {Nayak}, Omnarayani and {Naylor}, Bret and {Nickson}, Bryony and {Noriega-Crespo}, Alberto and {{\"O}stlin}, G{\"o}ran and {O'Sullivan}, Brian and {Ottens}, Richard and {Patapis}, Polychronis and {Penanen}, Konstantin and {Pietraszkiewicz}, Martin and {Ray}, Tom and {Regan}, Michael and {Roteliuk}, Anthony and {Royer}, Pierre and {Samara-Ratna}, Piyal and {Samuelson}, Bridget and {Sargent}, Beth A. and {Scheithauer}, Silvia and {Schneider}, Analyn and {Schreiber}, J{\"u}rgen and {Shaughnessy}, Bryan and {Sheehan}, Evan and {Shivaei}, Irene and {Sloan}, G.~C. and {Tamas}, Laszlo and {Teague}, Kelly and {Temim}, Tea and {Tikkanen}, Tuomo and {Tustain}, Samuel and {van Dishoeck}, Ewine F. and {Vandenbussche}, Bart and {Weilert}, Mark and {Whitehouse}, Paul and {Wolff}, Schuyler},
  title = {{The Mid-infrared Instrument for JWST and Its In-flight Performance}},
  journal = {\pasp},
  year = {2023},
  month = apr,
  volume = {135},
  number = {1046},
  eid = {048003},
  pages = {048003},
  doi = {10.1088/1538-3873/acbe66},
  adsurl = {https://ui.adsabs.harvard.edu/abs/2023PASP..135d8003W},
}

@ARTICLE{Hainline+25,
       author = {{Hainline}, Kevin N. and {Maiolino}, Roberto and {Juod{\v{z}}balis}, Ignas and {Scholtz}, Jan and {{\"U}bler}, Hannah and {D'Eugenio}, Francesco and {Helton}, Jakob M. and {Sun}, Yang and {Sun}, Fengwu and {Robertson}, Brant and et al.},
        title = "{An Investigation into the Selection and Colors of Little Red Dots and Active Galactic Nuclei}",
      journal = {\apj},
         year = 2025,
        month = feb,
       volume = {979},
       number = {2},
          eid = {138},
        pages = {138},
          doi = {10.3847/1538-4357/ad9920},
archivePrefix = {arXiv},
       eprint = {2410.00100},
 primaryClass = {astro-ph.GA},
       adsurl = {https://ui.adsabs.harvard.edu/abs/2025ApJ...979..138H}
}

@ARTICLE{BelldeJong01,
       author = {{Bell}, Eric F. and {de Jong}, Roelof S.},
        title = "{Stellar Mass-to-Light Ratios and the Tully-Fisher Relation}",
      journal = {\apj},
         year = 2001,
        month = mar,
       volume = {550},
       number = {1},
        pages = {212-229},
          doi = {10.1086/319728},
archivePrefix = {arXiv},
       eprint = {astro-ph/0011493},
 primaryClass = {astro-ph},
       adsurl = {https://ui.adsabs.harvard.edu/abs/2001ApJ...550..212B}
}

@ARTICLE{VandenBerk+04,
       author = {{Vanden Berk}, Daniel E. and {Wilhite}, Brian C. and {Kron}, Richard G. and {Anderson}, Scott F. and {Brunner}, Robert J. and {Hall}, Patrick B. and {Ivezi{\'c}}, {\v{Z}}eljko and {Richards}, Gordon T. and {Schneider}, Donald P. and {York}, Donald G. and {Brinkmann}, Jonathan V. and {Lamb}, Don Q. and {Nichol}, Robert C. and {Schlegel}, David J.},
        title = "{The Ensemble Photometric Variability of \raisebox{-0.5ex}\textasciitilde25,000 Quasars in the Sloan Digital Sky Survey}",
      journal = {\apj},
         year = 2004,
        month = feb,
       volume = {601},
       number = {2},
        pages = {692-714},
          doi = {10.1086/380563},
archivePrefix = {arXiv},
       eprint = {astro-ph/0310336},
 primaryClass = {astro-ph},
       adsurl = {https://ui.adsabs.harvard.edu/abs/2004ApJ...601..692V}
}

@ARTICLE{MacLeod+10,
       author = {{MacLeod}, C.~L. and {Ivezi{\'c}}, {\v{Z}}. and {Kochanek}, C.~S. and {Koz{\l}owski}, S. and {Kelly}, B. and {Bullock}, E. and {Kimball}, A. and {Sesar}, B. and {Westman}, D. and {Brooks}, K. and {Gibson}, R. and {Becker}, A.~C. and {de Vries}, W.~H.},
        title = "{Modeling the Time Variability of SDSS Stripe 82 Quasars as a Damped Random Walk}",
      journal = {\apj},
         year = 2010,
        month = oct,
       volume = {721},
       number = {2},
        pages = {1014-1033},
          doi = {10.1088/0004-637X/721/2/1014},
archivePrefix = {arXiv},
       eprint = {1004.0276},
 primaryClass = {astro-ph.CO},
       adsurl = {https://ui.adsabs.harvard.edu/abs/2010ApJ...721.1014M}
}

@ARTICLE{Zhang+15,
       author = {{Zhang}, Shaohua and {Zhou}, Hongyan and {Shi}, Xiheng and {Shu}, Xinwen and {Liu}, Wenjuan and {Ji}, Tuo and {Jiang}, Peng and {Sun}, Luming and {Zhou}, Junyan and {Pan}, Xiang},
        title = "{Discovery of Extremely Broad Balmer Absorption Lines in SDSS J152350.42+391405.2}",
      journal = {\apj},
         year = 2015,
        month = dec,
       volume = {815},
       number = {2},
          eid = {113},
        pages = {113},
          doi = {10.1088/0004-637X/815/2/113},
archivePrefix = {arXiv},
       eprint = {1511.03422},
 primaryClass = {astro-ph.GA},
       adsurl = {https://ui.adsabs.harvard.edu/abs/2015ApJ...815..113Z}
}

@ARTICLE{Hall07,
       author = {{Hall}, Patrick B.},
        title = "{A Quasar with Broad Absorption in the Balmer Lines}",
      journal = {\aj},
         year = 2007,
        month = apr,
       volume = {133},
       number = {4},
        pages = {1271-1274},
          doi = {10.1086/511272},
archivePrefix = {arXiv},
       eprint = {astro-ph/0611922},
 primaryClass = {astro-ph},
       adsurl = {https://ui.adsabs.harvard.edu/abs/2007AJ....133.1271H}
}

%%%%%%%%%%%%%%%%%%%%%%%%%%%%%%%%%%%%%%%%%%%%%%%%%%%%%%%%%%%%%%%
% Appendices must be placed after   \end{thebibliography}
% They will be placed automatically on a new page.
%%%%%%%%%%%%%%%%%%%%%%%%%%%%%%%%%%%%%%%%%%%%%%%%%%%%%%%%%%%%%%%
\appendix
 
\section{Removal of contaminating sources}
\label{app:gal-removal}
 
As discussed in Section~\ref{sec:photometry_extraction}, Ima-B is significantly contaminated by a foreground object located at a small projected separation from QSO1 ($\approx 1\arcsec$).
We therefore developed a procedure to model and remove the contaminating emission before performing the photometric extraction.
For consistency, the same procedure was applied to the cutouts of all three multiple images.
 
For each cutout, we first constructed a preliminary segmentation map using \textsc{sep} \citep{Barbary+16} and estimated the local background after masking the detected sources.
The background-subtracted image was then deconvolved using the Richardson--Lucy algorithm \citep{Richardson+72,Lucy+74}.
We adopted filter-dependent PSFs generated with \textsc{stpsf} \citep{Perrin+14} for the NIRCam and MIRI observations.
%For the HST images, for which an equivalent set of model PSFs was not employed in this procedure, we adopted a Gaussian kernel.
 
Before constructing the contaminant model, the pixels associated with QSO1 were identified from the segmentation map and replaced in the deconvolved image with values randomly drawn from source-free regions of the same cutout.
This prevented emission from QSO1 from being included in the model of the contaminating sources.
The resulting image was smoothed and reconvolved to obtain a model of the surrounding contaminating emission, which was then subtracted from the original background-subtracted cutout.
 
\begin{figure*}
    \centering
    \includegraphics[width=\textwidth]{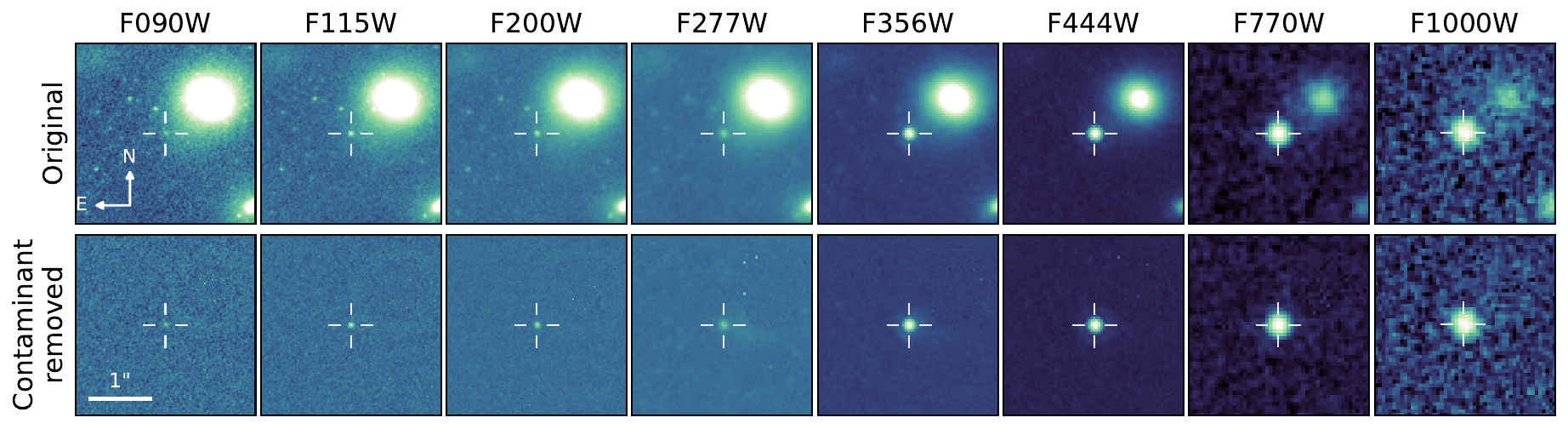}
    \caption{
    Examples of the contaminant-removal procedure applied to Ima-B of QSO1.
    The top panels show the original image cutouts for all NIRCam and MIRI wide bands available while the bottom panels show the corresponding residual images after the subtraction of all the models of nearby objects.
    The colourmap stretch and limits are kept consistent between top and bottom panels  for each filter.
    }
    \label{fig:gal-removal}
\end{figure*}

Figure~\ref{fig:gal-removal} presents representative examples of this procedure.
For each selected filter, we show the original background-subtracted cutout and the residual image.
 
\section{Combination of the multiple-image photometry}
\label{app:phot-combination}
 
After correcting the photometry of each multiple image for its corresponding lensing magnification, we found approximately wavelength-independent normalisation offsets between the three reconstructed SEDs.
As discussed in Section~\ref{sec:photometry_extraction}, we adopted Ima-C as the reference because it has the lowest magnification and is therefore expected to be less sensitive to uncertainties in the lens model.
 
For Ima-A and Ima-B, we determined a single multiplicative scaling factor, $s_i$, by minimising
\begin{equation}
    \chi^2(s_i) =
    \sum_j
    \frac{
    \left(s_i f_{i,j}-f_{{\rm C},j}\right)^2
    }{
    s_i^2\sigma_{i,j}^2+\sigma_{{\rm C},j}^2
    },
    \label{eq:image-rescaling}
\end{equation}
where $i$ denotes Ima-A or Ima-B, $j$ identifies the photometric band, and $f_{i,j}=f^{\rm obs}_{i,j}/\mu_i$ and $\sigma_{i,j}$ are the de-magnified flux density and its uncertainty, respectively.
The latter includes the uncertainty on the magnification, $\sigma_{i,j}^2=(\sigma^{\rm obs}_{i,j}/\mu_i)^2+(f_{i,j}\sigma_{\mu_i}/\mu_i)^2$, with $\mu_i\pm\sigma_{\mu_i}$ from the UNCOVER lens model v2 \citep{Furtak+25}. The same definition applies to Ima-C.
Both the flux densities and their uncertainties were multiplied by the resulting scaling factors, while Ima-C was left unchanged.
The scaling factors were constrained to be positive and were assumed to be independent of wavelength.
 
The final flux density in each band was calculated as the inverse-variance-weighted mean of the three rescaled measurements:
\begin{equation}
    \overline{f}_{\nu,j} =
    \frac{\displaystyle\sum_i w_{i,j}f_{i,j}}
         {\displaystyle\sum_i w_{i,j}},
    \qquad
    w_{i,j}=\frac{1}{\sigma_{i,j}^2}.
    \label{eq:weighted-photometry}
\end{equation}
The corresponding statistical uncertainty is
\begin{equation}
    \sigma_{{\rm stat},j} =
    \left(\sum_i w_{i,j}\right)^{-1/2}.
    \label{eq:weighted-photometry-error}
\end{equation}
 
To account for any residual dispersion between the three multiple-image measurements, we additionally estimated a band-dependent systematic uncertainty.
We first calculated the weighted inter-image variance,
\begin{equation}
    S_j^2 =
    \frac{
    \displaystyle\sum_i w_{i,j}
    \left(f_{i,j}-\overline{f}_{\nu,j}\right)^2
    }{
    \displaystyle\sum_i w_{i,j}
    }.
    \label{eq:weighted-scatter}
\end{equation}
Part of this scatter is already expected from the statistical uncertainties of the individual measurements.
We therefore defined the residual systematic variance as
\begin{equation}
    \sigma_{{\rm sys},j}^2 =
    \max\left[
    0,~
    S_j^2-
    \frac{N-1}{\displaystyle\sum_i w_{i,j}}
    \right],
    \label{eq:photometric-systematic}
\end{equation}
where $N=3$ is the number of multiple images.
The final uncertainty associated with the combined photometry was then obtained by adding the statistical and systematic contributions in quadrature:
\begin{equation}
    \sigma_{{\rm final},j} =
    \sqrt{
    \sigma_{{\rm stat},j}^2+
    \sigma_{{\rm sys},j}^2
    }.
    \label{eq:final-photometric-error}
\end{equation}
 
\begin{figure*}
    \centering
    \begin{minipage}[t]{0.48\textwidth}
        \includegraphics[width=\textwidth]{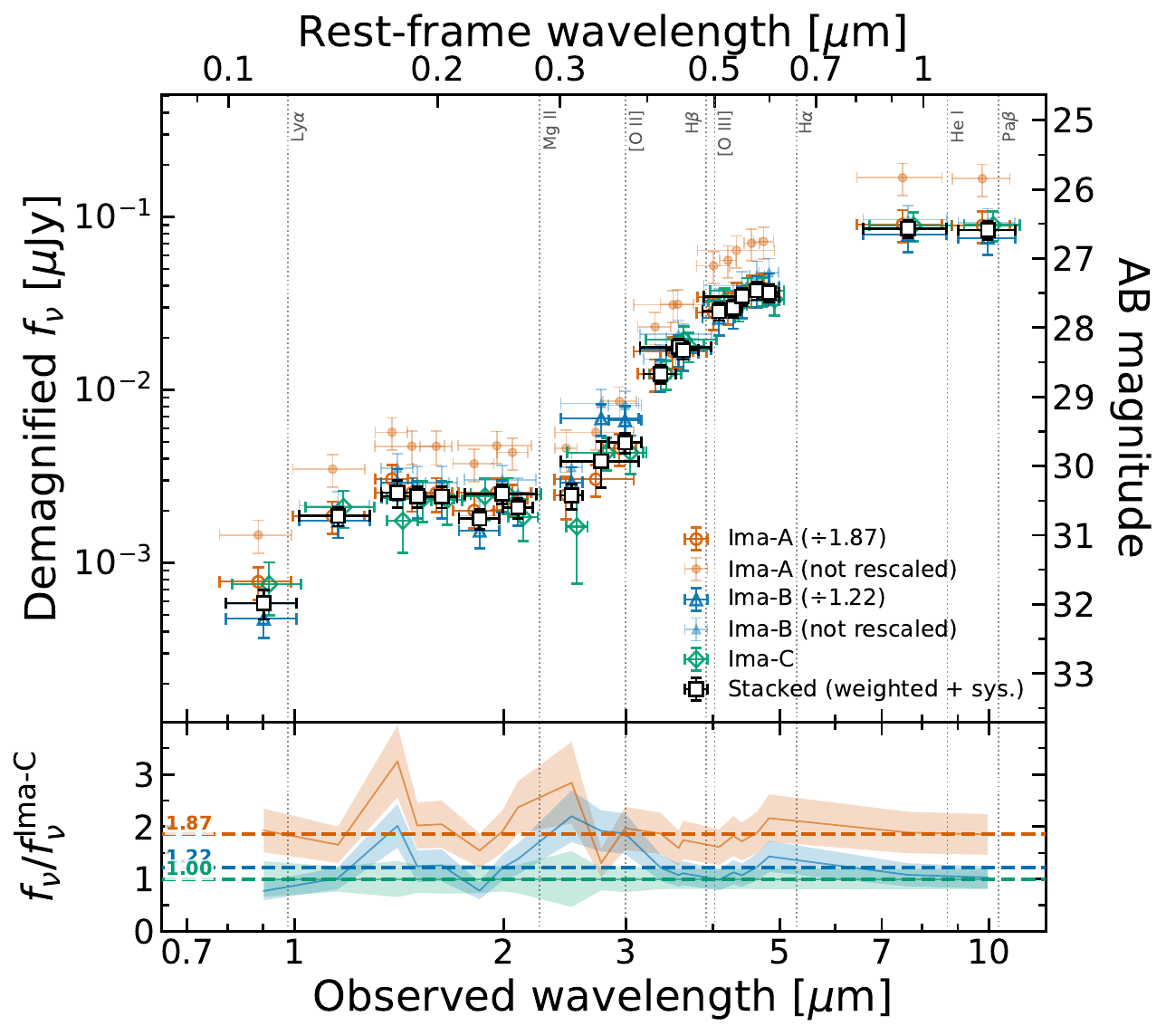}
    \caption{Comparison of the de-magnified photometry obtained from the three multiple images of QSO1 Ima-A (orange), Ima-B (blue) and Ima-C (green). For Ima-A and Ima-B we show the photometry before and after rescaling it for the wavelength-independent ratios $f_{\rm Ima-A}/ f_{\rm Ima-C} = 1.87$ and $f_{\rm Ima-B}/ f_{\rm Ima-C} = 1.22$. We also show the final inverse-variance-weighted photometry (white squares).
    }
    \label{fig:phot-rescaling-diagnostic}
    \end{minipage}\hfill
    \begin{minipage}[t]{0.5\textwidth}
    \includegraphics[width=\textwidth]{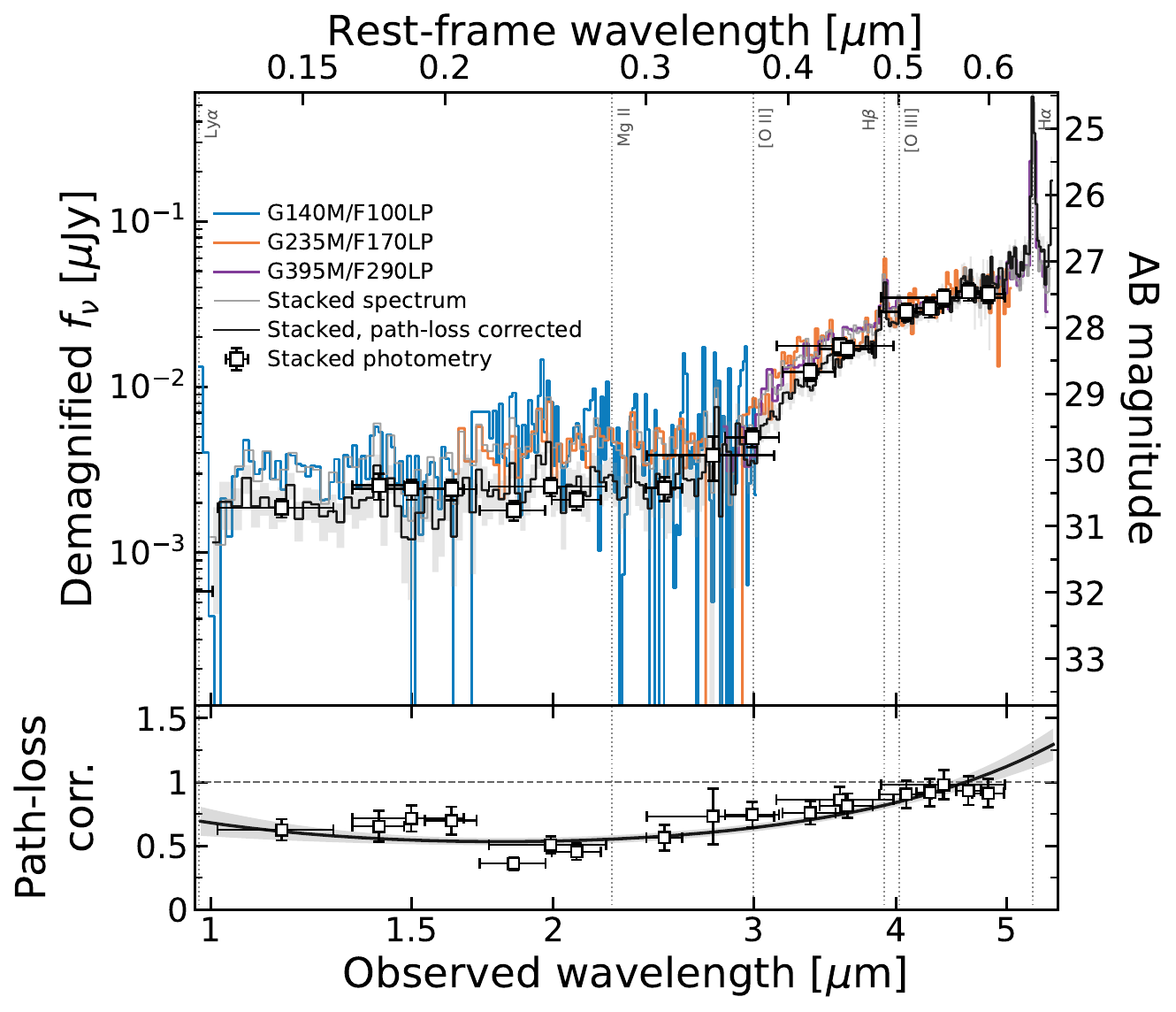}
    \caption{Spectrophotometric correction of the combined NIRSpec spectrum.
    Top: comparison between the combined photometry, the individual grating spectra, the stacked NIRSpec spectrum and the spectrum after applying the wavelength-dependent correction.
    Bottom: ratio between the imaging and synthetic photometry as a function of observed wavelength.
    The solid line shows the best-fitting second-order polynomial, while the shaded region represents the $1\sigma$ uncertainty derived from the Monte Carlo realisations.
    }
    \label{fig:pathloss-diagnostic}
    \end{minipage}
\end{figure*}

Figure~\ref{fig:phot-rescaling-diagnostic} shows the de-magnified photometry of the three images before and after applying the relative rescaling.
The lack of a clear wavelength dependence in the original offsets supports the use of a single multiplicative factor for each multiple image.
Following the rescaling, the individual measurements are consistent with the combined photometry within $3\sigma$ across the available passbands.
 
We report the observed photometry for the three individual multiple images of QSO1 and the magnification-corrected final stacked photometry in Table~\ref{tab:photometry}.

\begin{table*}
\centering
\caption{Photometry of QSO1. All flux densities are corrected for Galactic extinction using $f_{\rm ext}$ ($E(B-V)=0.0112$, \citealt{Gordon+23} curve). Ima-A, Ima-B and Ima-C fluxes are not corrected for lensing magnification. The MIRI aperture corrections use the in-flight encircled-energy curves of the CRDS reference files (Section~\ref{sec:photometry_extraction}). The stacked photometry is de-magnified with the magnifications of the UNCOVER lens model v2 \citep{Furtak+25}, including their uncertainties, and normalised to Ima-C (Section~\ref{sec:photometry_extraction} and Appendix~\ref{app:phot-combination}). The HST and ALMA reported upper-limits correspond to a $3\sigma$ limit. The MIPS and PACS limits are the mean $3\sigma$ depths of the Abell 2744 imaging, while the SPIRE values take into account confusion limits \citep{Rawle+14, Nguyen+10}. For both Spitzer and Herschel's upper-limits, we assumed independent and uniform image-plane noise. The stacked upper limits were obtained by taking one third of each $3\sigma$ depth as the $1\sigma$ uncertainty, de-magnifying and rescaling it to Ima-C as for the detections, and combining the three in inverse variance. For SPIRE, confusion noise is correlated over the beam, so the stacked limits are approximate.}
\label{tab:photometry}
\begin{tabular}{llc | cc | cc | cc | cc}
\hline
\hline
 &  &  & \multicolumn{2}{c}{Ima-A} & \multicolumn{2}{c}{Ima-B}  & \multicolumn{2}{c}{Ima-C}  & \multicolumn{2}{c}{Stacked}\\
Instrument & Filter & $f_{\rm ext}$ &
$\mu \cdot f_\nu$ & $\mu \cdot$ err$_{f_{\nu}}$ & $\mu \cdot f_\nu$ & $\mu \cdot$ err$_{f_{\nu}}$ & $\mu \cdot f_\nu$ & $\mu \cdot$ err$_{f_{\nu}}$ & $f_\nu$ & err$_{f_{\nu}}$ \\
 & &  & [$\mu$Jy] & [$\mu$Jy] & [$\mu$Jy] & [$\mu$Jy] & [$\mu$Jy] & [$\mu$Jy] & [$\mu$Jy] & [$\mu$Jy] \\
\hline
HST/WFC3\_UVIS & F225W & 1.091 & $<0.041$ & -- & $<0.045$ & -- & $<0.034$ & -- & $<0.00304$ & -- \\
HST/WFC3\_UVIS & F275W & 1.070 & $<0.027$ & -- & $<0.029$ & -- & $<0.024$ & -- & $<0.00197$ & -- \\
HST/WFC3\_UVIS & F336W & 1.054 & $<0.012$ & -- & $<0.013$ & -- & $<0.012$ & -- & $<0.00092$ & -- \\
HST/ACS\_WFC & F435W & 1.043 & $<0.005$ & -- & $<0.006$ & -- & $<0.005$ & -- & $<0.00042$ & -- \\
HST/ACS\_WFC & F606W & 1.031 & $<0.005$ & -- & $<0.006$ & -- & $<0.005$ & -- & $<0.00036$ & -- \\
JWST/NIRCam & F070W & 1.023 & $<0.004$ & -- & $<0.004$ & -- & $<0.004$ & -- & $<0.00027$ & -- \\
HST/ACS\_WFC & F814W & 1.019 & $<0.004$ & -- & $<0.005$ & -- & $<0.004$ & -- & $<0.00030$ & -- \\
JWST/NIRCam & F090W & 1.016 & 0.008 & 0.001 & 0.004 & 0.0005 & 0.002 & 0.001 & 0.00059 & 0.00011 \\
JWST/NIRCam & F115W & 1.010 & 0.019 & 0.001 & 0.016 & 0.001 & 0.006 & 0.001 & 0.00187 & 0.00023 \\
JWST/NIRCam & F140M & 1.007 & 0.031 & 0.002 & 0.026 & 0.001 & 0.005 & 0.001 & 0.00255 & 0.00046 \\
JWST/NIRCam & F150W & 1.006 & 0.026 & 0.002 & 0.022 & 0.003 & 0.006 & 0.001 & 0.00242 & 0.00034 \\
JWST/NIRCam & F162M & 1.005 & 0.025 & 0.002 & 0.022 & 0.003 & 0.006 & 0.001 & 0.00242 & 0.00034 \\
JWST/NIRCam & F182M & 1.004 & 0.020 & 0.001 & 0.014 & 0.001 & 0.007 & 0.001 & 0.00180 & 0.00024 \\
JWST/NIRCam & F200W & 1.004 & 0.026 & 0.001 & 0.023 & 0.001 & 0.007 & 0.001 & 0.00251 & 0.00031 \\
JWST/NIRCam & F210M & 1.004 & 0.024 & 0.001 & 0.019 & 0.001 & 0.005 & 0.001 & 0.00209 & 0.00028 \\
JWST/NIRCam & F250M & 1.003 & 0.025 & 0.005 & 0.027 & 0.003 & 0.004 & 0.002 & 0.00246 & 0.00041 \\
JWST/NIRCam & F277W & 1.002 & 0.031 & 0.001 & 0.063 & 0.003 & 0.012 & 0.001 & 0.00387 & 0.00114 \\
JWST/NIRCam & F300M & 1.002 & 0.046 & 0.002 & 0.061 & 0.003 & 0.012 & 0.002 & 0.00496 & 0.00064 \\
JWST/NIRCam & F335M & 1.002 & 0.125 & 0.006 & 0.113 & 0.006 & 0.033 & 0.002 & 0.01237 & 0.00144 \\
JWST/NIRCam & F356W & 1.001 & 0.168 & 0.008 & 0.157 & 0.008 & 0.053 & 0.003 & 0.01770 & 0.00207 \\
JWST/NIRCam & F360M & 1.001 & 0.169 & 0.008 & 0.149 & 0.007 & 0.048 & 0.002 & 0.01698 & 0.00198 \\
JWST/NIRCam & F410M & 1.001 & 0.282 & 0.014 & 0.239 & 0.012 & 0.087 & 0.004 & 0.02851 & 0.00334 \\
JWST/NIRCam & F430M & 1.001 & 0.303 & 0.015 & 0.261 & 0.013 & 0.083 & 0.004 & 0.02979 & 0.00348 \\
JWST/NIRCam & F444W & 1.001 & 0.347 & 0.017 & 0.300 & 0.015 & 0.101 & 0.005 & 0.03474 & 0.00406 \\
JWST/NIRCam & F460M & 1.001 & 0.381 & 0.019 & 0.345 & 0.017 & 0.101 & 0.005 & 0.03761 & 0.00439 \\
JWST/NIRCam & F480M & 1.001 & 0.390 & 0.019 & 0.358 & 0.018 & 0.090 & 0.005 & 0.03647 & 0.00429 \\
JWST/MIRI & F770W & 1.001 & 0.911 & 0.046 & 0.723 & 0.036 & 0.241 & 0.012 & 0.08570 & 0.01002 \\
JWST/MIRI & F1000W & 1.002 & 0.899 & 0.045 & 0.694 & 0.035 & 0.243 & 0.015 & 0.08404 & 0.00991 \\
Spitzer/MIPS & $24$\um\ & - & $<53$ & - &  $<53$ & - &  $<53$ & - & $<3.82$ & - \\
Herschel/PACS & $100$\um\ & - & $<2600$ & - &  $<2600$ & - &  $<2600$ & - & $<187$ & - \\
Herschel/PACS & $160$\um\ & - & $<5200$ & - &  $<5200$ & - &  $<5200$ & - & $<374$ & - \\
Herschel/SPIRE & $250$\um\ & - & $<16800$ & - &  $<16800$ & - &  $<16800$ & - & $<1209$ & - \\
Herschel/SPIRE & $350$\um\ & - & $<19200$ & - &  $<19200$ & - &  $<19200$ & - & $<1382$ & - \\
Herschel/SPIRE & $500$\um\ & - & $<19800$ & - &  $<19800$ & - &  $<19800$ & - & $<1425$ & - \\
ALMA           & Band 6 & - & $< 94.5$ &  - & $< 90.9$ & - & $< 84.0$ & - & $< 6.7$ & - \\
\hline
\hline
\end{tabular}
 
\end{table*}

\section{Construction and path-loss correction of the NIRSpec spectrum}
\label{app:pathloss}
 
We combined the NIRSpec spectra obtained with the G140M/F100LP, G235M/F170LP, and G395M/F290LP configurations.
Each spectrum was resampled onto a common wavelength grid extending from $0.8$ to $5.6\rm~\mu m$, with a sampling of $\Delta\lambda=0.005\rm~\mu m$, using the flux-conserving \texttt{spectres} algorithm \citep{Carnall+17}.
At wavelengths covered by more than one configuration, the combined flux density was calculated as
\begin{equation}
    f_\nu(\lambda) =
    \frac{
    \displaystyle\sum_k
    f_{\nu,k}(\lambda)w_k(\lambda)
    }{
    \displaystyle\sum_k w_k(\lambda)
    },
    \qquad
    w_k(\lambda)=\frac{1}{\sigma_k^2(\lambda)},
    \label{eq:spectral-stacking}
\end{equation}
where $k$ denotes the grating--filter configuration.
The corresponding uncertainty is
\begin{equation}
    \sigma_{\rm spec}(\lambda) =
    \left[
    \sum_k w_k(\lambda)
    \right]^{-1/2}.
    \label{eq:spectral-stacking-error}
\end{equation}
Only spectral elements with finite and positive uncertainties were included in the combination.
 
For each configuration, we also calculated the wavelength-dependent resolving power using the instrumental line-spread-function parameterisation presented by \citet{Shajib+25}.
The resulting resolution curves were interpolated onto the common wavelength grid and combined according to the relative inverse-variance contribution of the individual spectra.
This provided an effective spectral-resolution curve that accounts for the transitions and overlaps between the three configurations.
 
We corrected for residual wavelength-dependent path losses by comparing the combined NIRSpec spectrum with the photometry derived in Section~\ref{sec:photometry_extraction}.
We retrieved the NIRCam filter-transmission curves from the SVO FPS \citep[][]{Rodrigo+12,Rodrigo+20}.
For a spectrum expressed in terms of $f_\nu$, we calculated the synthetic flux density in each passband as
\begin{equation}
    f_{\nu,{\rm syn},j} =
    \frac{
    \displaystyle\int
    f_\nu(\lambda)T_j(\lambda)
    \frac{{\rm d}\lambda}{\lambda}
    }{
    \displaystyle\int
    T_j(\lambda)\frac{{\rm d}\lambda}{\lambda}
    },
    \label{eq:synthetic-photometry}
\end{equation}
where $T_j(\lambda)$ is the transmission curve of filter $j$.
Only filters sufficiently covered by the NIRSpec spectrum were included.
 
The uncertainty on the synthetic photometry was propagated from the uncertainties of the spectral elements contributing to the convolution.
We then calculated the ratio
\begin{equation}
    C_j =
    \frac{
    f_{\nu,{\rm phot},j}
    }{
    f_{\nu,{\rm syn},j}
    },
    \label{eq:pathloss-ratio}
\end{equation}
where $f_{\nu,{\rm phot},j}$ is the combined imaging photometry.
The uncertainty on this ratio includes both the photometric and spectroscopic contributions:
\begin{equation}
    \sigma_{C,j} =
    C_j
    \sqrt{
    \left(
    \frac{\sigma_{{\rm phot},j}}
         {f_{\nu,{\rm phot},j}}
    \right)^2+
    \left(
    \frac{\sigma_{{\rm syn},j}}
         {f_{\nu,{\rm syn},j}}
    \right)^2
    }.
    \label{eq:pathloss-ratio-error}
\end{equation}

We modelled the wavelength dependence of the photometric-to-synthetic flux-density ratios with a second-order polynomial in logarithmic space:
\begin{equation}
    \log_{10} C(\lambda) =
    a_0+
    a_1\log_{10}\lambda+
    a_2\left(\log_{10}\lambda\right)^2.
    \label{eq:pathloss-polynomial}
\end{equation}
The fit was weighted using the uncertainties given by Eq.~\eqref{eq:pathloss-ratio-error}.
The corrected spectrum was then obtained as
\begin{equation}
    f_{\nu,{\rm corr}}(\lambda) =
    C(\lambda)f_{\nu,{\rm spec}}(\lambda).
    \label{eq:pathloss-corrected-spectrum}
\end{equation}
 
Because the constant coefficient $a_0$ sets the absolute normalisation of the correction, the procedure simultaneously corrects for a wavelength-independent offset between the normalisation of the input spectrum and that of the de-magnified photometry.
Consequently, the final spectrum is tied directly to the absolute normalisation of the combined photometric SED, whereas the higher-order terms describe the wavelength-dependent component of the path losses.
 
We estimated the uncertainty associated with the correction through 1000 Monte Carlo realisations.
For each realisation, the values of $\log_{10}C_j$ were perturbed according to their uncertainties and the second-order polynomial was refitted.
The standard deviation of the resulting correction curves was adopted as the wavelength-dependent uncertainty $\sigma_C(\lambda)$.
The final uncertainty on the corrected spectrum was calculated as
\begin{equation}
    \sigma_{\rm corr}(\lambda) =
    \sqrt{
    \left[
    C(\lambda)\sigma_{\rm spec}(\lambda)
    \right]^2+
    \left[
    f_{\nu,{\rm spec}}(\lambda)\sigma_C(\lambda)
    \right]^2
    }.
    \label{eq:pathloss-final-error}
\end{equation}
 
Figure~\ref{fig:pathloss-diagnostic} shows the photometric-to-synthetic flux-density ratios and the best-fitting correction curve.
The comparison between the corrected synthetic photometry and the observed imaging measurements provides a direct assessment of the consistency of the final spectrophotometric calibration.

\section{Point-spread functions}
\label{sec:psf}
Testing whether QSO1 is resolved in MIRI requires a PSF accurate on the scale of the core ($r\lesssim0.3\arcsec$).
A default {\sc stpsf} model is inadequate, as it is computed for a single pointing and field position, sampled on native detector pixels and oriented along the detector axes, and therefore ignores how the dithered exposures were combined into our mosaics.
We built an ``ad-hoc'' effective PSF for each lensed image (A, B, C) and filter (F770W, F1000W), reproducing the construction of the mosaics: 20 exposures per filter combined with the JWST pipeline \texttt{resample} step (v1.20; \texttt{ivm} weighting, square kernel, \texttt{pixfrac}~=~1, $0.06\arcsec$ pixels), then reprojected onto a north-up grid aligned to NIRCam.
 
For every exposure covering the source with valid pixels (10, 13 and 11 exposures for images A, B and C), we computed an {\sc stpsf} (v2.0.0) model at the detector position of the source, including the sub-pixel phase, geometric distortion and MIRI detector effects (\texttt{DET\_DIST} product), for a blackbody spectrum of $T=4470~{\rm K}/(1+z)=556$~K (see Figure~\ref{fig:bb_slopes}).
%Source positions were mapped onto each frame after correcting for an offset of $(-0.083\arcsec,~-0.128\arcsec)$ between the frame WCS and the NIRCam-aligned mosaic, measured from field sources.
Each model was inserted into a copy of its exposure, preserving the original masks and variance, and all copies were resampled with the mosaic parameters onto a $0.06\arcsec$ grid centred on the source and normalised to unit flux.
Weights are thus those of the real data (effectively exposure-time weights, as all exposures are identical).
 
\begin{figure*}
    \centering
    \includegraphics[width=\textwidth]{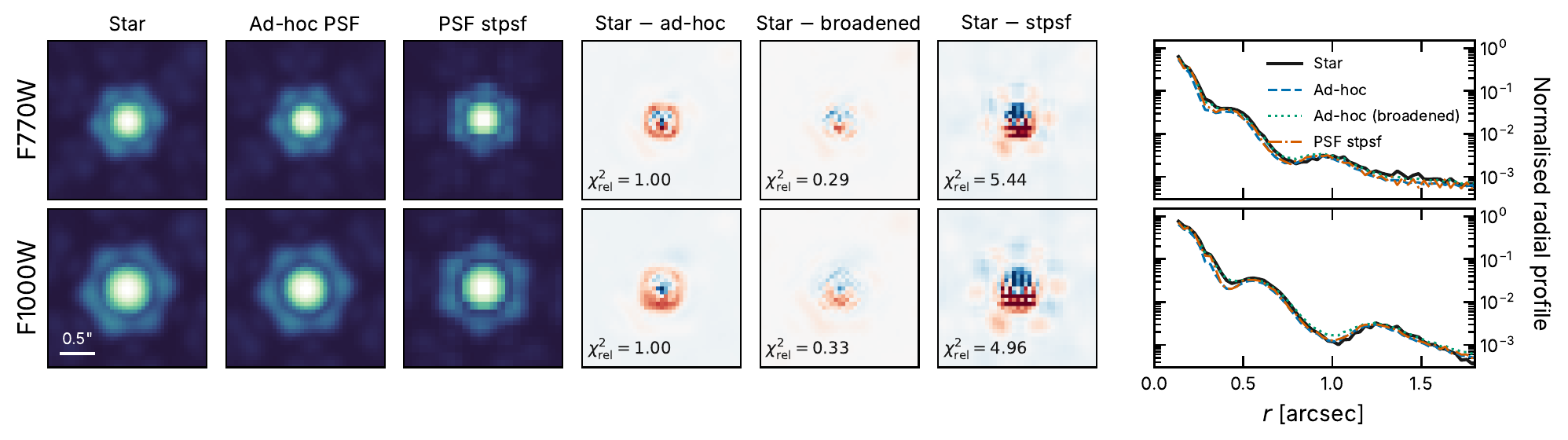}    \caption{Validation of the MIRI ad-hoc PSFs on an unsaturated field star (top: F770W; bottom: F1000W). From left to right: the star in the mosaic, the ad-hoc PSF, the default {\sc stpsf} PSF including detector effects (field centre, detector orientation, single pointing) and the residuals after fitting each model (and the ad-hoc PSF convolved with a Gaussian of $\sigma=0.042\arcsec$ and $0.054\arcsec$, ``broadened'') to the star within $r<1\arcsec$, shown within $\pm10\%$ of the stellar peak. $\chi^2_{\rm rel}$ is relative to the ad-hoc PSF. Panels are $2.5\arcsec$ wide. Right: azimuthally averaged profiles normalised to the peak.}
    \label{fig:psf}
\end{figure*}

We validated the procedure on an unsaturated field star ($\rm RA=3.56893~deg$, $\rm Dec=-30.40279~deg$; 10 exposures per filter), rebuilding its PSF in the same way (Figure~\ref{fig:psf}).
As a reference, we also considered a default {\sc stpsf} model including detector effects (\texttt{DET\_DIST}), computed at the centre of the field and resampled onto the same grid.
Both models reproduce the azimuthally averaged profile of the star, including the diffraction rings and the wings, but the ad-hoc PSF also matches its two-dimensional structure, i.e. the orientation of the hexagonal pattern and the centroid ($<0.005\arcsec$), with residuals 5.4 (F770W) and 5.0 (F1000W) times smaller ($\chi^2$) than the default model, whose single-pointing, native-pixel sampling leaves strong residuals in the core.
The stellar core is nevertheless slightly broader than the ad-hoc PSF: convolving the latter with a Gaussian of $\sigma=0.04\arcsec$ (F770W) and $0.05\arcsec$ (F1000W) reduces $\chi^2$ by a further factor of $\sim$3.
About one third of this excess is due to the final reprojection of the mosaic. We ascribe the rest to unmodelled detector effects (charge diffusion and possibly the brighter-fatter effect, which may affect the bright star more than QSO1).
The input spectrum is irrelevant (FWHM changes $\leq1\%$ between 556 and 5000~K).
As the true PSF of QSO1 should lie between the ad-hoc PSF and its broadened version, we performed all morphological fits with both and regard a source extent as robust only if required in both cases (see Section~\ref{sec:miri_morph}).
 
For aperture photometry, the core broadening is negligible, but the encircled energy at the adopted radii ($0.40$--$0.56\arcsec$) depends on the PSF wings.
We measured the growth curve of the field star out to $3\arcsec$ and corrected it to infinite radius with the {\sc stpsf} \texttt{DET\_DIST} model ($20\arcsec$ field of view).
At the adopted radii, the star encloses $65$--$73\%$ of its flux, in agreement within $1.5\%$ with the in-flight EE curves of the CRDS \texttt{apcorr} reference file (\texttt{jwst\_miri\_apcorr\_0014}), which we adopted (Section~\ref{sec:photometry_extraction}).
The {\sc stpsf} \texttt{DET\_DIST} model encloses $4$--$5\%$ (F770W) and $3$--$4\%$ (F1000W) more flux, and the {\sc stpsf} model without detector effects $\approx12\%$ and $4$--$6\%$ more, mostly because it lacks the detector scattering that moves light from the core to the wings, which is strongest at short wavelengths.
 
\section{MIRI morphology}
\label{app:morph}
We modelled the three lensed images in the contaminant-removed F770W and F1000W mosaics with {\sc socca}\footnote{\url{https://github.com/lucadimascolo/socca}} \citep{DiMascolo+26}, fitting $2.5\arcsec\times2.5\arcsec$ stamps centred on each image.
The drizzled MIRI mosaics have correlated noise (the rms of $2\times2$-binned blank regions is $\sim$1.4 times the value expected for independent pixels).
We therefore used a correlated Gaussian likelihood, with the pixel covariance taken from the empirical noise autocorrelation of each stamp, measured in pixels more than $1.5\arcsec$ from QSO1.
Each image and filter was fitted with its own ad-hoc PSF (Appendix~\ref{sec:psf}) and with the same PSF convolved with the Gaussian that best reproduces the field star ($\sigma=0.04\arcsec$ and $0.05\arcsec$ in F770W and F1000W; ``broadened'' PSF), the two bracketing the true PSF.
We quantified any extension with a profile likelihood: for each image, filter and PSF we fitted a model of the PSF convolved with a circular Gaussian of width $\sigma$, optimising flux and centroid on a grid of $\sigma=0$--$0.10\arcsec$, and derived 95\% upper limits from $\Delta\chi^2=3.84$ relative to the best fit.
As a cross-check, we computed Bayesian evidences ($\ln(\mathcal{B})$) for a point-source and a Gaussian model with {\sc nautilus} \citep{Lange+23}.
With the broadened PSF, all fits have $\sigma_{\rm best}=0$ and residuals consistent with noise.
With the ad-hoc PSF, the F770W fits of all three images and the F1000W fit of Ima-A prefer $\sigma\simeq0.025$--$0.035\arcsec$ ($\Delta\chi^2=18$--72; $\ln(\mathcal{B})\simeq7$--9 in favour of the Gaussian), an excess consistent with the PSF systematic discussed in Section~\ref{sec:miri_morph}.
 
We converted angular to physical sizes with the adopted cosmology ($5.21$~kpc~arcsec$^{-1}$ at $z=7.0364$) and magnifications \citep{Furtak+25}, taking $r_{\rm e}\simeq{\rm FWHM}/2$ \citep{Furtak+23}.
For Ima-A, \citet{Ma+25} provided the tangential magnification from the same lens model, $\mu_{\rm t}=3.21$, implying a radial magnification $\mu_{\rm r}=\mu_{\rm A}/\mu_{\rm t}\simeq1.68$.
A circular source of radius $r$ is thus imaged into an ellipse with semi-axes $\mu_{\rm t}r$ and $\mu_{\rm r}r$. Since our circular model is sensitive to the root mean square of the two axes, we converted the image-plane limits of Ima-A as $r_{\rm src}=r_{\rm img}/[(\mu_{\rm t}^2+\mu_{\rm r}^2)/2]^{1/2}=r_{\rm img}/2.56$.
For Ima-B and Ima-C, for which no shear information was available, we assumed isotropic magnification ($r_{\rm src}=r_{\rm img}/\sqrt{\mu}$).
 
A first, model-independent bound follows from the angular resolution alone: with effective PSFs of FWHM~$\simeq0.28\arcsec$ (F770W) and $0.36\arcsec$ (F1000W), an unresolved source must have $r_{\rm e}\lesssim0.14\arcsec$ in F770W, i.e. $\lesssim285$, 270 and 445~pc in the source plane for Ima-A, B and C.
The profile-likelihood fits improve on this by a factor of $\sim$4: with the ad-hoc PSF, the 95\% limits on the intrinsic F770W FWHM are $<0.063\arcsec$, $<0.073\arcsec$ and $<0.11\arcsec$ for Ima-A, B and C (F1000W gives weaker but consistent limits), corresponding to the radii quoted in Section~\ref{sec:miri_morph}.
With the broadened PSF the best-fit width is $\sigma=0$ in all fits and $\Delta\chi^2$ rises steeply from this boundary. The resulting limits ($r_{\rm e}<18$, 18 and 88~pc) reflect the mismatch between the broadened and the true PSF rather than a measurement, and we do not use them.
Since the PSF uncertainty ($\sigma\approx0.03$--$0.04\arcsec$) exceeds the statistical one, the ad-hoc-PSF limits should be regarded as the effective resolution floor of these data.
Dividing the image-plane size by $\mu_{\rm t}$ alone, as done by \citet{Furtak+24}, would give $r_{\rm e}<51$~pc for Ima-A.
 
\section{Line contamination and continuum fits}
\label{app:nir}
 
\subsection{Emission-line contribution to the MIRI bands}
Both MIRI filters contain hydrogen and helium recombination lines: F770W covers the Paschen lines from Pa$\delta$ to Pa10 (Pa$\gamma$ and He{\sc i}$~\lambda10830$ fall at its red edge, where the throughput is $\approx3\%$ and $\approx33\%$ of the peak, respectively), while F1000W covers Pa$\beta$.
To estimate their contribution, we adopted the broad and narrow H$\alpha$ fluxes measured by \citet{Deugenio+26b} for Ima-A, rescaled to our combined photometry.
For the narrow component, we assumed Case-B line ratios for $T=10^4$~K \citep{Storey+95} and no attenuation, consistent with the narrow-line Balmer decrement \citep{Deugenio+26b}.
For the broad component, we applied the same Paschen-to-H$\beta$ ratios to the H$\alpha$ flux corrected for the attenuation inferred from the broad-line Balmer decrement ($A_{V,{\rm b}}=2.9\pm0.2$~mag), adopting, as \citet{Deugenio+26b}, the SMC-Bar extinction curve of \citet{Gordon+03} and an intrinsic broad-line H$\alpha$/H$\beta=3.1$ \citep{Dong+08}, and re-attenuated each Paschen line accordingly.
Folding each line through the filter transmission curves, the hydrogen lines contribute $1.9\pm0.2\%$ and $6.9^{+1.0}_{-0.8}\%$ of the F770W and F1000W flux densities, respectively, almost entirely from the broad component.
For He{\sc i}$~\lambda10830$, we used the He{\sc i}/Pa$\gamma$ ratios measured in LRDs by \citet{Kapoor+26}, which are systematically lower ($\approx1.0$--1.6) than those of normal type-1 AGN ($>2.3$) and tentatively anti-correlate with the strength of the Balmer break, likely because of He{\sc i} self-absorption in high-column-density gas.
With a Balmer break $f_\nu(4050\textrm{\AA})/f_\nu(3650\textrm{\AA})\approx2.5$, comparable to the strongest breaks in their sample, QSO1 is expected to have He{\sc i}/Pa$\gamma\approx1.0$--1.3, implying a further contribution of $\approx0.8\%$ to F770W ($\lesssim1.5\%$ even for He{\sc i}/Pa$\gamma=2.3$). Possible He{\sc i} absorption would only lower it.
In total, line emission accounts for $\approx3\%$ and $\approx7\%$ of the F770W and F1000W flux densities.
These estimates are conservative: if the line contribution is instead scaled using the H$\alpha$ equivalent width reported by \citet{Deugenio+26b}, it decreases by a factor of $\approx2.7$. If the broad Balmer decrement is not due to dust, Case-B Paschen-to-Balmer ratios would lower it by a factor of $\approx3$--4. However, the Paschen lines of LRDs are boosted relative to Case~B \citep{Juodbalis+24a,Lin+26}, as expected for dense gas in which the Balmer lines are optically thick. Our adopted ratios correspond to a broad Pa$\beta$/H$\alpha\approx0.25$, about four times the Case-B value and comparable to the ratios observed in LRDs, so we adopt them independently of the origin of the decrement.
After correcting for the line emission, the slope between the two MIRI filters becomes $\alpha=-0.2\pm0.4$.
 
\subsection{Degeneracies of the continuum fits}
The blackbody temperature is degenerate with dust attenuation. A hotter, reddened blackbody provides a comparable description, e.g. $T_{\rm BB}=6874$~K with $A_V \simeq 1.5$~mag ($\chi^2_{\rm red}=2.4$), assuming the SMC-Bar extinction curve \cite[$R_V = 2.74$,][]{Gordon+03}.
The photometry alone cannot break this degeneracy: $\chi^2_{\rm red}$ increases only from $2.2$ for an unreddened $4450$~K blackbody to $2.4$ and $2.7$ for $T=7000$ and $30000$~K, with $A_V\approx1.6$ and $3.3$~mag, respectively.
Similarly, a dust-reddened power law ($f_\nu\propto\lambda^{\alpha}10^{-0.4A_\lambda}$) fitted to the same filters requires a steep intrinsic slope ($\alpha=-1.8$) and a large attenuation ($A_V\approx3.6$~mag; $\chi^2_{\rm red}=3.2$), whereas intrinsic slopes typical of type-1 AGN or of a standard accretion disc ($f_\nu\propto\nu^{-0.44}$ or $\nu^{1/3}$; \citealt{VandenBerk+01, ShakuraSunyaev73}, respectively) provide poor fits ($\chi^2_{\rm red}=10.2$ and $5.9$, respectively), as they cannot reproduce the turnover at $\approx1~$\um.

\section{Dust energy budget}
\label{app:dust_budget}
\textit{Absorbed luminosity.} We refitted the AGN-only model of Section~\ref{sec:ma25} (the \citealt{Temple+21} continuum without emission lines and hot dust, attenuated with the \citealt{Noll+09} curve with free $A_V$ and $\delta$, plus a free unattenuated scattered fraction) to the 18 NIRCam and MIRI bands, using the filter transmission curves.
The fit recovers the parameters of Table~\ref{tab:ma25_posteriors} ($\log_{10}(\lambda L_{3000}~[{\rm erg~s^{-1}}])=44.22\pm0.04$, $A_V=2.17^{+0.10}_{-0.08}$ mag, $\delta=-0.96^{+0.06}_{-0.03}$, scattered fraction $f_{\rm scat}=0.010\pm0.001$, $\chi^2_{\rm red}=1.70$).
We sampled its posterior with {\sc emcee} \citep{ForemanMackey+13} and, for each sample, integrated the difference between the intrinsic and attenuated continuum between $912$~\AA\ and $2.9$\um, obtaining $L_{\rm abs}=(3.9\pm0.3)\times10^{44}$~erg~s$^{-1}$.
The ionising continuum is not included in the template. We extended it with the incident continuum of the accretion disc used in our cocoon models (Appendix~\ref{app:cocoon}), in which the range 912~\AA--2.9\um\ carries 33.7\% of the luminosity, giving $L_{\rm AGN}=(1.23^{+0.11}_{-0.10})\times10^{45}$~erg~s$^{-1}$. Ionising photons are absorbed mainly by gas, especially at low metallicity, but are largely re-emitted as lines and recombination continua that are in turn absorbed by the dust. The upper value of $L_{\rm abs}$ therefore assumes complete reprocessing, while the lower value is independent of it. The bolometric correction of \citet{Runnoe+12}, $L_{\rm bol}=0.75\times5.15~\lambda L_{3000}$, gives $6.4\times10^{44}$~erg~s$^{-1}$, and would give $9.6\times10^{43}$~erg~s$^{-1}$ instead of $2.4\times10^{44}$~erg~s$^{-1}$ for the disc of the cocoon with the stellar-mass limit. The ionising continuum of LRDs is not constrained by observations, so this factor of $\approx2$--2.5 is a systematic uncertainty on all intrinsic luminosities.
 
\textit{Hot dust (NIRCam and MIRI).} We added to the AGN model an optically thin greybody, $L_\nu\propto\nu^\beta B_\nu(T_{\rm d})$ with $\beta=0$, 1 and 2, at fixed $T_{\rm d}$, and computed the profile likelihood of its normalisation, re-optimising all AGN parameters at each step.
Since F770W and F1000W are detections, we quote these limits at $\Delta\chi^2=1$ ($1\sigma$) relative to the fit without dust, whereas the far-infrared and millimetre limits are the $3\sigma$ upper limits of the non-detections. The limits are converted to luminosities by integrating the greybody over all frequencies.
 
\textit{Cold and warm dust (far-infrared and millimetre).} For the MIPS, PACS, SPIRE and ALMA bands, we normalised a greybody with $\beta=1.8$ to the stacked $3\sigma$ limit of each band (Table~\ref{tab:photometry}), dividing by the CMB contrast factor $1-B_\nu(T_{\rm CMB}(z))/B_\nu(T_{\rm d})$ \citep{daCunha+13b}, and adopted the most constraining band at each temperature. Here $T_{\rm d}$ is the dust temperature at $z=7.04$, including any CMB heating.
ALMA dominates below $T_{\rm d}\approx120$~K, \textit{Herschel}/PACS between $\approx130$ and 350~K and MIPS above.
 
\begin{table}
\centering
\caption{Upper limits on the luminosity of dust emission at a single temperature, in units of $10^{43}$~erg~s$^{-1}$: $1\sigma$ for the NIRCam and MIRI detections and $3\sigma$ for the far-infrared and millimetre non-detections.}
\label{tab:dust_limits}
\begin{tabular}{rcccc}
\hline
$T_{\rm d}$ [K] & $\beta=0$ & $\beta=1$ & $\beta=1.8$ & $\beta=2$ \\
\hline
\multicolumn{5}{l}{NIRCam+MIRI, joint fit with the reddened-disc model ($1\sigma$)} \\
500 & $1.3\times10^{4}$ & 2200 & 640 & 480 \\
600 & 640 & 130 & 44 & 35 \\
700 & 82 & 20 & 7.5 & 6.0 \\
800 & 19 & 5.1 & 2.2 & 1.8 \\
900 & 6.3 & 1.9 & 0.89 & 0.75 \\
1000 & 2.7 & 0.92 & 0.46 & 0.40 \\
1200 & 0.86 & 0.35 & 0.20 & 0.18 \\
1500 & 0.31 & 0.16 & 0.11 & 0.10 \\
2000 & 0.15 & 0.10 & 0.096 & 0.097 \\
\hline
\multicolumn{5}{l}{MIPS, \textit{Herschel} and ALMA ($3\sigma$)} \\
30 & 2.0 & 2.4 & 3.3 & 3.6 \\
40 & 2.1 & 3.4 & 5.9 & 6.9 \\
50 & 2.9 & 5.7 & 12 & 14 \\
80 & 7.1 & 23 & 68 & 92 \\
100 & 12 & 48 & 170 & 240 \\
150 & 33 & 200 & 480 & 440 \\
200 & 72 & 420 & 360 & 350 \\
300 & 220 & 440 & 510 & 540 \\
400 & 500 & 510 & 240 & 210 \\
500 & 350 & 140 & 77 & 68 \\
600 & 140 & 66 & 44 & 40 \\
800 & 58 & 36 & 30 & 29 \\
\hline
\end{tabular}
\end{table}
 
\section{Comparison with previous continuum models}
\label{app:ma25}
 
We reproduced the three model families of \citet{Ma+25} with their ingredients: {\sc fsps} \citep{Conroy+09} stellar populations (MIST, \citealt{Choi+16,Dotter+16}; MILES, \cite{Sanchez-Blazquez+06}; \citealt{Chabrier+03} IMF $\log_{10}(Z\rm~[Z_\odot])=-1$; delayed-$\tau$ history; nebular continuum), the line-free \citet{Temple+21} quasar continuum with or without its hot-dust component, and the \citet{Noll+09} attenuation law, with unattenuated fractions $f_{\rm nodust}$ (stars) and $f_{\rm scat}$ (AGN-only).
In the AGN+galaxy model, the AGN is attenuated by $A_V^{\rm gal}+A_V^{\rm AGN}$.
 
\begin{figure}
    \centering
    \includegraphics[width=\linewidth]{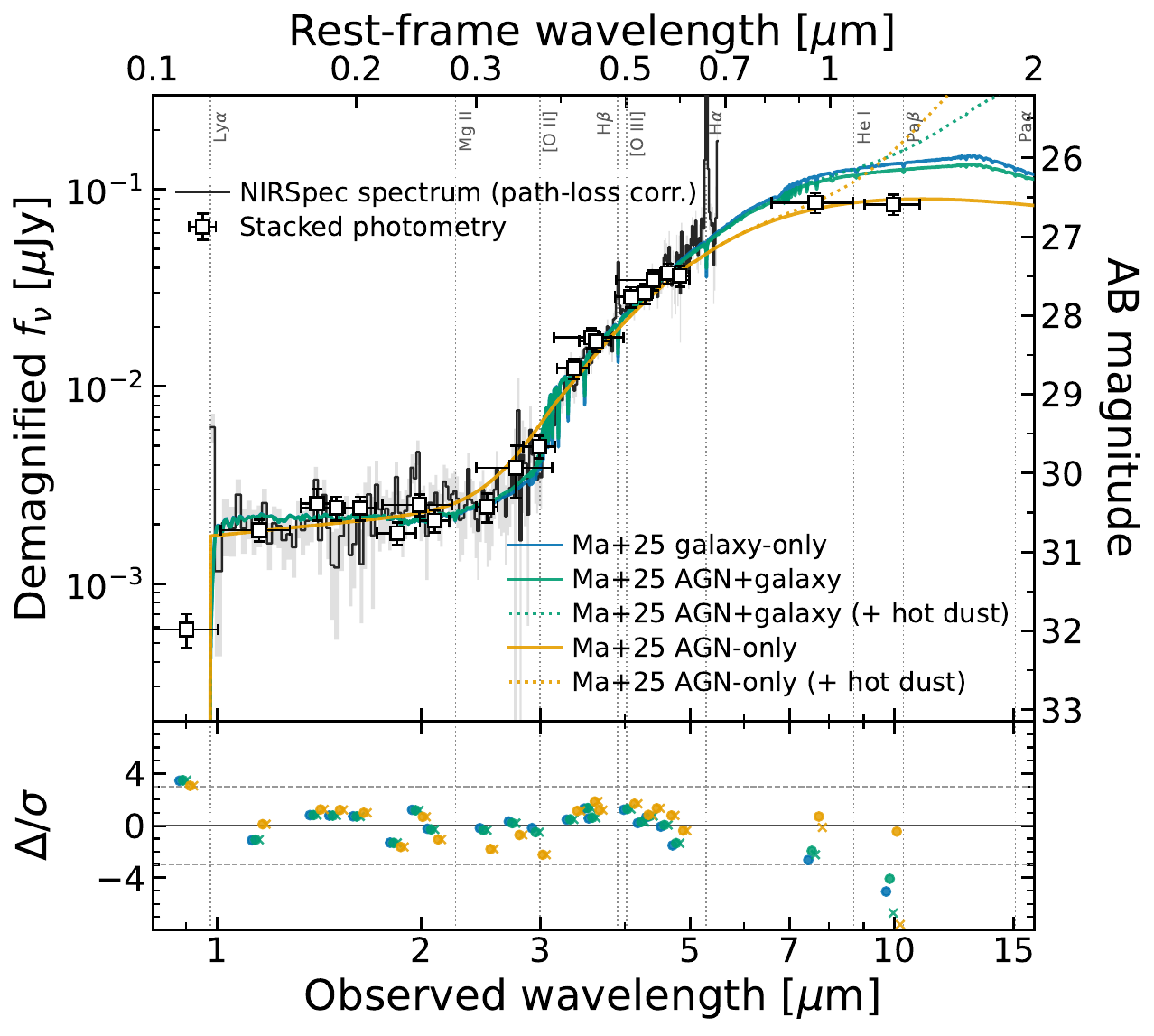}
    \caption{
    Best-fitting continuum models of \citet{Ma+25}, rebuilt from their Table~2 and extended to the MIRI bands, compared with the stacked photometry (white squares) and the NIRSpec spectrum (black) of QSO1.
    Each model is rescaled by a single factor fitted to the NIRCam continuum bands (F115W--F480M, excluding F410M and F444W).
    Dotted lines include the hot-dust component of the \citet{Temple+21} quasar template. Unlike Figure~\ref{fig:ma25_posteriors}, we show here the published solutions, not refits to our photometry.
    Bottom panel: residuals, $(f_{\rm obs}-f_{\rm model})/\sigma$, of the synthetic photometry of each model.
    }
    \label{fig:ma25_published}
\end{figure}

We first rebuilt their best-fitting models from their Table~2, with $\delta=-1$.
They reproduce the quoted $A_{1500}/A_V$ (9.4; effective 2.1) and the broad-line Balmer decrement (11.4) and their components within $\approx0.05$~dex.
Each model was rescaled by a single factor fitted to the NIRCam continuum (F115W--F480M, excluding F410M and F444W), which removes the 0.27~dex offset between their normalisation ($\mu_{\rm A}=5.4$, Ima-A) and ours.
The galaxy-only and AGN+galaxy models over-predict both MIRI bands by 23--60\% (Table~\ref{tab:ma25_miri}, Figure~\ref{fig:ma25_published}).
The AGN-only model is within $8\%$ without hot dust, but over-predicts F1000W by a factor of $\approx1.9$ with hot dust.
As \citet{Ma+25} fitted only up to rest-frame $0.66$\um, these are predictions of their models, not failures of their fits.
 
\begin{figure*}
    \centering
    \includegraphics[width=0.44\textwidth]{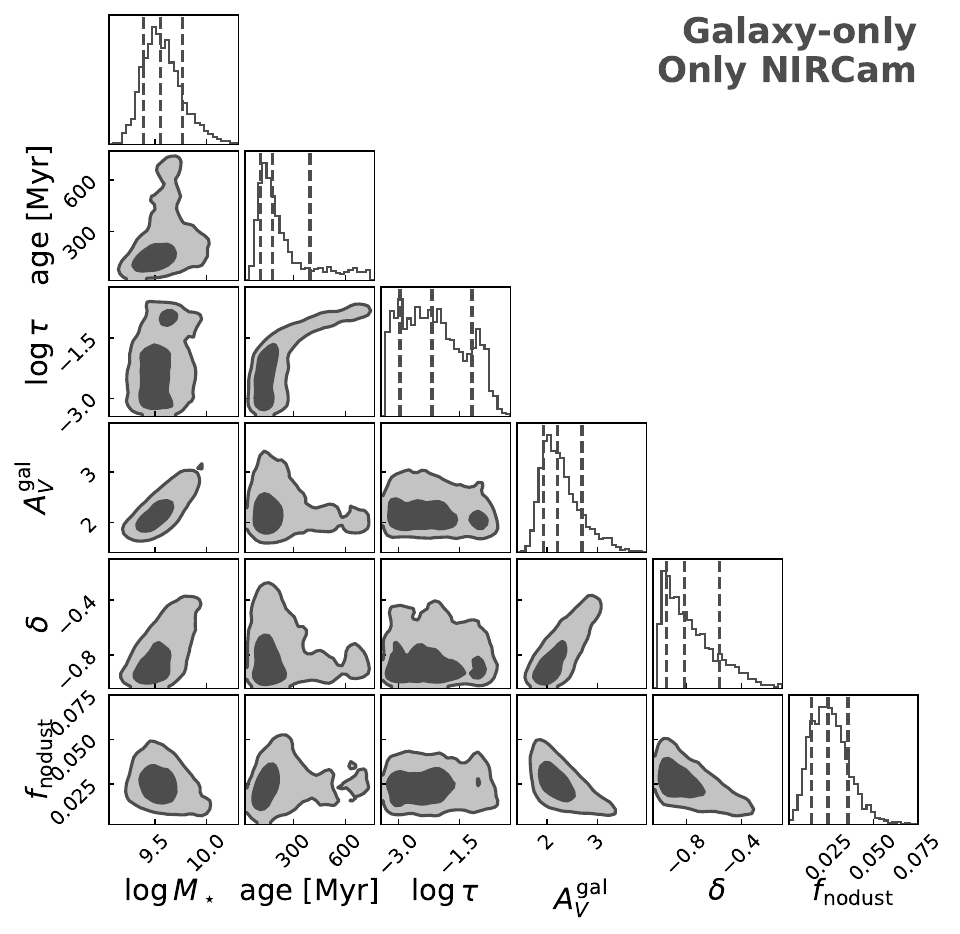}\hfill
    \includegraphics[width=0.44\textwidth]{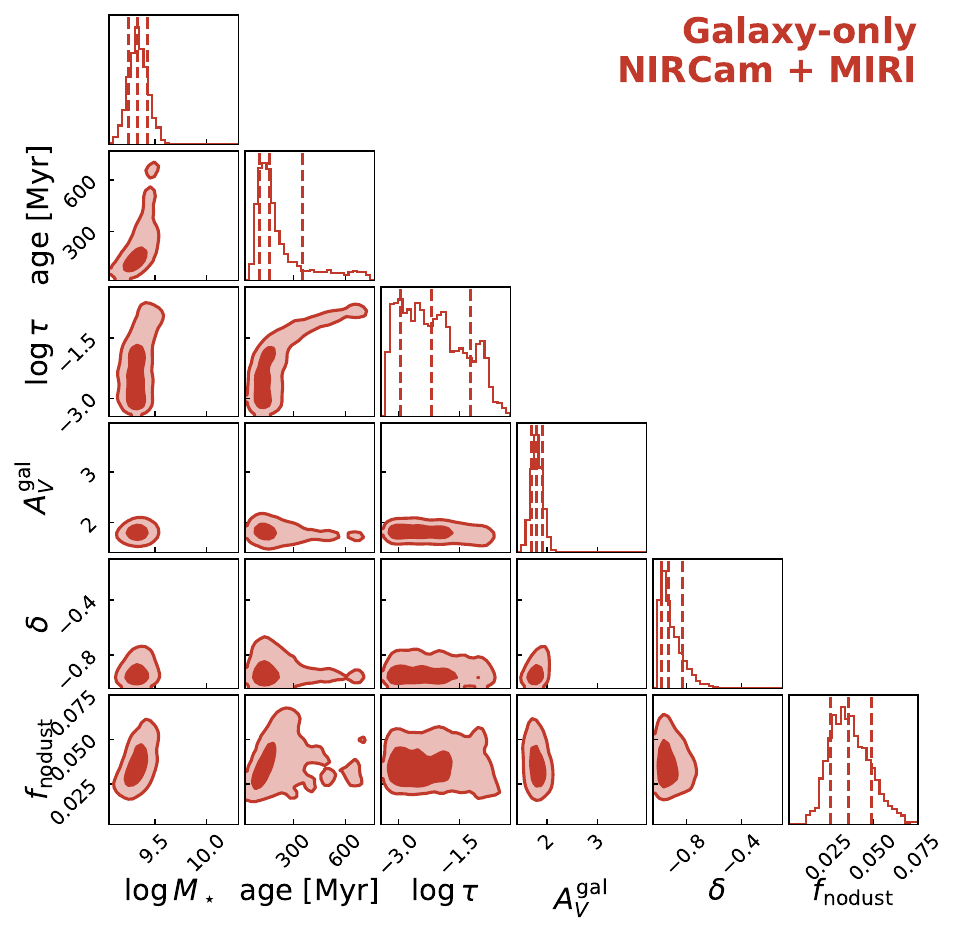}\\[4pt]
    \includegraphics[width=0.44\textwidth]{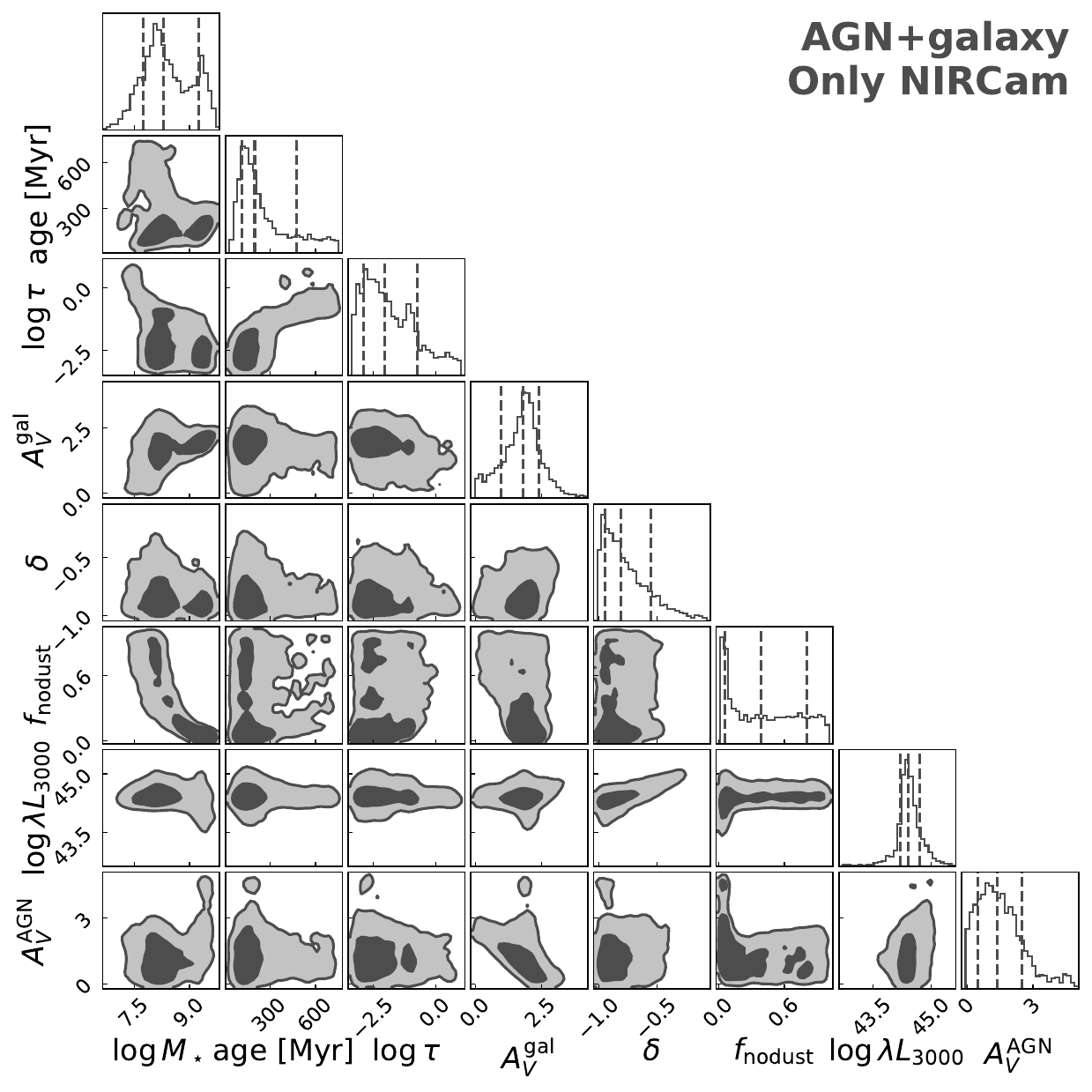}\hfill
    \includegraphics[width=0.44\textwidth]{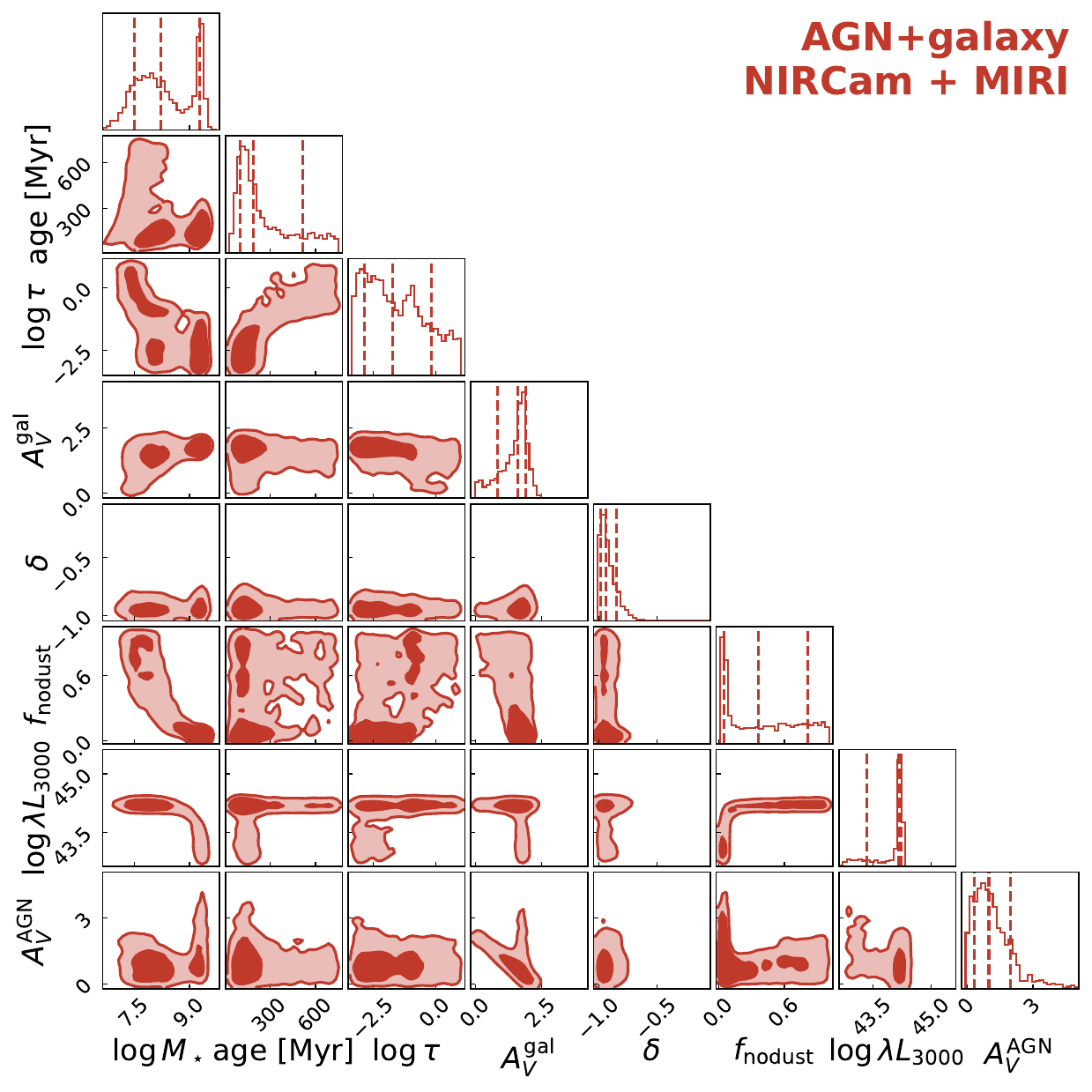}
    \caption{
    Posterior distributions of the galaxy-only (top) and AGN+galaxy (bottom) models of \citet{Ma+25}, fitted to the NIRCam photometry only (left, grey) and to NIRCam and MIRI (right, red).
    Contours enclose 39\% and 86\% of the posterior mass ($1\sigma$ and $2\sigma$ in two dimensions). Dashed lines in the 1D histograms mark the 16th, 50th and 84th percentiles (Table~\ref{tab:ma25_posteriors}).
    $M_\star$ is the surviving stellar mass in M$_\odot$ and $\lambda L_{3000}$ is in ergs$^{-1}$, both in our normalisation. The age is measured from the onset of star formation, and $\tau$ is in Gyr.
    Axis ranges are the same in the two panels of each row.
    }
    \label{fig:ma25_corners}
\end{figure*}

We then refitted the same families to our stacked photometry with \textsc{Nautilus} \citep{Lange+23}, using 1000 live points and an effective sample size of $10^4$.
The fits use the 16 NIRCam bands from F115W to F480M (excluding F410M and F444W), with or without F770W and F1000W, and uniform priors as in \citet{Ma+25} (Table~\ref{tab:ma25_posteriors}, Figures~\ref{fig:ma25_posteriors} and \ref{fig:ma25_corners}).
Fitted to NIRCam alone, the posteriors recover their solutions. For example, $A_V=2.21^{+0.49}_{-0.29}$ mag (galaxy-only) and $2.41^{+0.25}_{-0.18}$ mag (AGN-only), compared with their 2.12 mag and 2.08 mag.
All of these fits over-predict MIRI, so the tension is not limited to the published best fits.
With MIRI, the models without hot dust match F770W and F1000W within $13\%$, but only with a different optical solution: the galaxy-only model needs lower attenuation ($A_V=1.79^{+0.12}_{-0.11}$ mag) and $\log_{10}( M_\star~\rm[M_\odot])=9.33^{+0.10}_{-0.09}$, while the AGN+galaxy posterior is bimodal, between an AGN-dominated mode with a minor stellar component ($\log_{10}( M_\star\rm~[M_\odot])\approx7.5$--8) and the galaxy-only solution, with medians $\log_{10}( \lambda L_{3000}~[{\rm erg~s^{-1}}])=44.17^{+0.06}_{-0.82}$ and $\log_{10}( M_\star~\rm[M_\odot])=8.21^{+1.06}_{-0.70}$.
These two have similar evidence ($\Delta\ln(\mathcal{Z})=-0.7$), while AGN-only is disfavoured ($\Delta\ln(\mathcal{Z})=-5.4$).
The hot-dust component of the \citet{Temple+21} template is disfavoured.
The AGN-only model with hot dust is rejected ($\Delta\ln(\mathcal{Z})=-19.5$, $\chi^2_{\rm red}=3.66$).
In the AGN+galaxy model with hot dust, the AGN is suppressed by $\approx1.2$~dex, so that the model reverts to galaxy-only ($\Delta\ln(\mathcal{Z})=-2.4$). Fitted to NIRCam alone, the hot-dust models over-predict F1000W by up to a factor of $2.8$.
Fitted to NIRCam and MIRI, the galaxy-only and AGN+galaxy models are statistically acceptable ($\chi^2_{\rm red}=0.94$ and $1.20$), whereas the AGN-only model is not as good ($\chi^2_{\rm red}=1.70$). All place $\delta$ close to the lower edge of its prior ($-1$).
Removing the line contribution to MIRI (Section~\ref{sec:nir_continuum}) would only increase the tension with the NIRCam-only fits.
 
\begin{table*}
\centering
\caption{Predicted-to-observed MIRI flux-density ratios for the best-fit continuum models of \citet{Ma+25}, rebuilt from their best-fit parameters. Each model is rescaled by a single factor, either fitted to the NIRCam continuum bands (F115W--F480M, excluding F410M and F444W; ``global'') or matched to F480M. Numbers in parentheses give the deviation in units of the photometric uncertainty. Models labelled ``+ hot dust'' include the default hot-dust component of the \citet{Temple+21} quasar template.}
\label{tab:ma25_miri}
\begin{tabular}{lcccc}
\hline\hline
Model & \multicolumn{2}{c}{global} & \multicolumn{2}{c}{F480M-matched} \\
 & F770W & F1000W & F770W & F1000W \\
\hline
Galaxy-only                 & 1.31 ($-2.6\sigma$) & 1.60 ($-5.1\sigma$) & 1.11 ($-0.9\sigma$) & 1.35 ($-3.0\sigma$) \\
AGN+galaxy                  & 1.23 ($-1.9\sigma$) & 1.48 ($-4.1\sigma$) & 1.06 ($-0.5\sigma$) & 1.28 ($-2.4\sigma$) \\
AGN+galaxy + hot dust       & 1.26 ($-2.2\sigma$) & 1.79 ($-6.7\sigma$) & 1.09 ($-0.8\sigma$) & 1.55 ($-4.6\sigma$) \\
AGN-only                    & 0.92 ($+0.7\sigma$) & 1.05 ($-0.5\sigma$) & 0.88 ($+1.0\sigma$) & 1.01 ($-0.1\sigma$) \\
AGN-only + hot dust         & 1.02 ($-0.1\sigma$) & 1.89 ($-7.6\sigma$) & 0.97 ($+0.2\sigma$) & 1.81 ($-6.9\sigma$) \\
\hline\hline
\end{tabular}
 
\end{table*}
 
\begin{table*}
\centering
%\small\setlength{\tabcolsep}{3pt}
\caption{Posterior constraints for the \citet{Ma+25} model families fitted to the stacked photometry of QSO1, without and with the MIRI bands. Values are posterior medians with 16th--84th percentile ranges. Stellar masses and luminosities are de-magnified with the Ima-C-anchored stack adopted here (UNCOVER lens model v2, \citealt{Furtak+25}). They are 0.27dex lower than in the \citet{Ma+25} normalisation ($\mu_{\rm A}=5.4$). $M_\star$ is the surviving stellar mass in M$_\odot$ and $\lambda L_{3000}$ is in ergs$^{-1}$. $\chi^2_{\rm red}=\chi^2_{\rm min}/(N_{\rm bands}-N_{\rm par})$ of the maximum-likelihood model, with $N_{\rm bands}=16$ (NIRCam) or 18 (NIRCam+MIRI) and $N_{\rm par}=6$, 8 and 4 for the galaxy-only, AGN+galaxy and AGN-only families. $\Delta\ln(\mathcal{Z})$ is the log-evidence relative to the best model fitted to the same bands, and values for different band sets are not comparable. The last two columns give the predicted-to-observed MIRI flux ratios (for the NIRCam-only fits, these are extrapolations).}
\label{tab:ma25_posteriors}
\begin{tabular}{lcccccccc}
\hline\hline
Model & $\Delta\ln(\mathcal{Z})$ & $\chi^2_{\rm red}$ & $\log_{10} M_\star$ & $A_V^{\rm gal}$ & $\log_{10} \lambda L_{3000}$ & $A_V^{\rm AGN}$ & F770W & F1000W \\
 & & & & & & & pred/obs & pred/obs \\
\hline
\multicolumn{9}{l}{\textit{Fit to NIRCam only (F115W--F480M)}}\\
Galaxy-only & $+0.0$ & 0.91 & $9.56^{+0.21}_{-0.16}$ & $2.21^{+0.49}_{-0.29}$ & -- & -- & $1.27^{+0.28}_{-0.20}$ & $1.62^{+0.49}_{-0.33}$ \\
AGN+galaxy & $-0.7$ & 1.19 & $8.30^{+0.95}_{-0.56}$ & $1.81^{+0.59}_{-0.85}$ & $44.42^{+0.29}_{-0.20}$ & $1.37^{+1.13}_{-0.88}$ & $1.36^{+0.42}_{-0.25}$ & $1.76^{+0.84}_{-0.42}$ \\
AGN+galaxy (+hot dust) & $-0.7$ & 1.16 & $8.39^{+0.94}_{-0.60}$ & $1.86^{+0.54}_{-0.82}$ & $44.38^{+0.29}_{-0.32}$ & $1.34^{+1.11}_{-0.87}$ & $1.44^{+0.40}_{-0.23}$ & $2.78^{+1.36}_{-0.75}$ \\
AGN-only & $-5.9$ & 1.77 & -- & -- & $44.35^{+0.11}_{-0.09}$ & $2.41^{+0.25}_{-0.18}$ & $1.18^{+0.20}_{-0.15}$ & $1.43^{+0.30}_{-0.21}$ \\
AGN-only (+hot dust) & $-5.9$ & 1.77 & -- & -- & $44.35^{+0.12}_{-0.09}$ & $2.41^{+0.28}_{-0.18}$ & $1.33^{+0.22}_{-0.17}$ & $2.62^{+0.57}_{-0.40}$ \\
\hline
\multicolumn{9}{l}{\textit{Fit to NIRCam + MIRI (F115W--F1000W)}}\\
Galaxy-only & $+0.0$ & 0.94 & $9.33^{+0.10}_{-0.09}$ & $1.79^{+0.12}_{-0.11}$ & -- & -- & $0.95^{+0.06}_{-0.06}$ & $1.13^{+0.08}_{-0.09}$ \\
AGN+galaxy & $-0.7$ & 1.20 & $8.21^{+1.06}_{-0.70}$ & $1.61^{+0.29}_{-0.78}$ & $44.17^{+0.06}_{-0.82}$ & $1.00^{+0.98}_{-0.66}$ & $0.96^{+0.06}_{-0.06}$ & $1.13^{+0.09}_{-0.08}$ \\
AGN+galaxy (+hot dust) & $-2.4$ & 1.20 & $9.30^{+0.10}_{-0.10}$ & $1.75^{+0.12}_{-0.12}$ & $43.00^{+0.28}_{-0.19}$ & $2.42^{+1.81}_{-1.65}$ & $0.93^{+0.06}_{-0.06}$ & $1.16^{+0.08}_{-0.08}$ \\
AGN-only & $-5.4$ & 1.70 & -- & -- & $44.22^{+0.04}_{-0.04}$ & $2.17^{+0.10}_{-0.09}$ & $0.97^{+0.06}_{-0.06}$ & $1.13^{+0.09}_{-0.08}$ \\
AGN-only (+hot dust) & $-19.5$ & 3.66 & -- & -- & $44.03^{+0.03}_{-0.03}$ & $1.86^{+0.07}_{-0.07}$ & $0.76^{+0.05}_{-0.04}$ & $1.38^{+0.09}_{-0.09}$ \\
\hline
\end{tabular}
 
\vspace{2pt}\par\parbox{\textwidth}{\footnotesize \textbf{Note.} For AGN+galaxy, $A_V^{\rm AGN}$ is the additional attenuation of the AGN on top of $A_V^{\rm gal}$, as in \citet{Ma+25}. For AGN-only, it is the total attenuation. The Noll et al.\ slope $\delta$ lies close to the lower edge of its prior ($\delta\simeq-0.8$ to $-1$) in all fits. In the AGN+galaxy fits, the posteriors are broad or bimodal, and the quoted ranges span both the galaxy- and the AGN-dominated solutions.}
\end{table*}
 
\subsection{Comparison with accretion-flow envelope models}
\label{app:liu}
We used the 15 public continuum spectra of \citet{Liu+25}, computed for luminosities $L=L_{\rm Edd}(10^{l}~\mathrm{M}_\odot)$ with $l=5$, 6, 7 and five reference densities $\rho_{\rm ref}$ for each $l$, over rest-frame 1250--13000~\AA. We interpolated them linearly in $\log L_\nu$ over $(l, \log\rho_{\rm ref}-0.5~l)$, as in the notebook released by the authors, and extrapolated beyond $1.3$\um\ with a power law, which affects F1000W by $\leq1\%$. The models contain no emission lines and no flux below $\approx2500$~\AA, so we added a power law for the UV continuum. We fitted the stacked photometry from F115W to F1000W (excluding F410M and F444W) with the photometric uncertainties, on a grid of $l$, $\rho_{\rm ref}$ and UV slope, solving for the two normalisations. The shape fixes $\log\rho_{\rm ref}-0.5~l\approx-14.75\pm0.05$, while $l$ is degenerate with the normalisation. The solution whose luminosity matches the data without rescaling has $l\approx5.2$ and $\rho_{\rm ref}\approx10^{-12}~{\rm g~cm^{-3}}$.

\section{SED modelling}
\label{app:sedfit}

\subsection{Stellar populations and stellar-ionised gas}
For the stellar populations, we adopted the binary {\sc Bpass} v.~2.2 models \citep{Eldridge+17,Stanway+18}, with a Chabrier-like initial mass function extending to $300~\mathrm{M}_{\odot}$, and stellar metallicities of 0.05, 0.5, 5 and 10\% $Z_\odot$, bracketing the very low gas-phase metallicity of QSO1 \citep{Maiolino+26b}.
Each stellar population had a free formed mass ($\log_{10}(M_\star~[\mathrm{M}_\odot])\in[3,12]$, or $[3,7.5]$ in the test with the stellar-mass limit), its own colour excess and an age chosen from the native {\sc bpass} grid.
In the family with two ordinary-density populations, both ages were drawn from $\log_{10}(t~[\mathrm{Myr}])=0.0$--$2.2$ (1--160~Myr), with the second population older than the first.
In the dense-gas families, both populations could take any age in the same range, without ordering. In the cocoon families a single stellar population in the same age range was combined with the AGN component (Appendix~\ref{app:cocoon}).
 
Each {\sc bpass} spectrum was then used as ionising source for a grid of \textsc{Cloudy} (v25.00) photoionisation models to estimate the corresponding nebular emission.
The {\sc bpass} models were set in a spherical, ionisation-bounded, dust-free cloud (calculation stopped at an electron fraction of 1\%) with gas-phase metallicity tied to the stellar metallicity via the Galactic-concordance abundance pattern of \cite{Nicholls+17}.
The grid spanned $\log_{10}(n_\mathrm{H}~[\mathrm{cm^{-3}}])=0.5$--$12$ and $\log_{10}(U)=-4$ to $-1$ in steps of 0.5~dex, and stored the transmitted stellar light $F_\mathrm{trans}$, the diffuse nebular continuum $F_\mathrm{diffuse}$ and the absolute luminosities of the main emission lines $F_\mathrm{lines}$.
 
During fitting, each SSP was assigned a covering fraction $C_f$ such that its overall emission was $(1-C_f)~F_\mathrm{BPASS}+C_f~(F_\mathrm{trans}+F_\mathrm{diffuse}+F_\mathrm{lines})$. %, so the uncovered fraction played the role of a geometric Lyman-continuum escape channel ($f_\mathrm{esc}\simeq 1-C_f$)
In addition, stellar age and the pair $(\log_{10}(n_\mathrm{H}),\log_{10}(U))$ were kept as discrete parameters and no interpolation across the grid was performed.
Nodes spanning both ordinary ($\log_{10}(n_\mathrm{H})\le6$) and dense ($\log_{10}(n_\mathrm{H})\ge6.5$) gas were computed.
Ordinary-density populations were restricted to the former and had a free covering fraction.
In the dense-gas families (class ii in Section~\ref{sec:sedfit}), the second population was restricted to the dense nodes, where forbidden lines are collisionally suppressed and the emission is dominated by permitted lines and by the Balmer continuum and absorption.
Its covering fraction was free, and its lines had their own velocity dispersion, $\sigma_{v,\rm dense}\in[0,3000]~\rm{km~s^{-1}}$, independent of that of the ordinary-density gas ($\sigma_v\in[20,500]~\rm{km~s^{-1}}$).
For emission lines in the range covered by the NIRSpec spectrum, we integrated them exactly over each pixel using a Gaussian of width equal to the quadrature sum of the fitted intrinsic velocity dispersion and the instrumental dispersion at the line wavelength.
 
\subsection{Dust attenuation and emission}
The stellar light and its nebular emission were attenuated with the \cite{Calzetti+00} law, extended to the Lyman limit following \cite{Leitherer+02a}.
The dust screen carried a free slope offset, $A_\lambda\propto E(B-V)~k(\lambda)~(\lambda/0.55~\mu\mathrm{m})^{\delta}$, with $\delta\in[-1,0.4]$, to permit steeper or shallower ({\it greyer}) attenuation laws \cite[e.g.][]{Noll+09}.
In the families with independent attenuation, each component (each stellar population with its nebular emission, or the AGN cocoon) had its own colour excess. In the shared-screen families, all components were placed behind a single foreground screen with a common $E(B-V)$ and $\delta$.
Re-emission by dust was modelled with a modified blackbody including CMB heating and the CMB contrast at $z=7.04$ \citep{daCunha+13b}, plus a fixed-shape polycyclic aromatic hydrocarbons (PAH) template (Drude profiles, including the 3.3 and 3.4\um\ features) with a free total luminosity.
The intrinsic dust temperature had a log-uniform prior, $T_\mathrm{dust,0}\in[15,600]$~K, and the emissivity index and luminosity uniform priors, $\beta\in[1,2.5]$ and $\log_{10}(L_\mathrm{dust}~[L_\odot])\in[4,14]$.
$L_\mathrm{dust}$ was integrated over rest-frame 1--1000\um, so that it includes the full emission of warm dust.
The dust luminosity was tied through a Gaussian prior of width 0.3 dex in $\log_{10}(L_\mathrm{dust})$ to the luminosity absorbed by dust between 912~\AA\ and 3\um.
The absorbed luminosity included the stellar light leaving the photoionised gas (uncovered and transmitted), the nebular continuum and lines and, in the cocoon families, the cocoon emission.
The light absorbed by the gas itself is re-emitted as nebular emission and was therefore counted once.
The PAH luminosity was tied to $0.045~L_\mathrm{dust}$ with a Gaussian width of 0.3 dex (0.6 dex for the partially covering cocoon, since PAH emission is expected to be suppressed near AGN).
These were soft priors, so the long-wavelength data could pull the solution away from strict energy balance.
 
Finally, intergalactic medium (IGM) absorption was set to follow \cite{Inoue+14} as implemented in {\sc eazy}.
 
\subsection{Accretion disc embedded in a dense cocoon}
\label{app:cocoon}
In the cocoon families, an accretion disc is surrounded by a dense, dust-free envelope of gas, whose emission we computed with a dedicated grid of \textsc{Cloudy} (v25.00; \citealt{Gunasekera+25}) models.
The intrinsic disc spectrum was the native, dust-free (transparent) accretion-disc component of the {\sc skirtor} models \citep{Stalevski+12a,Stalevski+16}, a broken power law with $\lambda L_\lambda\propto\lambda^{1.2}$, $\lambda^{0}$, $\lambda^{-0.5}$ and $\lambda^{-3}$ below 0.01, between 0.01 and 0.1, between 0.1 and 5, and above $5~$\um, respectively.
We did not use the torus component of {\sc skirtor}.
The disc spectrum was extended to the extreme ultraviolet with a $T=3.3\times10^5$~K blackbody joined at 8~nm \citep{Zhang+26b}, and to the X-rays with a coronal power law with photon index $\Gamma_X=2.4$ and an exponential cut-off at 300~keV \citep{Yang+20}, joined at 0.1~keV and normalised through the optical-to-X-ray slope $\alpha_{\rm ox}=0.3838~\log_{10}[L_\nu(2~{\rm keV})/L_\nu(2500~\textrm{\AA})]=-1.5$ \citep{Tananbaum+79}. The resulting spectrum was renormalised to the integrated luminosity of the {\sc skirtor} disc.
In this continuum, $\approx64\%$ of the luminosity is emitted below 912~\AA\ ($\approx14\%$ above 0.1~keV), and ${\rm BC}_{3000}=L_{\rm AGN}/\lambda L_{3000}\approx9.6$.
 
The gas was modelled as a spherical, closed, dust-free shell fully surrounding the source, with constant hydrogen density $\log_{10}(n_\mathrm{H}~[\mathrm{cm^{-3}}])=8$, 9, 10, 11 and 12, ionisation parameter $\log_{10}(U)=-4$, $-3$, $-2$, $-1$ and total hydrogen column $\log_{10}(N_\mathrm{H}~[\mathrm{cm^{-2}}])=24.0$, 24.5, 25.0, 25.5, which spans the columns invoked by dense-envelope models of LRDs \citep{Inayoshi+25b,Ji+25,Naidu+25}.
Each calculation was stopped at the requested $\log_{10}(N_\mathrm{H})$.
The gas-phase metallicity was set equal to the stellar metallicity of each fit ($Z_\mathrm{gas}=10^{-5}$, $10^{-4}$, 0.001, 0.002 for \texttt{zem5}, \texttt{zem4}, \texttt{z001} and \texttt{z002}), with the Galactic-concordance abundance pattern of \citet{Nicholls+17}.
One node ($n_\mathrm{H}=10^{9}~\mathrm{cm^{-3}}$, $U=10^{-1}$, $N_\mathrm{H}=10^{25.5}~\mathrm{cm^{-2}}$) did not pass our convergence checks at $Z_\mathrm{gas}=10^{-5}$ and $10^{-4}$ and was excluded, leaving 79 nodes at these metallicities and 80 at the others.
Each model was run at a reference bolometric luminosity of $10^{11}~L_\odot$, with the corresponding ionising photon rate $Q(\mathrm{H})$ and the inner radius implied by $(Q(\mathrm{H}),U,n_\mathrm{H})$, so that the output is in absolute units.
For each model we stored, per unit bolometric luminosity of the disc, the transmitted continuum $F_\mathrm{trans}$, the outward diffuse continuum with the line emission removed (\textsc{Cloudy}'s outward diffuse continuum minus its outward-line contribution) $F_\mathrm{diffuse}$ and the emergent luminosity of each emission line $F_\mathrm{lines}$.
Lines were excluded from the saved diffuse continuum (\textsc{Cloudy}'s \texttt{set save line width suppress}), any residual outward-line contribution was subtracted explicitly, and duplicated and blended entries were removed from the line list, so that each line is counted exactly once.
 
In the fit, the cocoon emission was
\begin{equation}
L_\lambda=L_\mathrm{AGN}\left[(1-C)~F_\mathrm{inc}+C~(F_\mathrm{trans}+F_\mathrm{diffuse}+F_\mathrm{lines})\right],
\end{equation}
where $F_\mathrm{inc}$ is the intrinsic disc spectrum and $C$ the covering fraction, free in $[0,1]$.
The uncovered fraction thus reaches the observer as the unprocessed disc, while the covered fraction is replaced by the emission of the cocoon.
The bolometric luminosity, $\log_{10}(L_\mathrm{AGN}~[L_\odot])$, enters linearly and was solved for analytically at each likelihood evaluation. The cocoon node $(n_\mathrm{H},U,N_\mathrm{H})$ was a discrete parameter without interpolation, and the lines were broadened with a free velocity dispersion $\sigma_v\in[ 200,3000]~\rm km~s^{-1}$.
The cocoon emission was attenuated by a foreground screen with the \citet{Calzetti+00} curve, a free colour excess and a free slope offset $\delta$, as for the stellar populations. In the shared-screen families this screen was common to the stellar populations.
 
\textit{Limitations.} \textsc{Cloudy} solves the ionisation, thermal balance and non-LTE level populations of the gas self-consistently, including the $n=2$ population of hydrogen that shapes the Balmer break and the Balmer absorption, at a cost that allows a grid of 80 nodes per metallicity to be fitted with nested sampling. It assumes spherical symmetry and computes the shell along the radial direction only, and it transfers the lines with escape probabilities. In the closed geometry adopted here, scattering opacities do not attenuate the incident continuum, and the frequency redistribution of scattered photons is not followed. The ionising continuum is fixed ($\alpha_{\rm ox}=-1.5$, $\Gamma_X=2.4$). Given the X-ray non-detection of QSO1, an X-ray-weaker continuum could change the thermal and ionisation structure of the gas.
The electron-scattering optical depth computed by \textsc{Cloudy} is $\tau_{\rm es}\leq2.2$ for all nodes and $\lesssim0.3$ for the nodes selected by the fits, because the gas beyond the thin ionised layer is only partially ionised ($x_e\lesssim2\%$).
Scattering by neutral hydrogen is instead large in the UV. For the node of the cocoon with the stellar-mass limit, \textsc{Cloudy} gives a Rayleigh optical depth of 211 at 1300~\AA, yet transmits 40\% of the incident continuum at this wavelength. At $1$\um, both electron and Rayleigh scattering are negligible in this node ($\tau\lesssim0.1$), and the continuum shape is set by the absorption and emission of the gas.
These approximations affect the UV transmitted through the cocoon, the shape of the broad lines and the structure of optically thick gas. Monte Carlo radiative-transfer codes such as \textsc{Sirocco} \citep{Matthews+25,Sneppen+26c} treat electron and resonance scattering in multi-dimensional geometries and should be used to test the regime selected by our fits.
 
\begin{figure}
    \centering
    \includegraphics[width=\columnwidth]{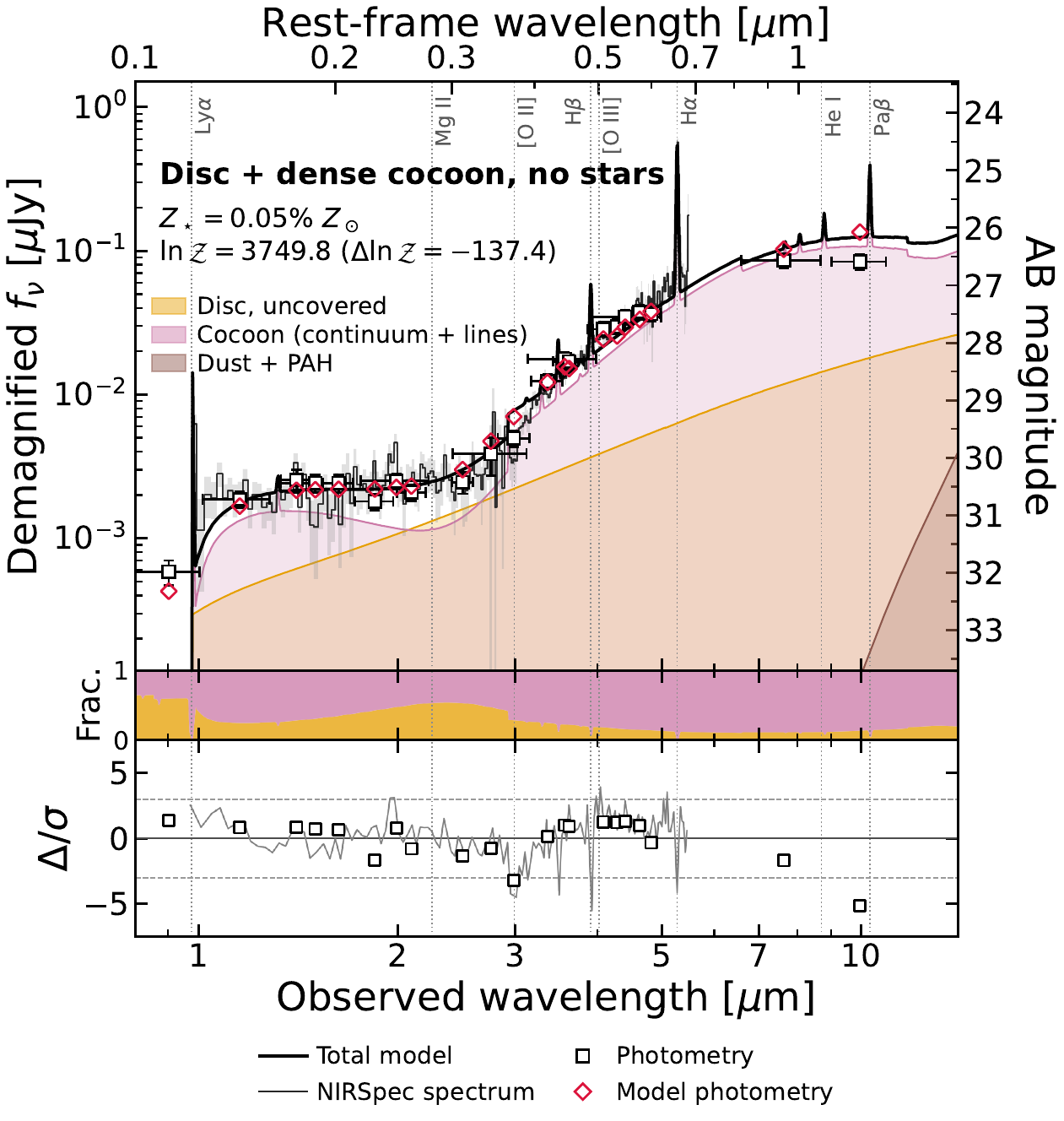}
    \caption{Maximum-likelihood solution of the model without stars (Appendix~\ref{app:extra_families}, \texttt{zem5}), in which an accretion disc is seen directly and through the dense cocoon. Panels as in Fig.~\ref{fig:sed_models}, with $\Delta\ln(\mathcal{Z})$ relative to the highest-evidence model of Table~\ref{tab:sedfit_evidence}. Without a separate UV component, the model misses the Balmer break ($3\sigma$ in F300M) and over-predicts F1000W by $5\sigma$.}
    \label{fig:sedfit_nostars}
\end{figure}
 
\subsection{Test models}
\label{app:extra_families}
The last block of Tables~\ref{tab:sedfit_families} and \ref{tab:sedfit_evidence} lists two models that test the components of the cocoon solution.
In the model without stars, the stellar population was switched off by fixing its mass to $10^{3}~\mathrm{M}_\odot$, its covering fraction to zero and its colour excess to zero, so that it contributes no measurable light. The disc then has to produce the UV continuum as well.
The disc is seen both directly and through the cocoon, with a free covering fraction (Appendix~\ref{app:cocoon}).
In the model with $E(B-V)_\star=0$, the colour excess of the stellar population and of its nebular emission is fixed to zero, its mass is not limited, and the disc and cocoon keep their own screen.
All other components, priors, data and sampler settings are those of the other families, so the evidences are directly comparable.
Without stars, the cocoon solution (\texttt{zem5}) has $\log_{10}(L_{\rm AGN}~[L_\odot])=10.71\pm0.06$ ($\approx2\times10^{44}$~erg~s$^{-1}$), a cocoon with $C=0.49\pm0.04$, $n_{\rm H}=10^{11}~\mathrm{cm^{-3}}$, $U=10^{-4}$ and $N_{\rm H}=10^{25.5}~\mathrm{cm^{-2}}$, lines with $\sigma_v\approx910$~km~s$^{-1}$, and a screen with $A_V=1.6\pm0.1$~mag. Twice as much of the disc is covered as in the solution with a young stellar population ($C\approx0.22$), because the uncovered disc must now account for the UV alone.
 
\subsection{Fits to the individual images}
\label{app:images}
The stacked photometry combines three epochs separated by up to $\approx19$~yr in the observed frame and depends on the relative magnifications of the lens model, while the spectrum samples Ima-B only. To test whether the model comparison of Section~\ref{sec:sedfit} depends on how the images are combined, we evaluated the posterior samples of the joint fits against the photometry of each image separately. For each of the six models in Table~\ref{tab:images}, we took the parameter sets sampled in the joint fit to the stacked photometry and the spectrum (up to 6000 per model, including the highest-likelihood points) and computed their synthetic photometry. We then compared it with the photometry of each image (Table~\ref{tab:photometry}), using the photometric uncertainties and the upper limits of that image and no spectrum. A free overall normalisation was solved for at each point, so that the test is sensitive only to the shape of the SED and not to the magnification. The best point of each model is the best solution for that image within the region of parameter space allowed by the spectrum. It is not a posterior of the photometry alone, which would explore a wider region.
 
For every model, the fitted normalisations reproduce the image ratios of the stack ($1.77$--$1.86$, $1.16$--$1.23$ and $1.02$--$1.06$ for Ima-A, B and C, against 1.87, 1.22 and 1), so the three images have the same SED shape and achromatic offsets. Without the magnification and inter-image terms of the stack, the uncertainties are smaller and the $\chi^2$ values are larger than for the stack (Table~\ref{tab:images}). The largest residuals, however, are in the same bands for every model (F140M and F277W in Ima-A, F182M and F277W in Ima-B, F480M in Ima-C). They trace differences between the images, such as the varying broad-line equivalent widths and the residual contamination of Ima-B, rather than differences between the models, and do not affect their ranking.
 
In each image the ranking is the same as for the stack. The best solution with a massive stellar population fits better than the cocoon with the stellar-mass limit by $\Delta\chi^2=107$, 62 and 31 for Ima-A, B and C (13 for the stack). The cocoon over-predicts F1000W by 7.2, 8.8 and $4.6\sigma$, while the stellar solutions reproduce it within $\approx1\sigma$ in Ima-A and Ima-C. Without the MIRI bands, the difference drops to $\Delta\chi^2=51$, $-6$ and 9. The NIRCam photometry of Ima-B and Ima-C alone therefore does not separate the two classes, and the NIRCam-only cocoon fits predict F1000W fluxes 5--$15\sigma$ above the observed ones. In Ima-B all models over-predict F1000W by 3--$5\sigma$, as expected from its lower MIRI-to-NIRCam ratio (Section~\ref{sec:disc_variability}). The cocoon with $E(B-V)_\star=0$ behaves as the cocoon with the stellar-mass limit, and the model without stars gives the poorest fits in every image.
 
\begin{table*}
\centering
\caption{$\chi^2$ of the best solution of each model for the photometry of the individual images and of the stack, without the spectrum and with a free normalisation (Appendix~\ref{app:images}), using all 21 detected bands and the upper limits (``all'') or excluding F770W and F1000W (19 bands, ``no MIRI''). The stack column uses the uncertainties of the stacked photometry. The lowest value of each column is in bold.}
\label{tab:images}
\small
\setlength{\tabcolsep}{4pt}
\begin{tabular}{lcccccccc}
\hline\hline
 & \multicolumn{2}{c}{Ima-A} & \multicolumn{2}{c}{Ima-B} & \multicolumn{2}{c}{Ima-C} & \multicolumn{2}{c}{Stack} \\
Model & all & no MIRI & all & no MIRI & all & no MIRI & all & no MIRI \\
\hline
\texttt{2ssp\_nodense\_independent} (\texttt{zem4}) & \textbf{95.1} & 94.2 & 182.6 & 165.7 & 69.3 & 68.5 & 11.7 & 11.5 \\
\texttt{normal\_dense\_independent} (\texttt{zem5}) & 95.3 & \textbf{91.9} & 171.5 & 161.3 & \textbf{62.5} & \textbf{60.1} & \textbf{10.9} & \textbf{10.5} \\
\texttt{normal\_accretion\_cocoon\_independent} (\texttt{zem4}) & 121.2 & 117.7 & \textbf{169.3} & 135.8 & 65.5 & 64.7 & 13.9 & 13.3 \\
\texttt{accretion\_cocoon\_independent}, $M_\star\leq10^{7.5}~\mathrm{M}_\odot$ (\texttt{zem4}) & 202.2 & 143.0 & 231.2 & 129.9 & 94.0 & 69.4 & 24.3 & 14.3 \\
\texttt{accretion\_cocoon\_independent}, $E(B-V)_\star=0$ (\texttt{zem4}) & 196.1 & 145.6 & 253.5 & 142.6 & 92.6 & 68.5 & 24.0 & 15.4 \\
\texttt{accretion\_cocoon\_only} (\texttt{zem5}) & 319.5 & 270.2 & 240.2 & \textbf{111.7} & 116.4 & 70.6 & 33.9 & 19.9 \\
\hline\hline
\end{tabular}
\end{table*}

\begin{table*}
\centering
\caption{Model families fitted to QSO1. The family with $M_\star\leq10^{7.5}~\mathrm{M}_\odot$ is the cocoon fitted with the stellar mass limited to the upper limit on the host stellar mass. Every family except the one without stars contains a young, ordinary-density {\sc bpass} population for the UV continuum, with its own \textsc{Cloudy} nebular emission and covering fraction. The second column gives the component added for the red optical continuum. In the highest-evidence unconstrained cocoon this role is taken by the stellar population, and the disc produces the UV (Section~\ref{sec:sedfit}). ``Independent'' attenuation gives each component its own colour excess, ``shared screen'' places all components behind one foreground screen while ``single screen'' is for families with one single component. Each family was fitted at stellar metallicities of 0.05, 0.5, 5 and 10\%~$Z_\odot$ (\texttt{zem5}, \texttt{zem4}, \texttt{z001}, \texttt{z002}). $N_{\rm par}$ is the number of sampled parameters. In the cocoon families the AGN luminosity is solved for analytically and is not counted. The last block lists the test models of Appendix~\ref{app:extra_families}. For the model without stars, $N_{\rm par}$ counts only the parameters that affect the model.}
\label{tab:sedfit_families}
\begin{tabular}{llcc}
\hline\hline
Family & Red-continuum component & Attenuation & $N_{\rm par}$ \\
\hline
\texttt{2ssp\_nodense\_independent}                    & ordinary-density SSP ($t\geq10$~Myr)         & independent & 15 \\
\hline
\texttt{normal\_dense\_independent}                    & dense-gas SSP            & independent   & 16 \\
\texttt{normal\_dense\_screen}    & dense-gas SSP          & shared screen & 15 \\
\hline
\texttt{normal\_accretion\_cocoon\_independent}              & accretion disc + dense cocoon                    & independent   & 15 \\
\texttt{normal\_accretion\_cocoon\_screen}           & accretion disc + dense cocoon                 & shared screen & 13 \\
\hline
\texttt{accretion\_cocoon\_independent}, $M_\star\leq10^{7.5}~\mathrm{M}_\odot$ & accretion disc + dense cocoon & independent & 15 \\
\texttt{accretion\_cocoon\_independent, $E(B-V)_\star=0$} & accretion disc + dense cocoon & independent & 13\\
\texttt{accretion\_cocoon\_only} & disc + cocoon, no stars &	single screen &	9  \\
\hline\hline
\end{tabular}
 
\end{table*}

\begin{table*}
\centering
\caption{Bayesian evidence, $\ln(\mathcal{Z})$, of every family--metallicity combination fitted to QSO1, with $\Delta\ln(\mathcal{Z})$ relative to the highest-evidence model (bold) in parentheses.
The row with $M_\star\leq10^{7.5}~\mathrm{M}_\odot$ is the cocoon fitted with the stellar mass limited to the upper limit on the host stellar mass. The last block lists the cocoon with an unattenuated stellar population and no mass limit and the model without stars.}
\label{tab:sedfit_evidence}
\small
\setlength{\tabcolsep}{4pt}
\begin{tabular}{lcccc}
\hline\hline
Family & \texttt{zem5} & \texttt{zem4} & \texttt{z001} & \texttt{z002}  \\
 & (0.05\%~$Z_\odot$) & (0.5\%~$Z_\odot$) & (5\%~$Z_\odot$) & (10\%~$Z_\odot$) \\
\hline
\texttt{2ssp\_nodense\_independent} & 3739.1 ($-148.1$) & 3747.4 ($-139.8$) & 3721.3 ($-165.9$) & 3705.2 ($-182.0$) \\
\hline
\texttt{normal\_dense\_independent} & 3874.5 ($-12.7$) & 3858.2 ($-29.0$) & 3866.6 ($-20.6$) & 3857.1 ($-30.1$) \\
\texttt{normal\_dense\_screen} & 3600.3 ($-286.9$) & 3776.9 ($-110.3$) & 3767.0 ($-120.1$) & 3724.1 ($-163.1$) \\
\hline
\texttt{normal\_accretion\_cocoon\_independent} & 3873.4 ($-13.8$) & \textbf{3887.2 (0)} & 3875.0 ($-12.2$) & 3833.0 ($-54.2$) \\
\texttt{normal\_accretion\_cocoon\_screen} & 3776.4 ($-110.8$) & 3769.4 ($-117.8$) & 3750.3 ($-136.9$) & 3738.7 ($-148.5$) \\
\hline
\texttt{accretion\_cocoon\_independent}, $M_\star\leq10^{7.5}~\mathrm{M}_\odot$ & 3853.2 ($-34.0$) & 3854.4 ($-32.8$) & 3840.2 ($-46.9$) & 3833.4 ($-53.8$) \\
\texttt{accretion\_cocoon\_independent, $E(B-V)_\star=0$} & 3848.5 ($-38.7$) & 3850.9 ($-36.3$) & 3839.3 ($-47.9$) & 3832.2 ($-55.0$) \\
\texttt{accretion\_cocoon\_only} & 3749.8 ($-137.4$) & 3731.8 ($-155.4$) & 3706.8 ($-180.4$) & 3693.2 ($-194.0$) \\
\hline\hline
\end{tabular}
\end{table*}
 
\subsection{Likelihood and sampling}
The likelihood combined the photometry and the NIRSpec spectrum, which was rebinned to three pixels per resolution element and compared with models convolved to the instrumental resolution, without masking the emission lines. A fractional error floor (jitter) of 5\% was added to the spectrum, the same for all families.
For every combination of model family and stellar metallicity we ran the nested sampler {\sc Nautilus} with 800 live points, a target effective sample size of 4000, a live-evidence fraction $f_\mathrm{live}=0.01$ and at most $4\times10^6$ likelihood calls.
Priors were uniform over the ranges above and over the allowed nodes for discrete parameters.
The sampler returned the Bayesian evidence $\ln(\mathcal{Z})$ and posterior samples for each run.
 
\begin{figure*}[h]
    \centering
    \includegraphics[width=0.45\textwidth]{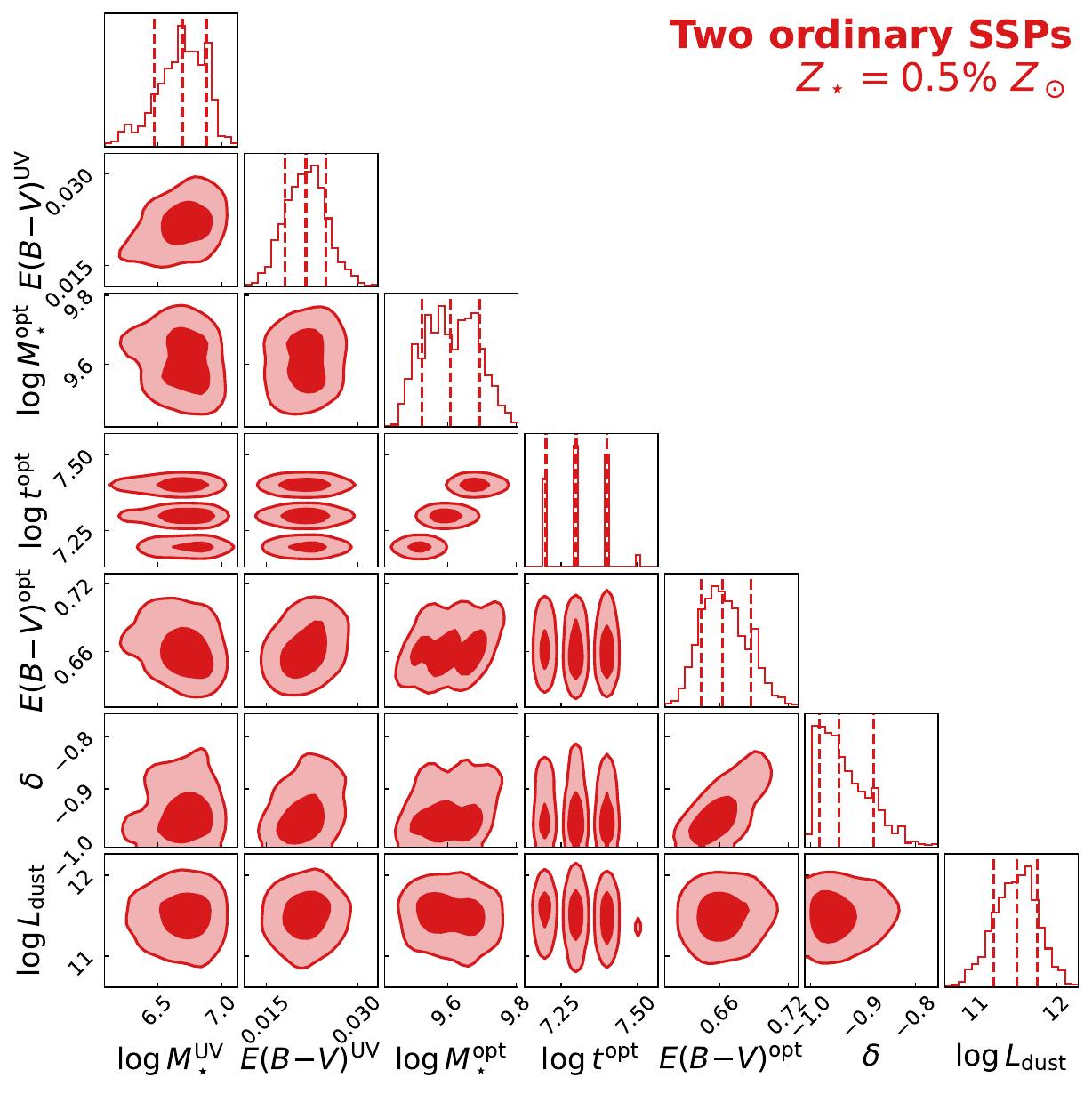}\hfill
    \includegraphics[width=0.45\textwidth]{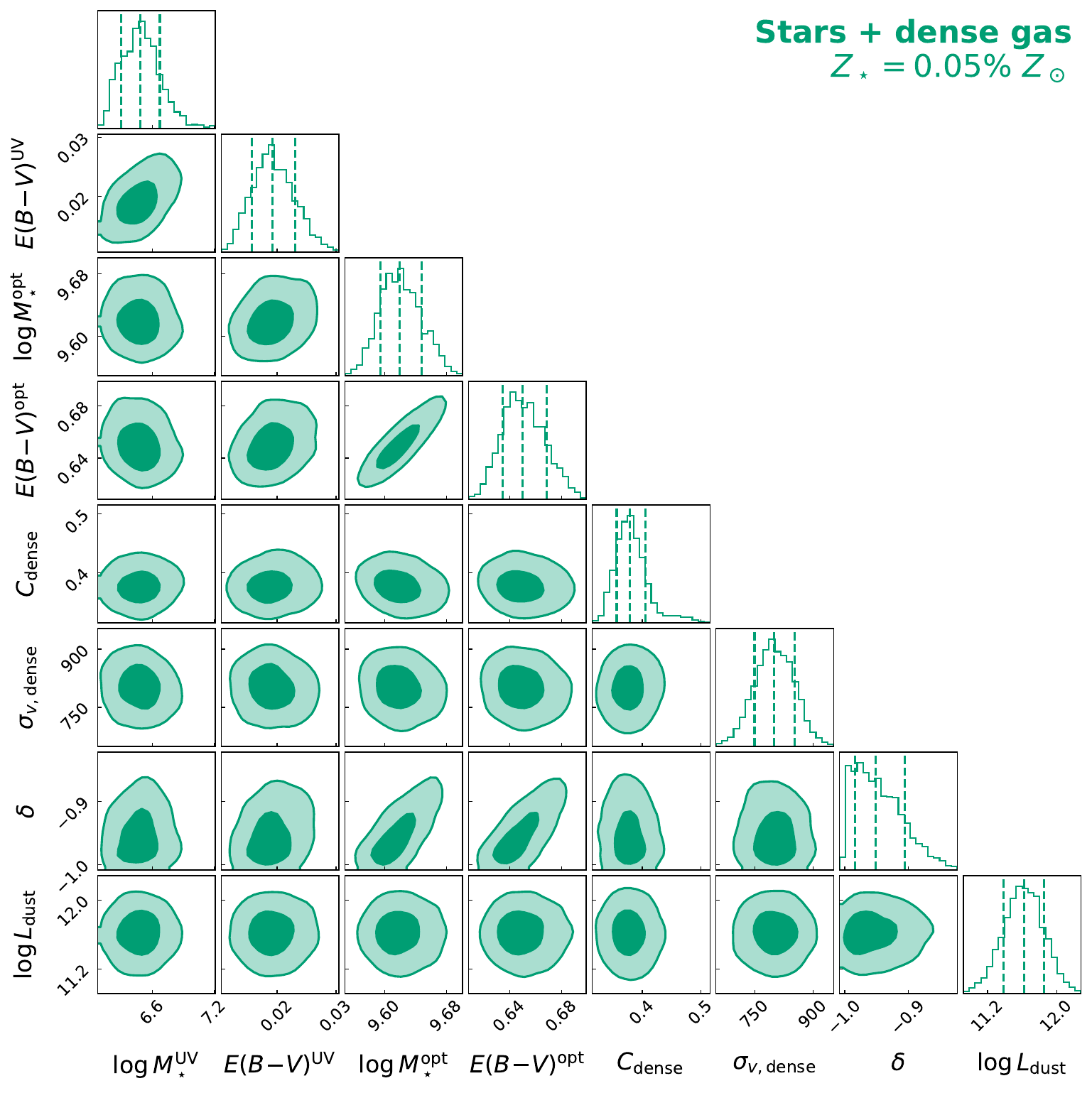}\\[4pt]
    \includegraphics[width=0.45\textwidth]{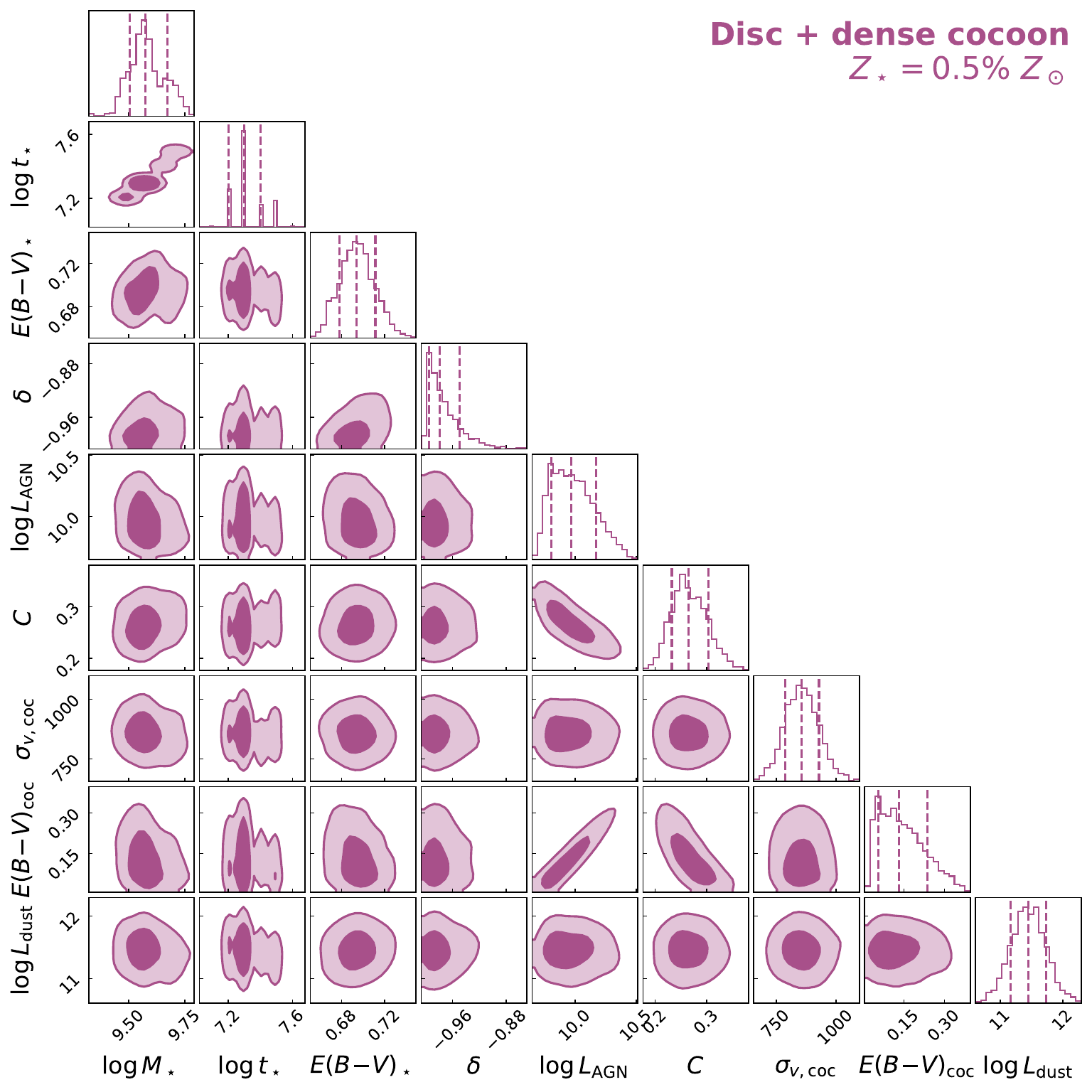}\hfill
    \includegraphics[width=0.45\textwidth]{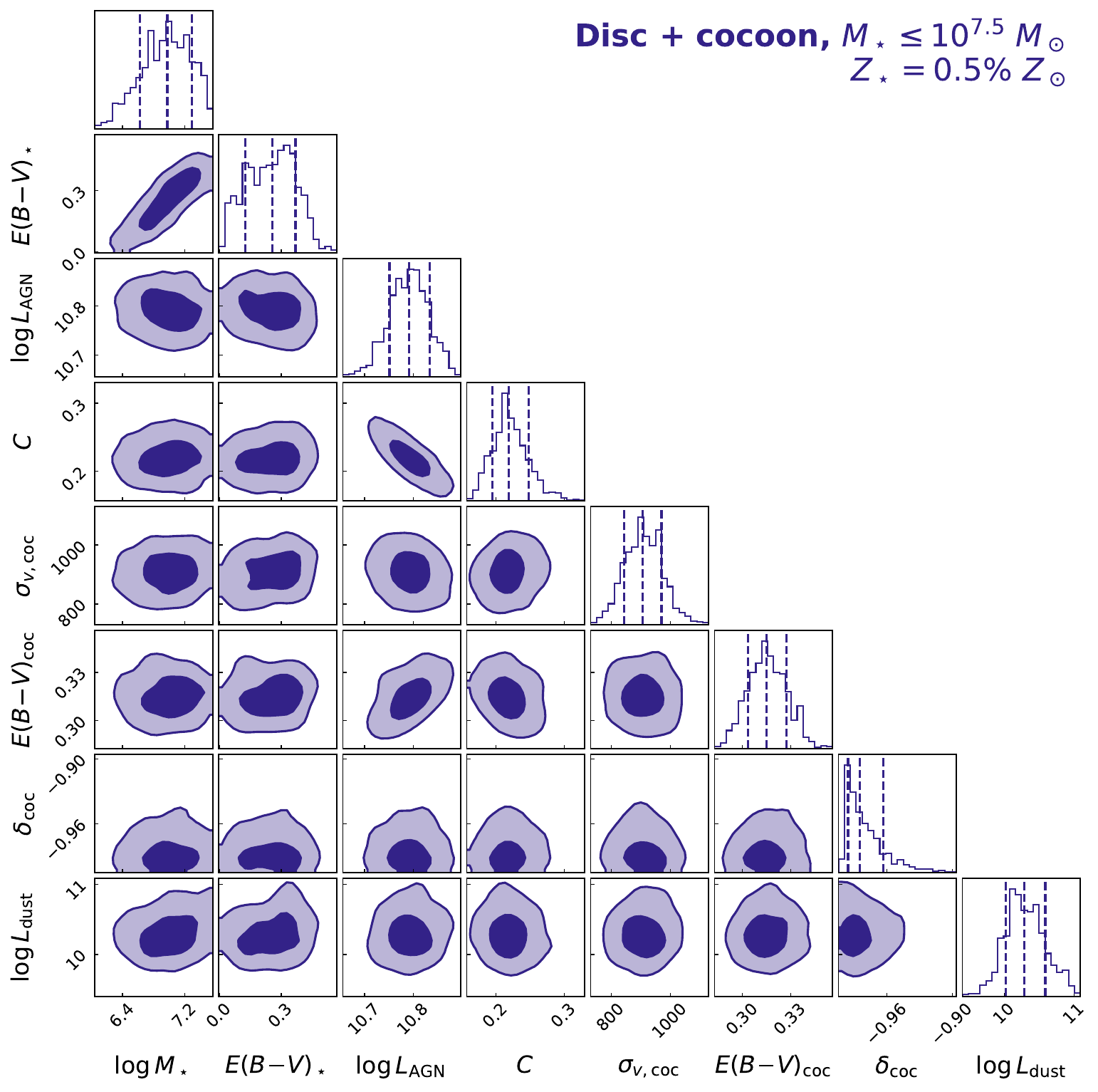}
    \caption{Posterior distributions of the main parameters of the four solutions in Fig.~\ref{fig:sed_models}: two ordinary-density populations (\texttt{zem4}, top left), stellar populations with dense gas (\texttt{zem5}, top right), the accretion disc in a dense cocoon (\texttt{zem4}, bottom left) and the same model with $M_\star\leq10^{7.5}~\mathrm{M}_\odot$ (\texttt{zem4}, bottom right). Contours enclose 39\% and 86\% of the posterior, and dashed lines mark the 16th, 50th and 84th percentiles. Stellar masses are formed masses, $\log_{10}(M_\star~[\mathrm{M}_\odot])$, and ages $\log_{10}(t~[\mathrm{yr}])$ on the 0.1~dex {\sc bpass} grid. Luminosities are $\log_{10}(L~[L_\odot])$ and velocity dispersions in km~s$^{-1}$. ``UV'' and ``opt.'' denote the components dominating the rest-frame UV and optical continua. In the unconstrained cocoon, $M_\star$, $t_\star$ and $E(B-V)_\star$ refer to the stellar population that produces the optical continuum, and in the cocoon with the mass limit to the one that produces the UV. The subscript ``coc'' refers to the attenuation of the cocoon. In the dense-gas model the optical population has $\log_{10}(t~[\mathrm{yr}])=7.2$ in more than 99.9\% of the posterior and its age is not shown.}
    \label{fig:sedfit_corners}
\end{figure*}

All runs completed the exploration phase ($f_\mathrm{live}<0.01$).
The cocoon with the stellar-mass limit was run with the same settings, except for a lower target effective sample size.
Figure~\ref{fig:sedfit_corners} shows the posteriors of the four solutions in Fig.~\ref{fig:sed_models}.
The formal statistical uncertainty on $\ln(\mathcal{Z})$ returned by the sampler ($\lesssim0.2$) does not include the run-to-run scatter.
The NIRSpec spectrum was flux-calibrated with a smooth polynomial derived from the same photometry (Appendix~\ref{app:pathloss}), whose uncertainty was propagated pixel by pixel. The covariance this introduces between spectral pixels and between the spectrum and the photometry is not included in the likelihood.
We therefore regard differences of $\Delta\ln(\mathcal{Z})\sim100$ between classes as robust, and differences of $\Delta\ln(\mathcal{Z})\sim10$--30, such as those between the best stellar and cocoon solutions ($\simeq13$) and between these and the cocoon with the stellar-mass limit ($\simeq20$--33), as model-dependent rather than decisive, since they may be affected by spectral covariance, discrete-grid sampling and run-to-run variance. The latter difference is supported by the MIRI and emission-line residuals (Section~\ref{sec:disc_host}).
Several solutions sit at the edge of a grid or prior: the slope of the attenuation curve ($\delta\simeq-1$), the column of the cocoon in the solution with the stellar-mass limit ($N_{\rm H}=10^{25.5}~\mathrm{cm^{-2}}$), and, in the dense-gas and capped cocoon solutions, the photoionised gas of the UV-emitting population ($n_{\rm H}=10^{6}~\mathrm{cm^{-3}}$ and $U=10^{-1}$, the upper limits for ordinary-density gas).
  
%\clearpage
\bsp	% typesetting comment
\label{lastpage}
\end{document}